\documentclass[11pt]{article}

\usepackage{subfigure}
\usepackage{graphicx}
\usepackage{amsfonts}
\usepackage{amsmath}
\usepackage{framed}
\usepackage{algorithmic}
\usepackage{enumerate,bm}
\usepackage{amsthm}

\newcommand{\setlinespacing}[1]%
           {\setlength{\baselineskip}{#1 \defbaselineskip}}

\newlength{\defbaselineskip}
\begin{document}

%%%
%%%\begin{titlepage}

\title{Localization on low-order eigenvectors of data matrices} 

\author{
Mihai Cucuringu
\thanks{
Program in Applied and Computational Mathematics,
Princeton University,
Princeton, NJ 08544.
Email: mcucurin@math.princeton.edu
}
\and 
Michael W. Mahoney
\thanks{
Department of Mathematics, 
Stanford University, 
Stanford, CA 94305. 
Email: mmahoney@cs.stanford.edu
}
}

\date{}
\maketitle

%{\bf
%\emph{
%\begin{center}
%Note:  This is still a rough draft - from \today. \\
%Please do NOT distribute.
%\end{center}
%}
%}

\begin{abstract}
Eigenvector localization refers to the situation when most of the components
of an eigenvector are zero or near-zero.
This phenomenon has been observed on eigenvectors associated with extremal 
eigenvalues, and in many of those cases it can be meaningfully interpreted 
in terms of ``structural heterogeneities'' in the data.
For example, the largest eigenvectors of adjacency matrices of large 
complex networks often have most of their mass localized on high-degree 
nodes; and the smallest eigenvectors of the Laplacians of such networks are 
often localized on small but meaningful community-like sets of nodes.
Here, we describe localization associated with low-order eigenvectors, 
\emph{i.e.}, eigenvectors corresponding to eigenvalues that are not extremal 
but that are ``buried'' further down in the spectrum.
Although we have observed it in several unrelated applications, this 
phenomenon of low-order eigenvector localization defies common intuitions 
and simple explanations, and it creates serious difficulties for the 
applicability of popular eigenvector-based machine learning and data 
analysis tools.
After describing two examples where low-order eigenvector localization 
arises, we present a very simple model that qualitatively reproduces several 
of the empirically-observed results.
This model suggests certain coarse structural similarities among the 
seemingly-unrelated applications where we have observed low-order 
eigenvector localization, and it may be used as a diagnostic tool to help 
extract insight from data graphs when such low-order eigenvector 
localization is present. 
\end{abstract}

%%%\end{titlepage}
%%%

\section{Introduction}
\label{sxn:intro}

The problem that motivated the work described in this paper had to do with 
using eigenvector-based methods to infer meaningful structure from 
graph-based or network-based data.
Methods of this type are ubiquitous.
For example, Principal Component Analysis and its variants have been 
widely-used historically.
More recently, nonlinear-dimensionality reduction methods, spectral 
partitioning methods, spectral ranking methods, etc. have been used in 
increasingly-sophisticated ways in machine learning and data analysis.

Although they can be applied to any data matrix, these eigenvector-based 
methods are generally most appropriate when the data possess some sort of 
linear redundancy structure (in the original or in some 
nonlinearly-transformed basis) and when there is no single data point or 
no small number of data points that are particularly important or 
influential~\cite{hast-tibs-fried}.
The presence of linear redundancy structure is typically quantified by the 
requirement that the rank of the matrix is small relative to its size,
\emph{e.g.}, that most of the Frobenius norm of the matrix is captured by a 
small number of eigencomponents.
The lack of a small number of particularly-influential data points is 
typically quantified by the requirement that the eigenvectors of the data 
matrix are delocalized.  
For example, matrix coherence and statistical leverage capture this 
idea~\cite{ChatterjeeHadi88}.

Localization in eigenvectors arises when most of the components of an 
eigenvector are zero or near-zero~\cite{HS10}.
(Thus, eigenvector delocalization refers to the situation when most or all 
of the components of an eigenvector are small and roughly the same magnitude.
Below we will quantify this idea in two different ways.)
While creating serious difficulties for recently-popular eigenvector-based 
machine learning and data analysis methods, such a situation is far from 
unknown.
Typically, though, this phenomenon occurs on eigenvectors associated with 
extremal eigenvalues.
For example, the largest eigenvectors of adjacency matrices of large 
complex networks often have most of their mass localized on high-degree 
nodes~\cite{CF06_survey}.
Alternatively, the smallest eigenvectors of the Laplacian of such networks 
are often localized on small but meaningful community-like sets of 
nodes~\cite{LLDM09_communities_IM}.
More generally, this phenomenon arises on extremal eigenvectors in 
applications where extreme sparsity is coupled with randomness or 
quasi-randomness~\cite{FDBV01,GKK01,DGMS03,MT09}.
In these cases, as a rule of thumb, the localization can often be 
interpreted in terms of a ``structural heterogeneity,'' \emph{e.g.}, that 
the degree (or coordination number) of a node is significantly higher or 
lower than average, in the data~\cite{FDBV01,GKK01,DGMS03,MT09}.

In this paper, the phenomenon of localization of low-order eigenvectors in 
Laplacian matrices associated with certain classes of data graphs is 
described for several real-world data sets and analyzed with a simple model.
By \emph{low-order eigenvectors}, we mean eigenvectors associated with 
eigenvalues that are not extremal (in the sense of being the largest or 
smallest eigenvalues), but that are ``buried'' further down in the 
eigenvalue spectrum of the data matrix.
As a practical matter, such localization is most interesting in two cases: 
first, when it occurs in eigenvectors that are below, \emph{i.e.}, 
associated with smaller eigenvalues than, other eigenvectors that are 
significantly more delocalized; and 
second, when the localization occurs on entries or nodes that are 
meaningful, \emph{e.g.}, that correspond to meaningful clusters or other 
structures in the data, to a downstream~analyst.

We have observed this phenomenon of low-order eigenvector localization in 
several seemingly-unrelated applications (including, but not limited to, 
the \textsc{Congress} and the \textsc{Migration} data discussed in this 
paper, DNA single-nucleotide polymorphism data, spectral and hyperspectral 
data in astronomy and other natural sciences, etc.). 
Moreover, based on informal discussions with both practitioners and 
theorists of machine learning and data analysis, it has become clear that 
this phenomenon defies common intuitions and simple explanations.
For example, the variance associated with these low-order eigenvectors is 
much less than the variance associated with ``earlier'' more-delocalized 
eigenvectors.
Thus, these low-order eigenvectors must satisfy the global requirement of 
exact orthogonality with respect to all of the earlier delocalized 
eigenvectors, and they must do so while keeping most of their components 
zero or near-zero in magnitude.
This requirement of exact orthogonality is responsible for the usefulness 
of eigenvector-based methods in machine learning and data analysis, but it
often leads to non-interpretable vectors---recall, \emph{e.g.}, the 
characteristic ``ringing'' behavior of eigenfaces associated with low-order 
eigenvalues~\cite{TP91,MMH04} as well as the issues associated with 
eigenvector reification in the natural and social sciences~\cite{CUR_PNAS}.
For this and related reasons, it is often the case that by the time that 
most of the variance in the data is captured, the residual consists mostly 
of relatively-delocalized noise.
Indeed, eigenvector-based models and methods in machine learning and data 
analysis typically simply assume that this is the case.

In this paper, our contributions are threefold:
first, we will introduce the notion of low-order eigenvector localization; 
second, we will describe several examples of this phenomenon in two real 
data sets; and 
third, we will present a very simple model that qualitatively reproduces 
several of the empirical observations.
Our model is a very simple two-level tensor product construction in which 
each level can be ``structured'' or ``unstructured.''
Aside from demonstrating the existence of low-order eigenvector localization
in real data, our empirical results will illustrate that meaningful very 
low variance parts of the data can---in some cases---be extracted in an 
unsupervised manner by looking at the localization properties of low-order 
eigenvectors. 
In addition, our simple model will suggest certain coarse structural 
similarities among seemingly-unrelated applications, and it may be used as a 
diagnostic tool to help extract meaningful insight from real data graphs 
when such low-order eigenvector localization is present. 
We will conclude the paper with a brief discussion of the implications of 
our results in a broader context.

\section{Data, methods, and related work} 
\label{sxn:background}

In this section, we will provide a brief background on two classes of data 
where we have observed low-order eigenvector localization, and we will 
describe our methods and some related work.

\subsection{The two data sets we consider}
\label{sxn:background:data}

The main data set we consider, which will be called the \textsc{Congress} 
data set, is a data set of roll call voting patterns in the U.S. Senate 
across time~\cite{PR97,WPFMP09_TR,multiplex_Mucha}.
We considered Senates in the $70^{th}$ Congress through the $110^{th}$ 
Congress, thus covering the years $1927$ to $2008$.
During this time, the U.S. went from $48$ to $50$ states, and thus the 
number of senators in each of these $41$ Congresses was roughly the same.
After preprocessing, there were $n=735$ distinct senators in these $41$ 
Congresses.
We constructed an $n \times n$ adjacency matrix $A$, where each 
$A_{ij} \in [0,1]$ represents the extent of voting agreement between 
legislators $i$ and $j$, and where identical senators in adjacent Congresses 
are connected with an inter-Congress connection strength.
We then considered the Laplacian matrix of this graph, constructed in 
the usual way; see~\cite{WPFMP09_TR} for more details.

We also report on a data set, which we will call the \textsc{Migration} 
data set, that was recently considered in~\cite{CVB11_DRAFT}.
This contains data on county-to-county migration patterns in the U.S., 
constructed from the $2000$ U.S. Census data, that reports the number of 
people that migrated from every county to every other county in the 
mainland U.S. during the 1995-2000 time 
frame~\cite{censusB,Sla08_TR,CVB11_DRAFT}.
We denote by $M = (M_{ij})_{1 \leq i,j \leq N}$ the total number of people 
who migrated from county $i$ to county $j$ or from county $j$ to county $i$ 
(so $M_{ij} = M_{ji}$), where $N=3107$ denotes the number of counties in the 
mainland U.S.; and we let $P_i$ denote the population of county $i$.
We then build the similarity matrix $W_{ij} = \frac{ M_{ij}^2 }{P_i P_j}$ and 
the diagonal scaling matrix $ D_{ii} = \sum_{j=1}^{N} w_{ij}$; and we 
considered the usual random walk matrix, $D^{-1}W$, associated with this graph. 
We refer the reader to~\cite{CVB11_DRAFT} for a discussion of variants of 
this similarity matrix.

\subsection{The methods we will apply}

In both of these applications, we will look at eigenvectors of matrices 
constructed from the data graph.
Recall that given a weighted graph $G=(V,E,W)$, one can define the Laplacian 
matrix as $L=D-W$, where $W$ is a weighted adjacency matrix, and where $D$ 
is a diagonal matrix, with $i^{th}$ entry $D_{ii}$ equal to the degree (or 
sum of weights) of the $i^{th}$ node.  
Then consider the solutions to the generalized eigenvalue problem
$Lx = \lambda D x$.
These are related to the solutions of the eigenvalue problem 
$Px = \lambda x$, where $P=D^{-1} W$ is a row-stochastic matrix that can be
interpreted as the transition matrix of a Markov chain with state space 
equal to the nodes in $V$ and where $P_{ij}$ represents the transition 
probability of moving from node $i$ to node $j$ in one step.
In particular, if $(\lambda,x)$ is an eigenvalue-eigenvector solution to 
$Px=\lambda x$, then $(1-\lambda,x)$ is a solution to $Lx=\lambda D x$.

The top (resp. bottom) eigenvectors of the Markov chain (resp. generalized
Laplacian eigenvalue) problem define the coarsest modes of variation or 
slowest modes of mixing, and thus these eigenvectors have a natural 
interpretation in terms of diffusions and random walks.
As such, they have been widely-studied in machine learning and data analysis 
to perform such tasks as partitioning, ranking, clustering, and visualizing 
the data~\cite{spielman96_spectral,ShiMalik00_NCut,RS00,MS01,BN03,CLLMNWZ05a}.
We are interested in localization, not on these top eigenvectors, but on 
lower-order eigenvectors---for example, on the $41^{st}$ eigenvector (or 
$43^{rd}$ or $\ldots$ out of a total of hundreds of eigenvectors) in the 
\textsc{Congress} data below.
(As a matter of convention, we will refer to eigenvectors that are associated 
with eigenvalues that are not near the top part of the spectrum of the 
Markov chain matrix as \emph{low-order eigenvectors}---thus, they actually
correspond to larger eigenvalues in the generalized eigenvalue problem 
$Lx = \lambda D x$.)

We will consider several measures to quantify the idea of localization in 
eigenvectors as arising when most of the components of an eigenvector are 
zero or near-zero.
Perhaps most simply, we will consider histograms of the entries of the 
eigenvectors.
More generally, let $V$ be a matrix consisting of the eigenvectors of 
$P=D^{-1}W$, ordered from top to bottom; let $V^{(j)}$ denote the $j^{th}$ 
eigenvector and $V_{ij}$ the $i^{th}$ element of this $j^{th}$ vector; and 
let $\mathcal{N}=\sum_{i=1}^{n}V_{ij}^{2}=1$.
Then:
\begin{itemize}
\item
Then the \emph{$j$-componentwise-statistical-leverage} (CSL) of node $j$ is 
an $n$-dimensional vector with $i^{th}$ element given by 
$V_{ij}^{2}/\mathcal{N}$.
Thus, this measure is a score over nodes that describes how localized is a 
given node along a particular eigendirection. 
\item
The \emph{$j$-inverse participation ratio} (IPR) is the number 
$\sum_{i=1}^{n}V_{ij}^{4}/\mathcal{N}$.
Thus, this measure is a score over eigendirections that describes how 
localized is a given eigendirection.
\end{itemize}
To gain intuition for these two measures, consider their behavior in the 
following limiting cases.
If the $j^{th}$ eigenvector is $(1/\sqrt{n},\ldots,1/\sqrt{n})$, 
\emph{i.e.}, is very delocalized everywhere, then every element of the 
$j$-CSL is $1/n$, and the $j$-IPR is $1/n$.
That is, they are both ``small.''
On the other hand, if the $j^{th}$ eigenvector is $(1,0,\ldots,0)$, 
\emph{i.e.}, is very localized, then the $j$-CSL is $(1,0,\ldots,0)$, and 
the $j$-IPR~is~$1$.
Thus, for both measures, higher values indicate the presence of 
localization, while smaller values indicate delocalization.

\subsection{Related work in machine learning and data analysis}

The $j$-CSL is based on the idea of statistical leverage, which has been 
used to characterize localization on the top eigenvectors in statistical 
data analysis~\cite{CUR_PNAS}; while the $j$-IPR originated in quantum 
mechanics and has been applied to study localization on the top 
eigenvectors of complex networks~\cite{FDBV01}. 
Depending on whether one is considering the adjacency matrix or the 
Laplacian matrix, localized eigenvectors have been found to correspond to 
structural inhomogeneities such as very high degree nodes or very small 
cluster-like sets of 
nodes~\cite{FDBV01,GKK01,DGMS03,MT09,LLDM09_communities_IM}.
More generally, localization on the top eigenvectors often has an 
interpretation in terms of the ``centrality'' or ``network value'' of a 
node~\cite{CF06_survey}, two ideas which are of use in applications such as
viral marketing and immunizing against infectious agents. 
Localization on extremal eigenvectors has also found application in 
a wide range of problems such as distributed control and estimation 
problems~\cite{BH06} as well as asymptotic space localization in sensor 
networks~\cite{JLHB07}.

There have been a great deal of work on clustering and community detection 
that rely on the eigenvectors of graphs.
Much of this work finds approximations to the best global partition of the 
data~\cite{pothen90partition,ShiMalik00_NCut,newman2006finding,spielman07_spectral}.
More recent work, however, has focused on local versions of the global 
spectral partitioning method~\cite{Spielman:2004,andersen06local,MOV09_TR}; 
and this work can be interpreted as partitioning with respect to a 
locally-biased vector computed from a locally-biased seed set.
Random walks have been of interest in machine learning and data analysis, 
both because of their usefulness in nonlinear dimensionality reduction
methods such as Laplacian Eigenmaps and the related diffusion 
maps~\cite{BN03,CLLMNWZ05a,BDRPVO04} as well as for the connections with 
spectral methods more 
generally~\cite{weiss99_segmentation,MS01,NJW01_spectral,luxburg05_survey}.
One line of work related to this but from which ours should be 
differentiated has to do with looking at the smallest eigenvectors of a graph 
Laplacian~\cite{newman2006finding,Tre09}.
These eigenvectors are not ``buried'' in the middle of the spectrum---they 
are associated with extremal eigenvalues and they typically have to do with 
identifying bipartite structure in the graph.

There is a large body of work in mathematics and physics on the localization 
properties of the continuous Laplace operator, nearly all of which studies 
the localization properties of eigenfunctions associated with extremal 
eigenvalues, and there is also a rich literature on the relationship between 
the spectrum and the geometry of the domain.
Only recently, however, has work advocated studying localized eigenfunctions 
associated to lower-order eigenvalues~\cite{HS10}.
Also recently, it was noticed that low-order localization exists in two 
spatially-distributed networks (the \textsc{Migration} data we report on here 
and a data set of mobile phone calls between cities in Belgium) and that this 
localization correlated with geographically-meaningful 
regions~\cite{CVB11_DRAFT}.

\section{Motivating empirical results}
\label{sxn:empirical}

In this section, we will illustrate low-order eigenvector localization for
the two data sets described in Section~\ref{sxn:background}, and we will 
show that in both cases the localization highlights interesting properties 
of the data.

\subsection{Overview of empirical results}

\begin{figure}[t]
\begin{center}
\subfigure[Two-dimensional grid.]{\includegraphics[width=0.22\columnwidth]{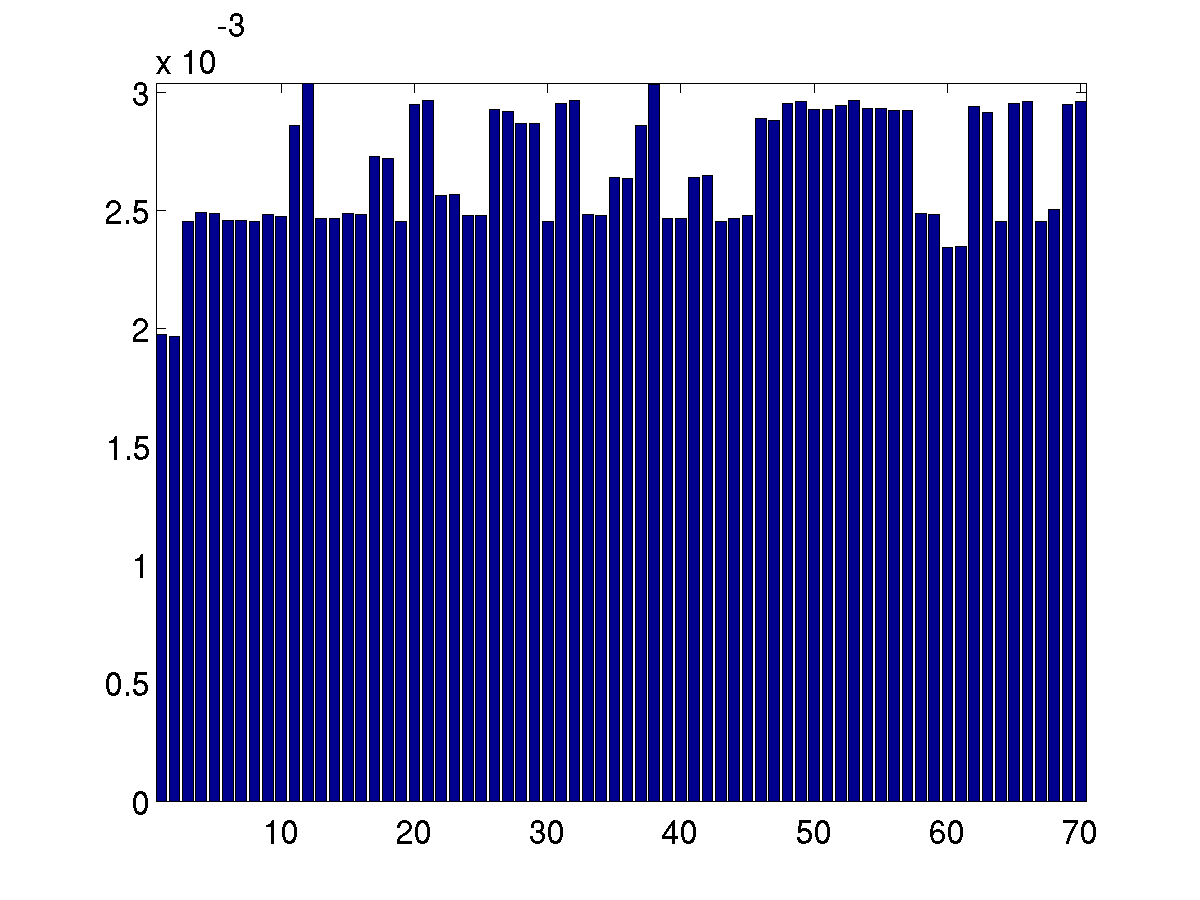} \label{fig:ipr-2dg} }
\subfigure[Random graph, $G(n,p=0.01)$.]{\includegraphics[width=0.22\columnwidth]{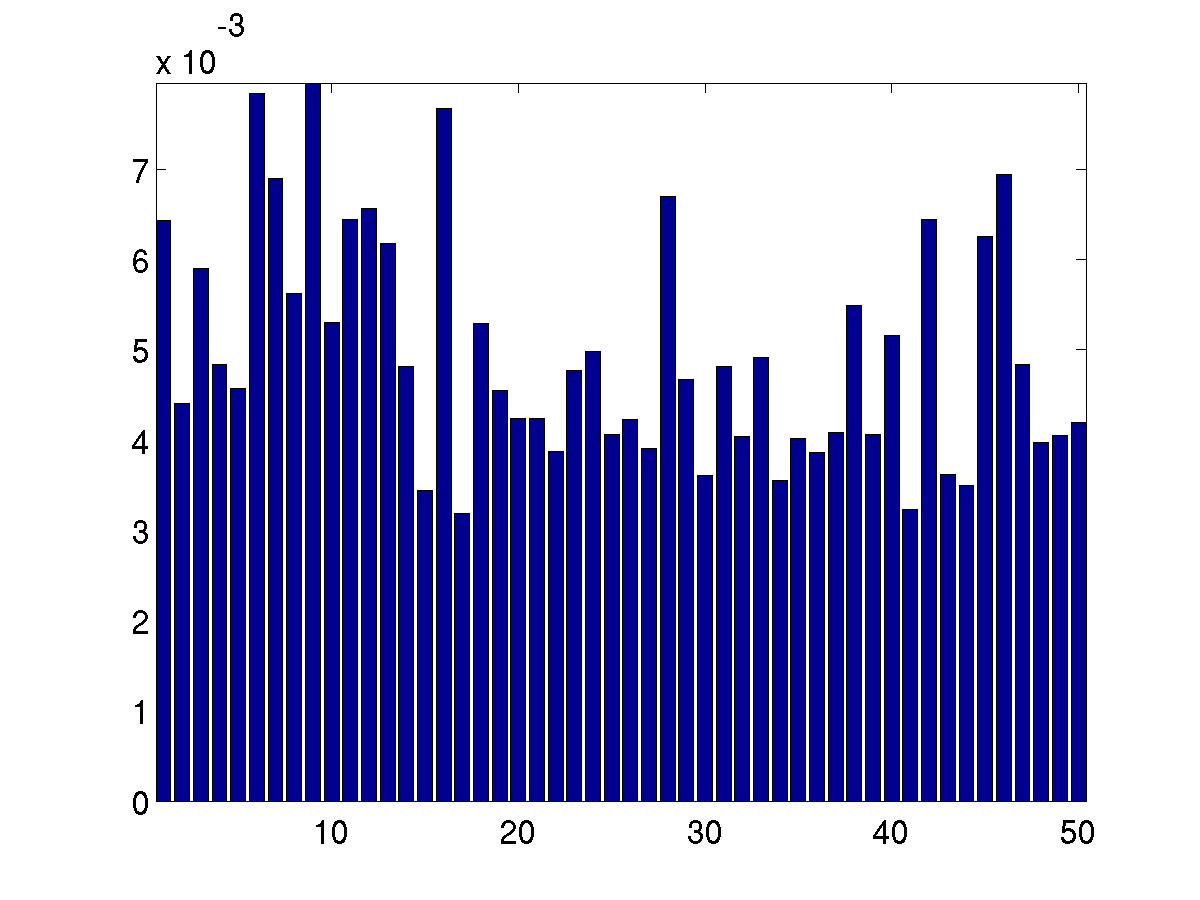} \label{fig:ipr-gnp1} }
\subfigure[Random graph, $G(n,p=0.03)$.]{\includegraphics[width=0.22\columnwidth]{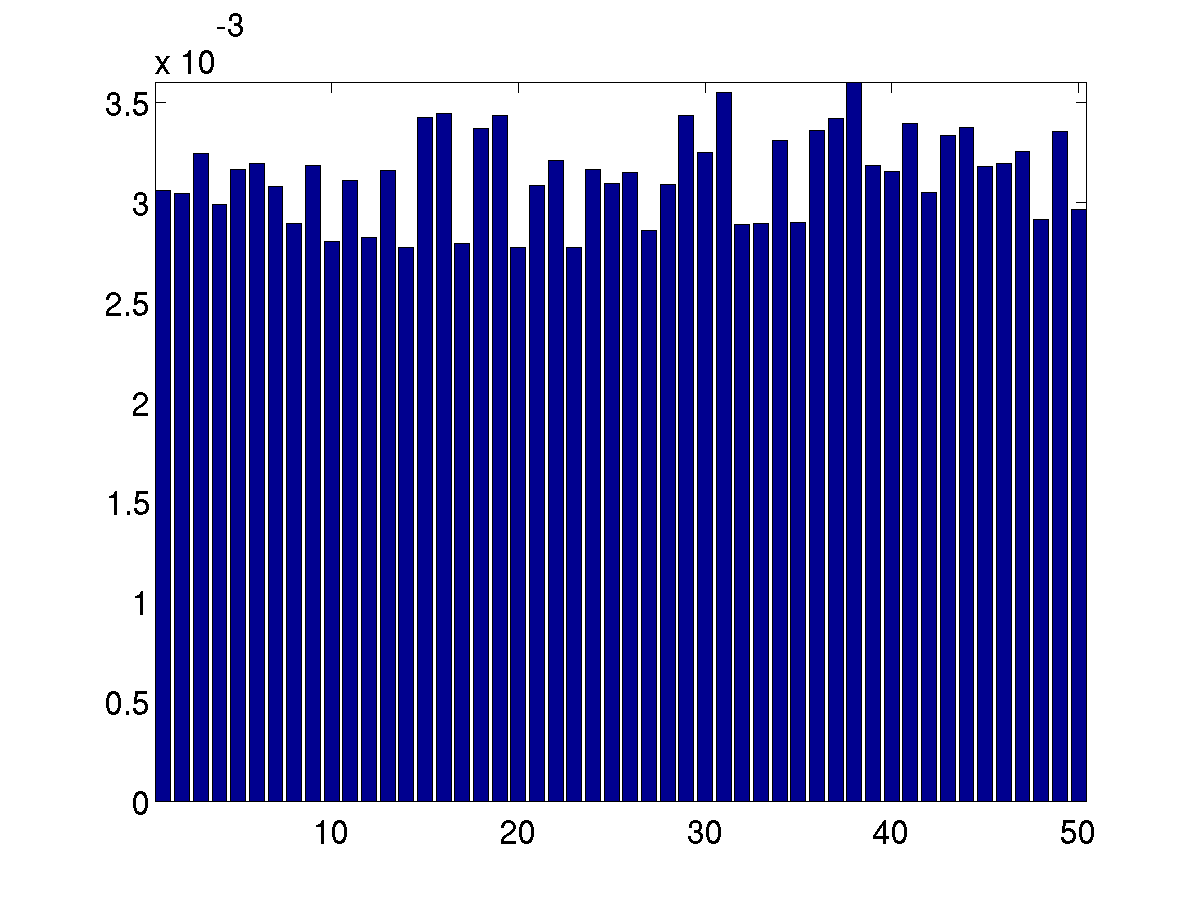} \label{fig:ipr-gnp2} }
\subfigure[\textsc{Congress} $\epsilon=0.01$]{\includegraphics[width=0.22\columnwidth]{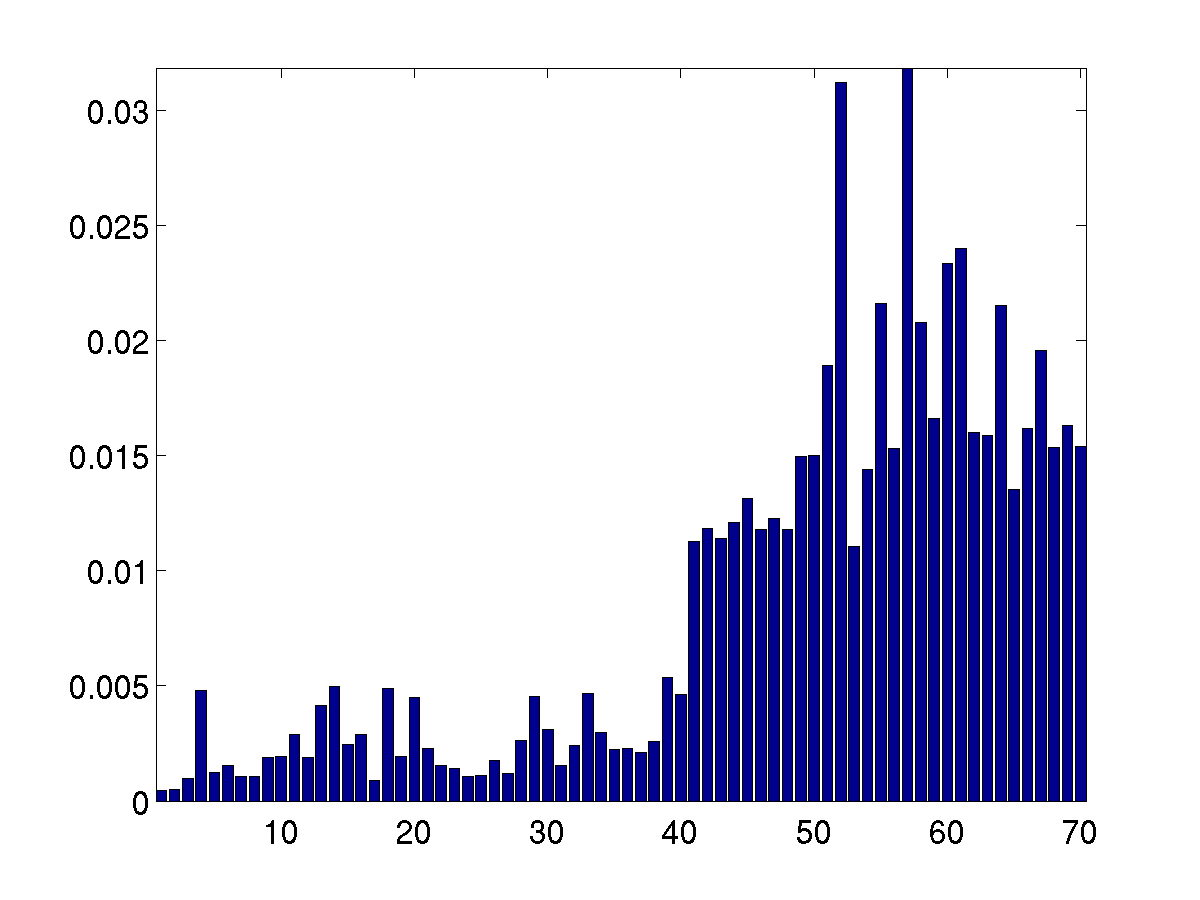} \label{fig:ipr-cong0} } 
\subfigure[\textsc{Congress}, $\epsilon=0.1$.]{\includegraphics[width=0.22\columnwidth]{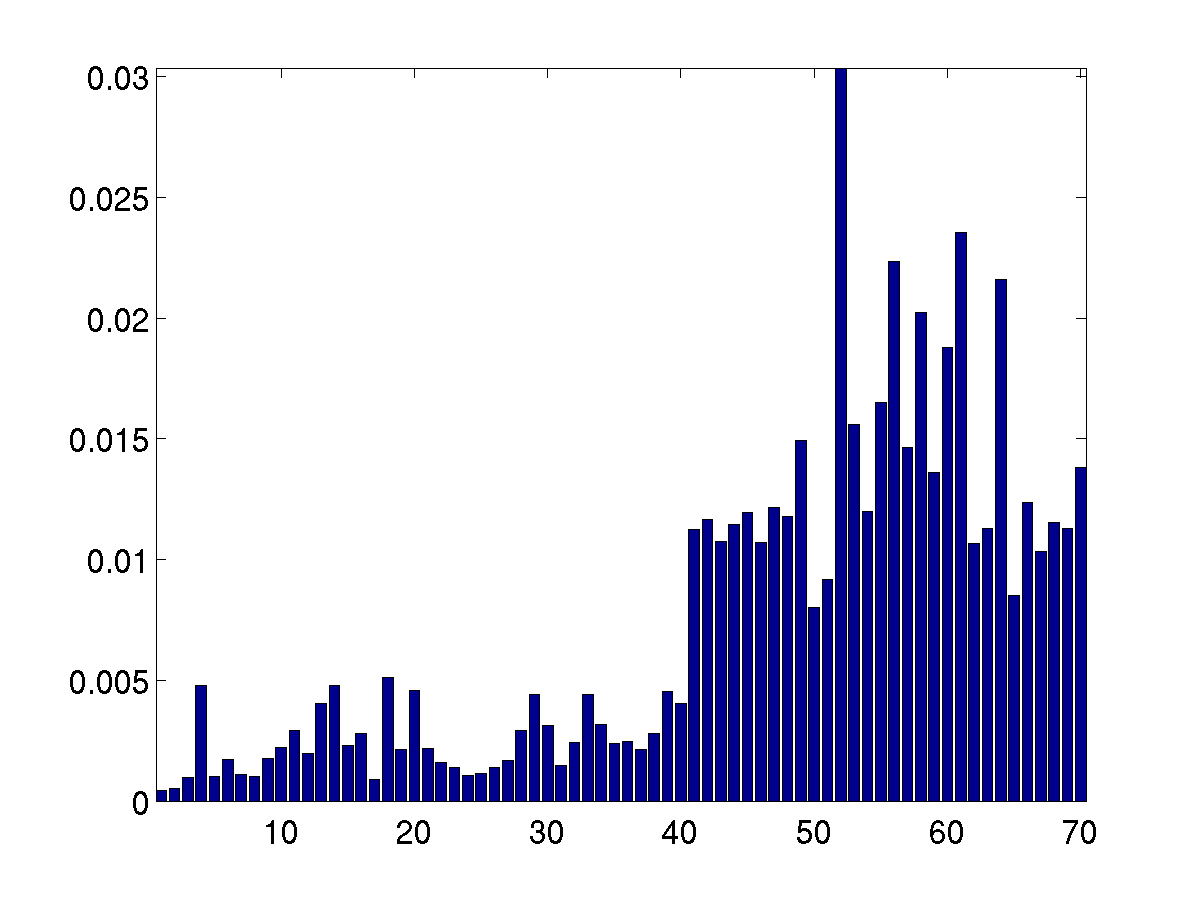} \label{fig:ipr-cong1} }
\subfigure[\textsc{Congress}, $\epsilon=1$.]{\includegraphics[width=0.22\columnwidth]{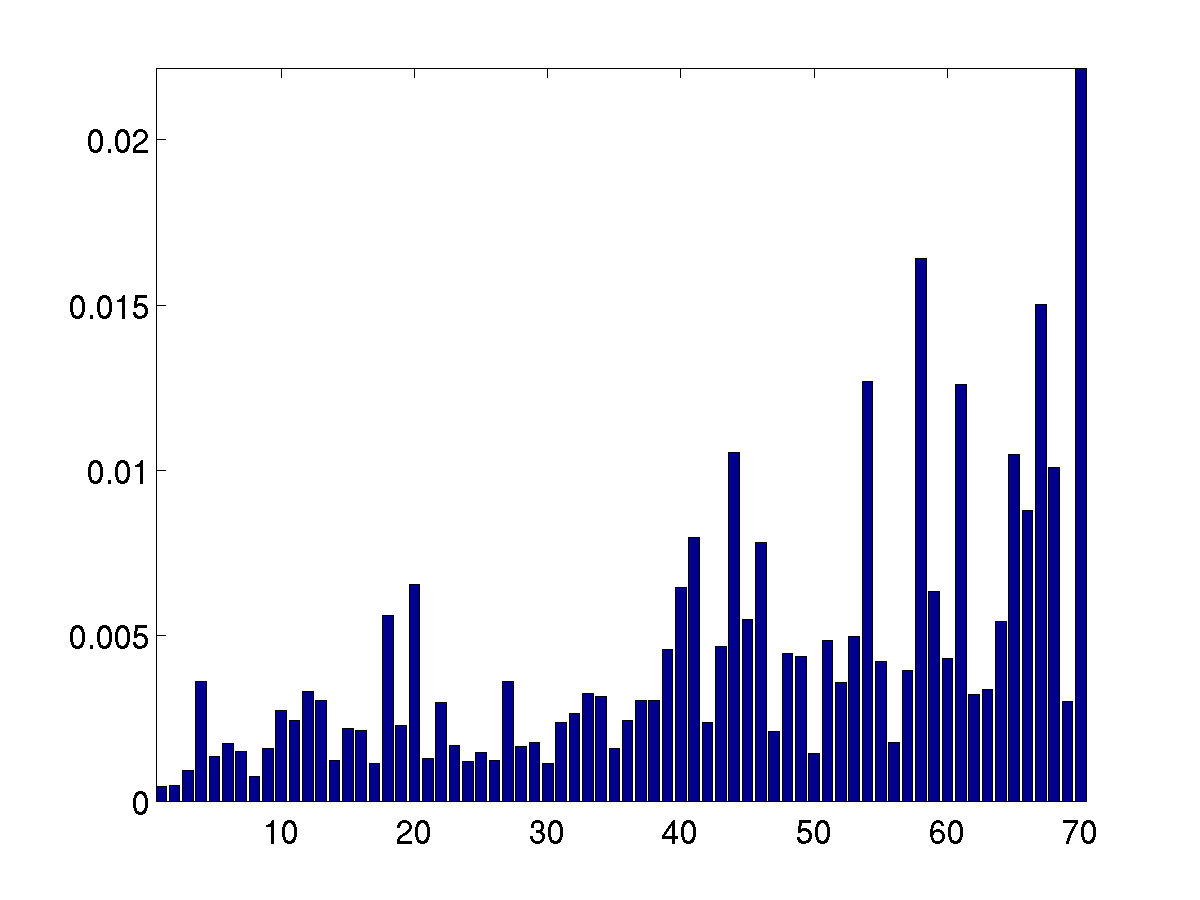} \label{fig:ipr-cong2} }
\subfigure[\textsc{Congress}, $\epsilon=10$.]{\includegraphics[width=0.22\columnwidth]{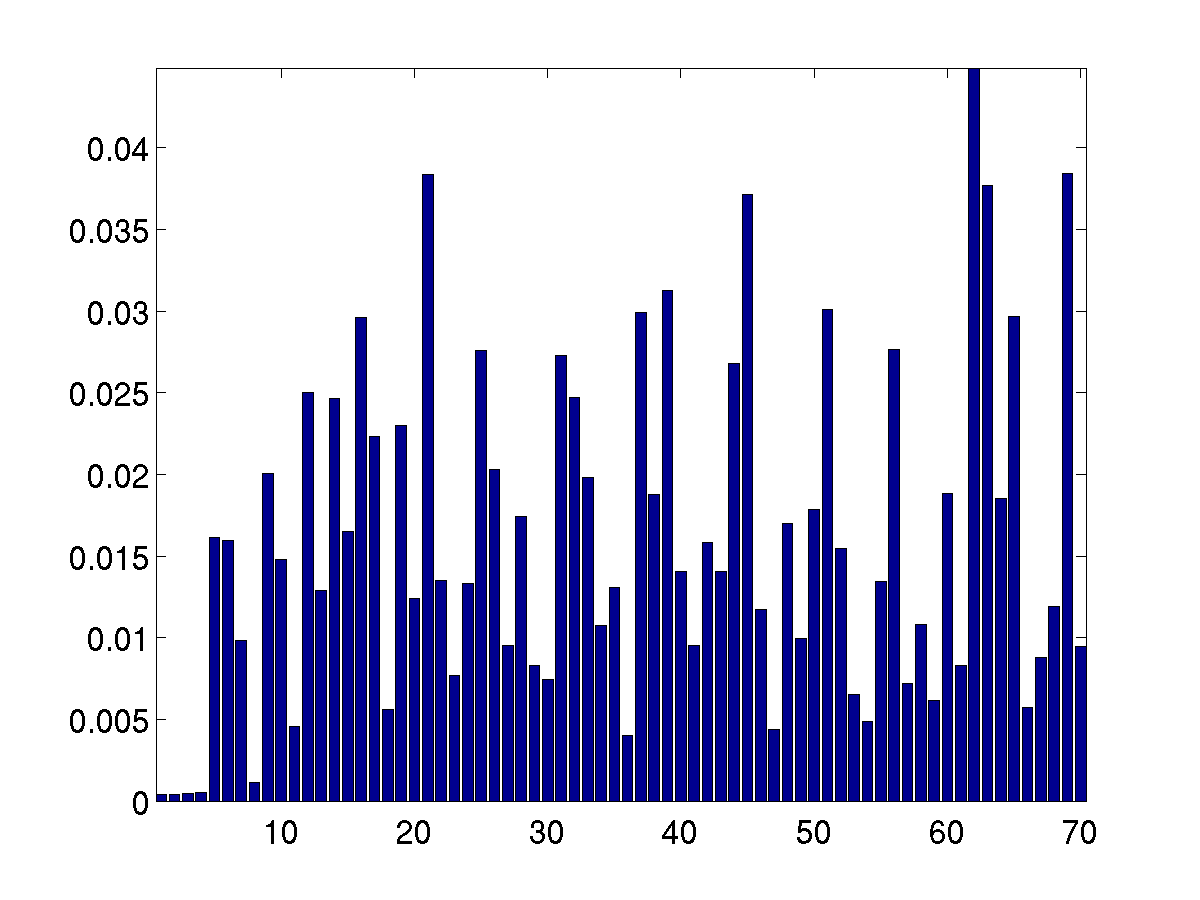} \label{fig:ipr-cong3} }    
\subfigure[\textsc{Migration} data.]{\includegraphics[width=0.22\columnwidth]{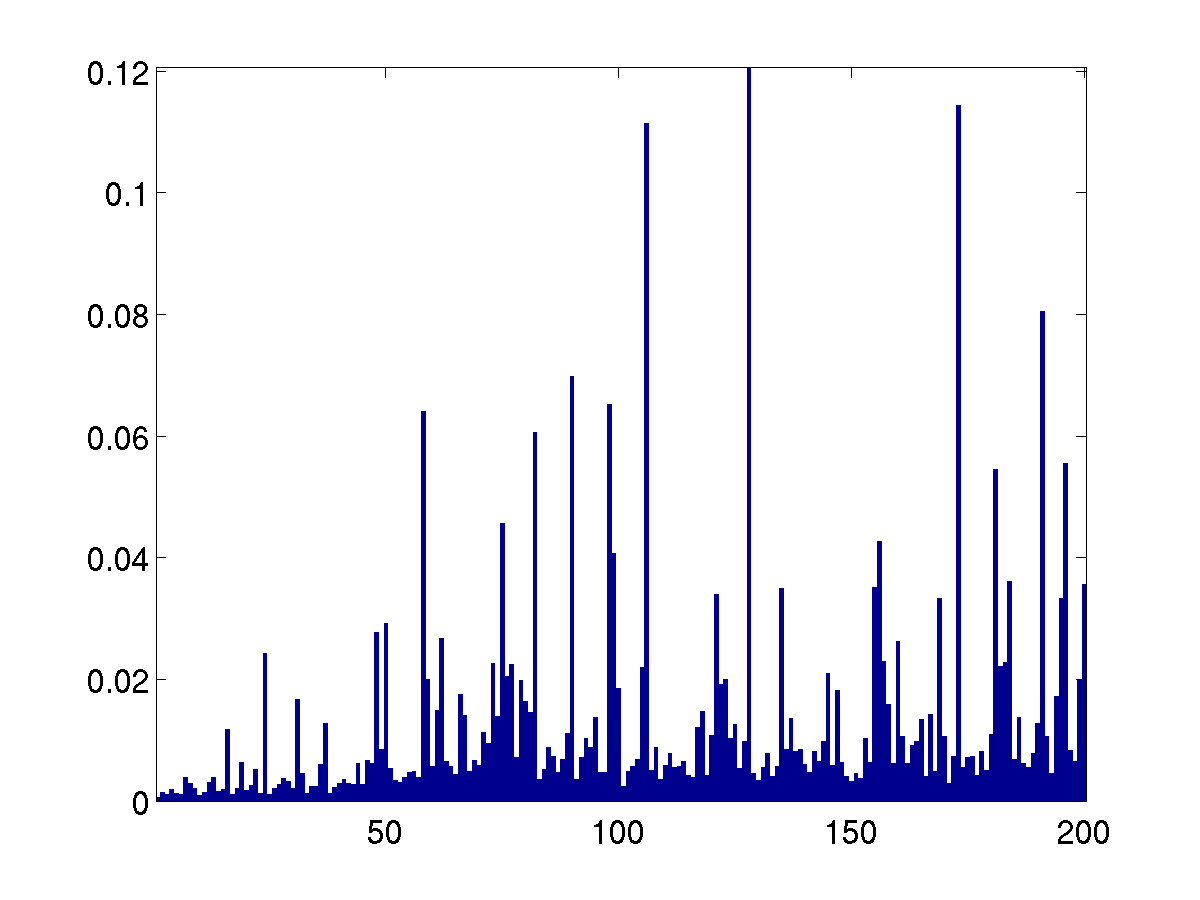} \label{fig:ipr-migr} }
\end{center}
\vspace{-4mm}
\caption{Inverse Participation Ratio (IPR), as a function of the rank of the
corresponding eigendirection, for several data graphs. 
For grids and other well-formed low-dimensional meshes as well as for 
not-extremely-sparse random graphs, all eigenvectors are fairly delocalized 
and the IPR is relatively flat.
For \textsc{Congress} and \textsc{Migration}, there is substantial 
localization on low-order eigenvectors.}
\label{fig:ipr}
\end{figure}

To start, consider Figure~\ref{fig:ipr}.
This figure illustrates the IPR for several toy data sets, for
\textsc{Congress} for several values of the connection parameter, and  for
\textsc{Migration}.
In each case, the IPR is plotted as a function of the rank of the 
corresponding eigenvector.
Figure~\ref{fig:ipr-2dg} shows this plot for a discretization of a 
two-dimensional grid; and Figures~\ref{fig:ipr-gnp1} and~\ref{fig:ipr-gnp2} 
show this plot for a not-too-sparse $G_{np}$ random graph, where $G_{np}$
refers to the Erd\H{o}s-R\'{e}nyi random graph model on some number $n$ of
nodes, where $p$ is the connection probability between each pair of 
nodes~\cite{Bollobas85}.
These toy synthetic graphs represent limiting cases where measure 
concentration occurs and where delocalized eigenvectors are known to 
appear.
More generally, the same delocalization holds for discretizations of 
other low-dimensional spaces, as well as low-dimensional manifolds under 
the usual assumptions made in machine learning, \emph{i.e.}, those without 
bad ``corners'' and without pathological curvature or other pathological 
distributional properties.
Not surprisingly, similar results are 
seen for other similar toy data sets that have been used to validate 
eigenvector-based algorithms in machine learning.
Two things should be noted about these results:
first, even in these idealized cases, the IPR is not perfectly uniform, even 
for large values of the rank parameter, although the nonuniformity due to 
the random noise is relatively modest and seemingly-unstructured; and
second, when the data are sparser, \emph{e.g.}, when the connection 
probability $p$ is smaller in the random graph model, the nonuniformity due 
to noise is somewhat more pronounced.

Next, Figures~\ref{fig:ipr-cong0}, \ref{fig:ipr-cong1}, \ref{fig:ipr-cong2}, 
and~\ref{fig:ipr-cong3} illustrate the IPR for the \textsc{Congress} 
data for several different values of the parameter defining the strength of 
interactions between successive Congresses, and Figure~\ref{fig:ipr-migr} 
illustrates the IPR for the \textsc{Migration} data. 
In all these cases, the IPR indicates that many of the low-order 
eigenvectors are significantly more localized than earlier eigenvectors.
Moreover, the localization is robust in the sense that similar (but often 
noisier) results are obtained if the details of the kernel connecting 
different counties is changed or if the connection probability between 
individuals in successive Congresses is modified within reasonable ranges. 
This is most prominent in the \textsc{Congress} data.
For example, when the connection probability is small, \emph{e.g.}, 
$\epsilon=0.1$, as it was in the original 
applications~\cite{WPFMP09_TR,multiplex_Mucha}.
there is a significant localization-delocalization transition talking place 
between the $40^{th}$ and $41^{st}$ eigenvector.
(The significance of this will be described below, but recall that the data 
consists of $41$ Congresses.
If the \textsc{Congress} data set is artificially truncated to consist of 
some number of Congresses other than $41$, then this transition would have 
taken place at some other location in the spectrum, and we would have 
illustrated those eigenvectors.)
Note, however, that when the connection parameter is increased from 
$\epsilon=0.1$ to $\epsilon=1$ and above, the low-order localization becomes 
much less structured.
In addition, unpublished results clearly indicate that in this case the 
structures highlighted by the low-order localization are much more noisy and 
much less meaningful to the domain scientist. 

%TMP% 

\subsection{The \textsc{Congress} data}

\begin{figure}[t]
\begin{center}
%% \subfigure[Congress Data]{ \includegraphics[width=0.32\columnwidth]{Plots/congress/eps_0p1/Network.png}}
%% \subfigure[Variance-Score of the eigenvalues]{\includegraphics[width=0.32\columnwidth]{Plots/congress/eps_0p1/Variance_Cont.png}}
%% \subfigure[IPR]{\includegraphics[width=0.32\columnwidth]{Plots/congress/eps_0p1/IPR.png}} \\
\includegraphics[width=0.24\columnwidth]{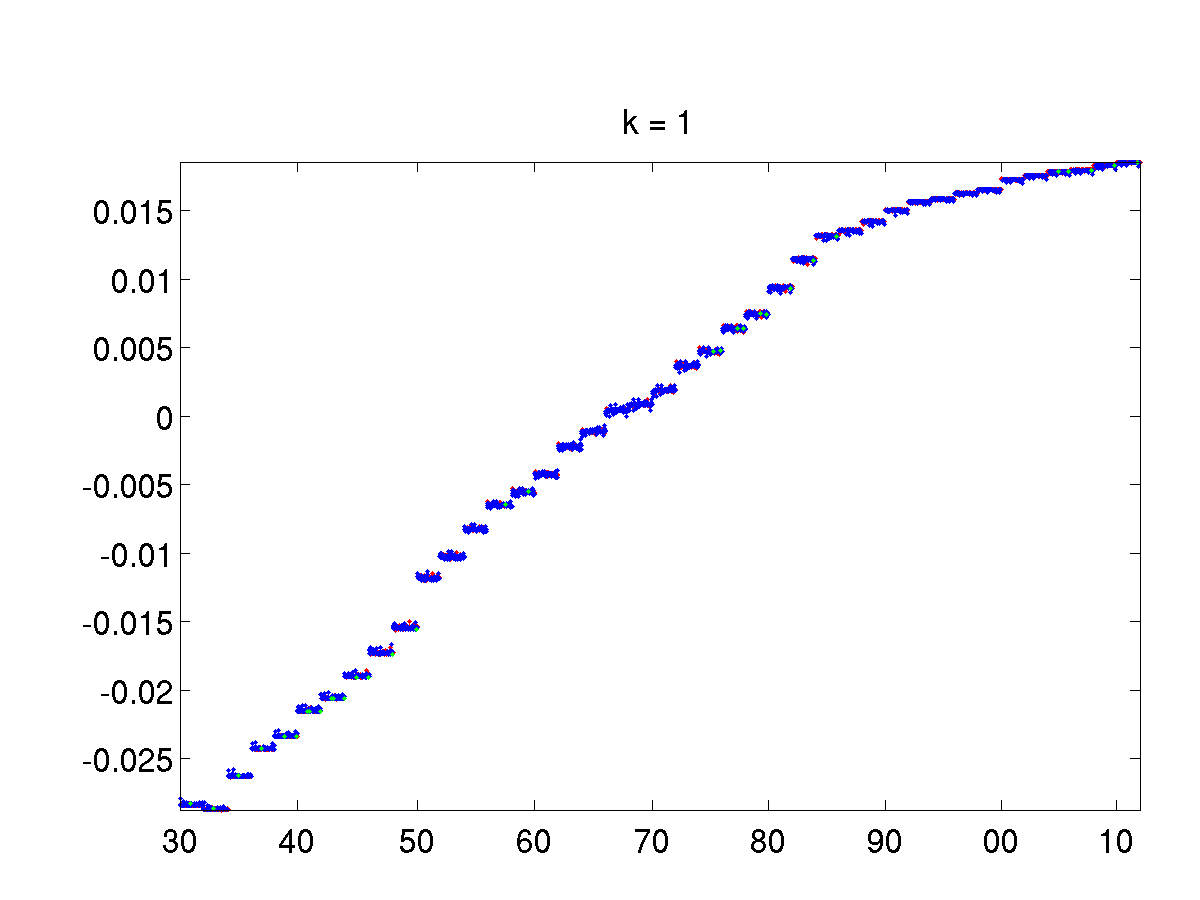}
\includegraphics[width=0.24\columnwidth]{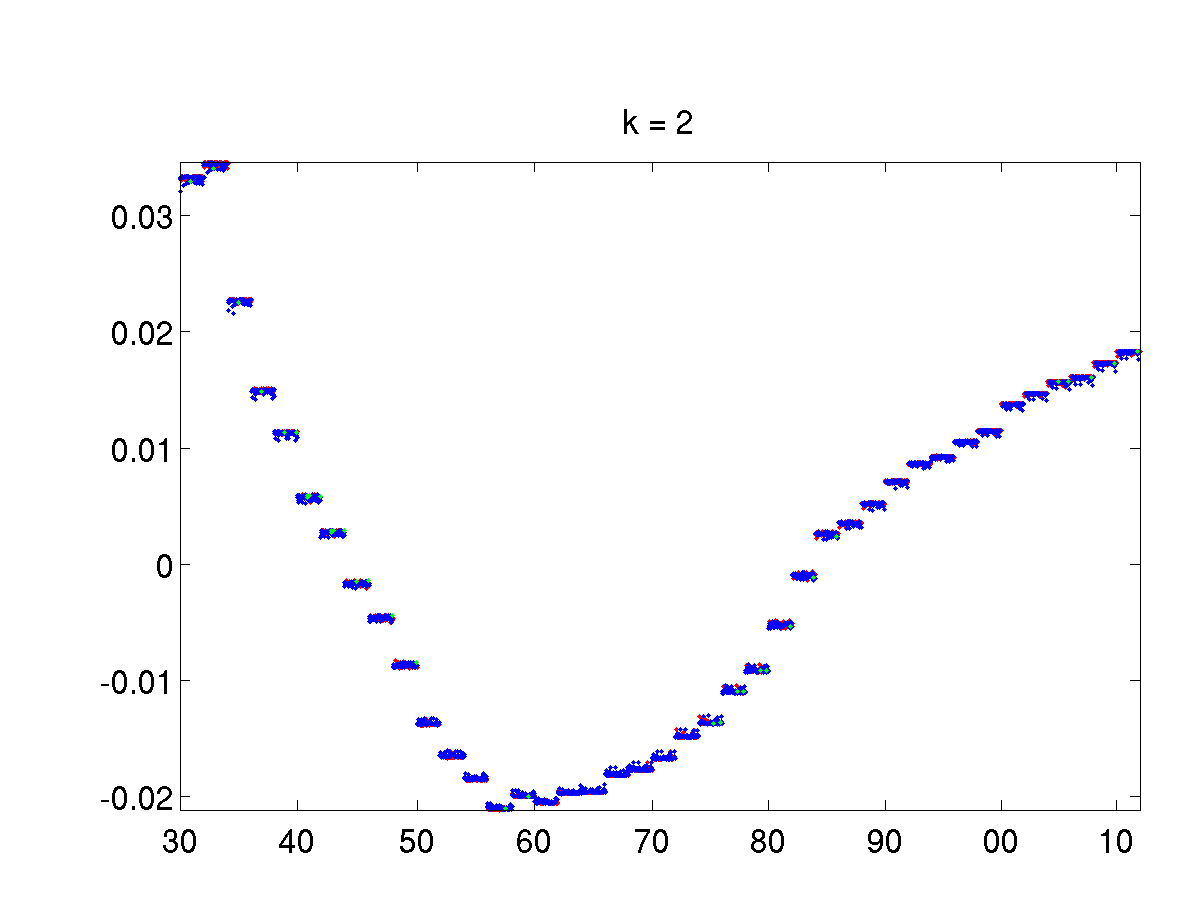}
\includegraphics[width=0.24\columnwidth]{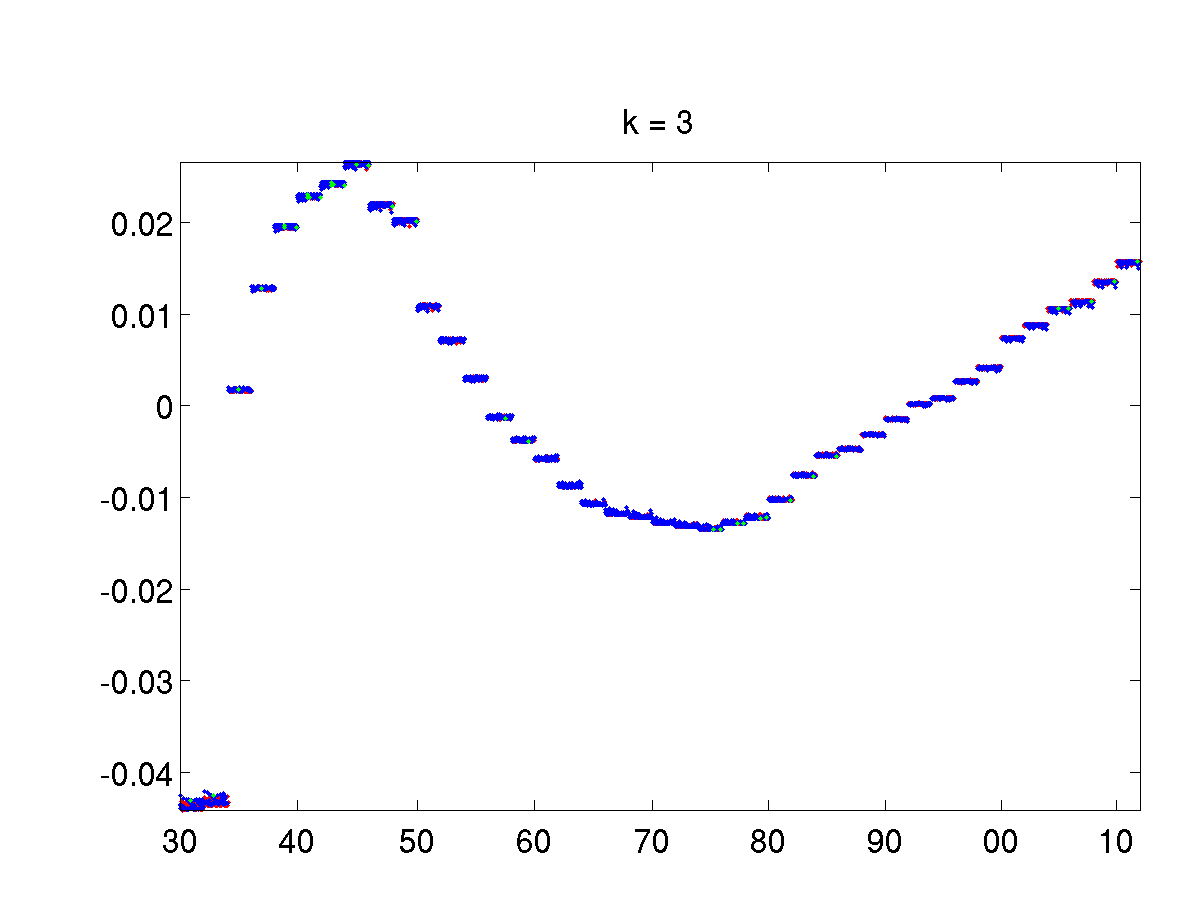}
\includegraphics[width=0.24\columnwidth]{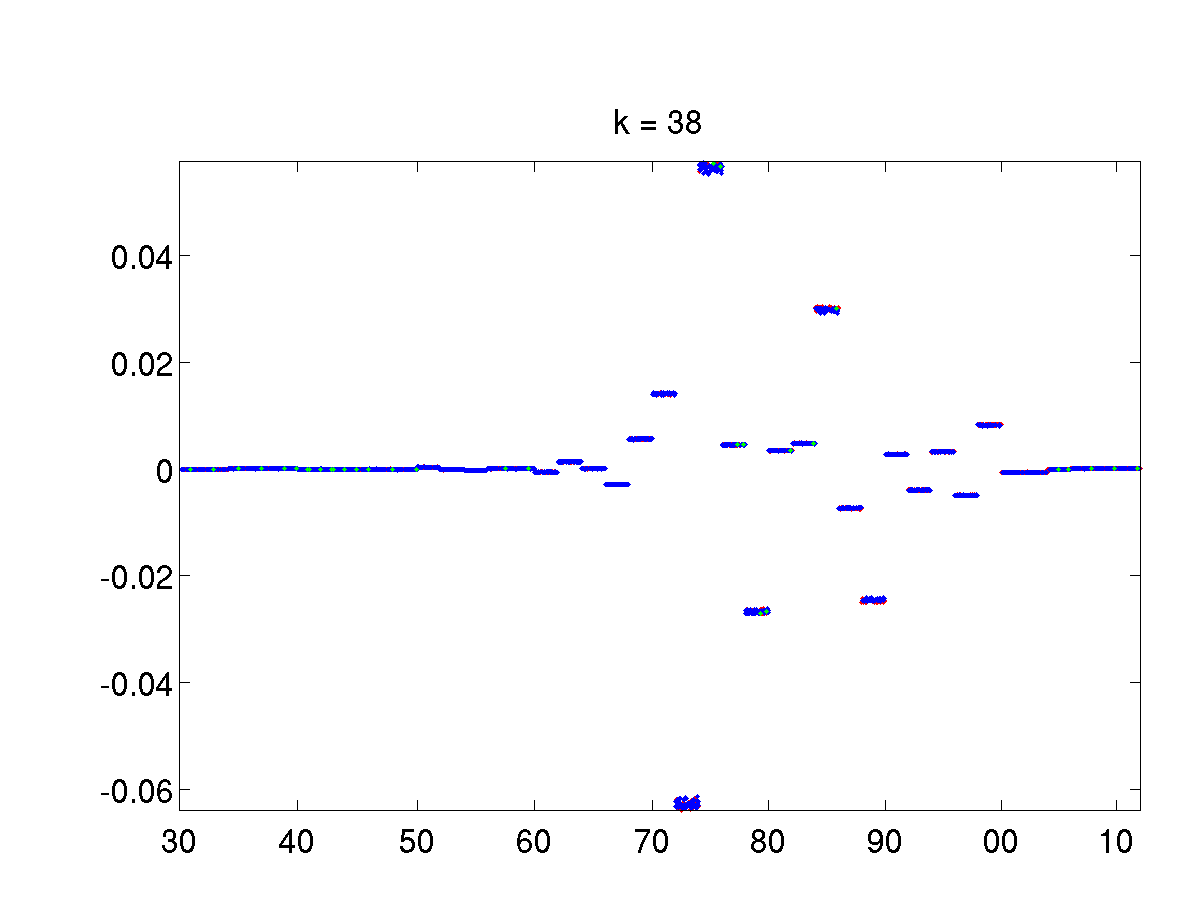}
\includegraphics[width=0.24\columnwidth]{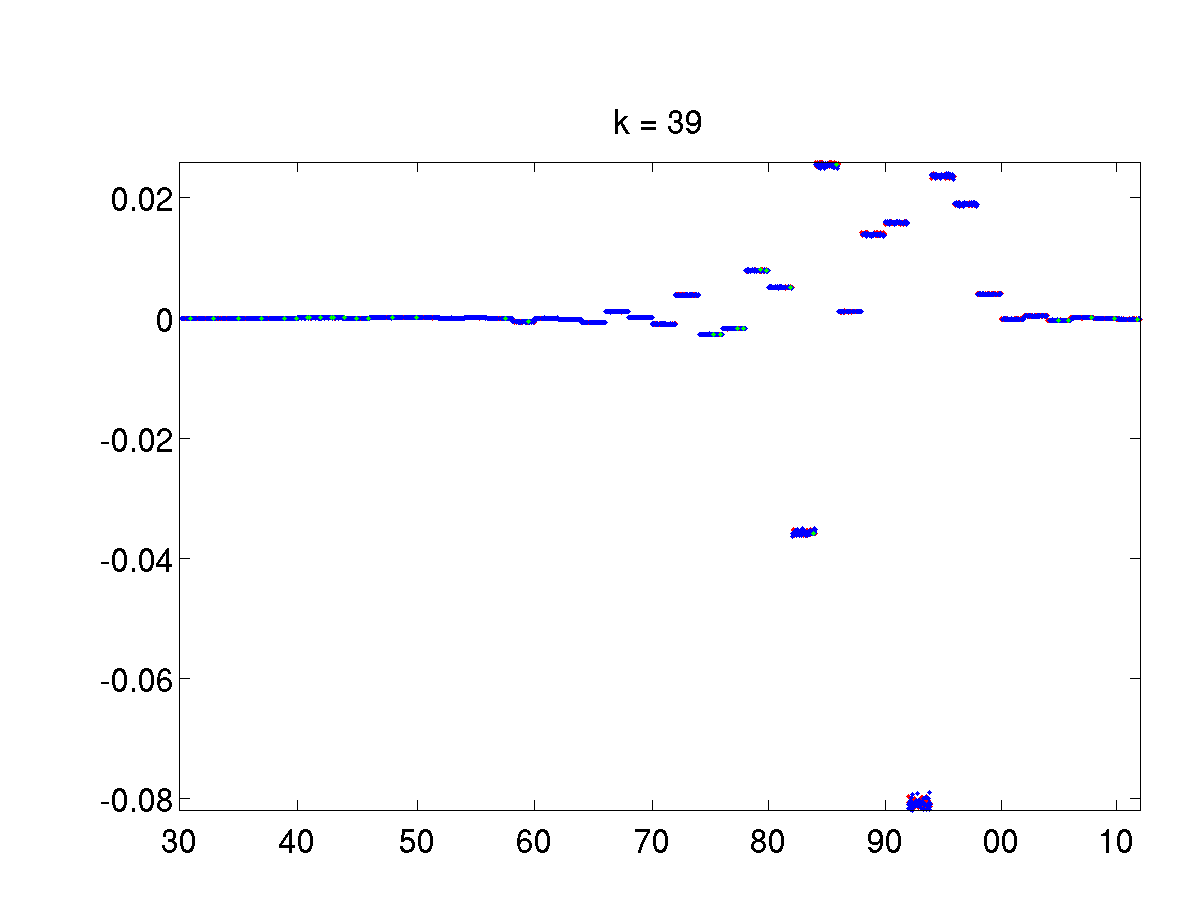}
\includegraphics[width=0.24\columnwidth]{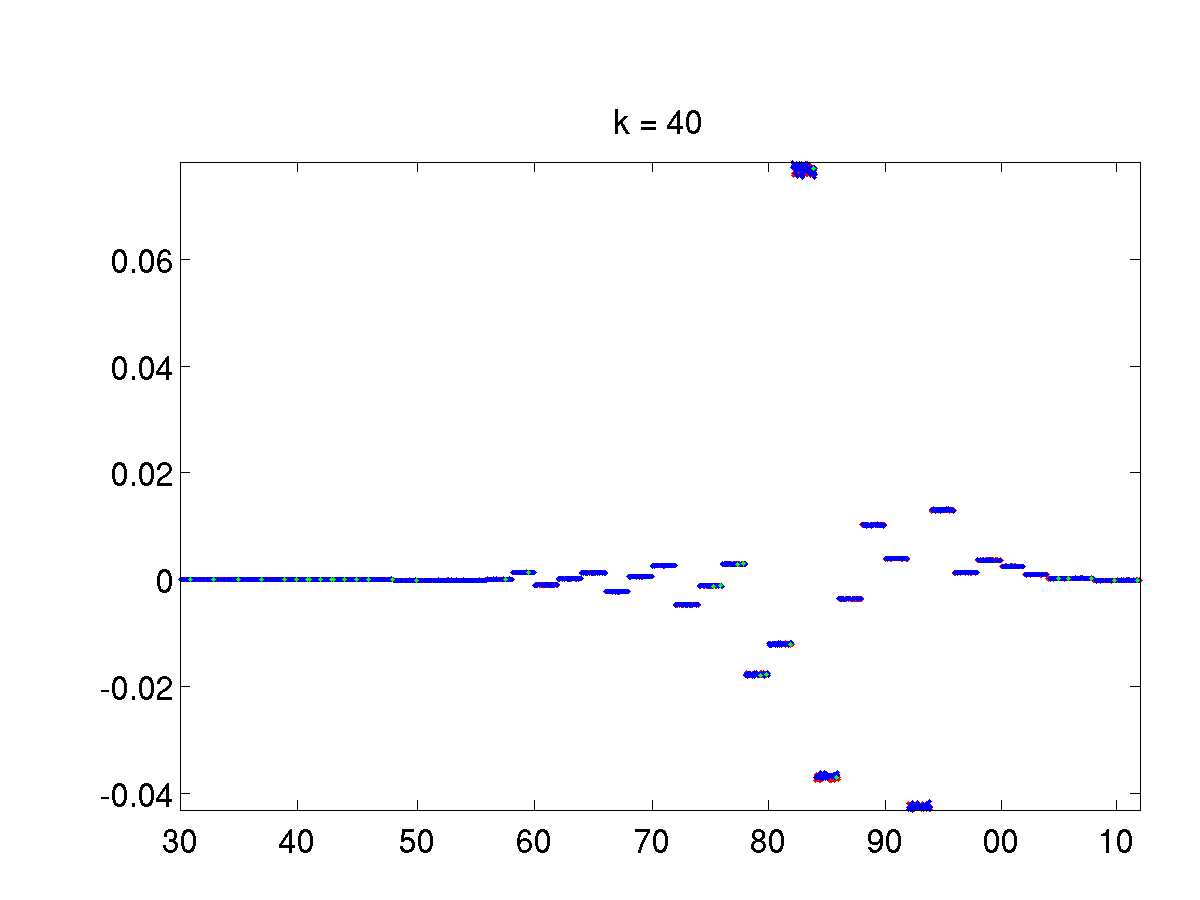}
\includegraphics[width=0.24\columnwidth]{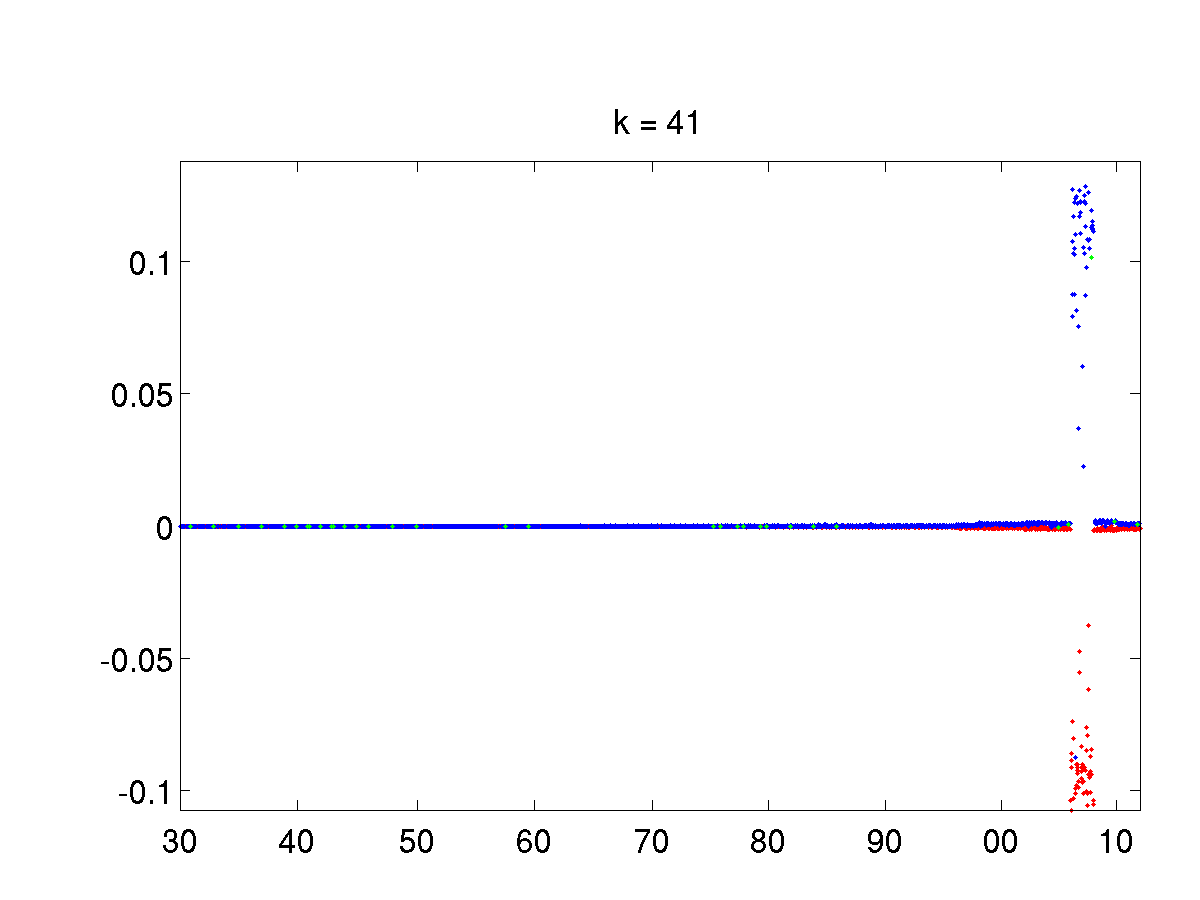}
\includegraphics[width=0.24\columnwidth]{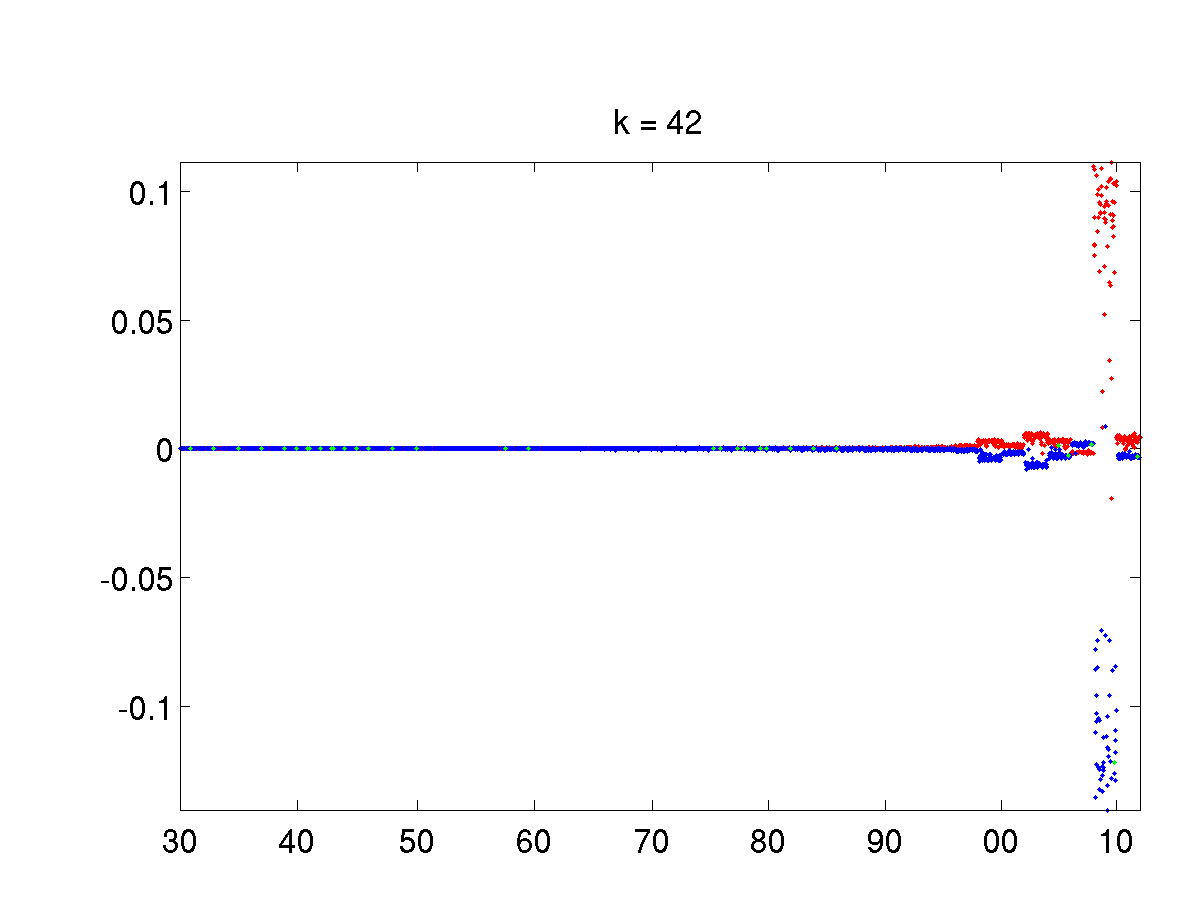}
\includegraphics[width=0.24\columnwidth]{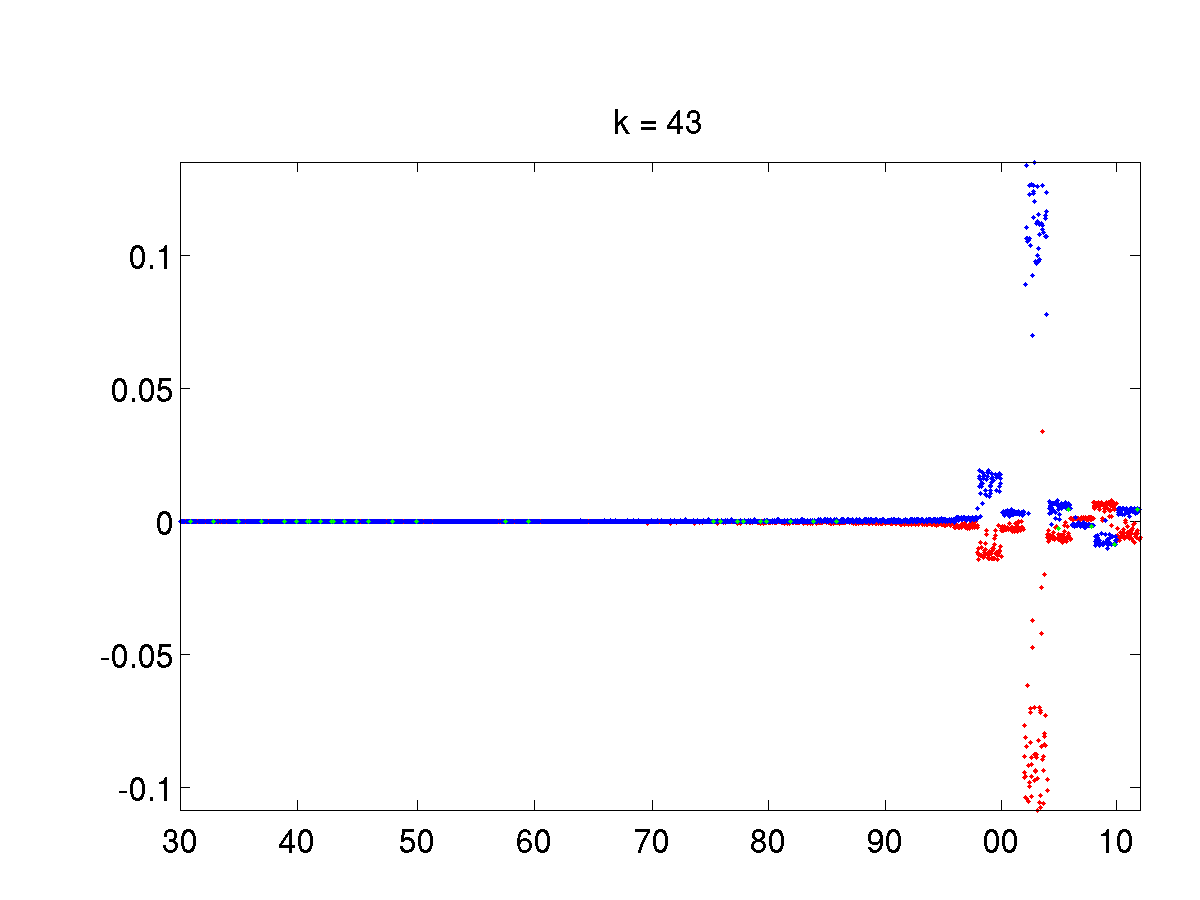}
\includegraphics[width=0.24\columnwidth]{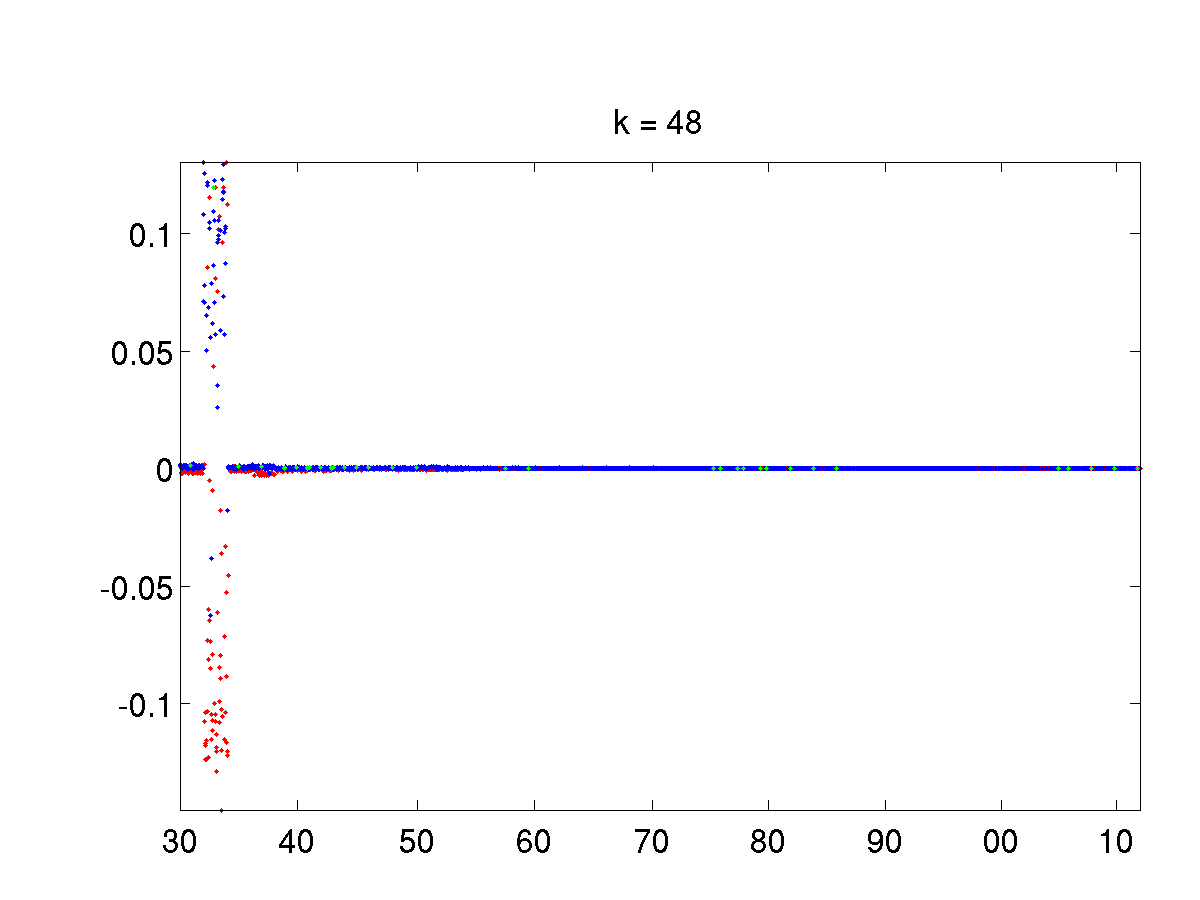}
\includegraphics[width=0.24\columnwidth]{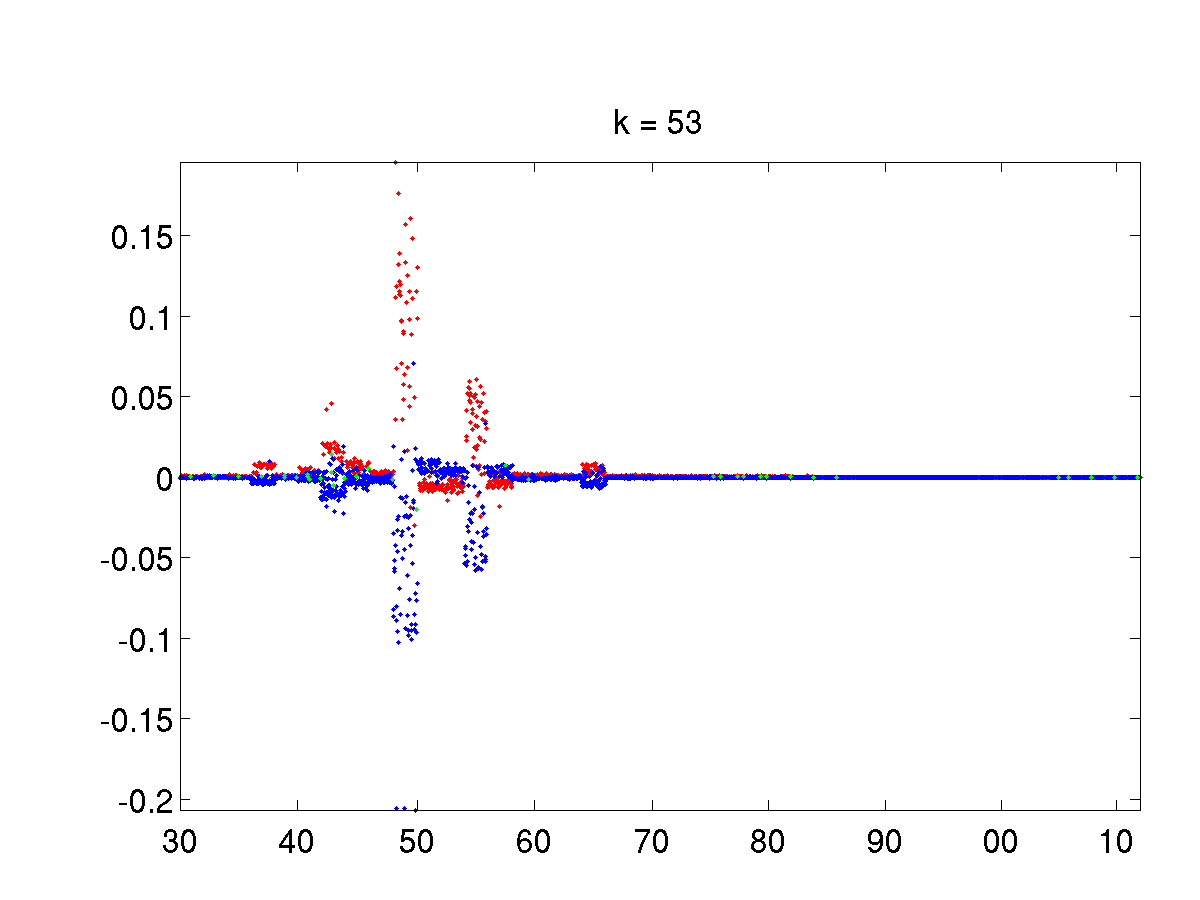}
\includegraphics[width=0.24\columnwidth]{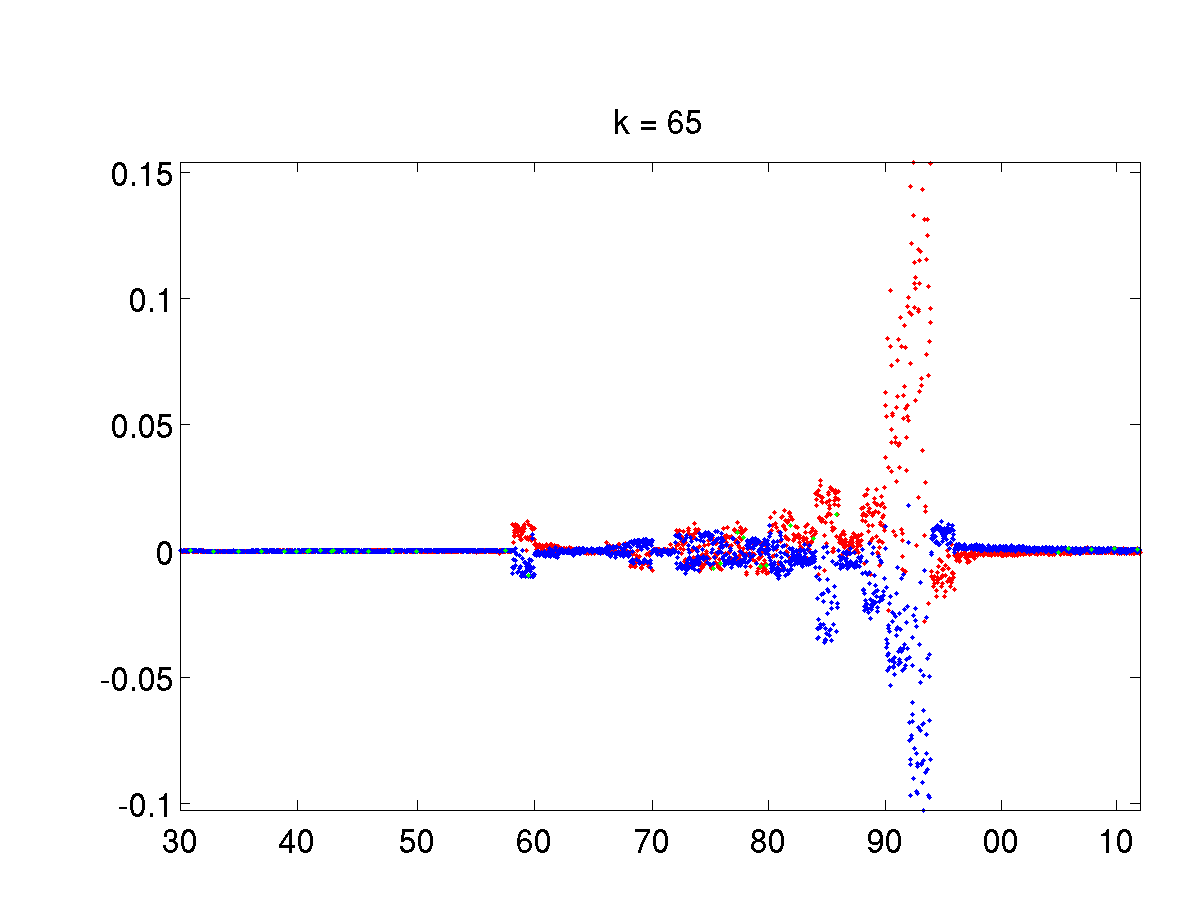}
\end{center}
\caption{The \textsc{Congress} data: illustration of several of the 
eigenvectors, when the inter-Congress coupling is set to $\epsilon = 0.1$.
(Recall that the X-axis essentially corresponds to time.)
Shown are the top eigenvectors and several of the lower-order 
eigenvectors that exhibit varying degrees of localization.}
\label{fig:cong-vect}
\end{figure}

\begin{figure}[t]
\begin{center}
\includegraphics[width=0.24\columnwidth]{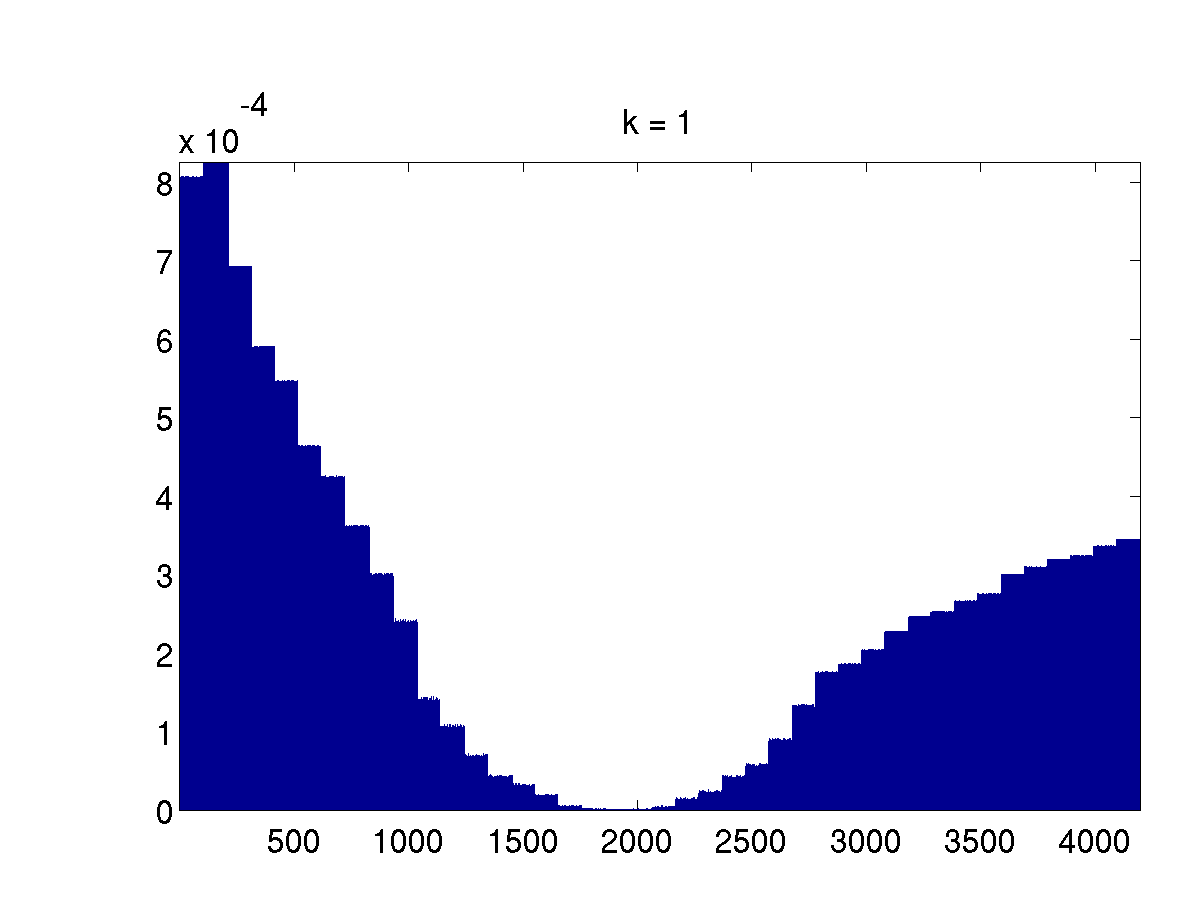}
\includegraphics[width=0.24\columnwidth]{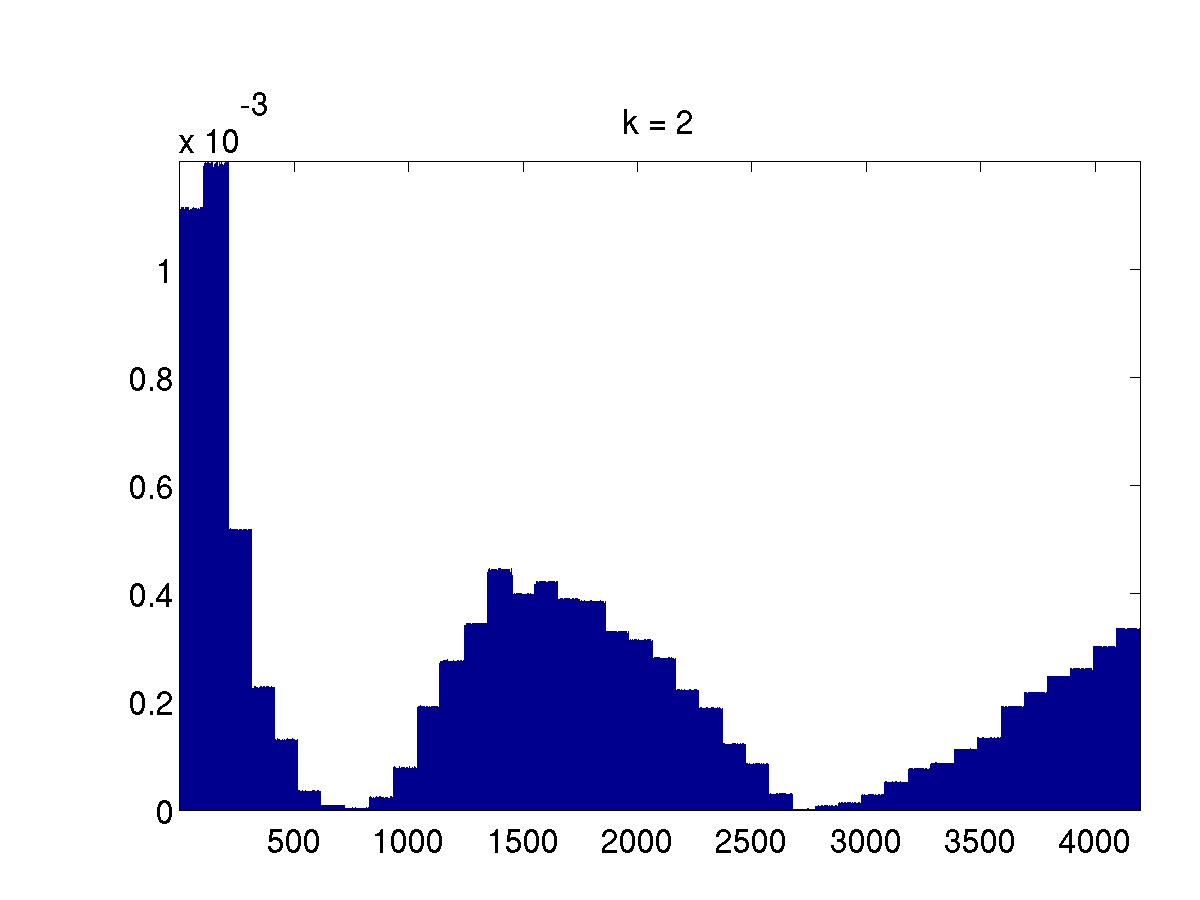}
\includegraphics[width=0.24\columnwidth]{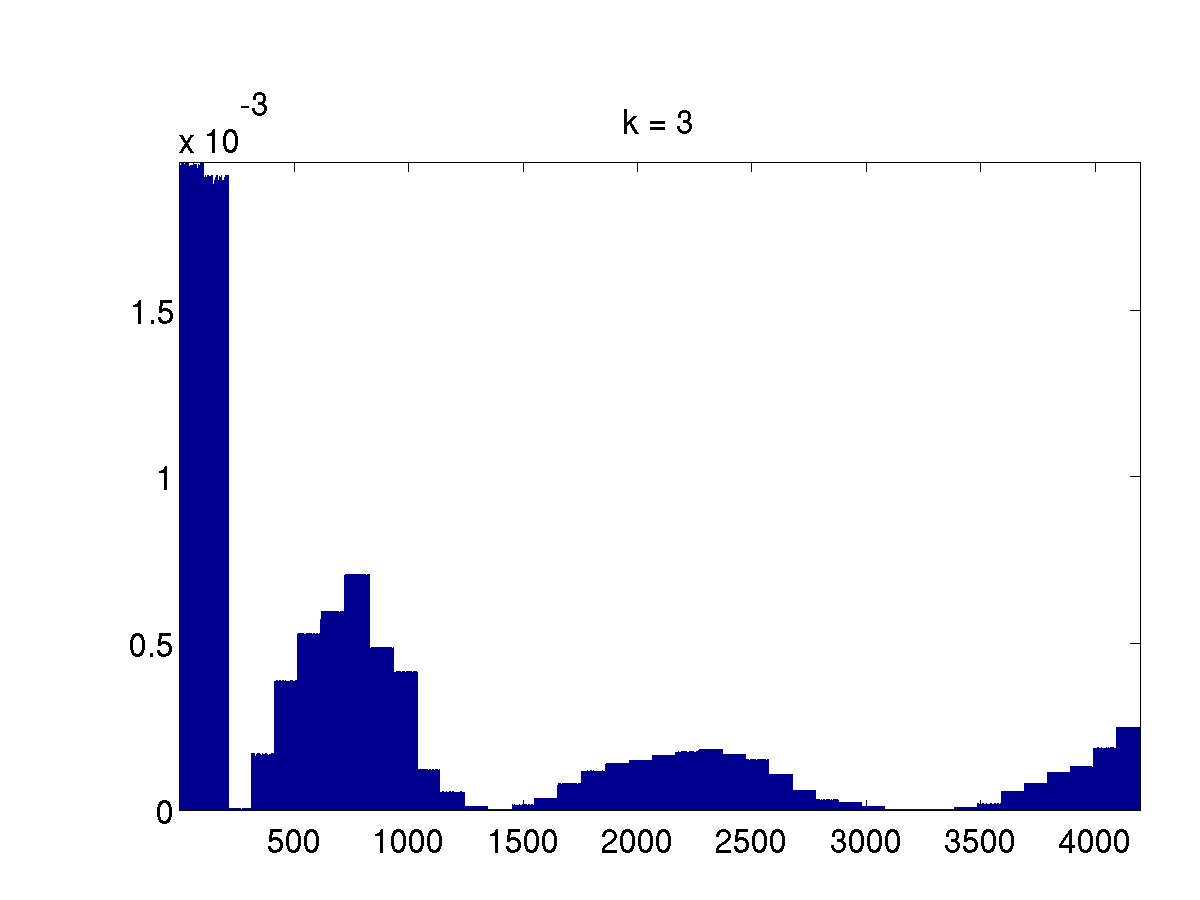}
\includegraphics[width=0.24\columnwidth]{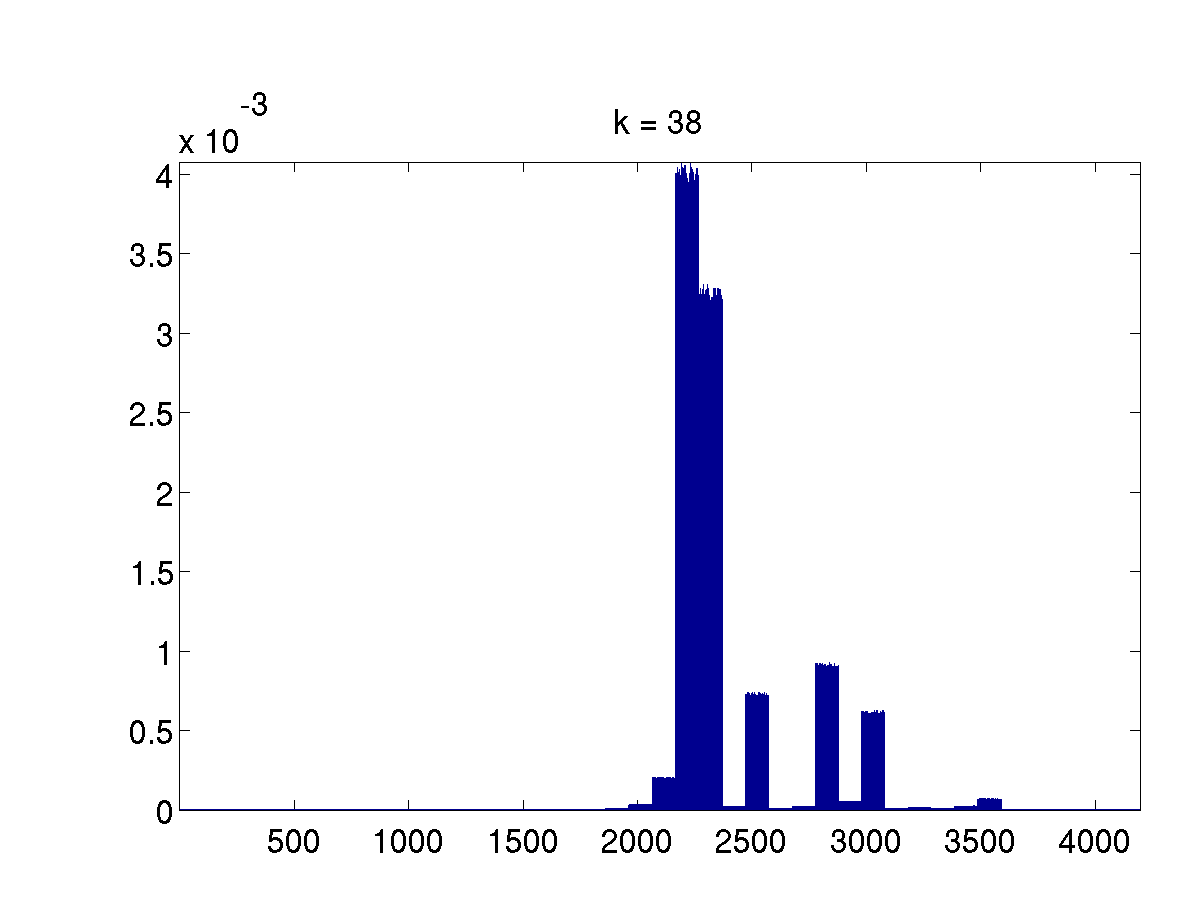}
\includegraphics[width=0.24\columnwidth]{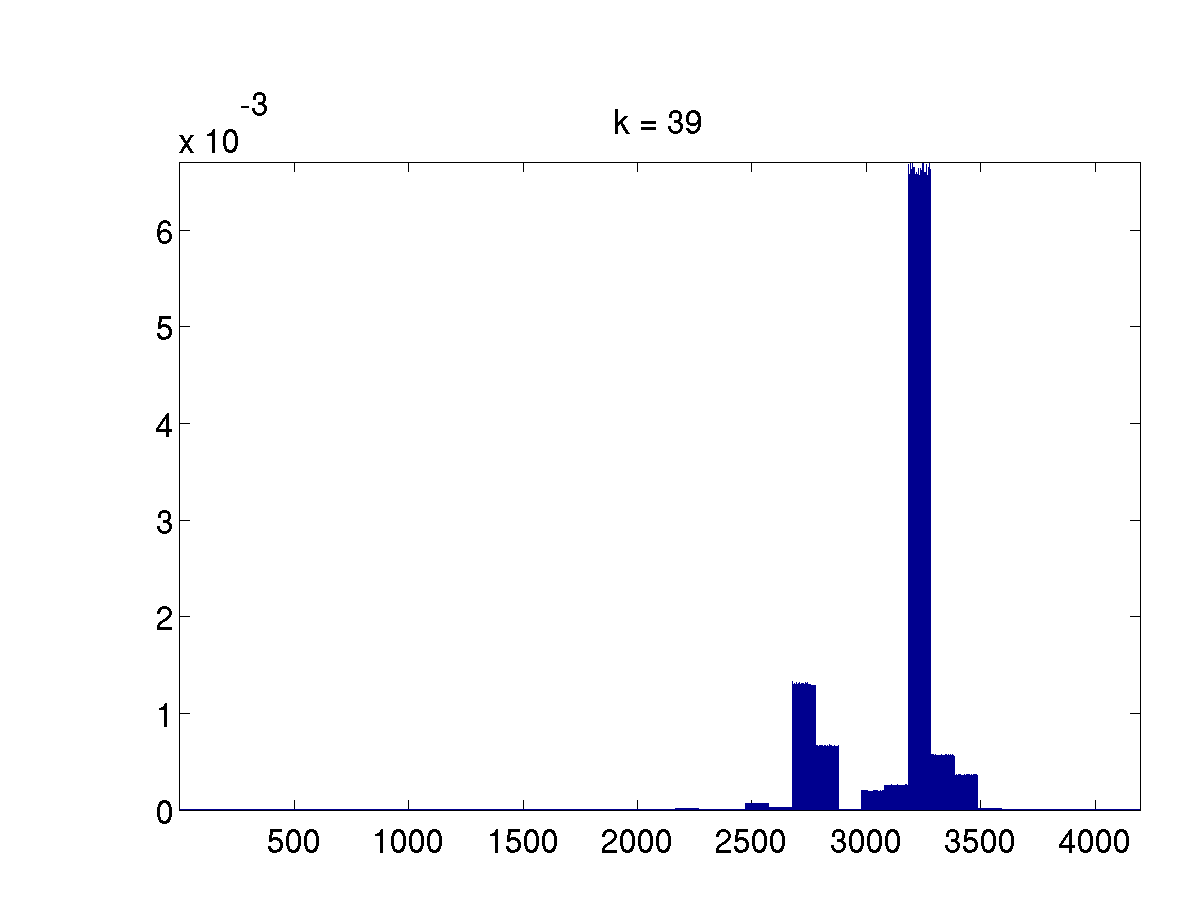}
\includegraphics[width=0.24\columnwidth]{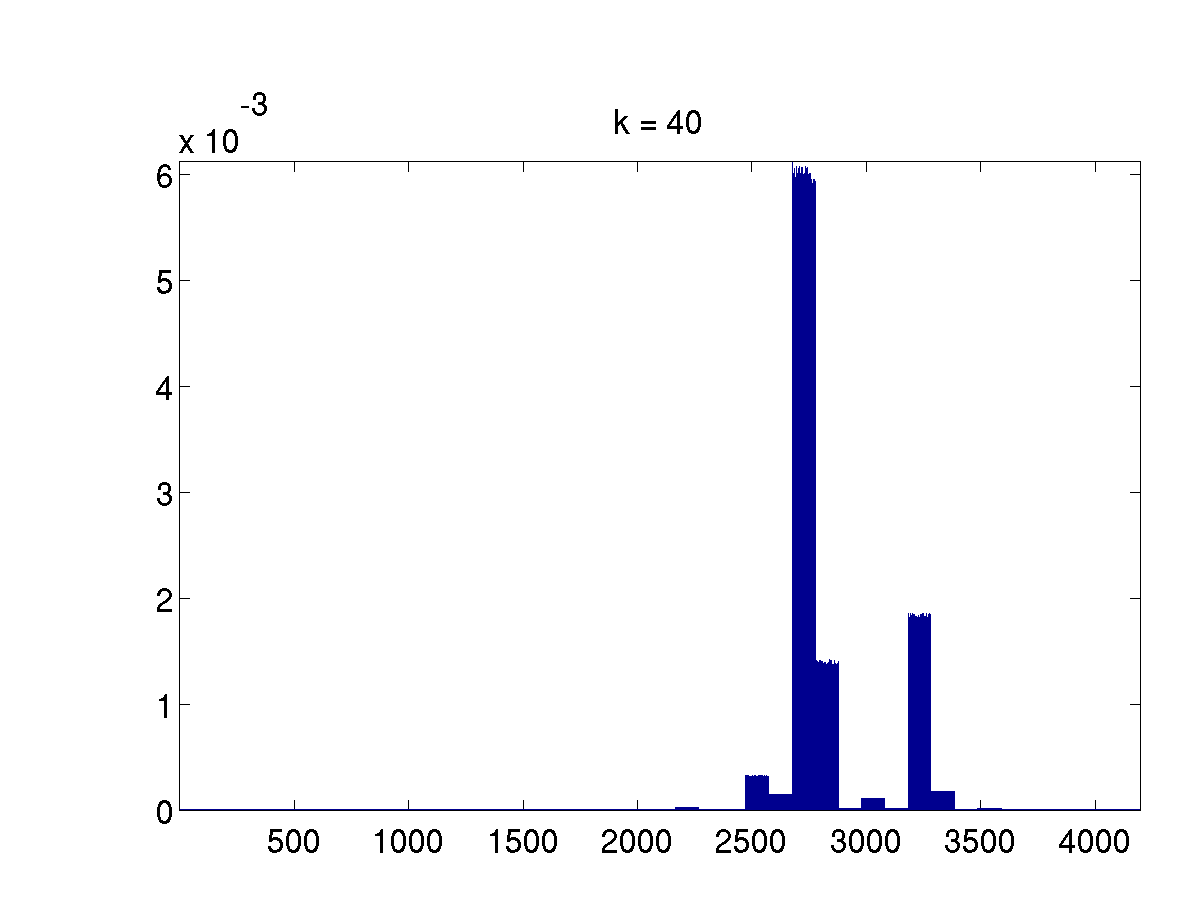}
\includegraphics[width=0.24\columnwidth]{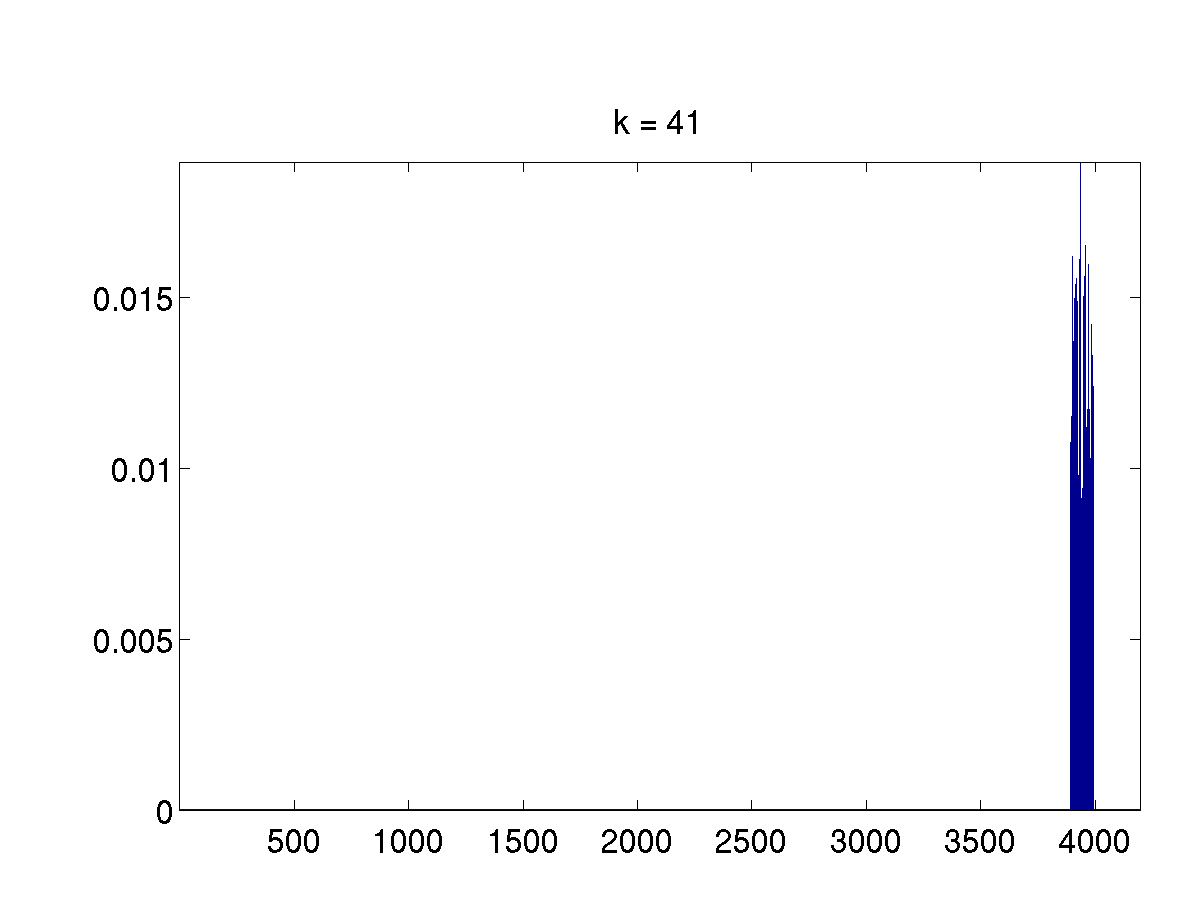}
\includegraphics[width=0.24\columnwidth]{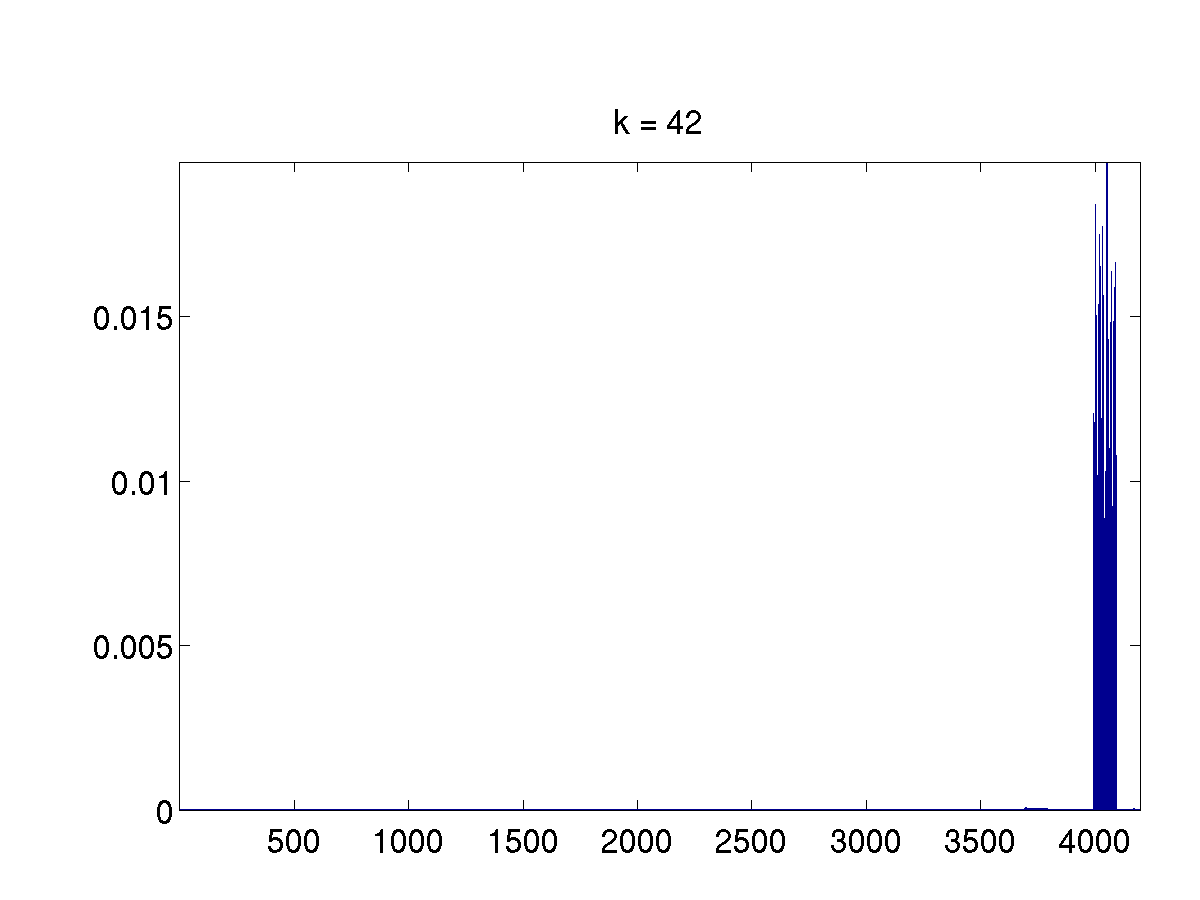}
\includegraphics[width=0.24\columnwidth]{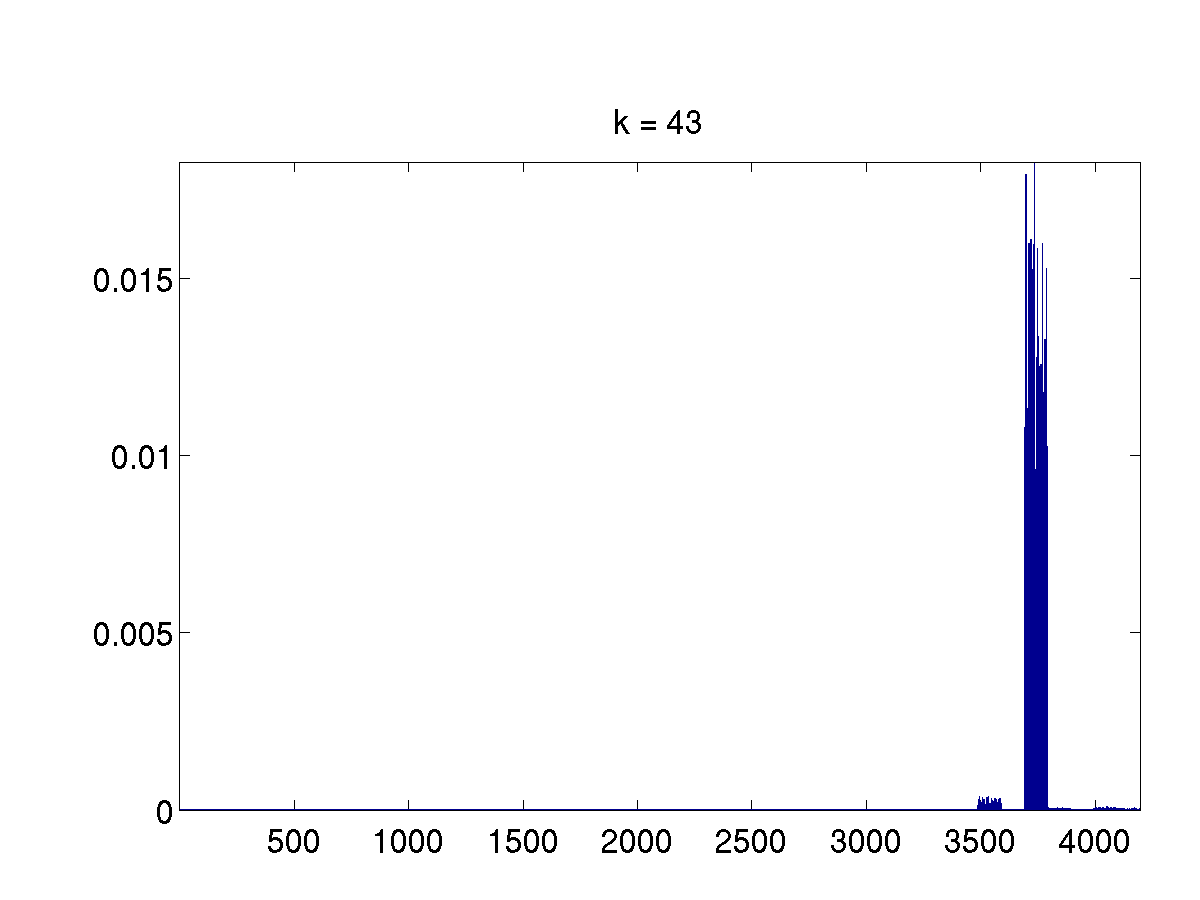}
\includegraphics[width=0.24\columnwidth]{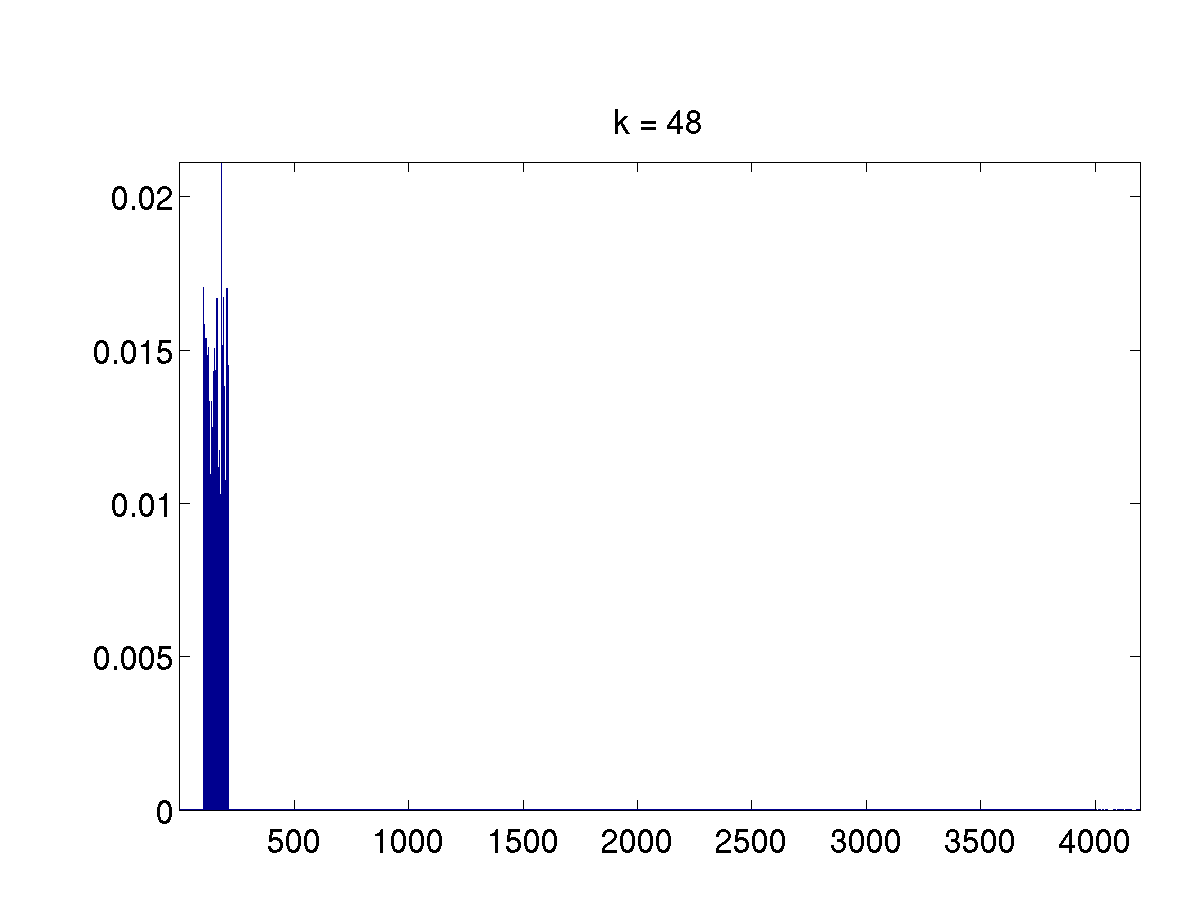}
\includegraphics[width=0.24\columnwidth]{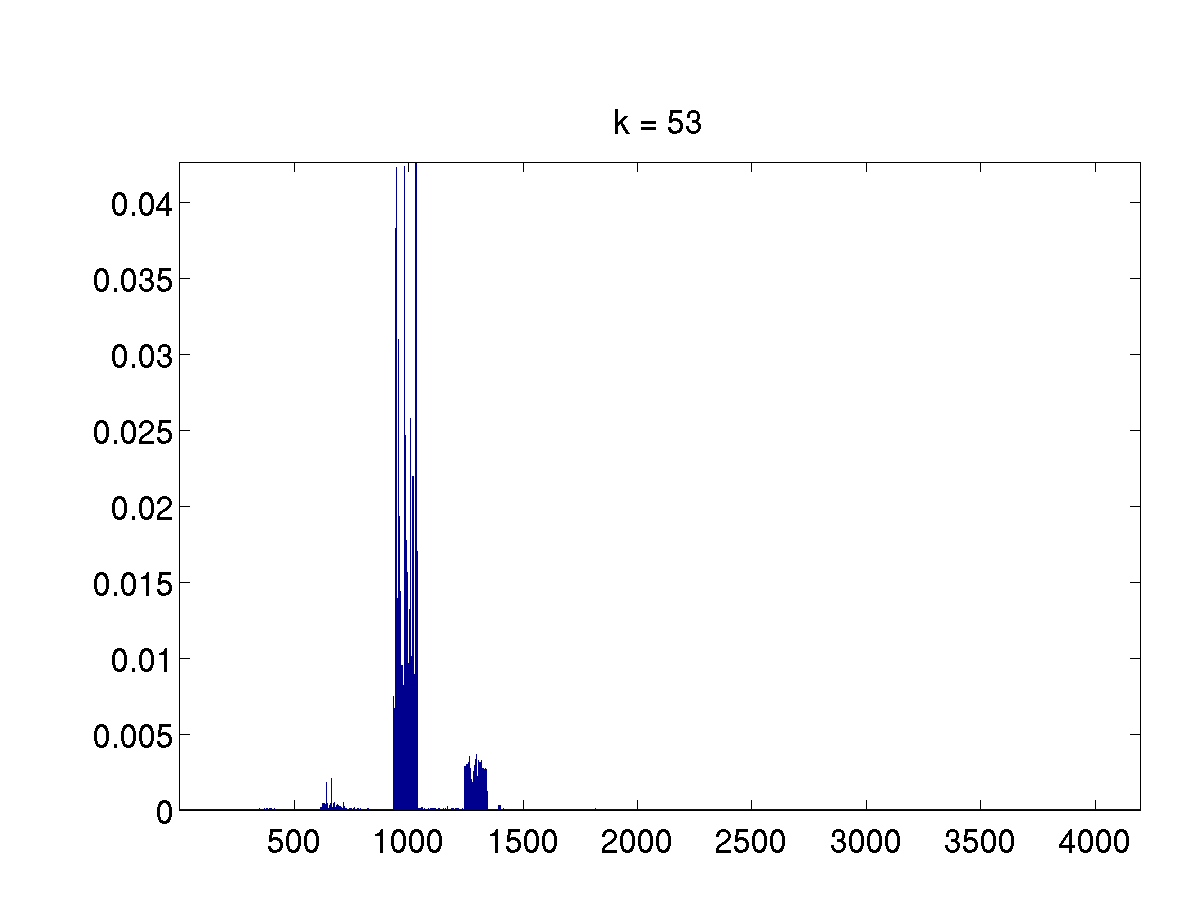}
\includegraphics[width=0.24\columnwidth]{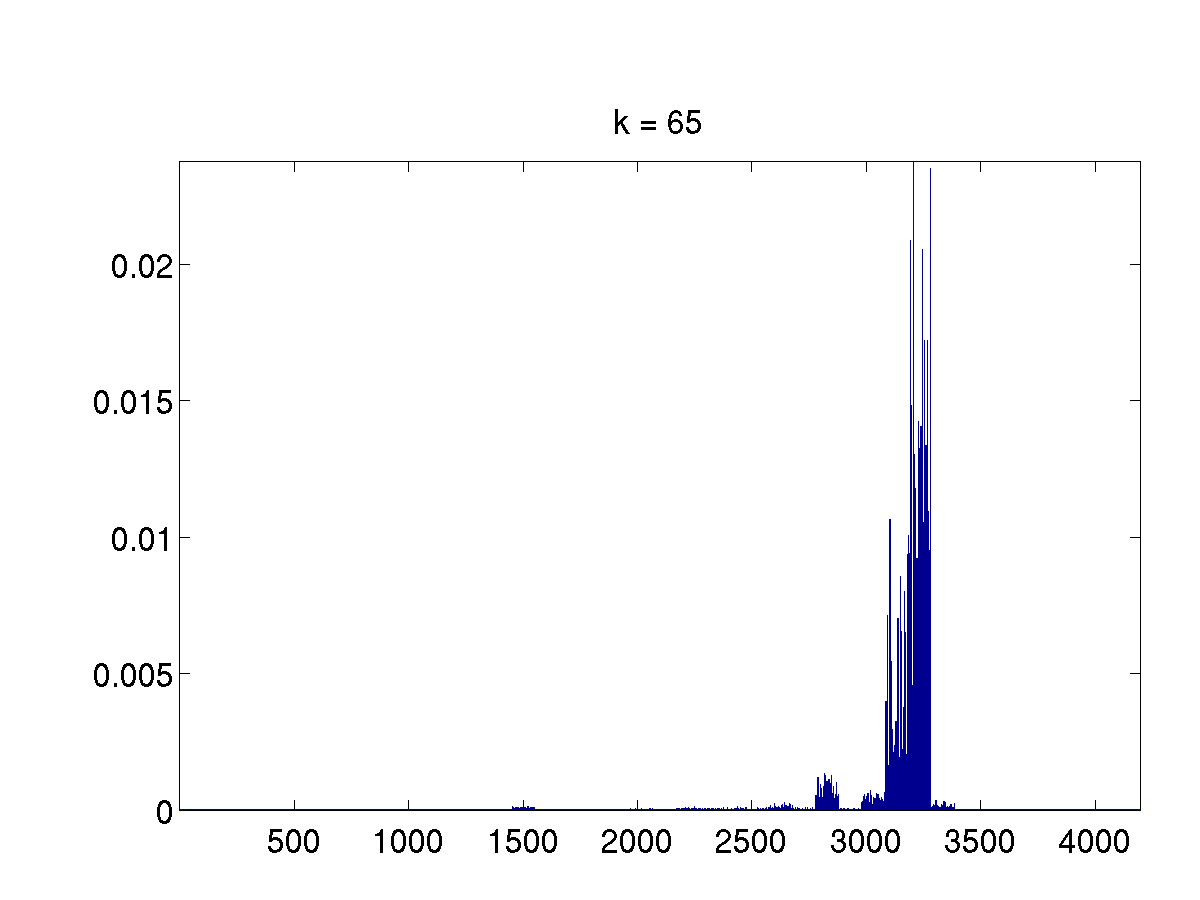}
\end{center}
\caption{The \textsc{Congress} data: the CSL scores of the eigenvectors that 
were shown in Figure~\ref{fig:cong-vect}, clearly indicating strong 
localization on some of the low-order eigenvectors.}
\label{fig:cong-sls}
\end{figure}

\begin{figure}[t]
\begin{center}
\includegraphics[width=0.24\columnwidth]{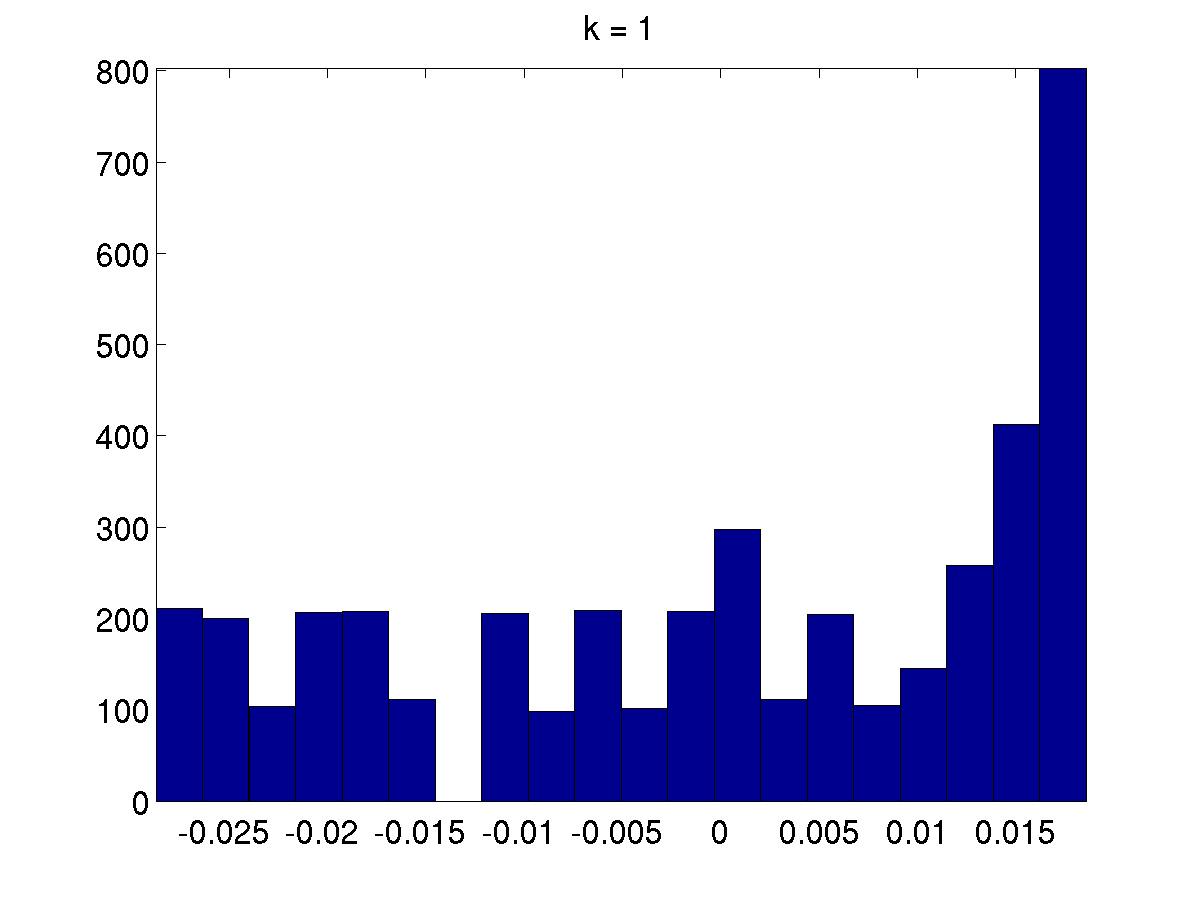}
\includegraphics[width=0.24\columnwidth]{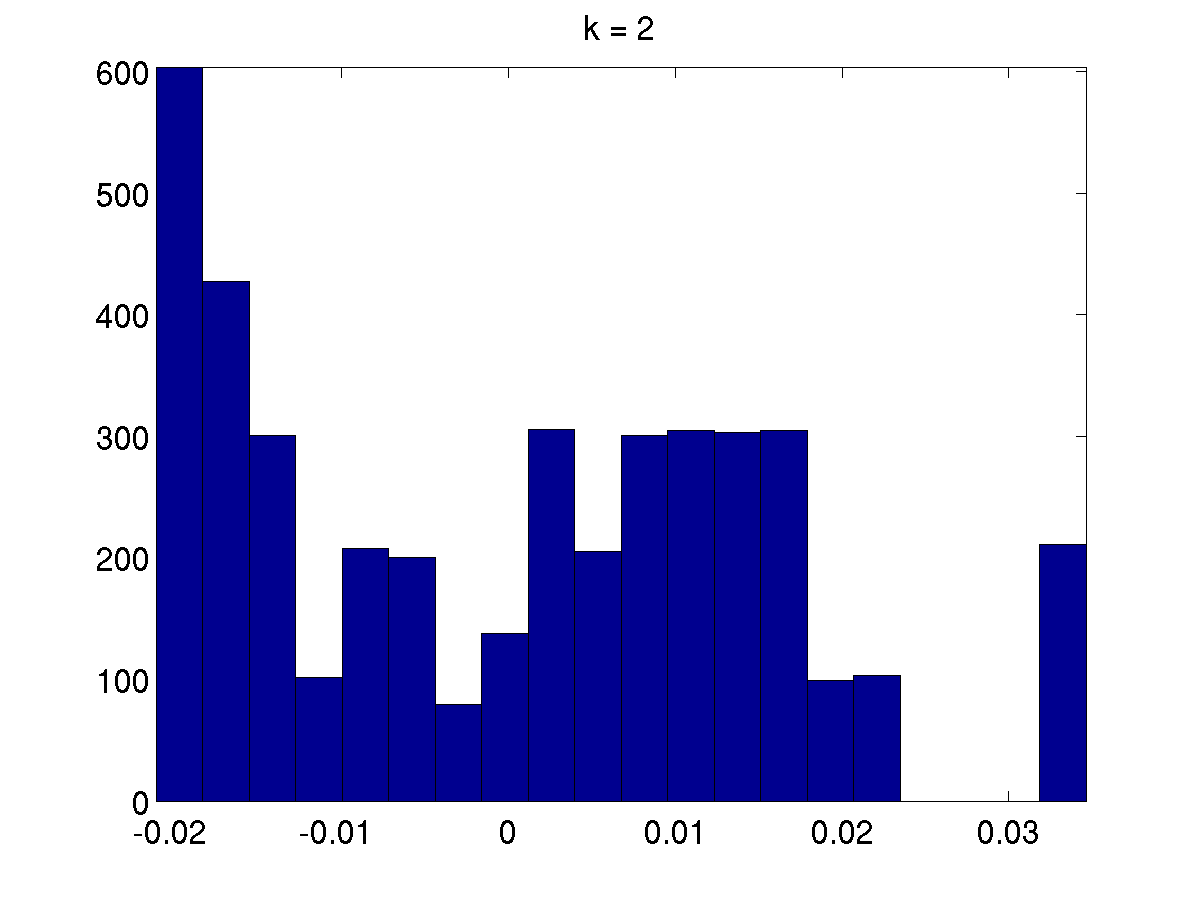}
\includegraphics[width=0.24\columnwidth]{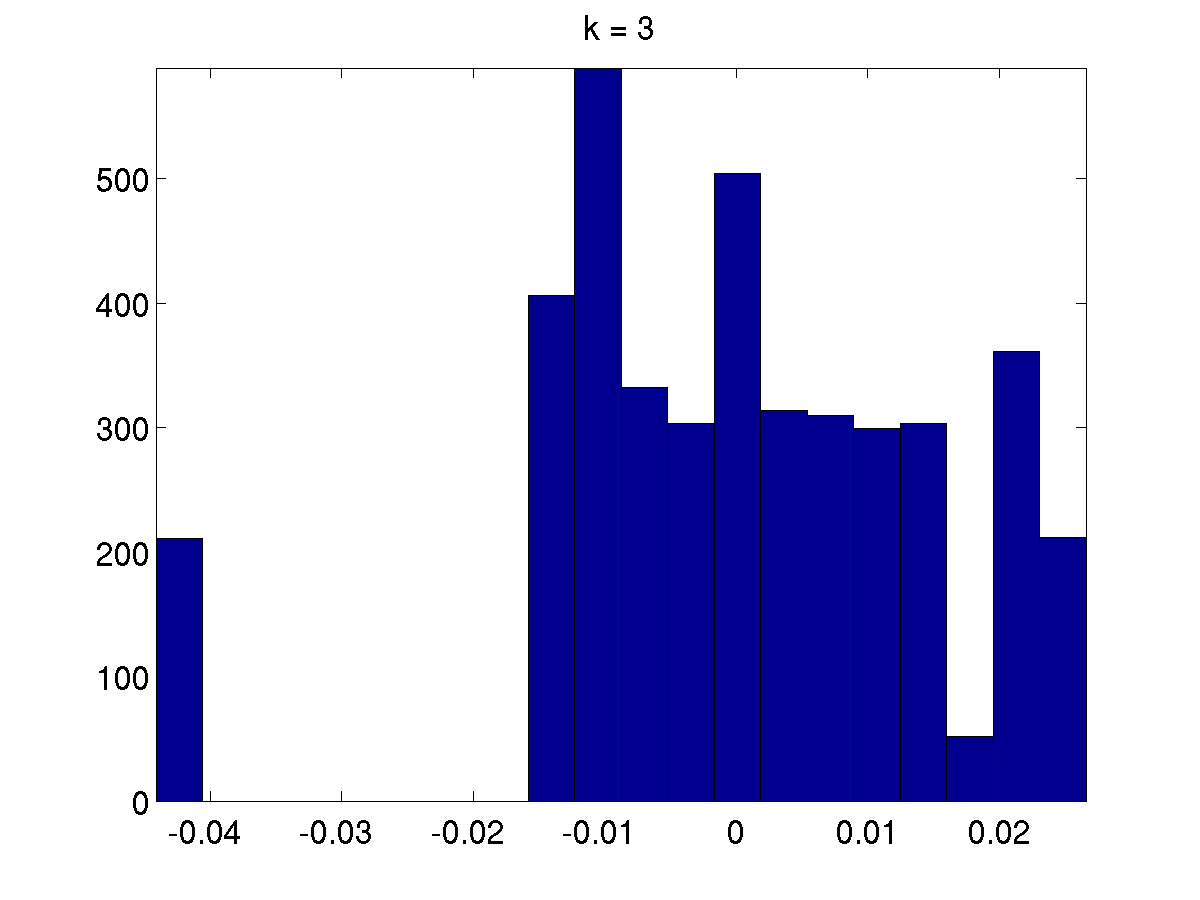}
\includegraphics[width=0.24\columnwidth]{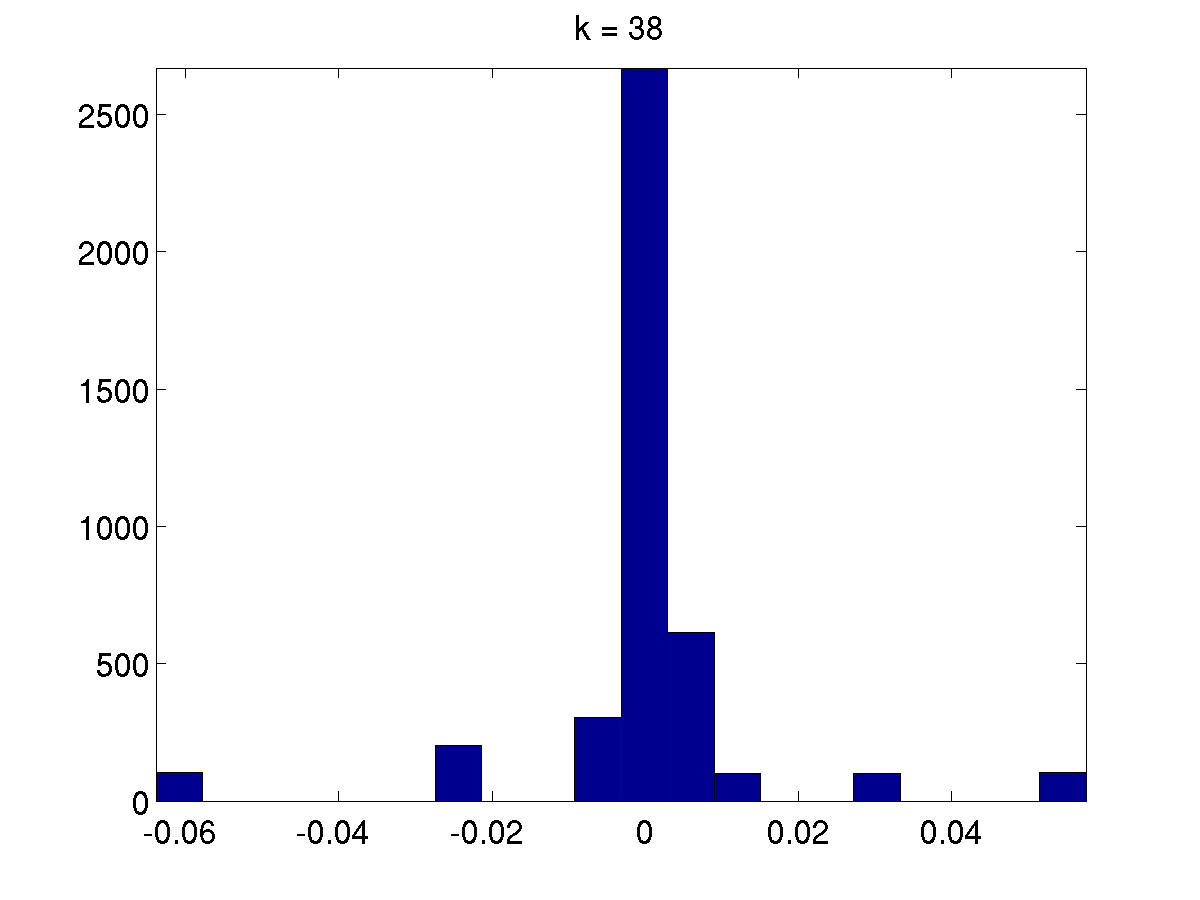}
\includegraphics[width=0.24\columnwidth]{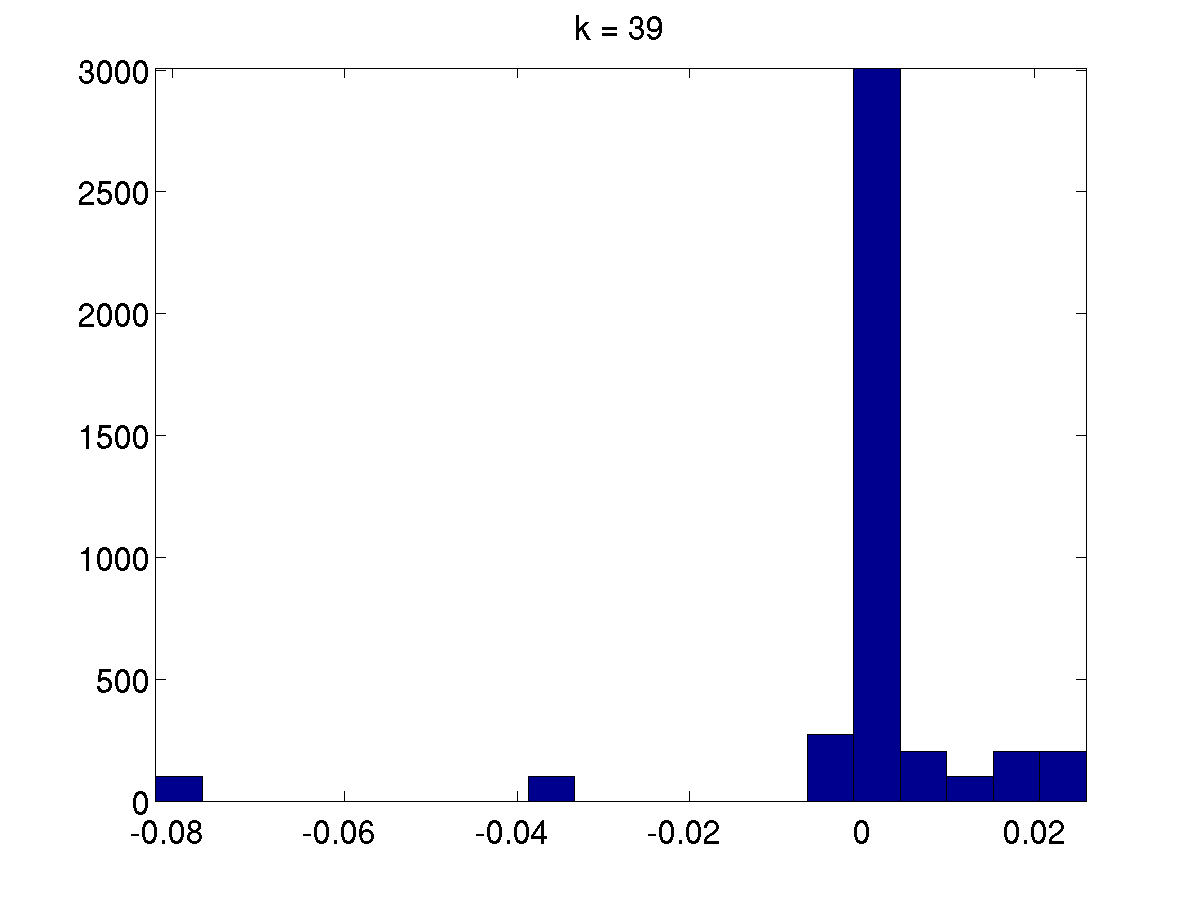}
\includegraphics[width=0.24\columnwidth]{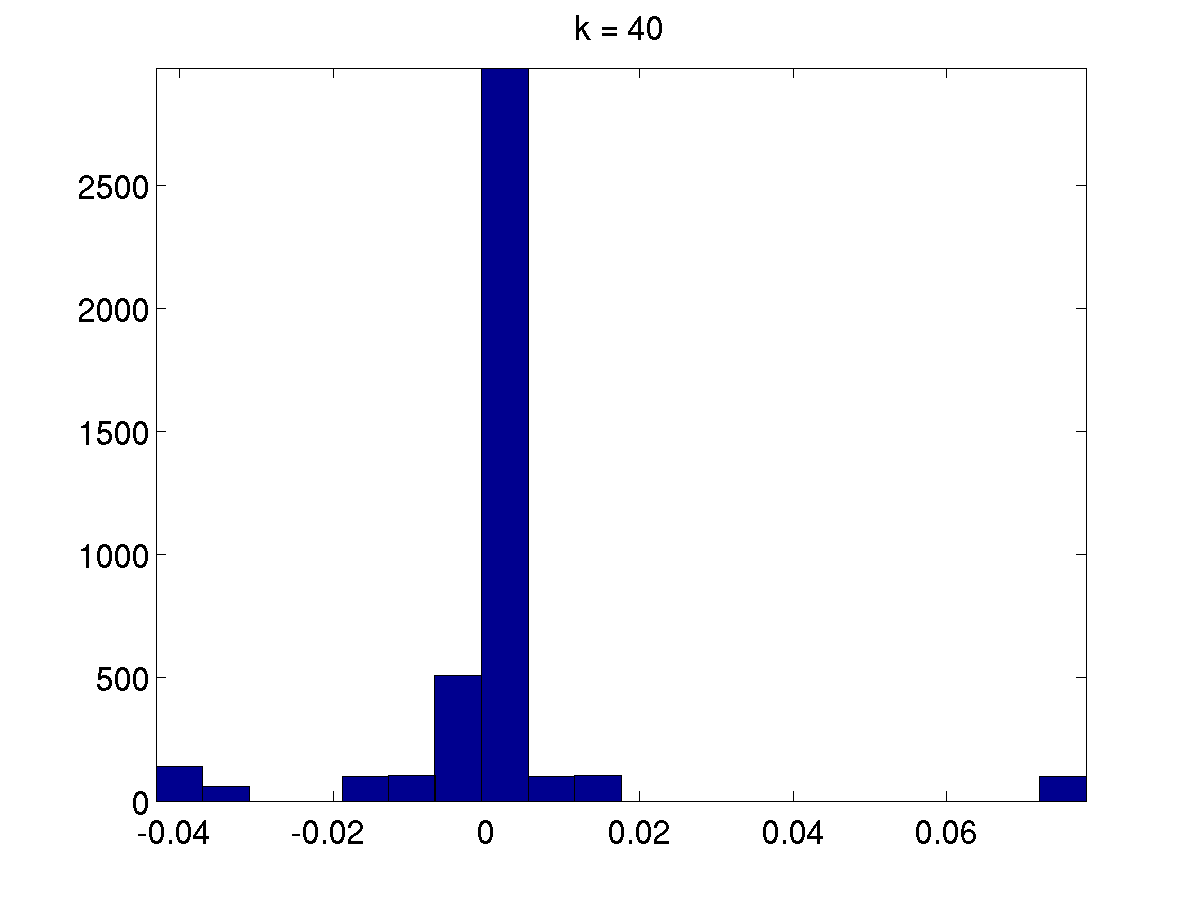}
\includegraphics[width=0.24\columnwidth]{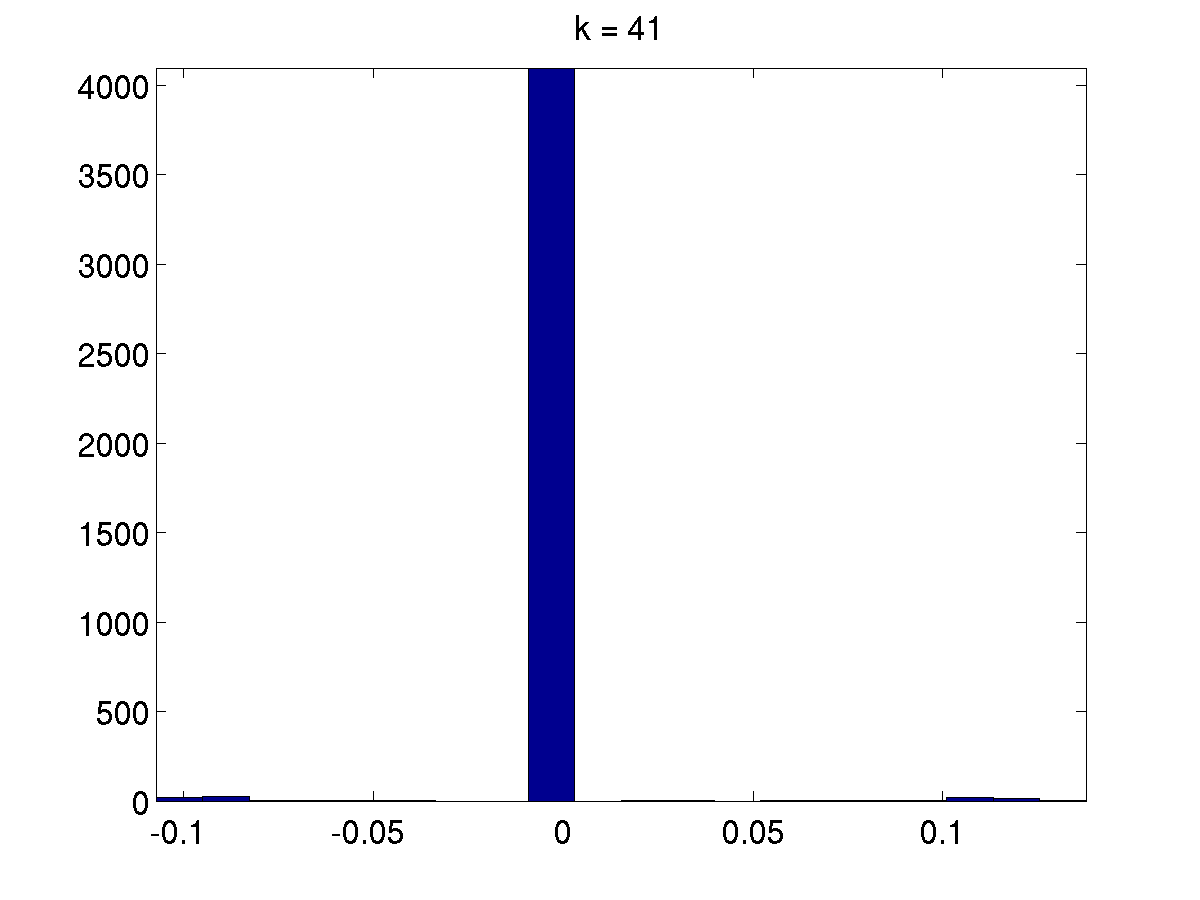}
\includegraphics[width=0.24\columnwidth]{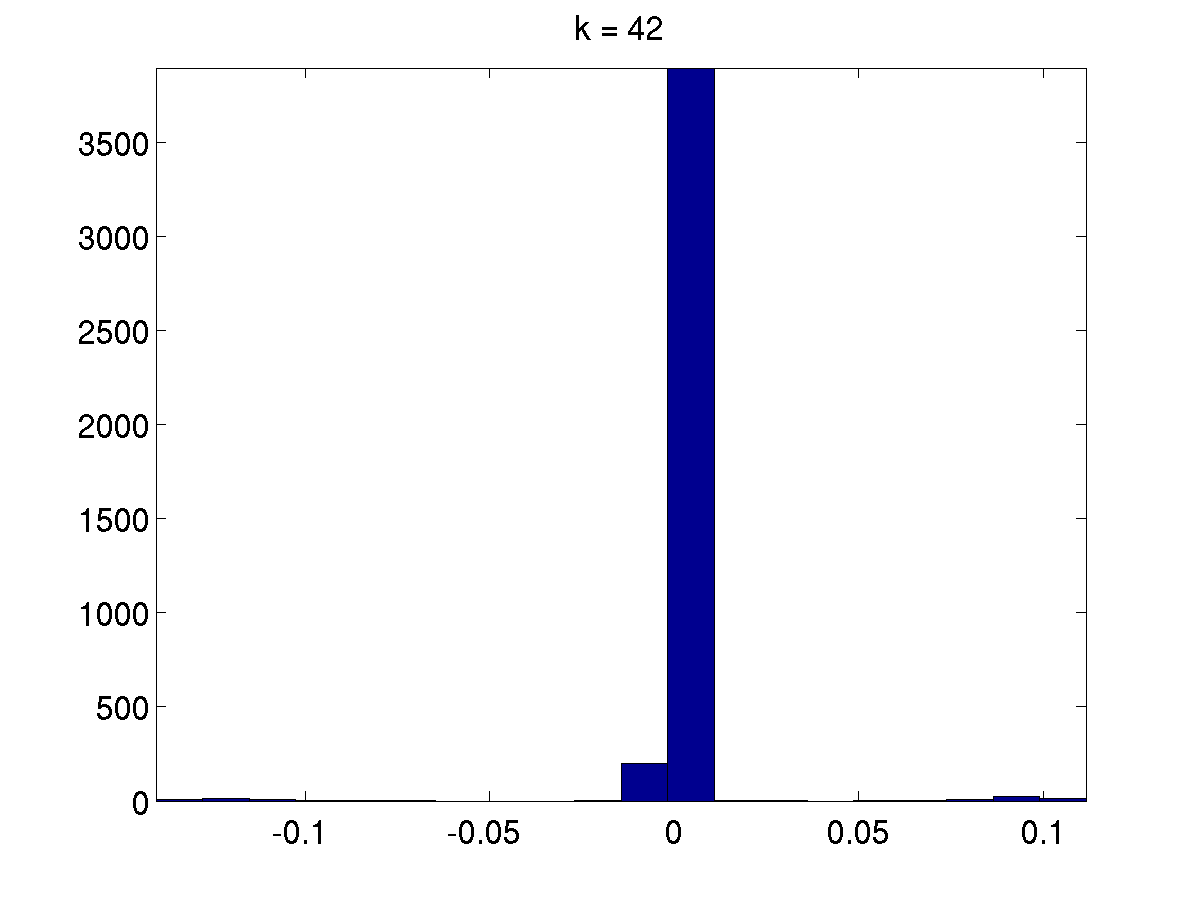}
\includegraphics[width=0.24\columnwidth]{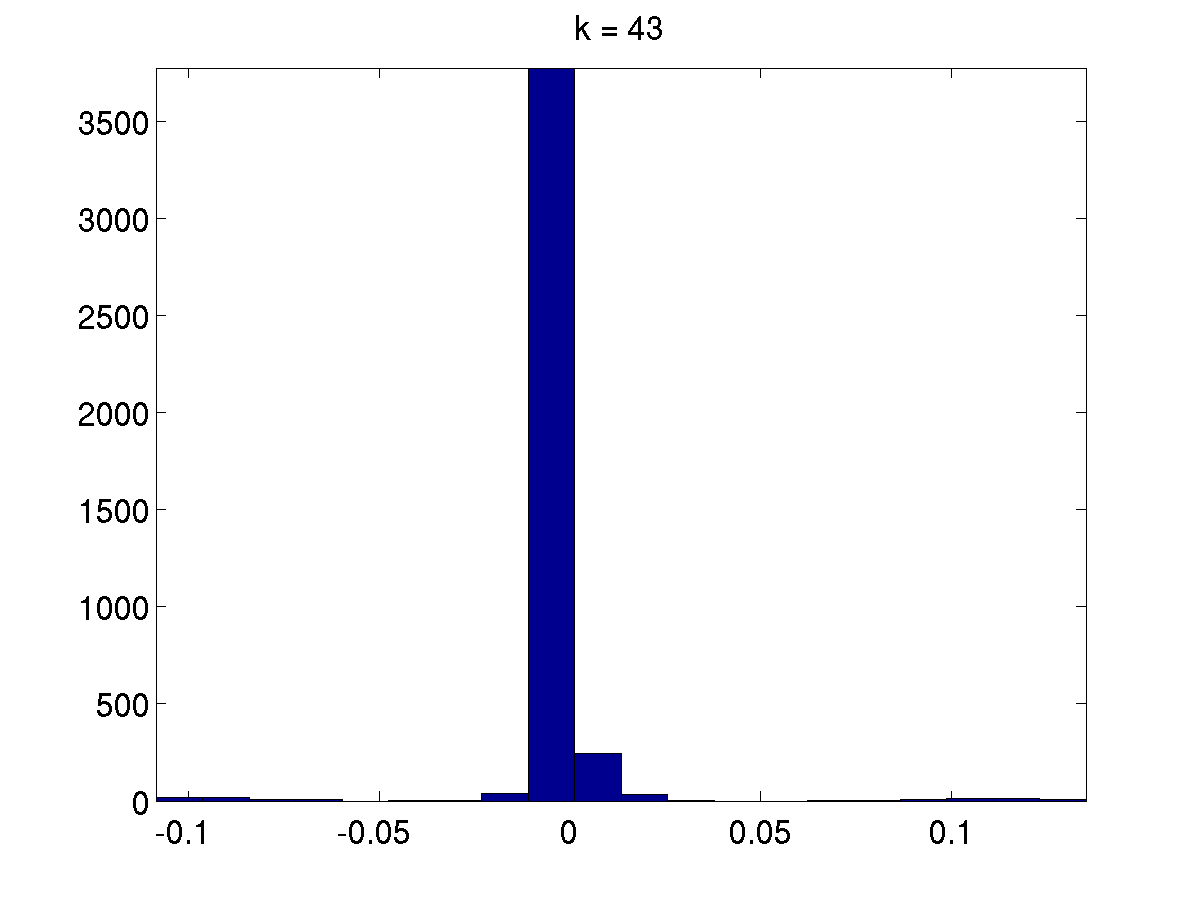}
\includegraphics[width=0.24\columnwidth]{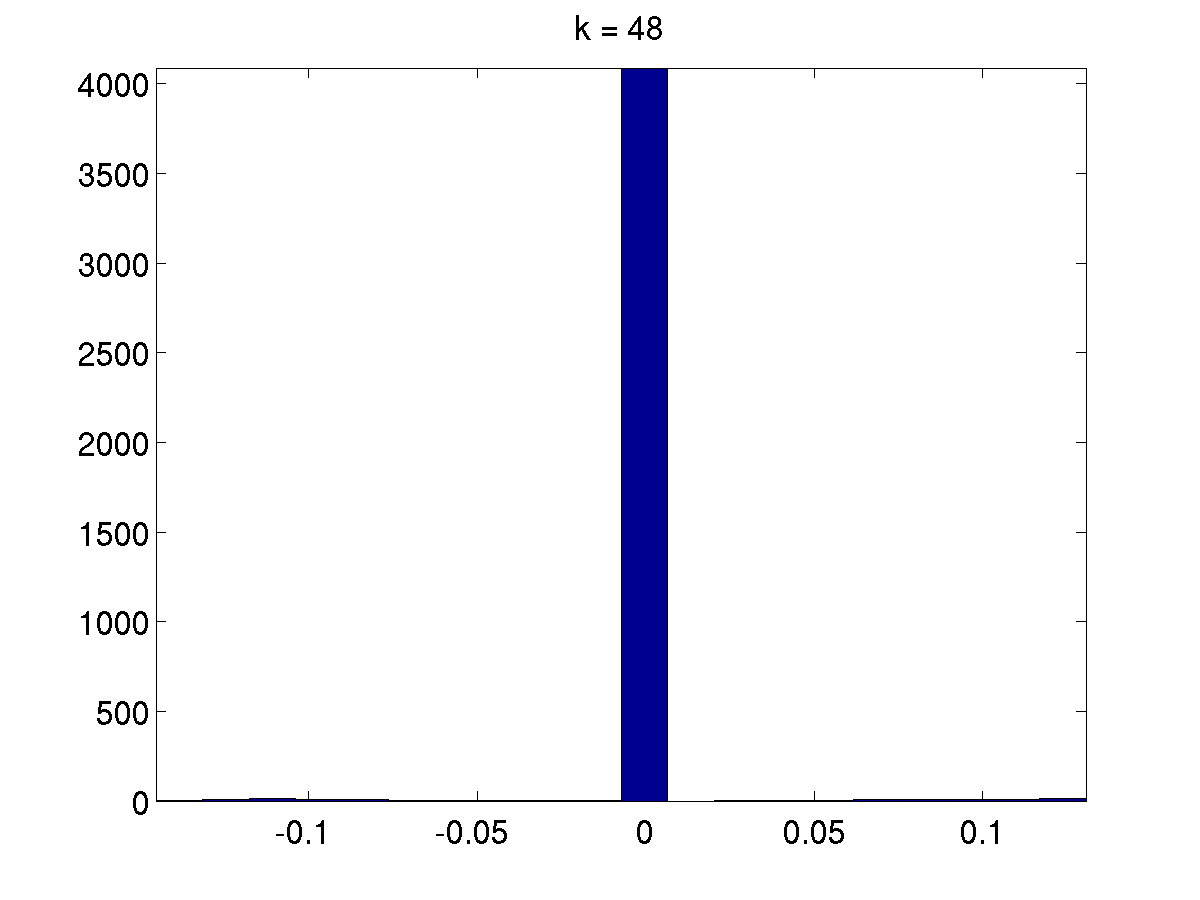}
\includegraphics[width=0.24\columnwidth]{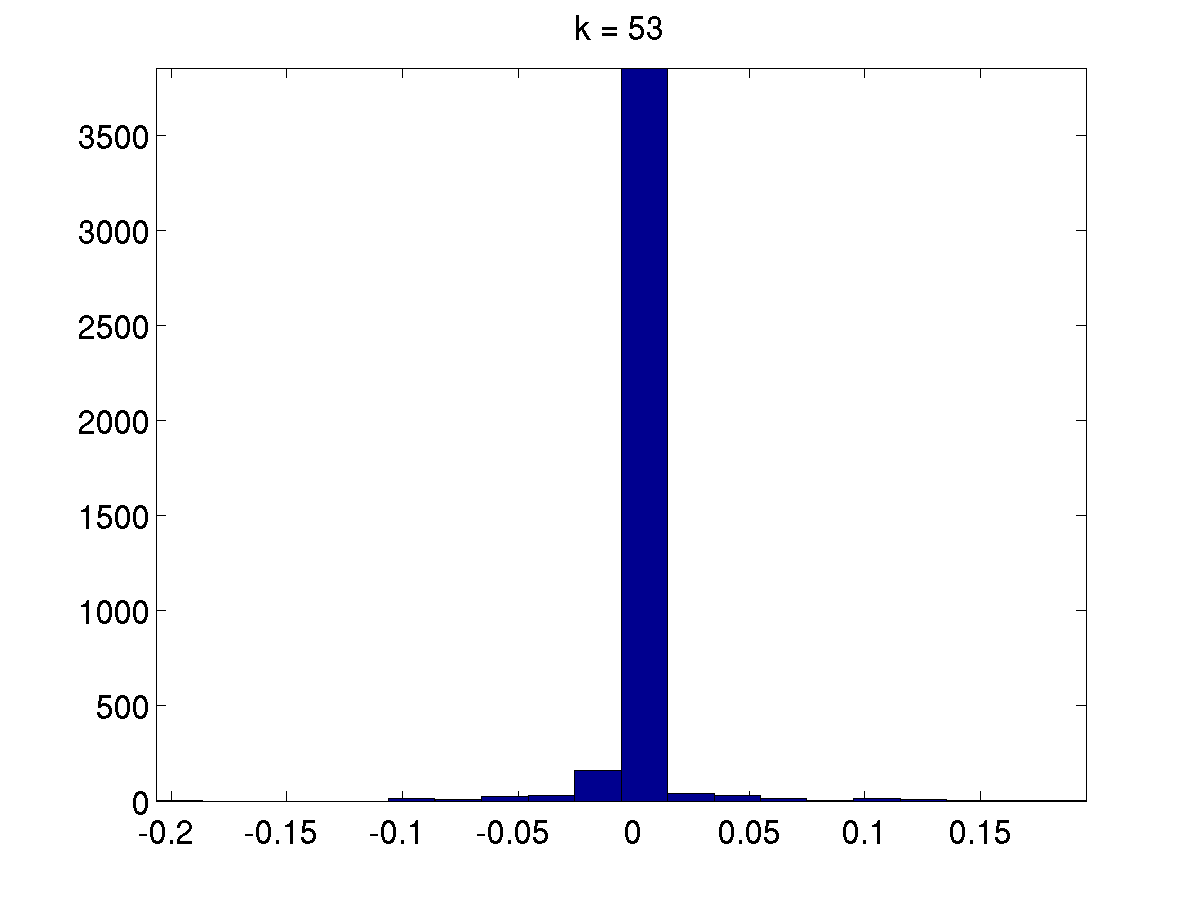}
\includegraphics[width=0.24\columnwidth]{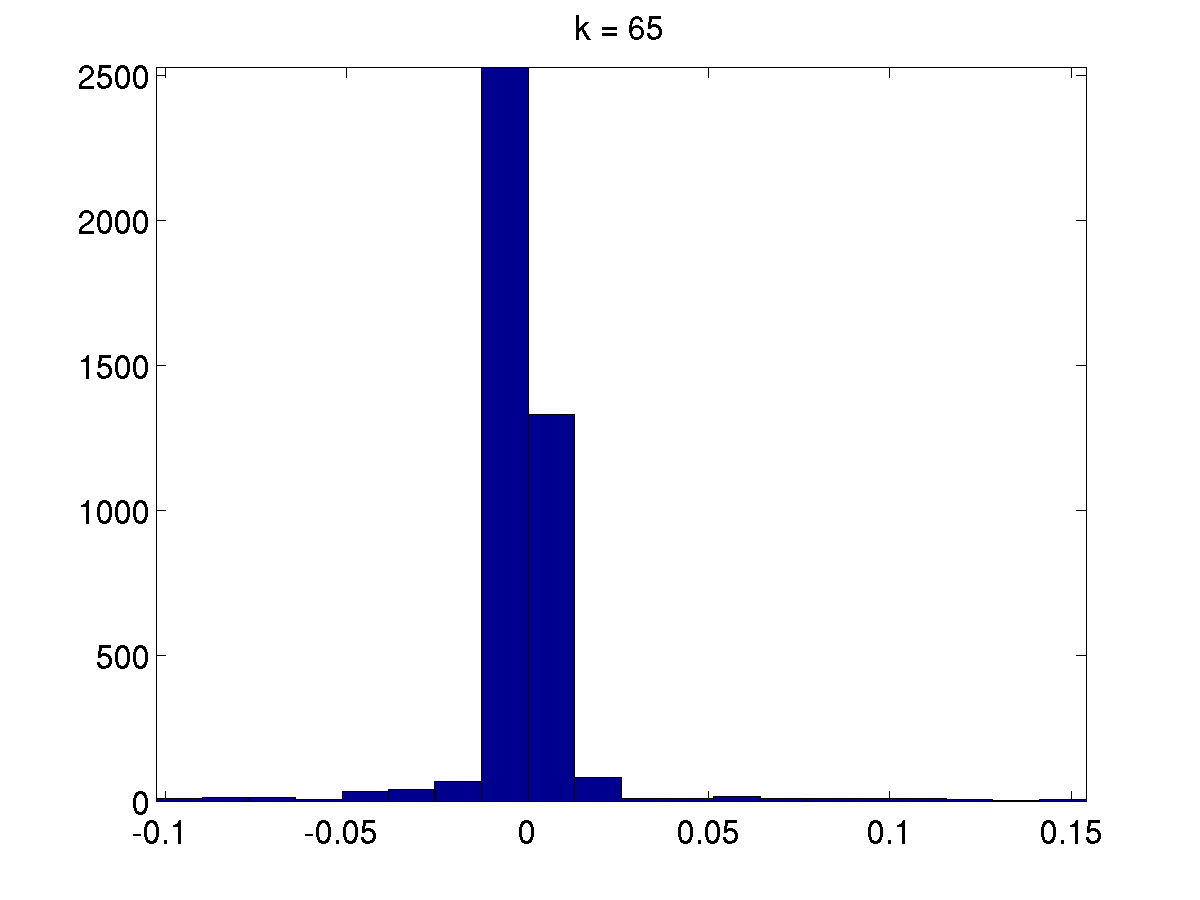}
\end{center}
\caption{The \textsc{Congress} data: histograms of the entries of the 
eigenvectors that were shown in Figure~\ref{fig:cong-vect}, clearly 
indicating strong localization on some of the low-order eigenvectors.}
\label{fig:cong-hist}
\end{figure}

For a more detailed understanding of the localization phenomenon for the 
\textsc{Congress} data (when $\epsilon=0.1$), consider 
Figures~\ref{fig:cong-vect}, \ref{fig:cong-sls}, and~\ref{fig:cong-hist}.
Figure~\ref{fig:cong-vect} presents a pictorial illustration of the top
several eigenvectors and several of the lower-order eigenvectors.
(Note that the numbering starts with the first nontrivial eigenvector.)
These particular eigenvectors have been chosen to illustrate: 
the top three directions defining the coarsest modes of variation in the data; 
the three eigenvectors above and the three eigenvectors below the low-order 
localization-delocalization transition; and
three eigenvectors further down in the spectrum.
The first three eigenvectors are fairly delocalized and exhibit global 
oscillatory behavior characteristic of sinusoids that might be expected for 
data that ``looked'' coarsely one-dimensional.
Eigenvectors $38$ to $40$ are quite far down in the spectrum; 
interestingly, they exhibit some degree of localization, perhaps more than 
one would na\"{i}vely expect, but  are still fairly delocalized relative 
to subsequent eigenvectors.
Starting with the $41^{st}$ eigenvectors, and continuing with many more 
eigenvectors that are not illustrated, one sees a remarkable 
transition---although they are quite far down in the spectrum, these 
eigenvectors exhibit a remarkable degree of localization, very often on a 
single Congress or a few temporally-adjacent Congresses.
(Note that in these and other figures the Y-axis is often different from 
subfigure to subfigure.
While creating difficulties for comparing different plots, the alternative 
would involve losing the resolution along the Y-axis for all but the most 
localized eigenvectors.)
Figure~\ref{fig:cong-sls} shows the SLS for these twelve eigenvectors, and 
Figure~\ref{fig:cong-hist} shows a histogram of the entries for each of 
these twelve eigenvectors.
By both of these measures, very pronounced localization is clearly observed, 
complementing the observations in the previous figure.

%TMP% 

%TMP% 

%TMP% 

As an illustration of the significance of the structure highlighted by these
low-order eigenvectors, note that only $0.73\%$ of the spectrum is captured
by the $41^{st}$ eigenvector and that over $99.9\%$ of the (L2) ``mass''
(and $92.3\%$ of the L1 mass) of $41^{st}$ eigenvector is on individuals who 
served in the $108^{th}$ Congress.
Similarly, only $0.42\%$ of the spectrum is captured by the $43^{rd}$
eigenvector and $98.5\%$ of the (L2) ``mass'' (and $71.7\%$ of the L1 mass)
of $43^{rd}$ eigenvector is on individuals who served in the $106^{th}$ 
Congress.
Similar results are seen for many (but certainly not all) of the low-order
eigenvectors.
That is, in many cases, although these low-order eigenvectors account for 
only a small fraction of the variance in the data, they are often strongly 
localized on a single Congress (or, as Figure~\ref{fig:cong-vect} 
illustrates, a small number of temporally-adjacent Congresses), \emph{i.e.}, 
at a single time step of the time series of voting data.
In part because of this, these low-order eigenvectors can in some cases be 
used to perform common machine learning and data analysis tasks.

\begin{figure}[t]
\begin{center}
\subfigure[ 
Illustration of \textsc{Congress} in the form of a ``spy'' plot.
]{ 
\includegraphics[width=0.31\columnwidth]{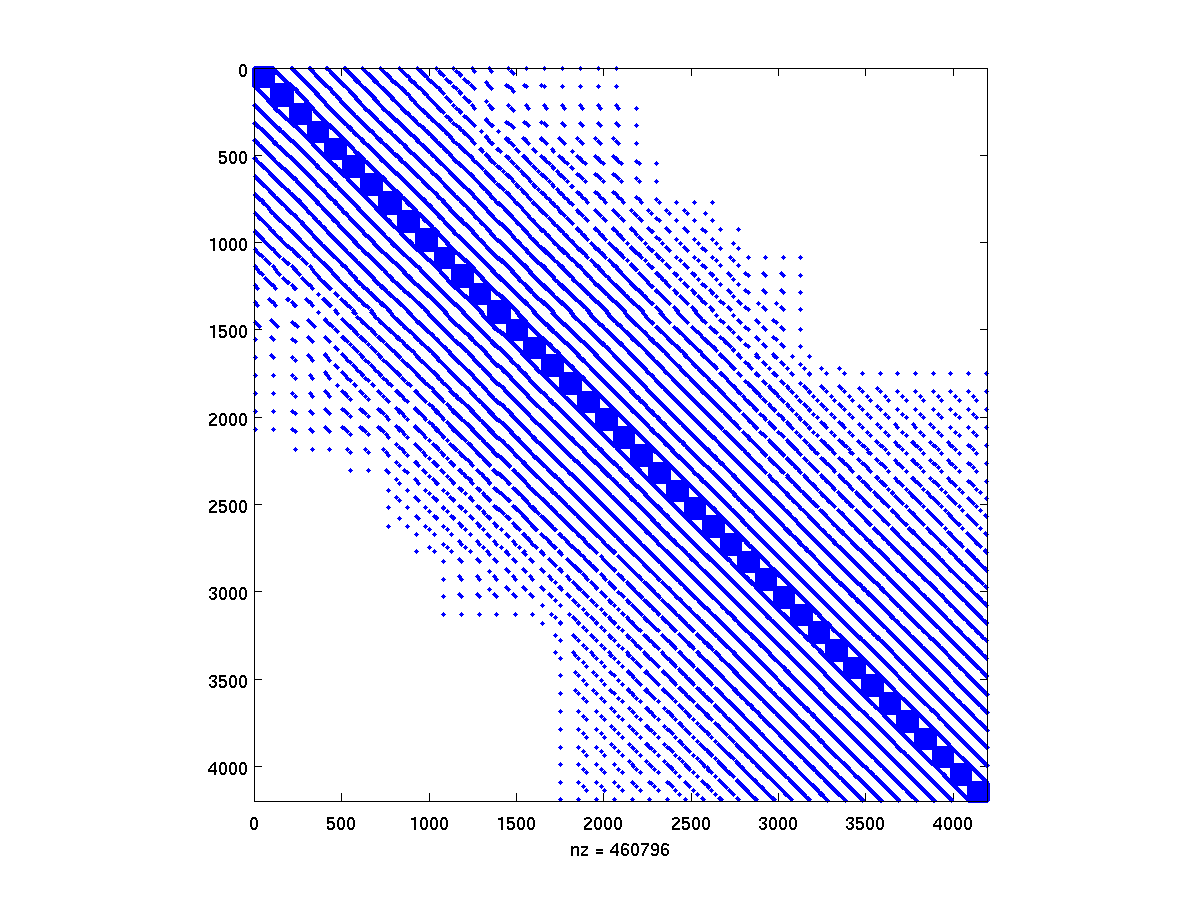}
}
%\;\;\;
\subfigure[
Normalized square spectrum.
]{
\includegraphics[width=0.31\columnwidth]{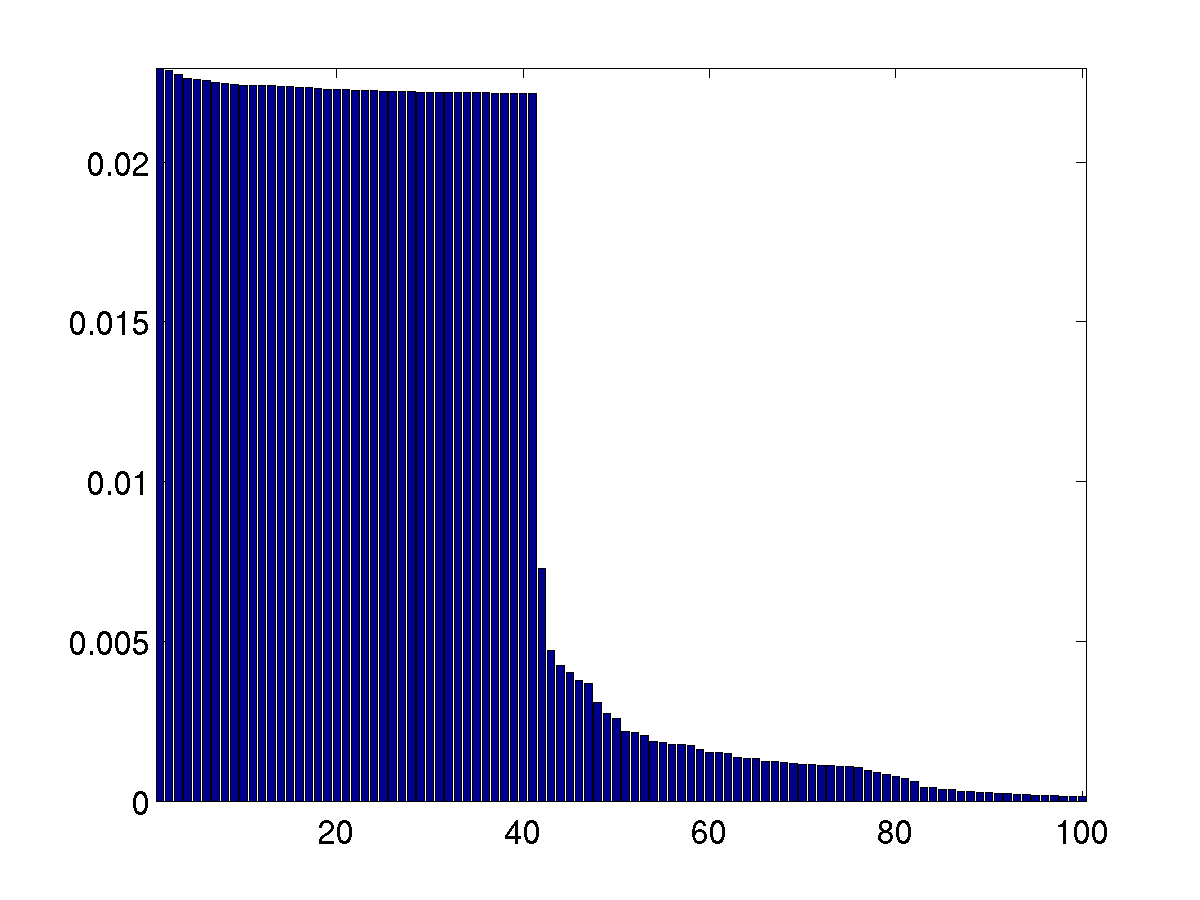}
}
\subfigure[
Partitioning based on the $41^{st}$ eigenvector.
]{
\includegraphics[width=0.31\columnwidth]{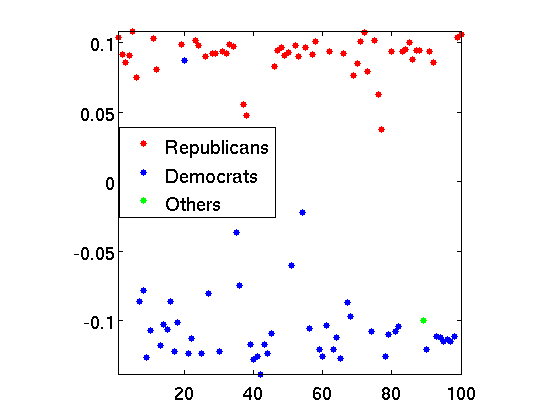}
}
\end{center}
\caption{
First panel:
A ``spy'' plot of the \textsc{Congress} data.  The blocks on the diagonal 
correspond to the voting patterns in each of the $41$ Congresses, and the 
off-diagonal entries take the value $\epsilon=0.1$ when a single individual 
served in two successive Congresses.
Second panel:
Barplot of the normalized square spectrum of the Congress matrix, 
\emph{i.e.}, $\frac{\lambda_i^2}{\sum_{j=1}^{n} \lambda_j^2}$, for 
$i=1,\ldots,100$, indicating that the low-order eigenvalues account for a
relatively-small fraction of the variance in the data. 
Third panel:
Plot of spectral clustering based on the first nontrivial eigenvector 
$v^{(1)}_{G_{2006}}$ of the matrix $G_{2006}$, where $G_{2006}$ denotes the 
full \textsc{Congress} restricted to the senators from the $110^{th}$ 
Congress (which includes the years $2006$ and $2007$).  If we let 
$v^{(41)}_{2006}$ denote the restriction of the (localized) $41^{st}$ 
eigenvector of full \textsc{Congress} data to the the senators in the 
$110^{th}$ Congress, then 
$ | v^{(1)}_{G_{2006}} -  v^{(41)}_{2006} | \leq 3 \times 10 ^{-3}$, and an
identical partition and plot (at the level of resultuion of this figure) is 
generated by partitioning by performing a sweep cut along $v^{(41)}_{2006}$.
}
\label{fig:cong-details}
\end{figure}

Consider, for example, spectral clustering, which involves partitioning the
data by performing a ``sweep cut'' over an eigenvector computed from the 
data.
The first eigenvector shown in Figure~\ref{fig:cong-vect} clearly 
illustrates that a sweep cut over the first nontrivial eigenvector of the 
Laplacian of the full data set will partition the network based on time, 
\emph{i.e.}, into a temporally-earlier cluster and a temporally-later cluster.
Not surprisingly, low-order eigenvectors can highlight very different 
structures in the data.
For example, by performing a sweep cut over the first nontrivial eigenvector 
of the Laplacian of the subnetwork induced by the nodes in the $110^{th}$ 
Congress, one obtains the same partition (basically, a partition along party 
lines~\cite{PR97,WPFMP09_TR,multiplex_Mucha}) as when the sweep cut is 
performed on the $41^{st}$ eigenvector of the Laplacian of the full data set.
This is illustrated in Figure~\ref{fig:cong-details}.
Clearly, there is a strong correlation, as that low-order eigenvector is 
effectively finding the partition of the $110^{th}$ Congress into two parties.
(Indeed, the color-coding in Figure~\ref{fig:cong-vect} corresponds to party 
affiliation.)
As a consequence, other clustering and classification tasks lead to 
similar or identical results, whether one considers the second eigenvector 
of the Laplacian of the subnetwork induced by the nodes in $110^{th}$ 
Congress or the $41^{st}$ eigenvector of the Laplacian of the full data set.
Similar results hold for many of the other low-order eigenvectors, especially 
when the localization is very pronounced.

%TMP% 

\subsection{The \textsc{Migration} data}

\begin{figure}[t]
\begin{center}
\includegraphics[width=0.24\columnwidth]{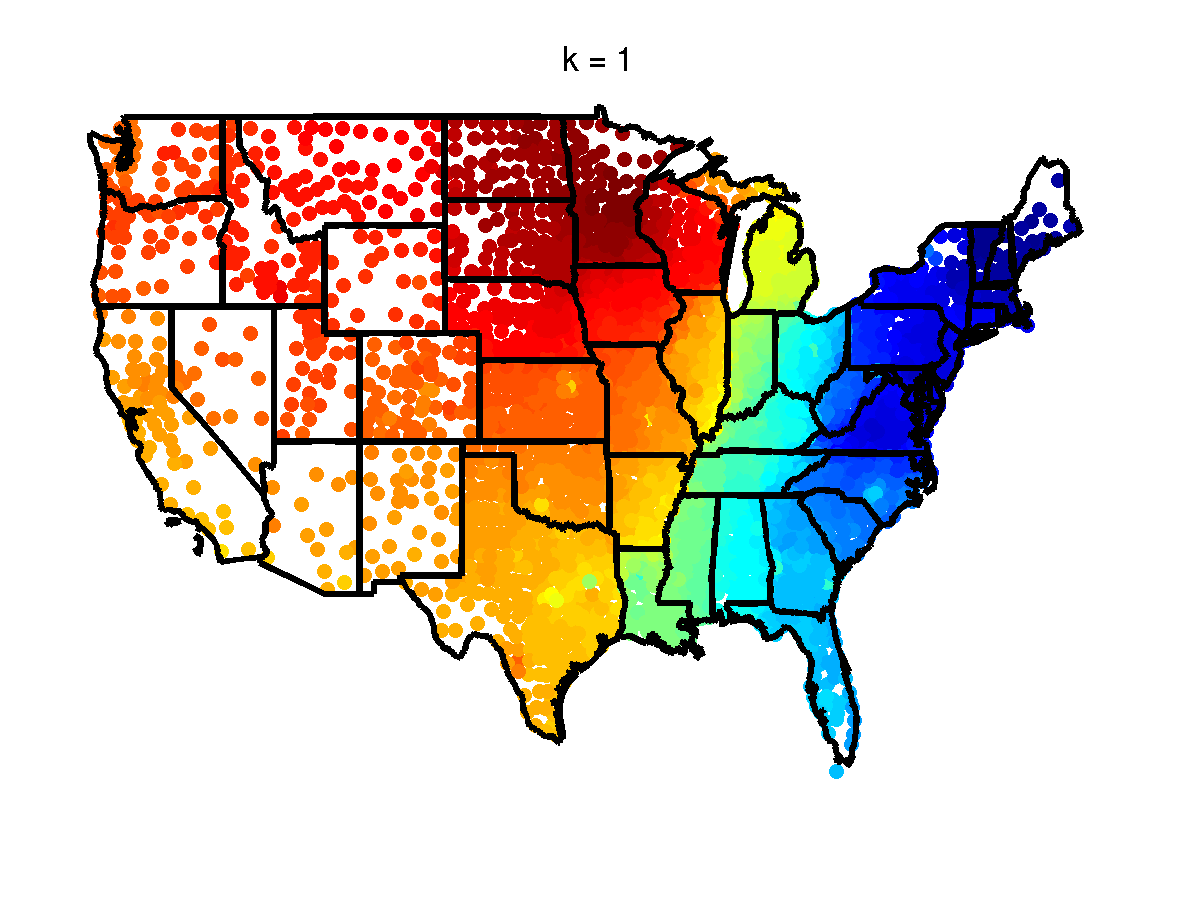}
\includegraphics[width=0.24\columnwidth]{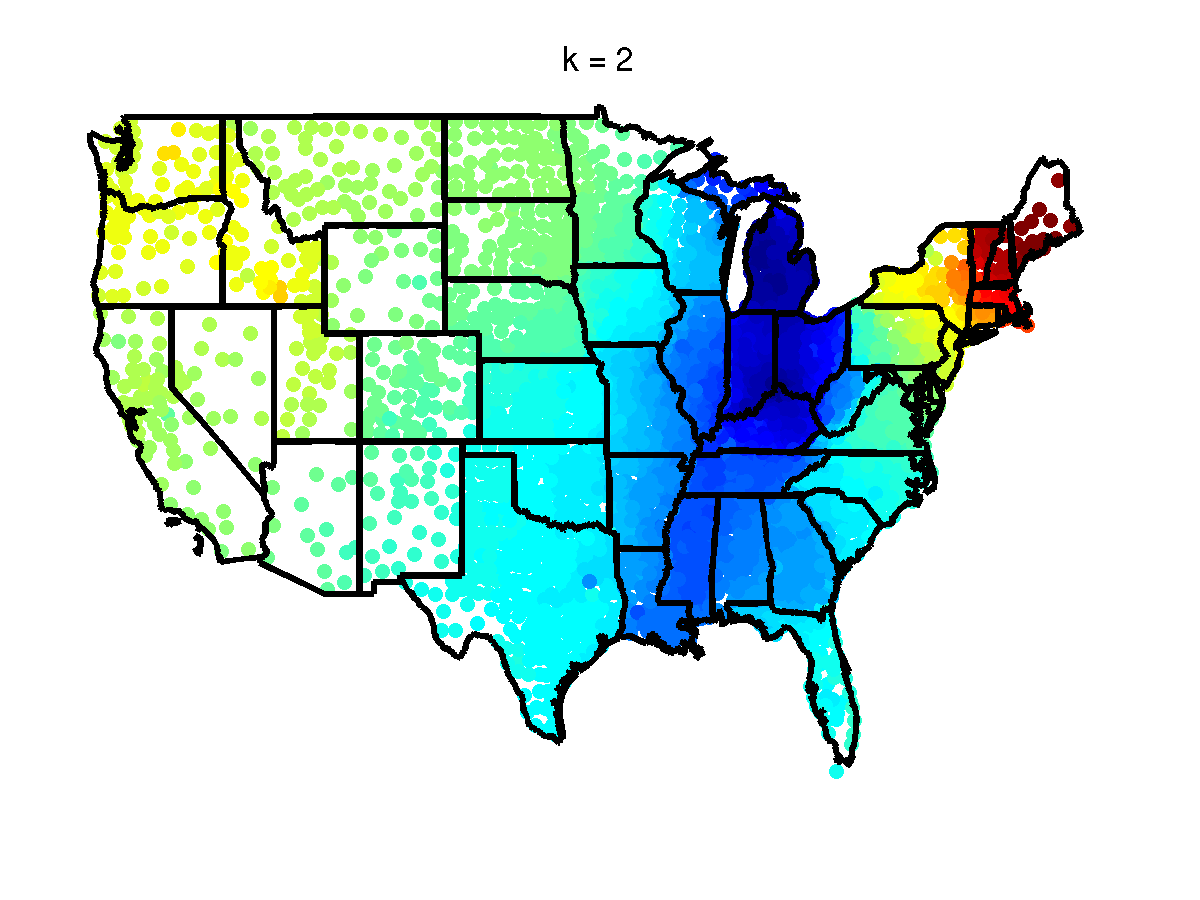}
\includegraphics[width=0.24\columnwidth]{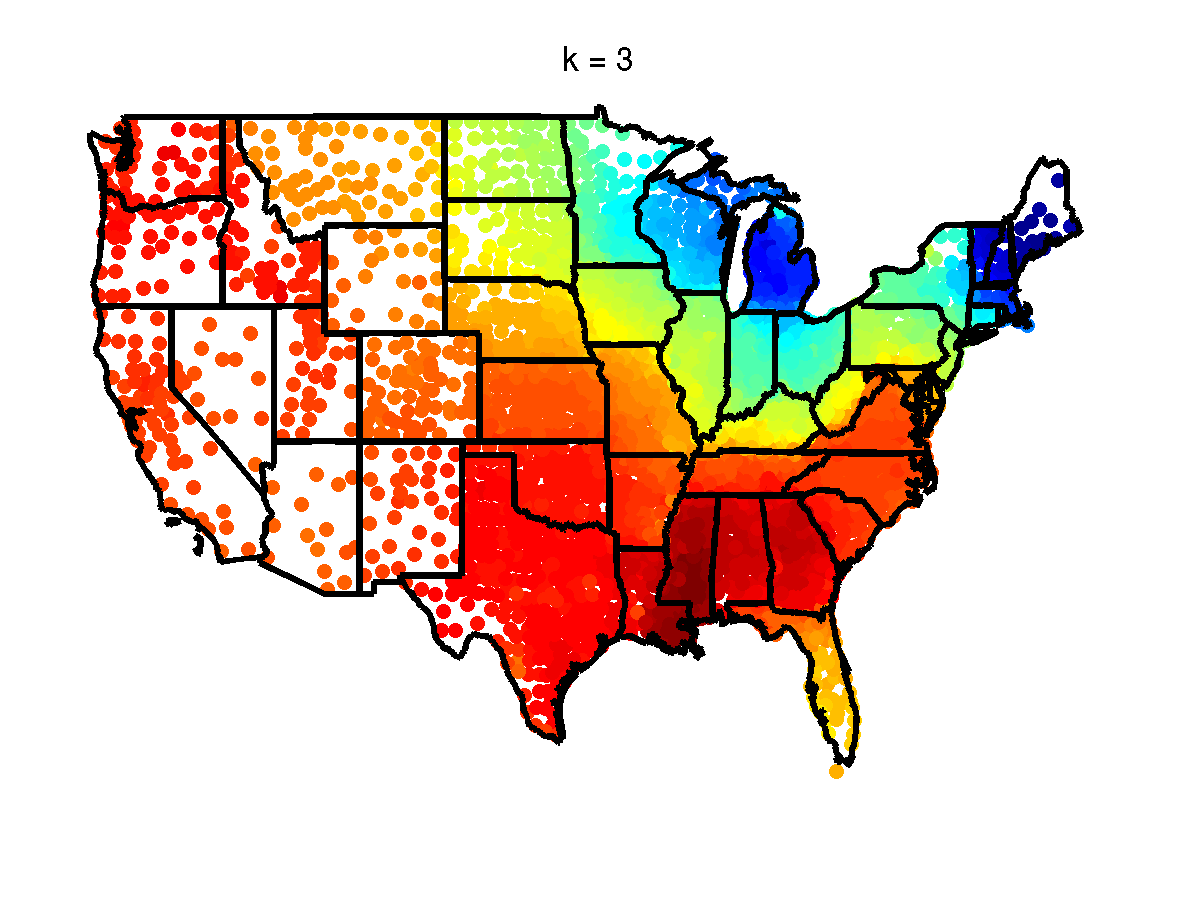}
\includegraphics[width=0.24\columnwidth]{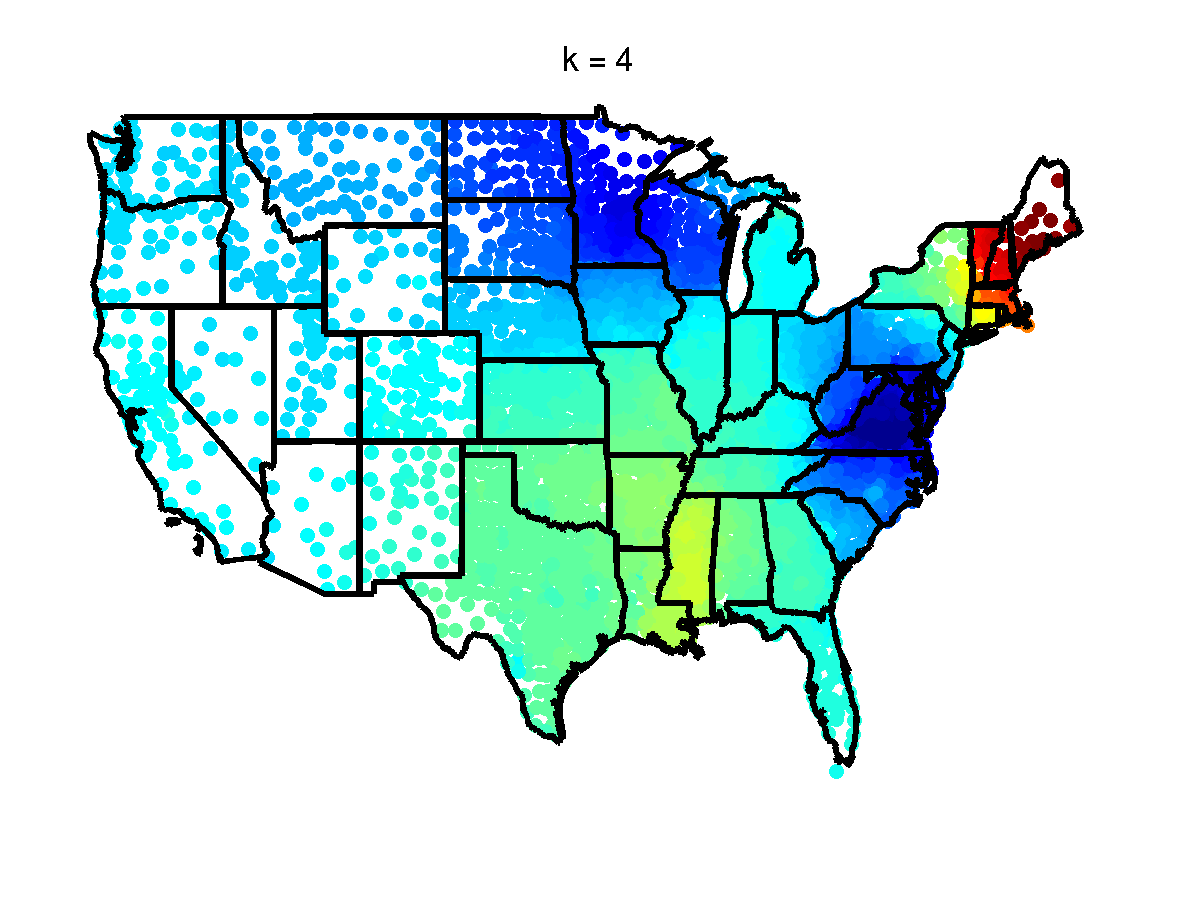}
\includegraphics[width=0.24\columnwidth]{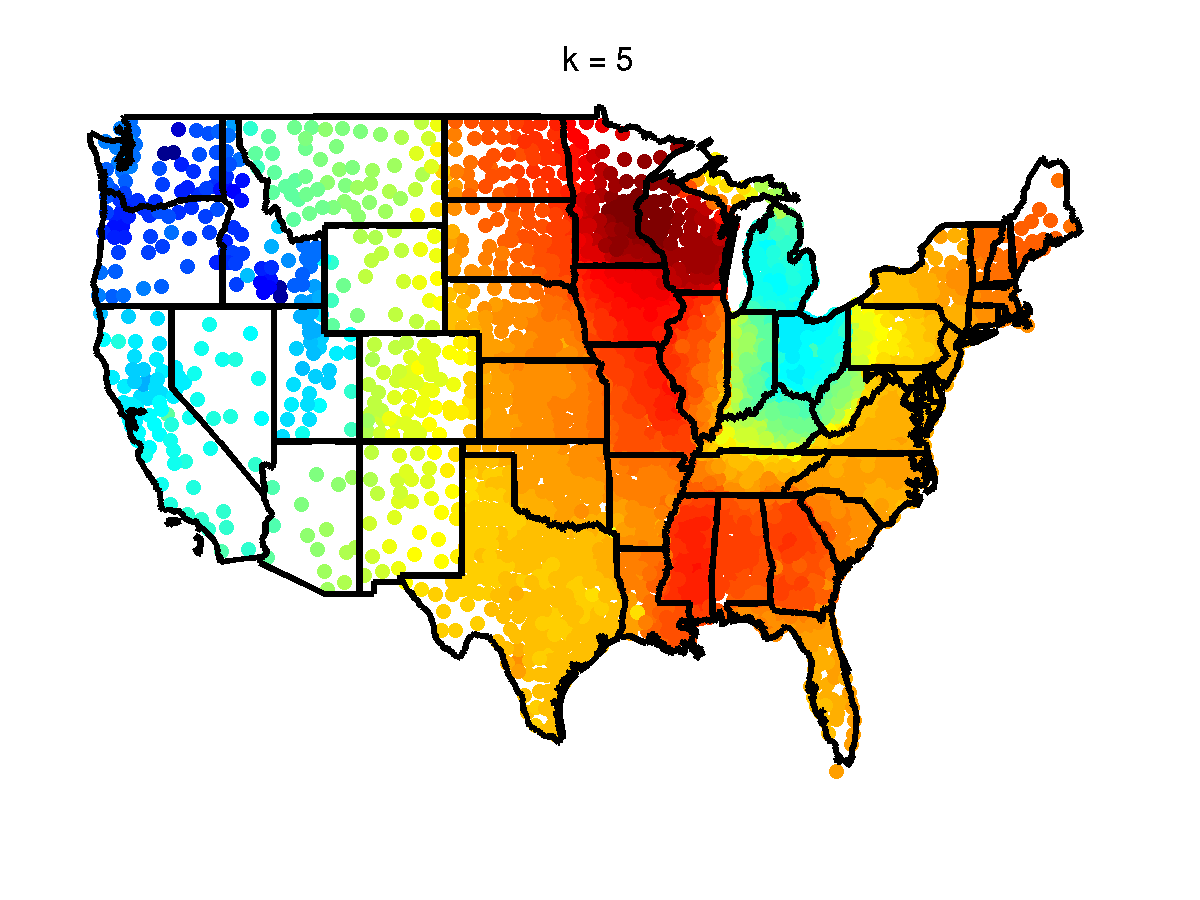}
\includegraphics[width=0.24\columnwidth]{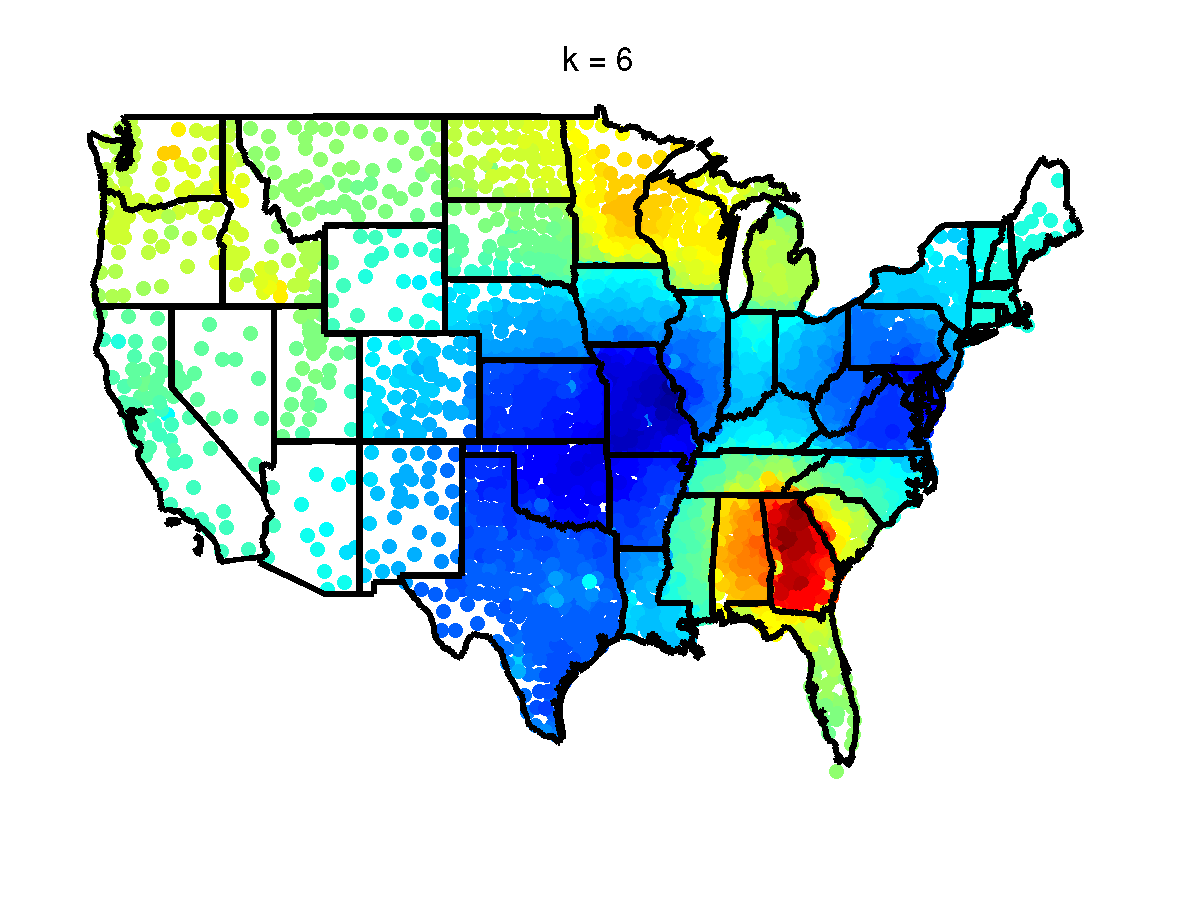}
\includegraphics[width=0.24\columnwidth]{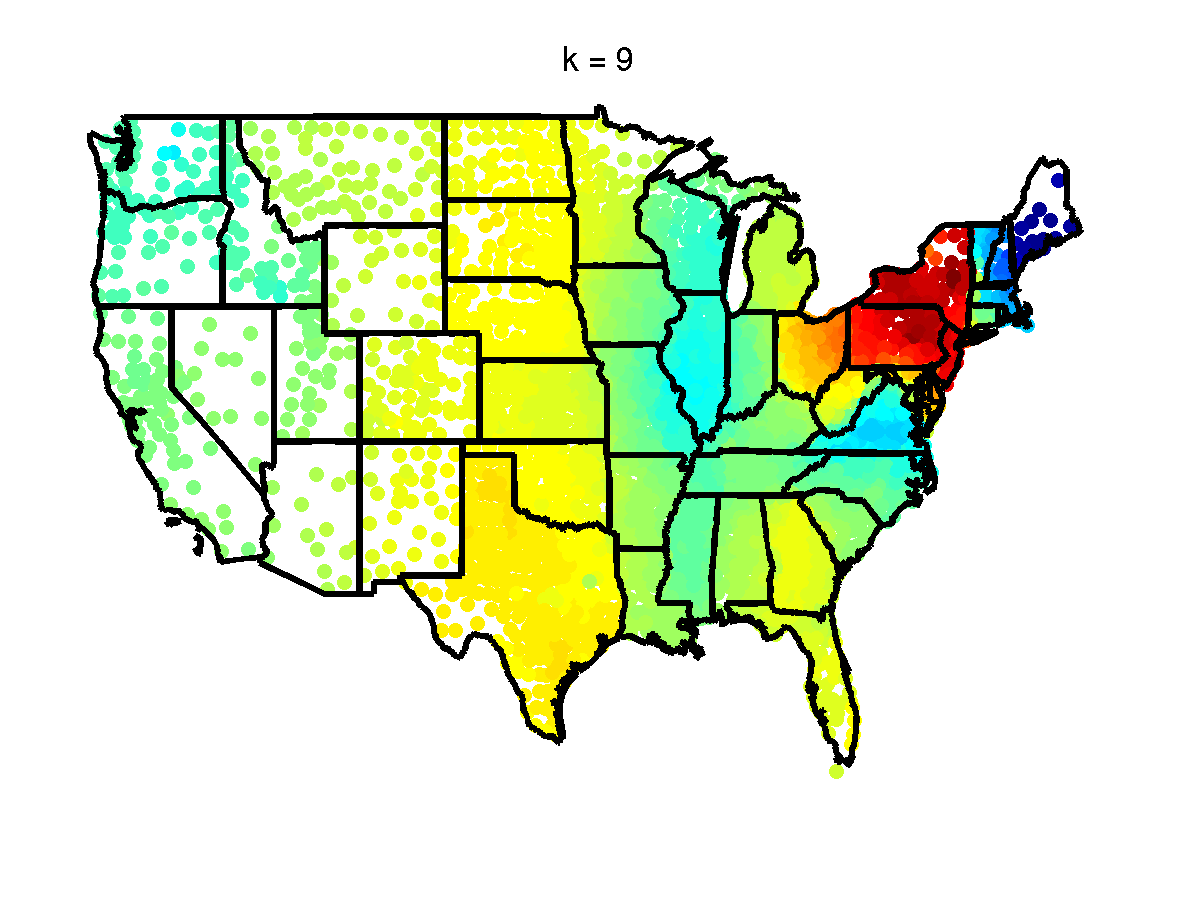}
\includegraphics[width=0.24\columnwidth]{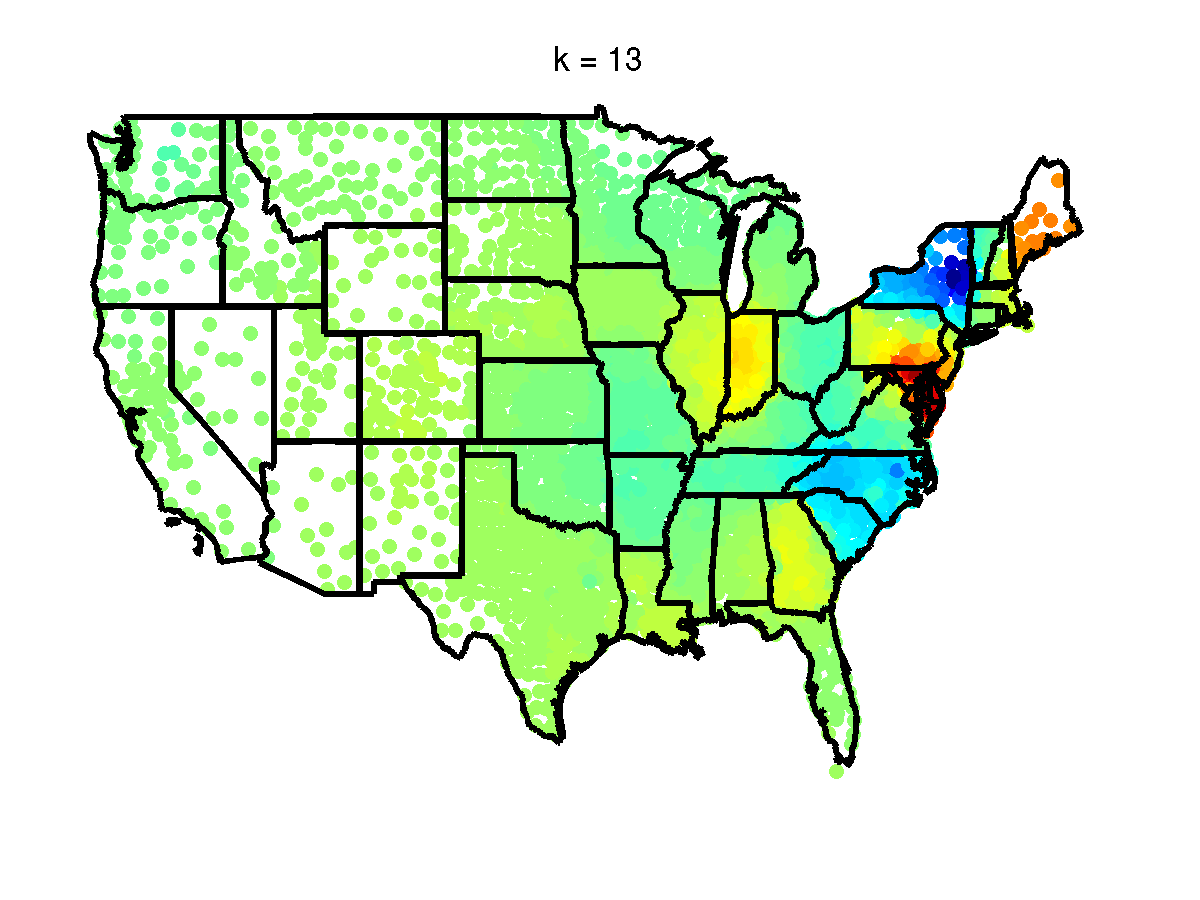}
\includegraphics[width=0.24\columnwidth]{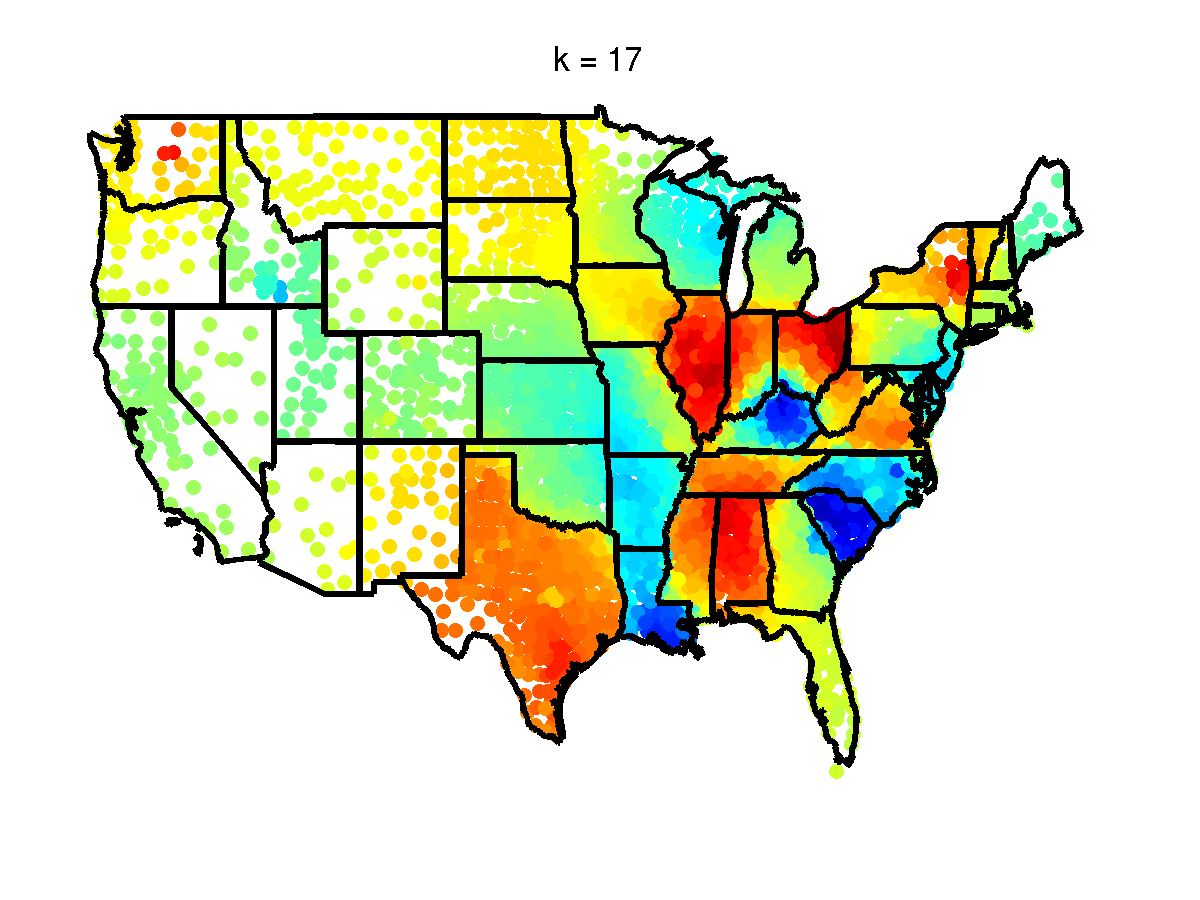}
\includegraphics[width=0.24\columnwidth]{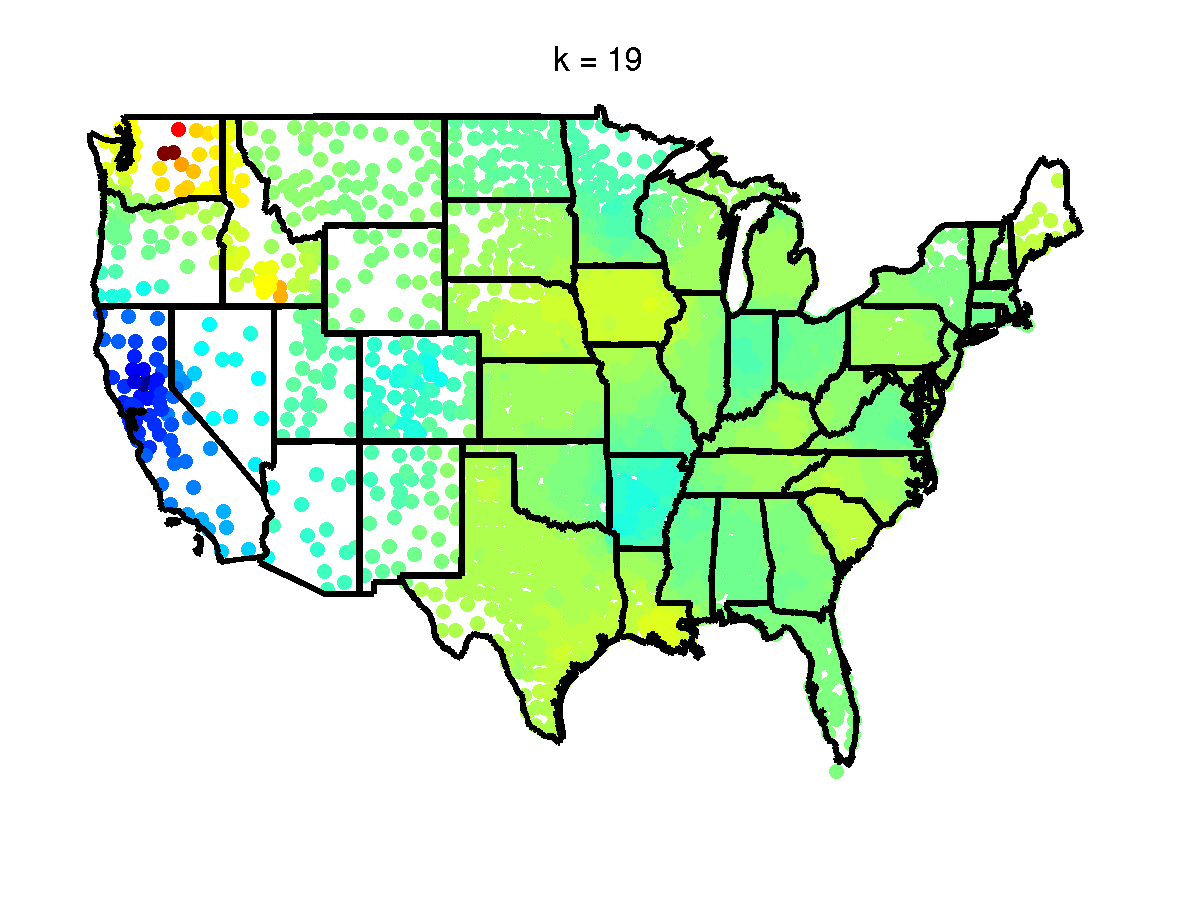}
\includegraphics[width=0.24\columnwidth]{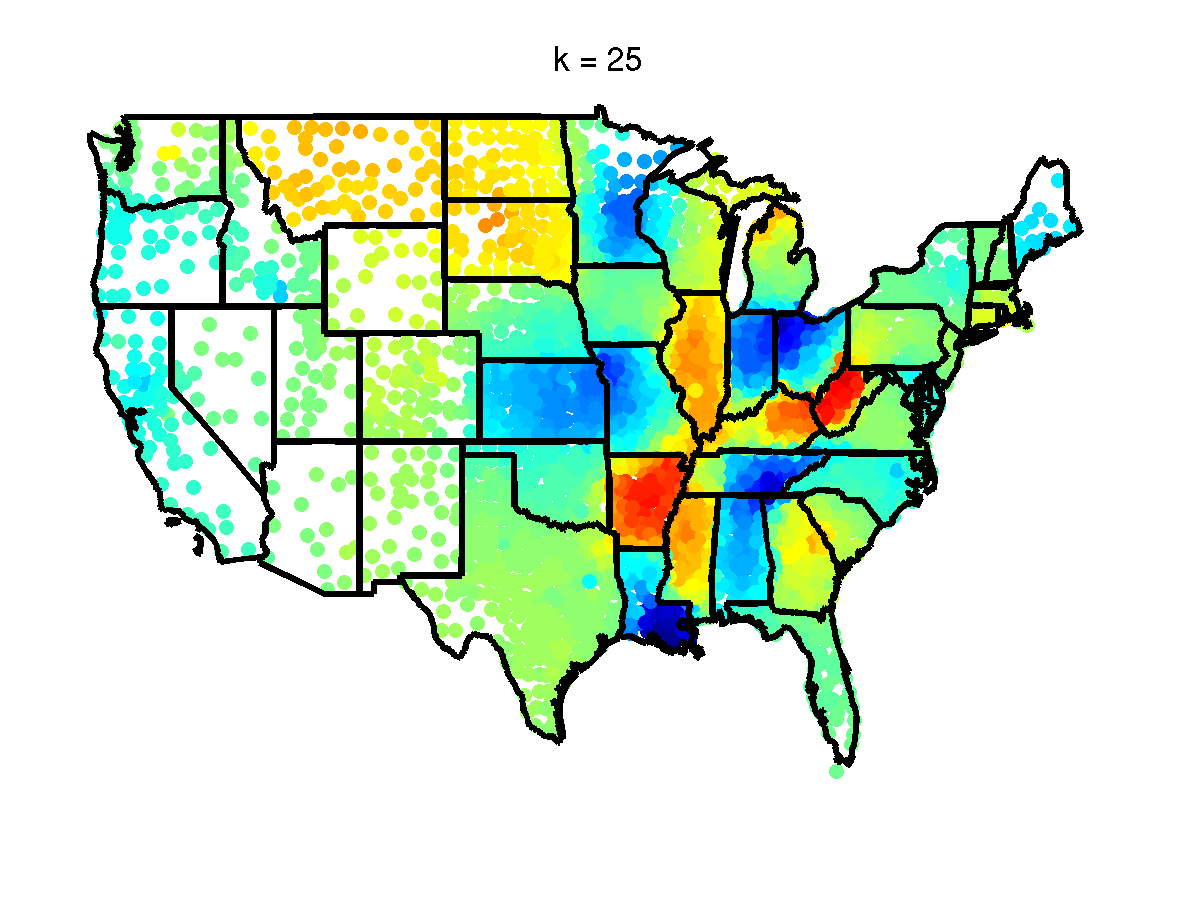}
\includegraphics[width=0.24\columnwidth]{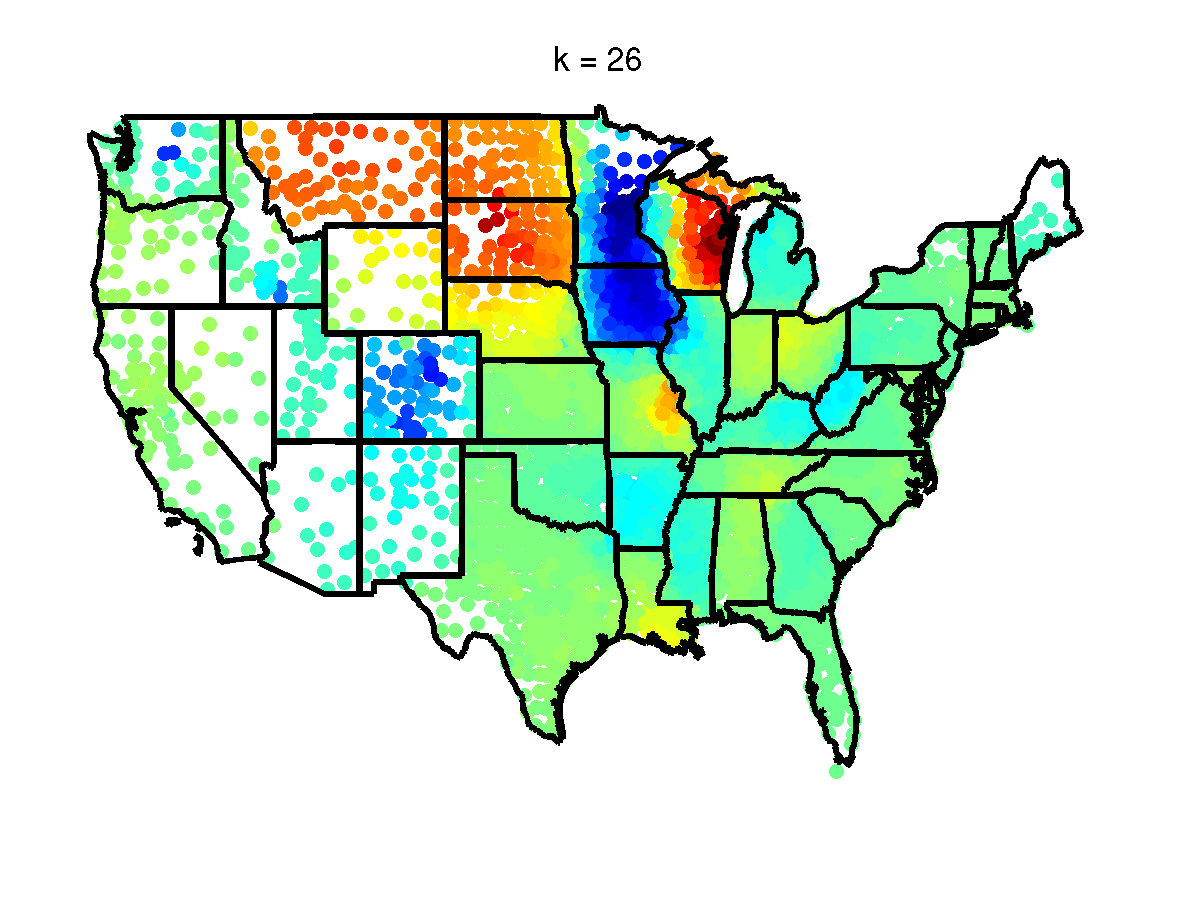}
\includegraphics[width=0.24\columnwidth]{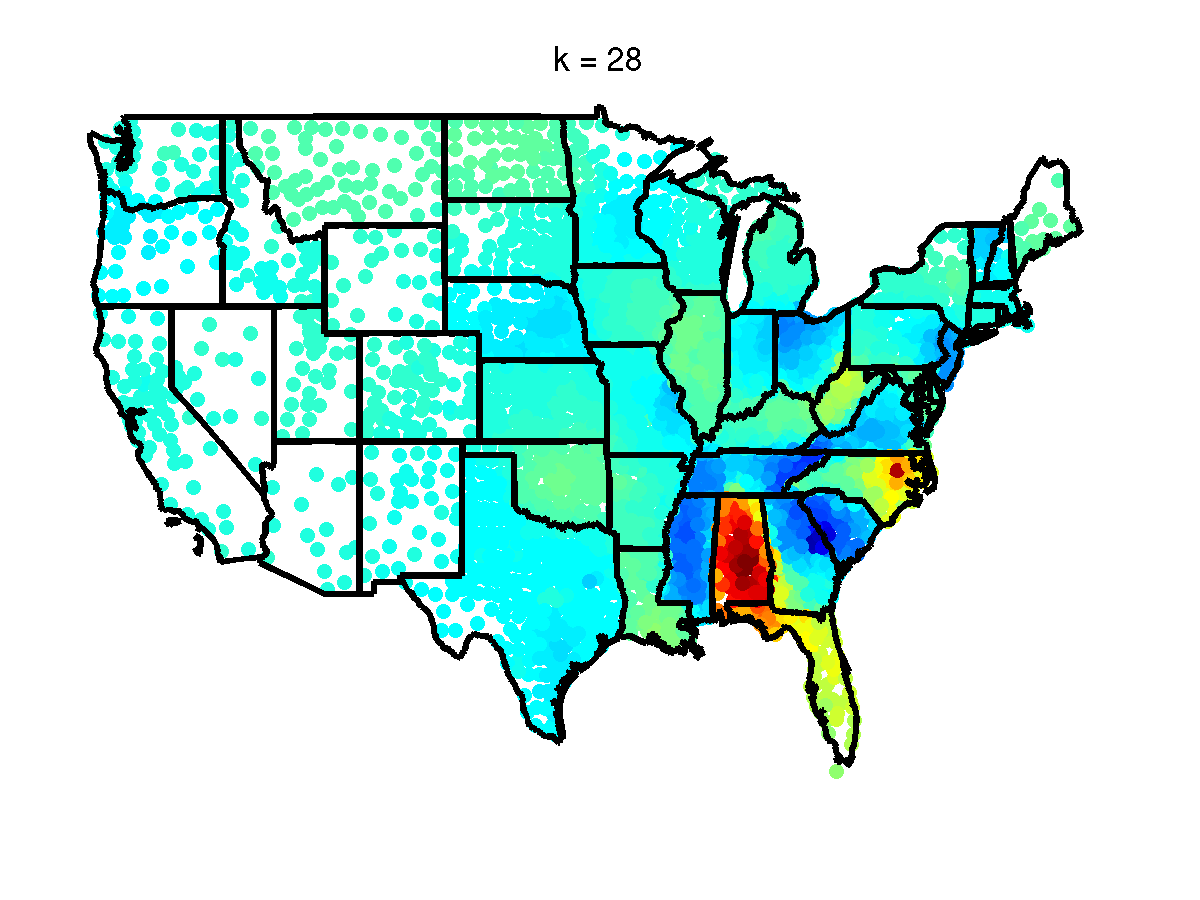}
\includegraphics[width=0.24\columnwidth]{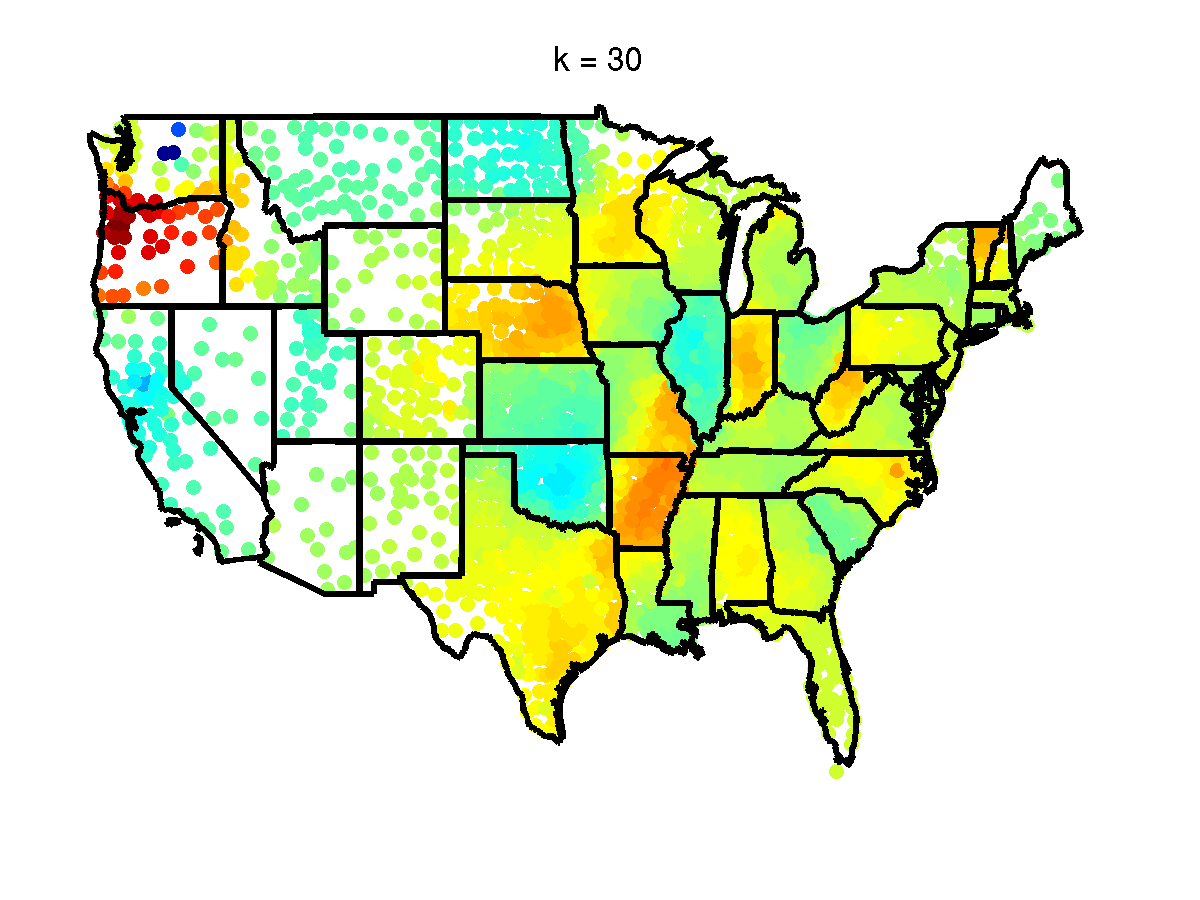}
\includegraphics[width=0.24\columnwidth]{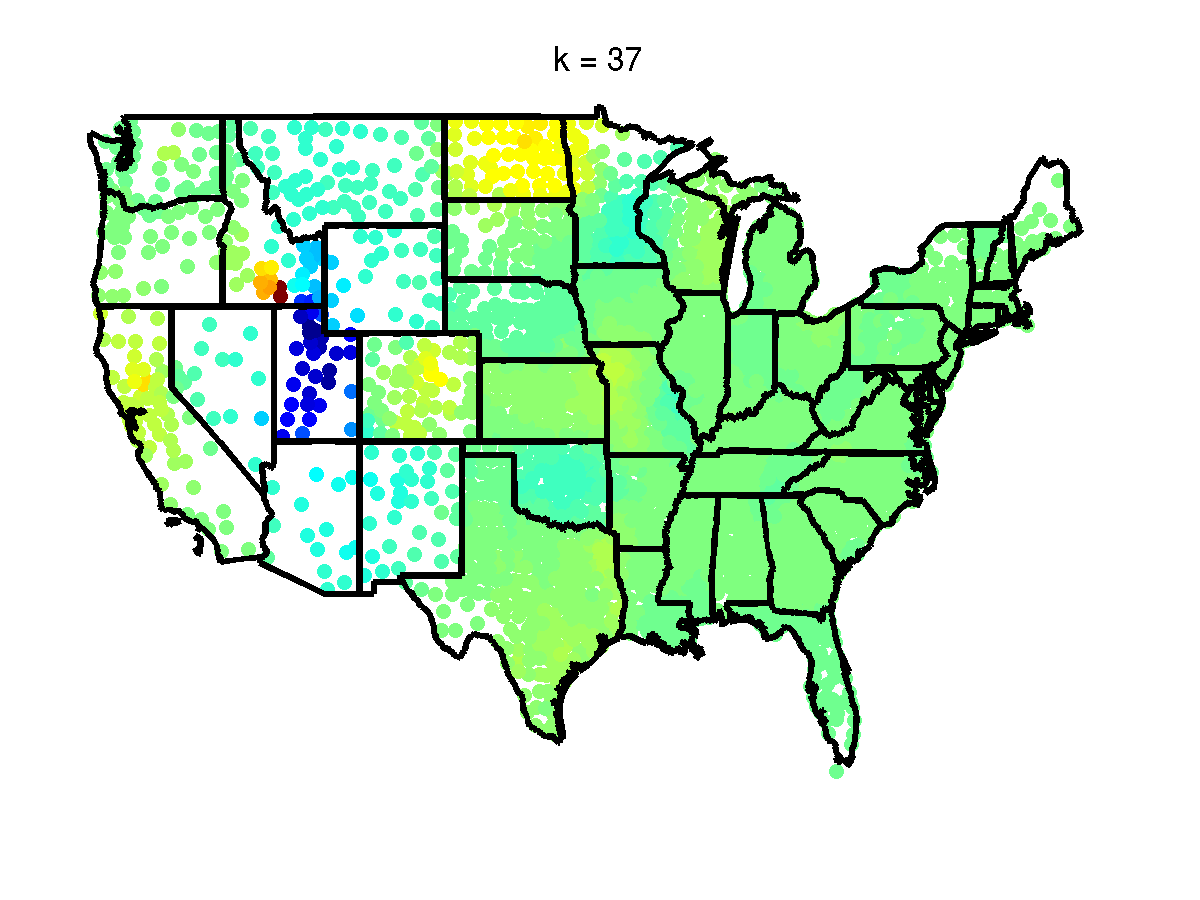}
\includegraphics[width=0.24\columnwidth]{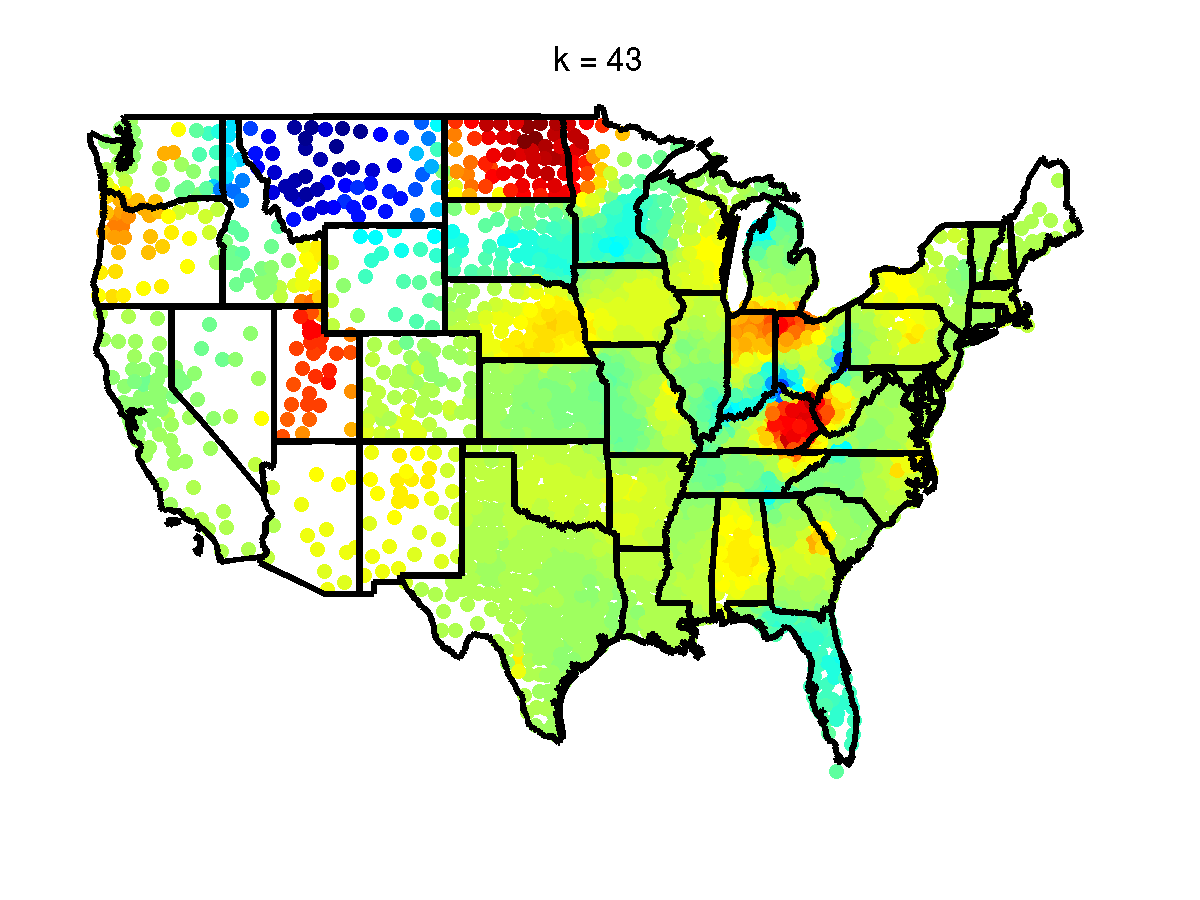}
\end{center}
\caption{
The \textsc{Migration} data:
pictorial illustration of several of the eigenfunctions.
Shown are the top eigenfunctions and several of the lower-order 
eigenfunctions that exhibit varying degrees of localization.
}
\label{fig:migr-vect}
\end{figure}

\begin{figure}[t]
\begin{center}
\includegraphics[width=0.24\columnwidth]{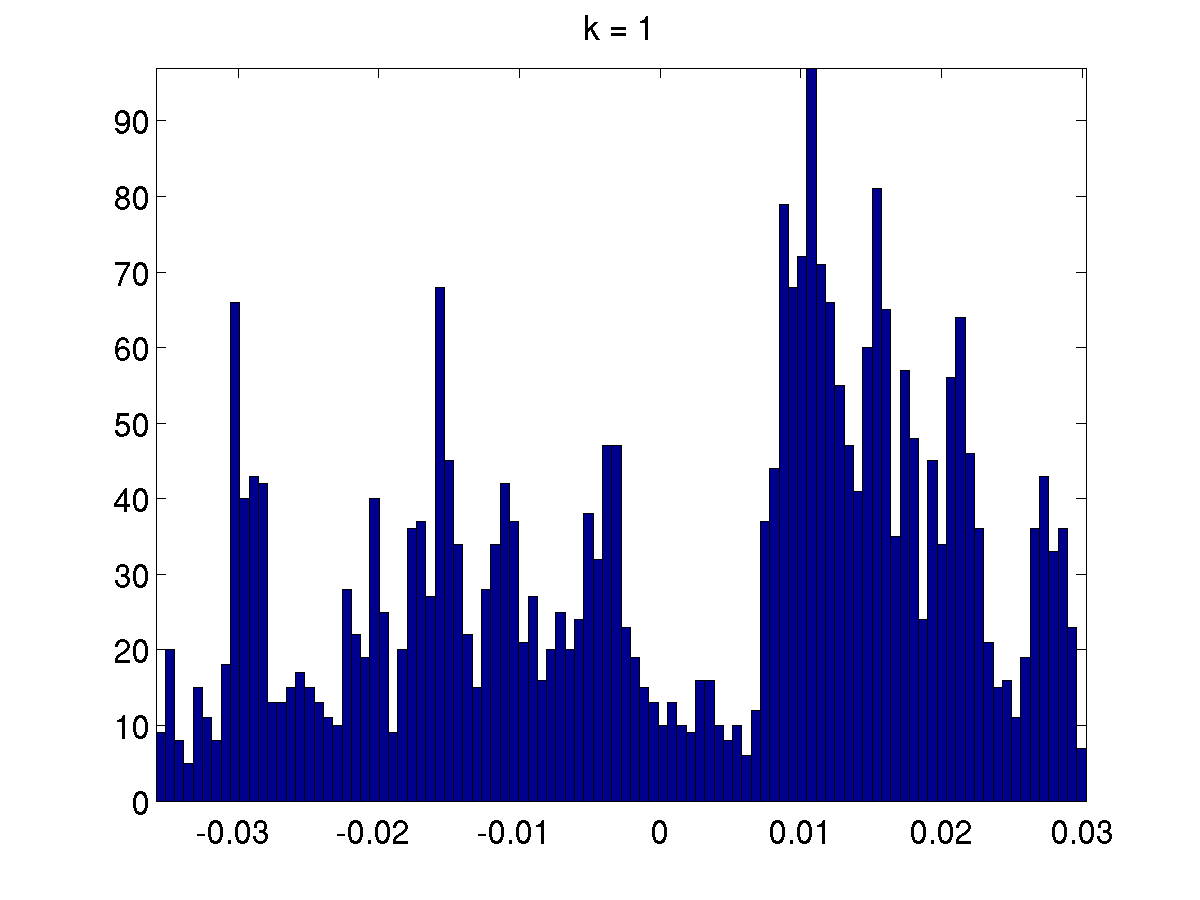}
\includegraphics[width=0.24\columnwidth]{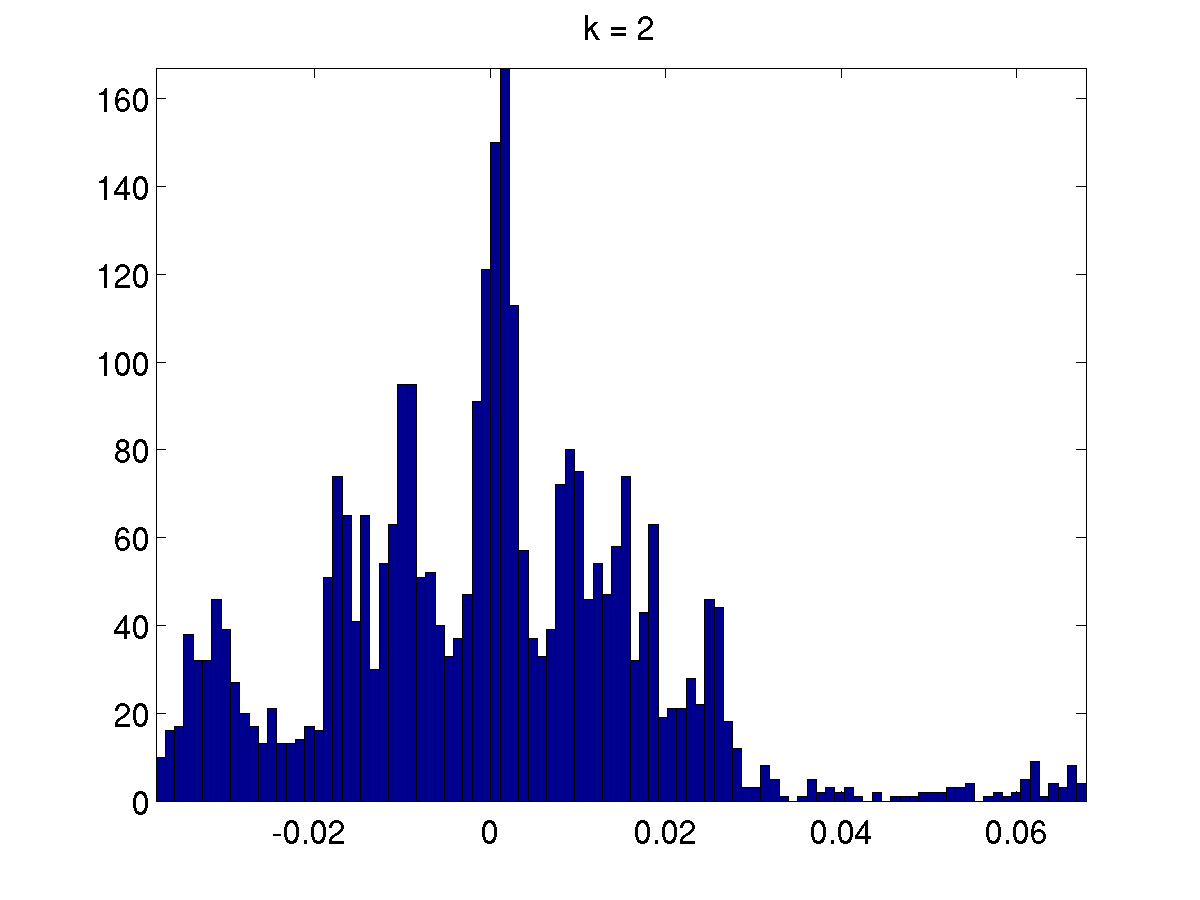}
\includegraphics[width=0.24\columnwidth]{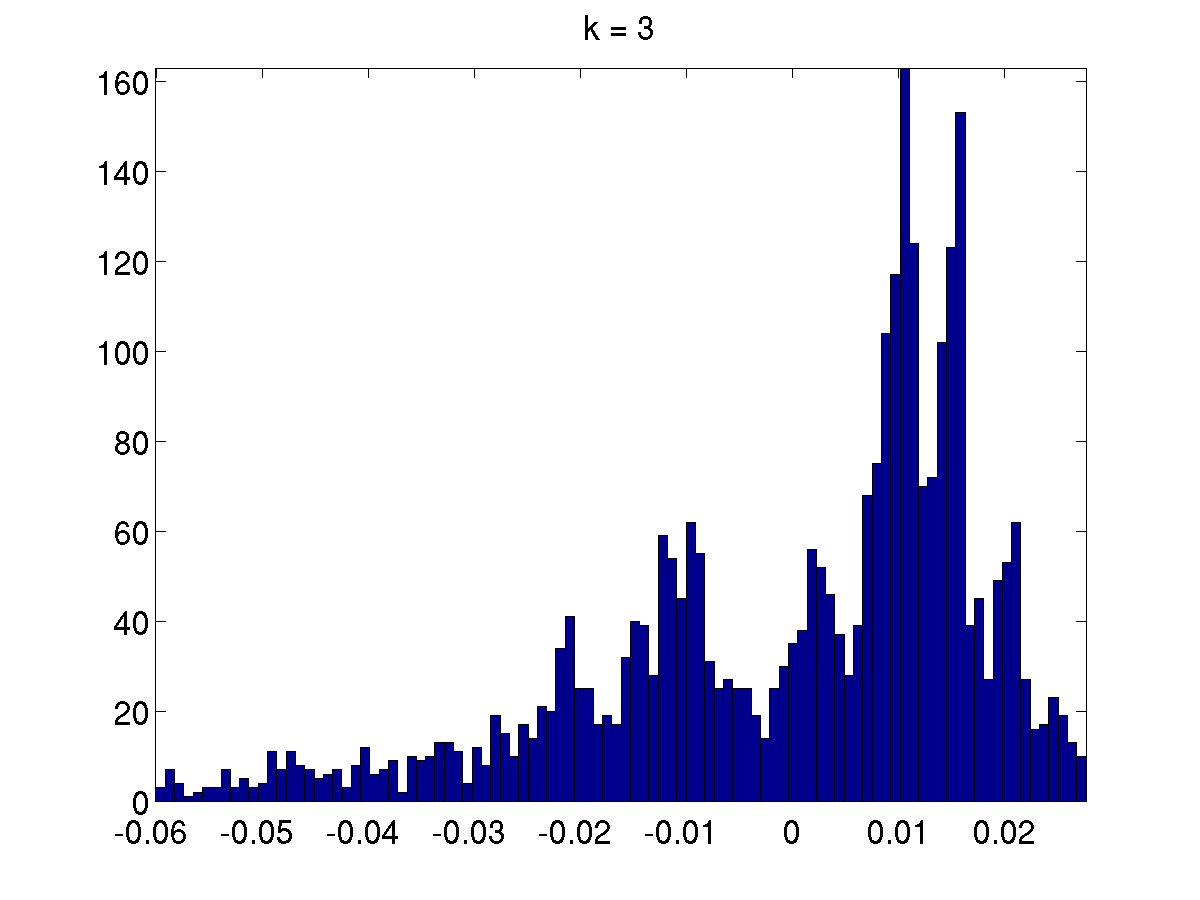}
\includegraphics[width=0.24\columnwidth]{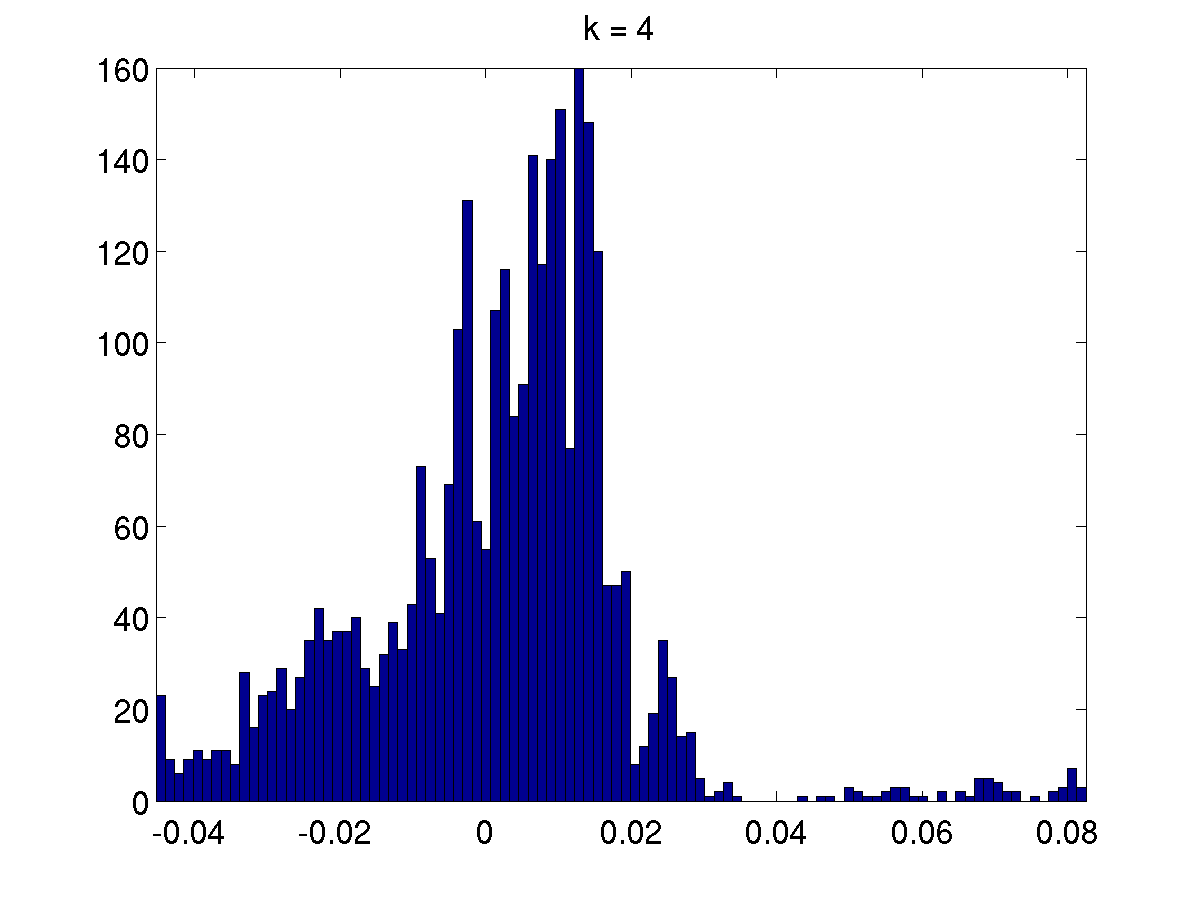}
\includegraphics[width=0.24\columnwidth]{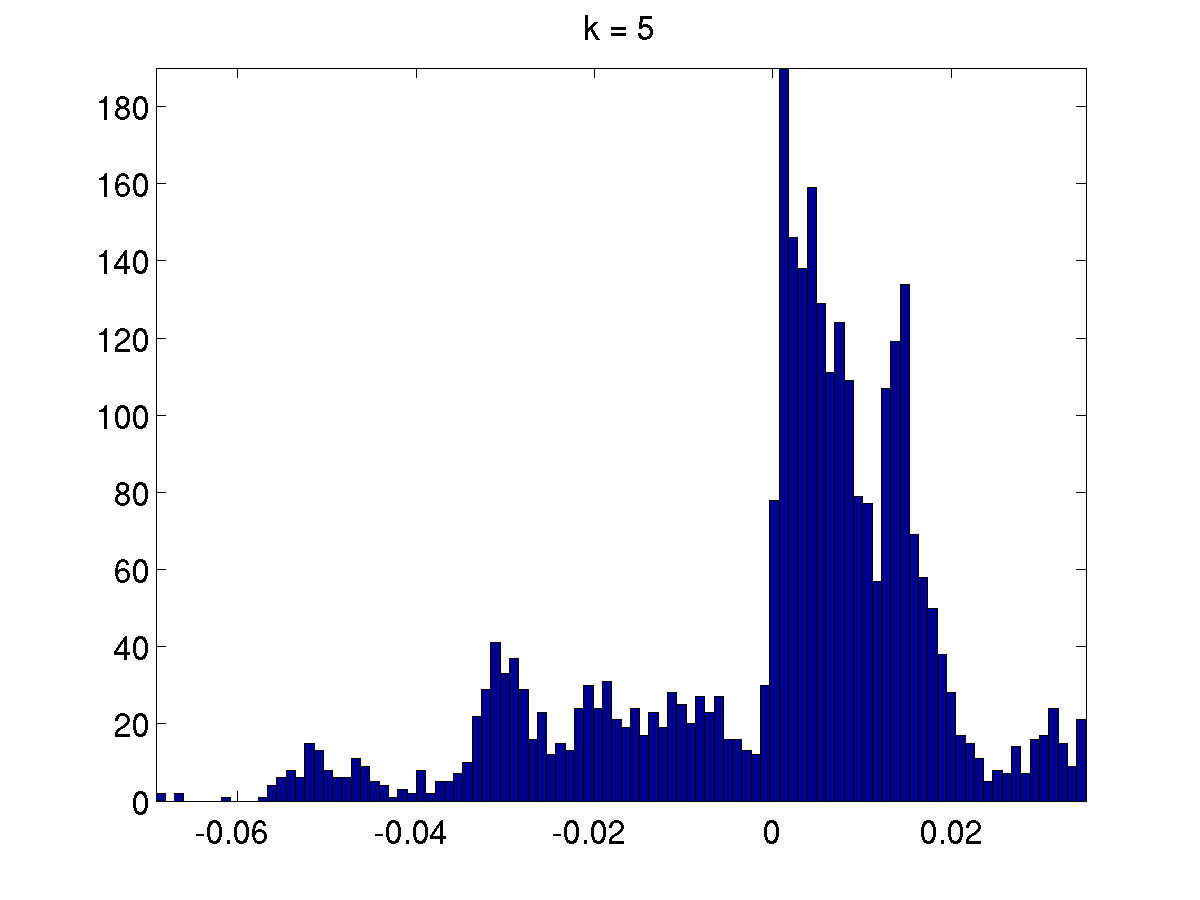}
\includegraphics[width=0.24\columnwidth]{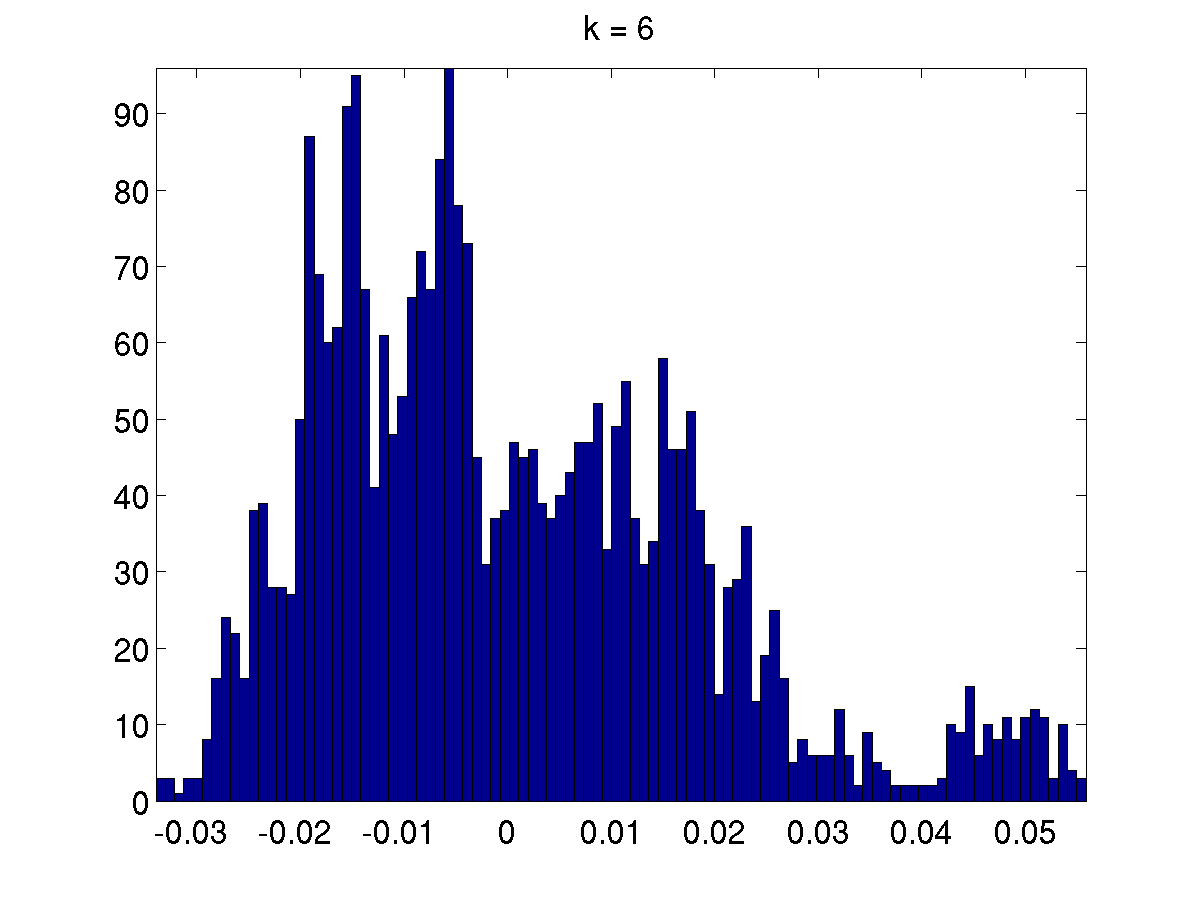}
\includegraphics[width=0.24\columnwidth]{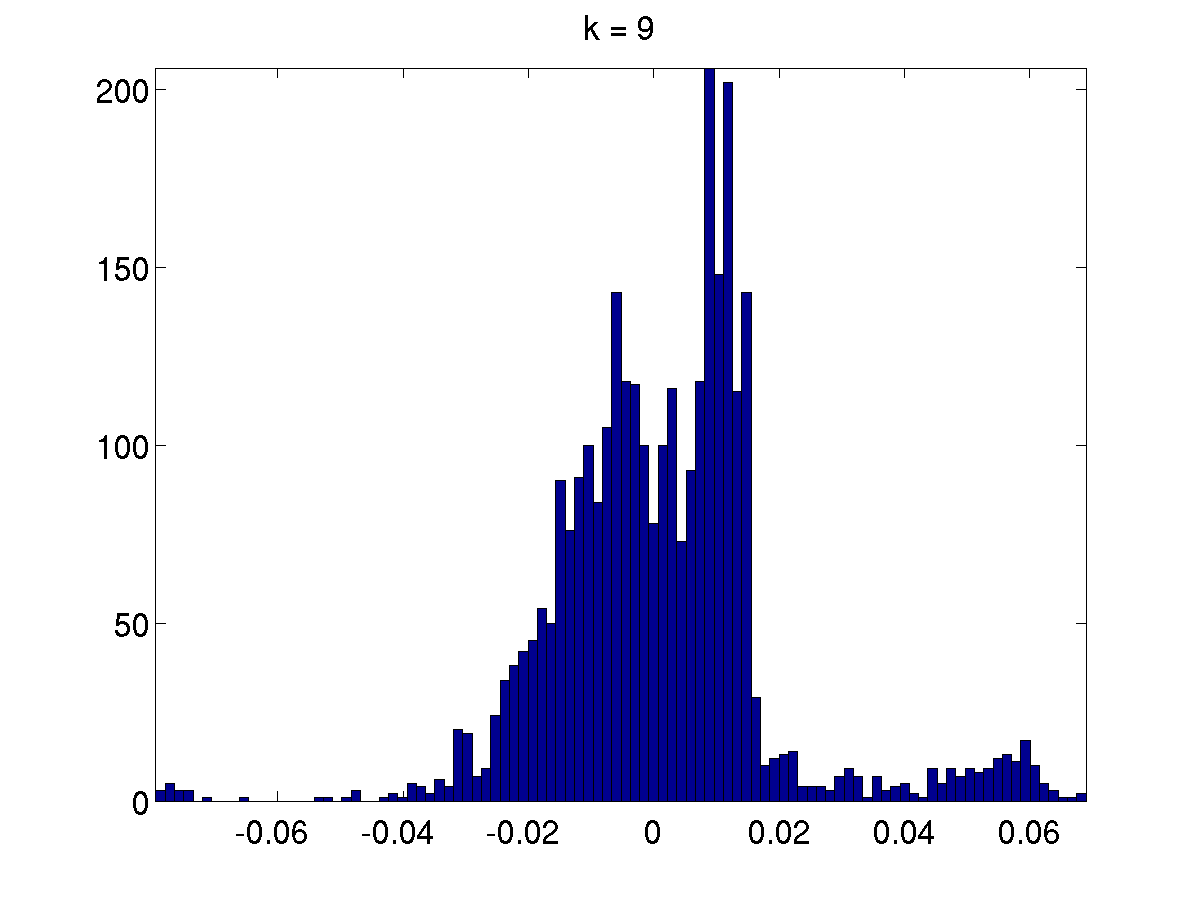}
\includegraphics[width=0.24\columnwidth]{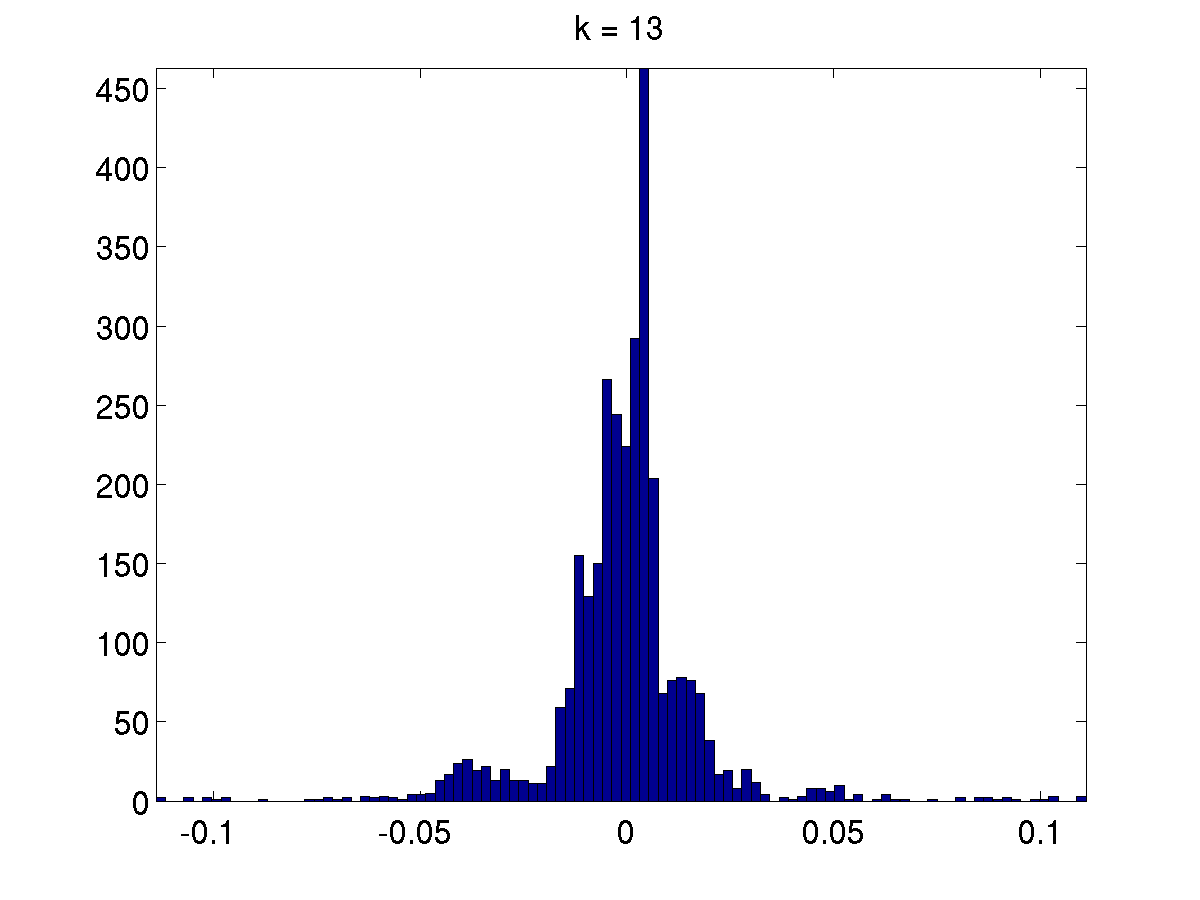}
\includegraphics[width=0.24\columnwidth]{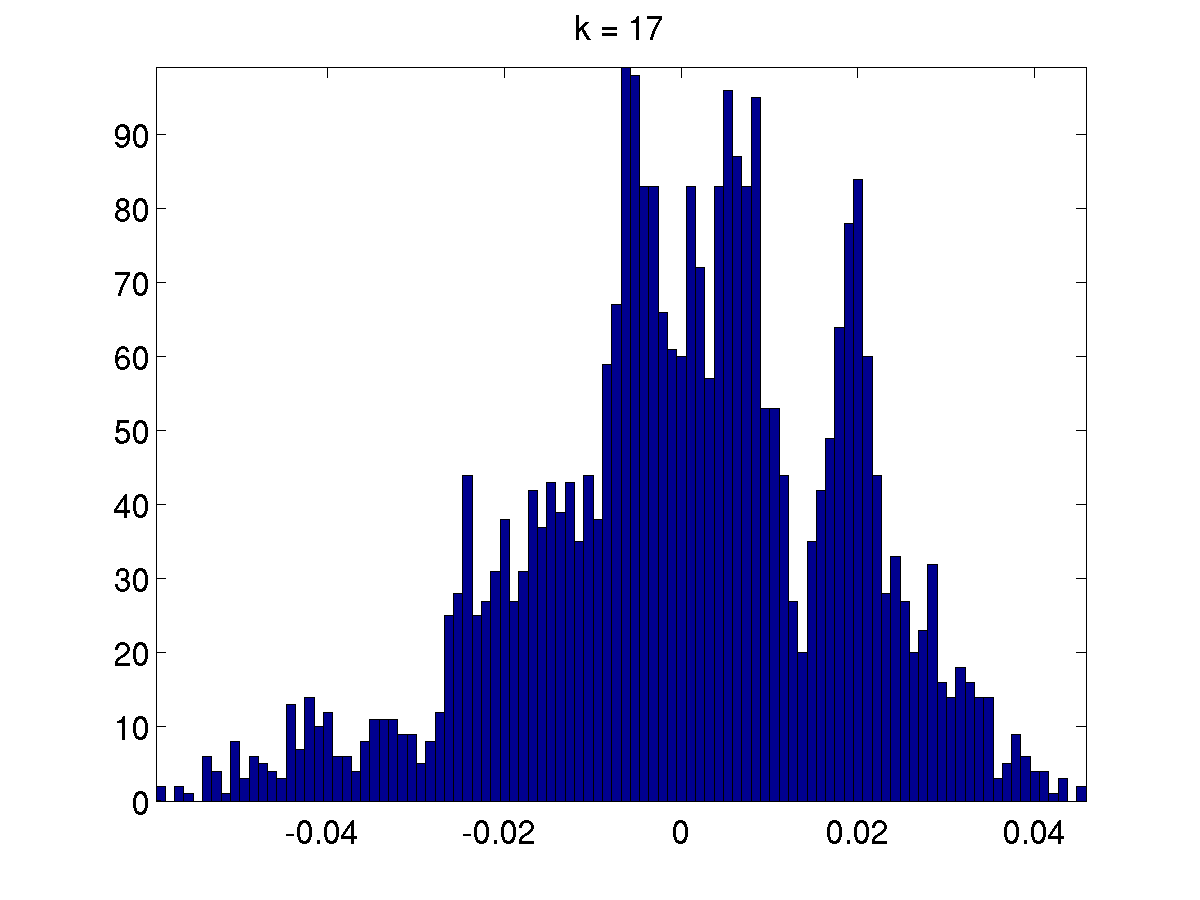}
\includegraphics[width=0.24\columnwidth]{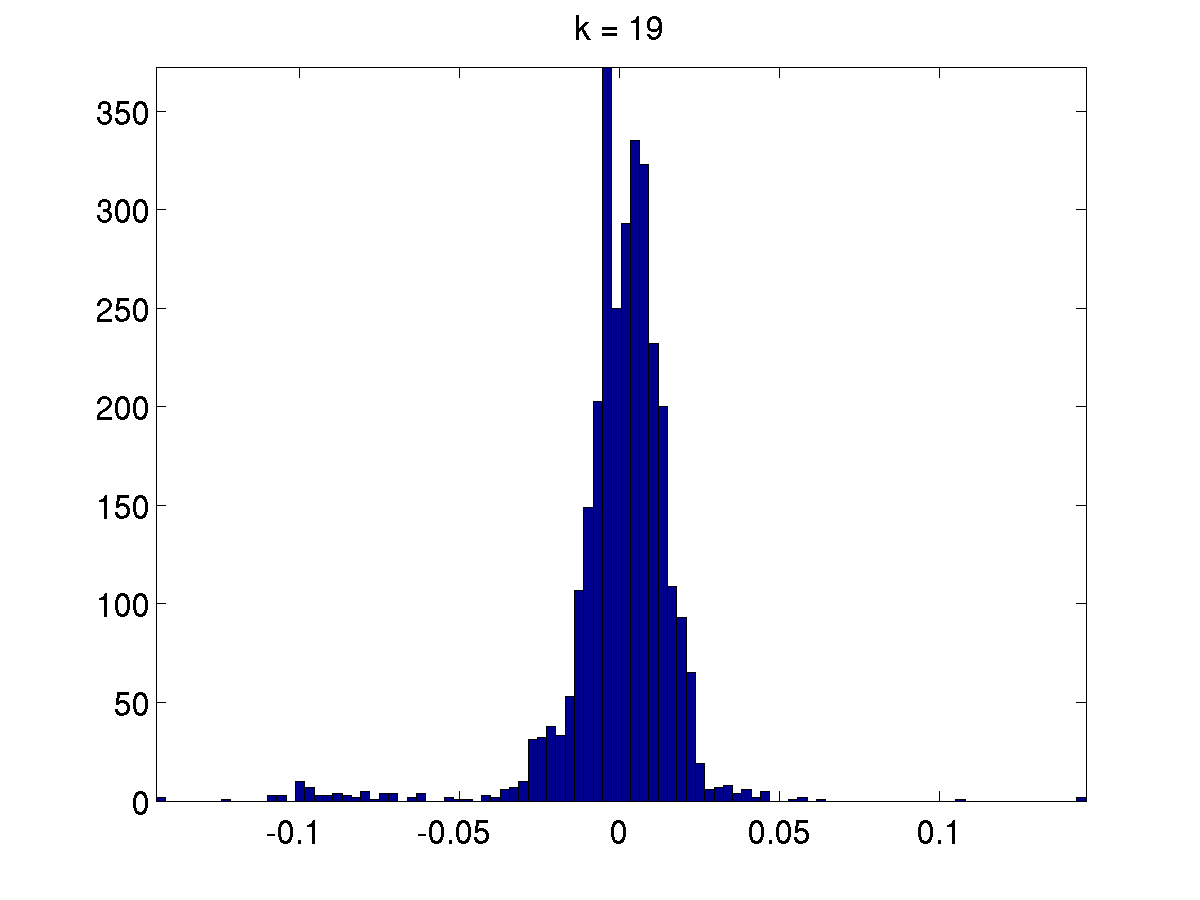}
\includegraphics[width=0.24\columnwidth]{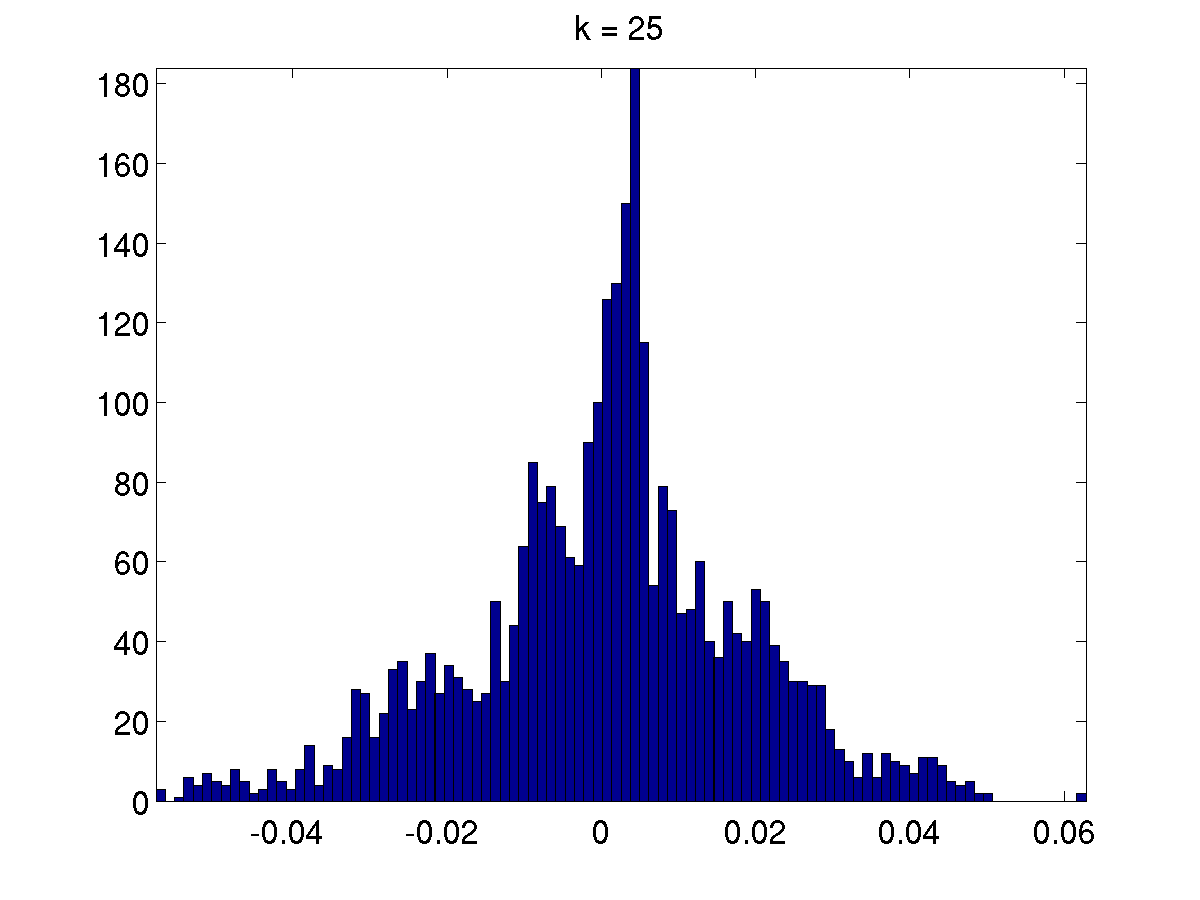}
\includegraphics[width=0.24\columnwidth]{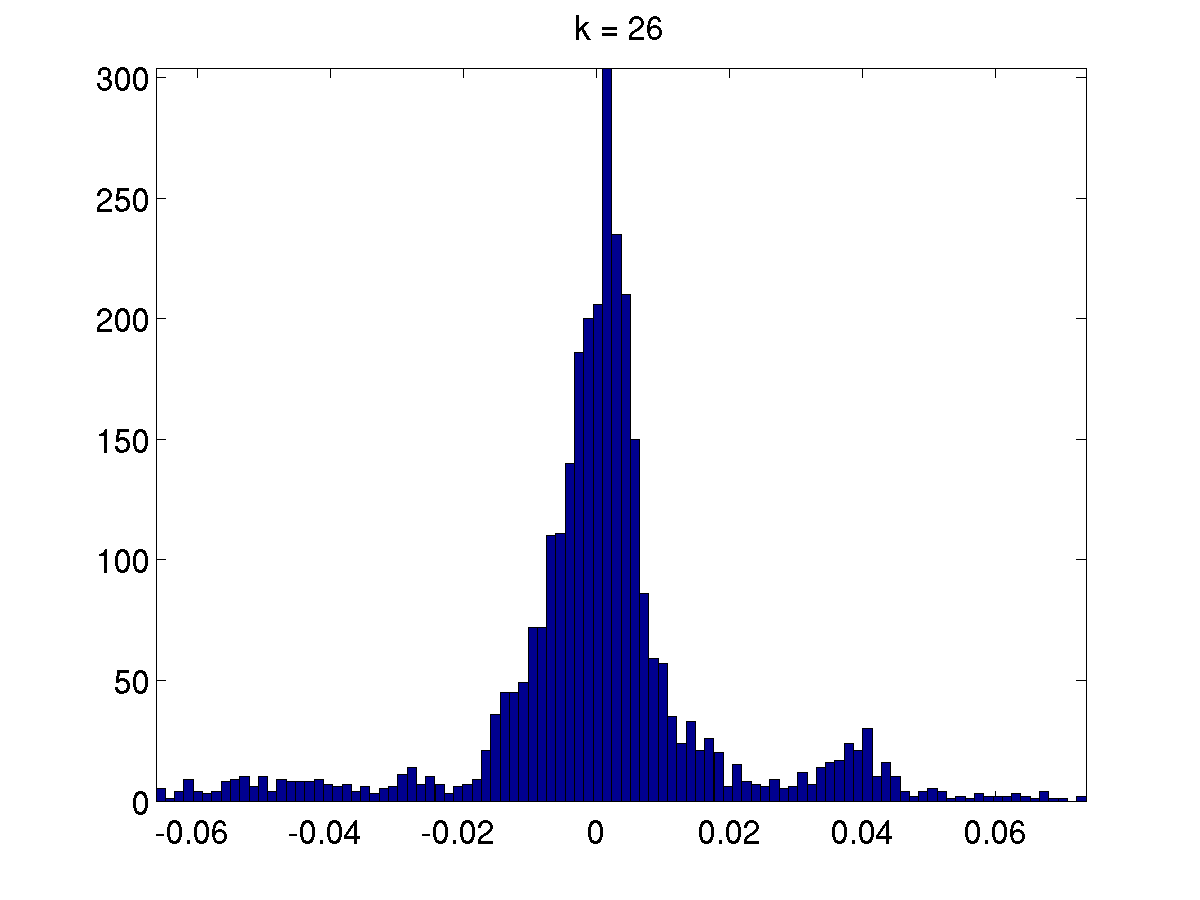}
\includegraphics[width=0.24\columnwidth]{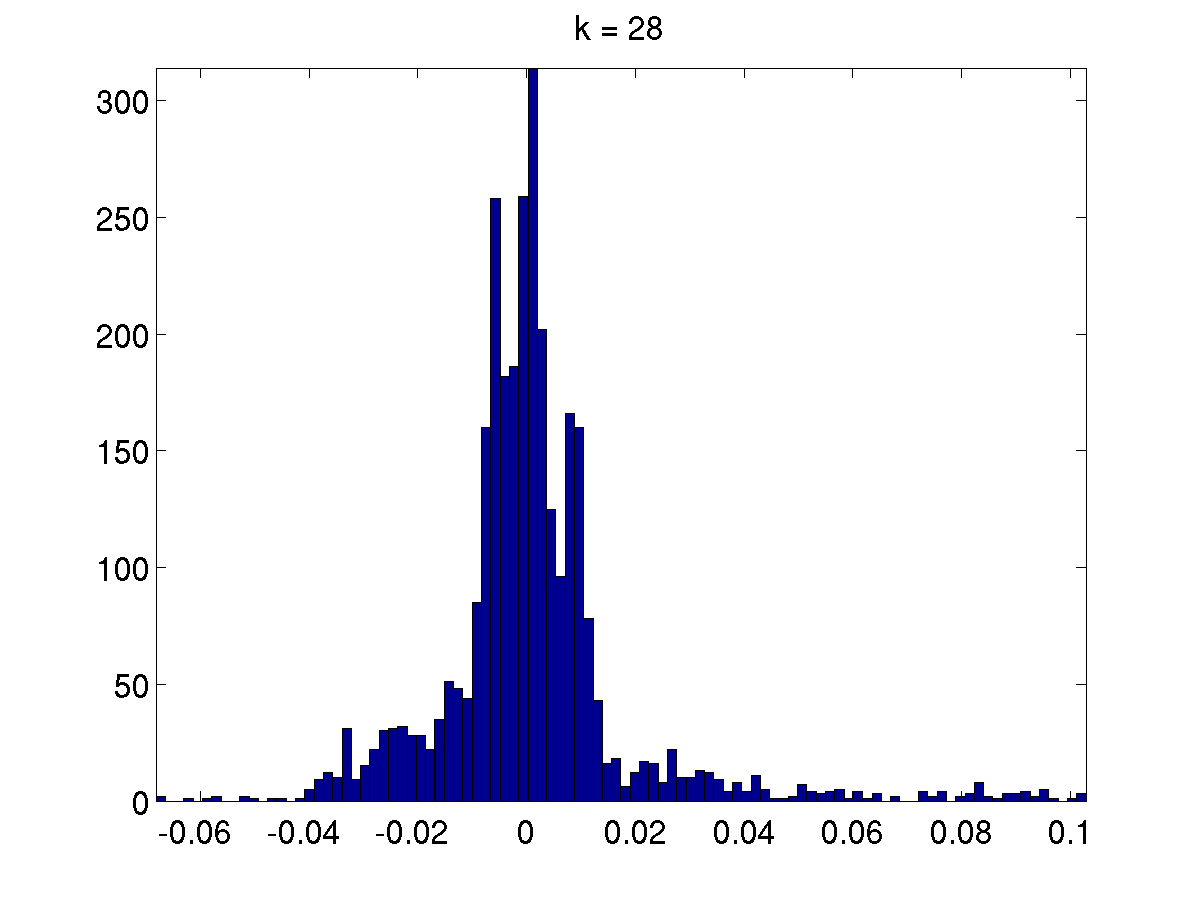}
\includegraphics[width=0.24\columnwidth]{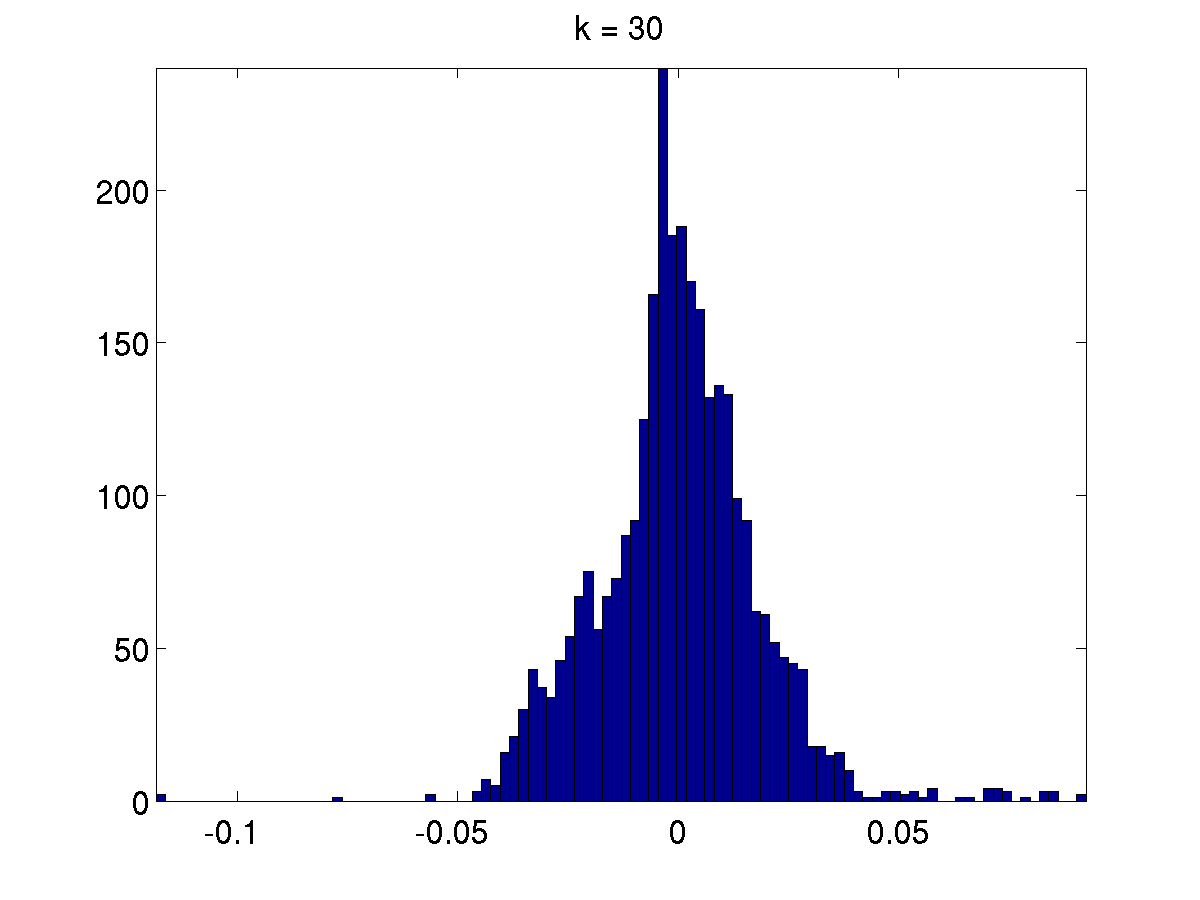}
\includegraphics[width=0.24\columnwidth]{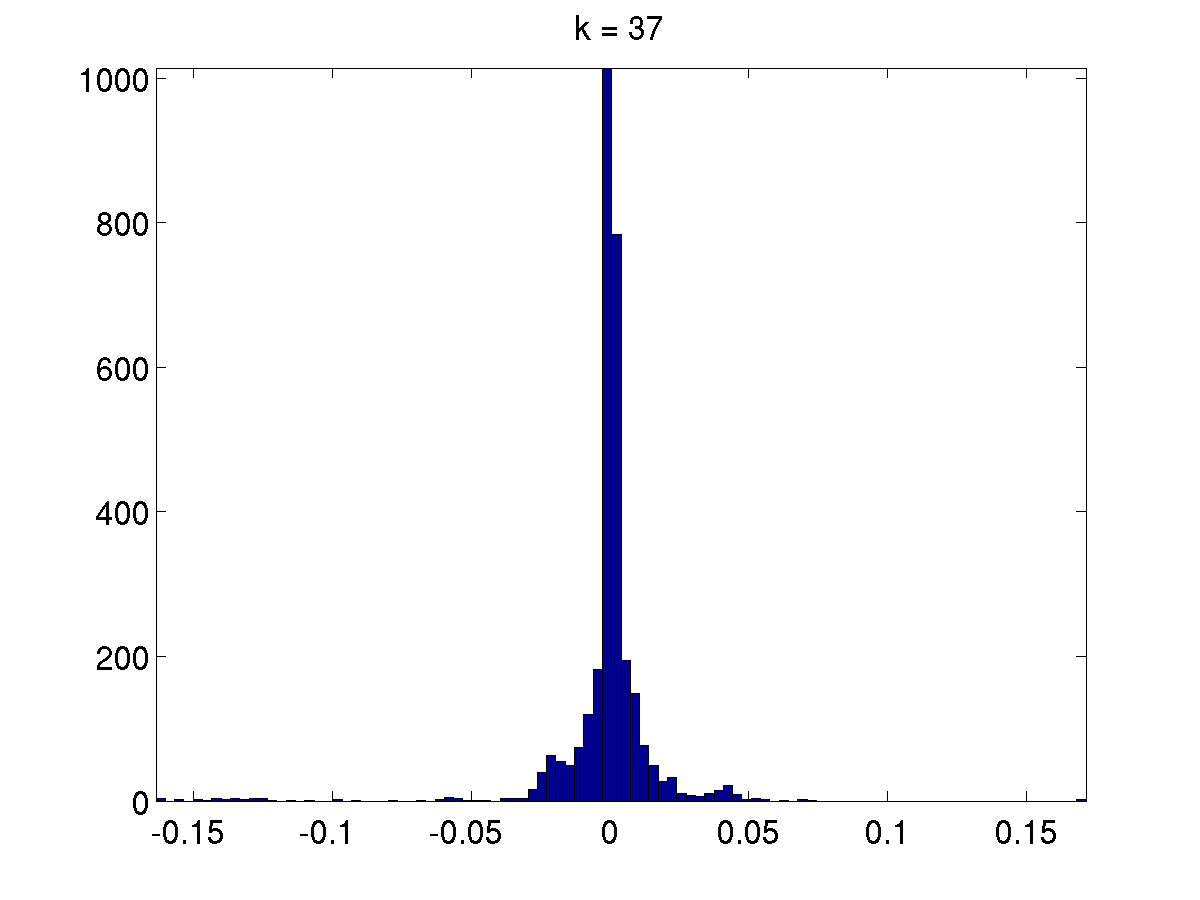}
\includegraphics[width=0.24\columnwidth]{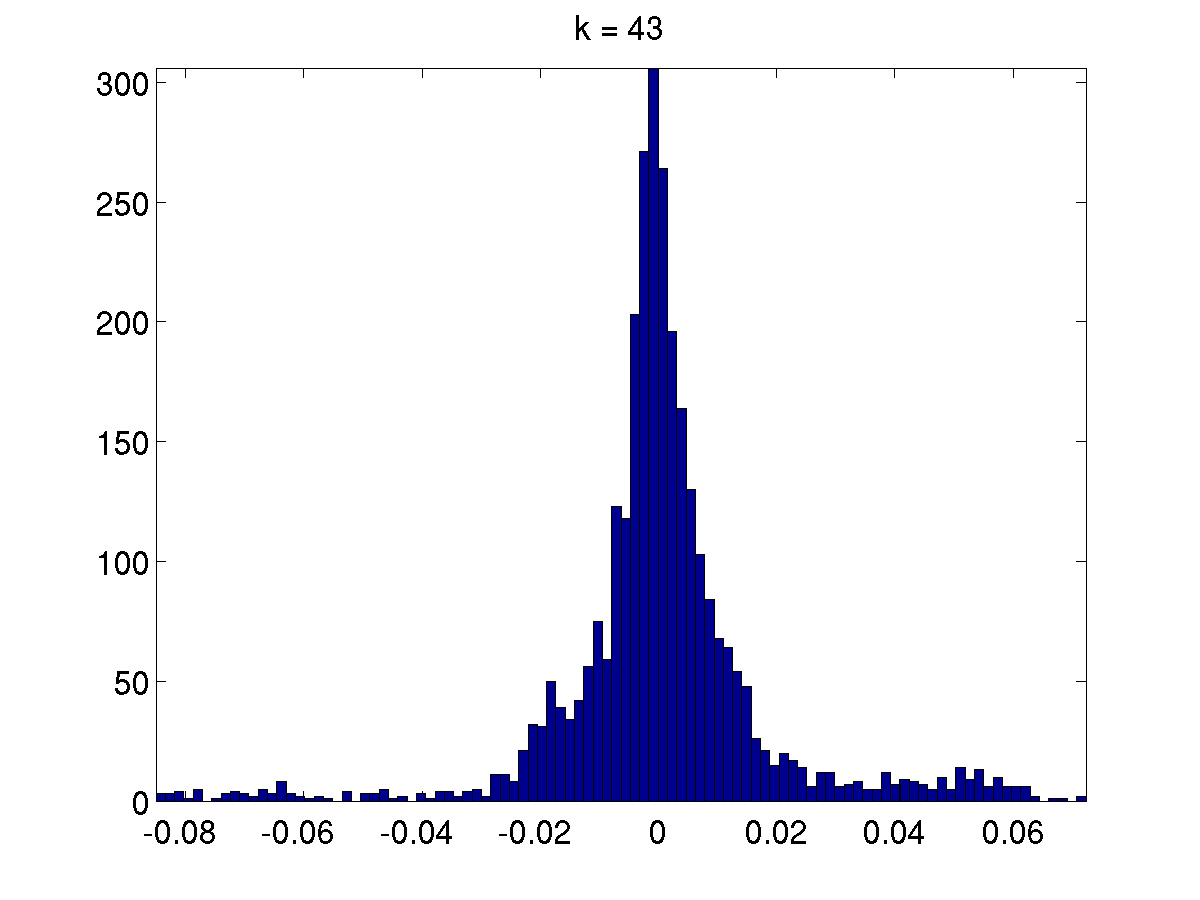}
\end{center}
\caption{The \textsc{Migration} data: histograms of the entries of the 
eigenfunctions that were shown in Figure~\ref{fig:migr-vect}, clearly
indicating localization on some of the low-order eigenfunctions.
}
\label{fig:migr-hist}
\end{figure}

For a more detailed understanding of the localization phenomenon for the 
\textsc{Migration} data, consider Figures~\ref{fig:migr-vect}
and~\ref{fig:migr-hist}~\cite{CVB11_DRAFT}.
Figure~\ref{fig:migr-vect} provides a pictorial illustration of the top
eigenvectors as well as several of the lower-order eigenvectors of the 
county-to-county migration matrix.
As with the \textsc{Congress} data, the \textsc{Migration} data demonstrates 
characteristic global oscillatory behavior on the the top three 
eigenvectors; and many of the low-order eigenvectors are fairly localized in 
way that seems to correspond to interesting domain-specific characteristics.
In particular, some of the low order eigenvectors that localize very well 
seem to reveal small geographically cohesive regions that correlate 
remarkably well with political and administrative boundaries.
In addition, Figure~\ref{fig:migr-hist} shows a histogram of the entries for 
each of these eigenvectors, quantifying the degree of localization.
Recent work on analyzing migration patterns using this data set
highlight cosmopolitan or hub-like regions, as well as isolated regions 
that emerge when there is a high measure of separation between a cluster 
and its environment, some of which are discovered by the localization 
properties of low-order eigenvectors~\cite{Sla08_TR}.
Our observations are also consistent with previous observations on the 
localization properties of the \textsc{Migration} data~\cite{CVB11_DRAFT}. 

%TMP% 

%TMP% 

Clearly, in both the \textsc{Congress} data and in the \textsc{Migration} 
data, there is more going on in the spectrum than we have discussed, and it 
is not obvious the extent to which these represent real properties of the 
data or are simply artifacts of noise.  
For example, there is a fairly strong tendency in the \textsc{Congress} 
data for localization to occur on very early Congresses or very late 
Congresses; when this happens, there is a tendency for eigenvectors with 
localization on recent Congresses to account for a larger fraction of the 
variance of the data than eigenvectors with localization on much older 
Congresses.; etc.
In addition, there are also many other low-order eigenvectors in these two 
data sets that are delocalized, noisy, and seemingly-meaningless in terms of 
the domain from which the data are drawn.
We will discuss these and other issues below.
Our point here is simply to illustrate that there can exist a substantial 
degree of localization on certain low-order eigenvectors; this this 
localization can highlight properties of the data---temporally-local 
information such as a party-line partition of a single Congress or small 
geographically cohesive regions that have experienced nontrivial migration 
patterns---of interest to the domain scientist; and that these properties 
are not highlighted among the coarsest modes of variation of the data when 
the data are viewed globally.

\section{A simple model}
\label{sxn:model}

In this section, we will describe a simple model that exhibits low-order
eigenvector localization.
This model qualitatively reproduces several of the results that were 
empirically observed in Section~\ref{sxn:empirical}, and it can be used as 
a diagnostic tool to help extract insight from data graphs when such 
low-order eigenvector localization is present.

\subsection{Description of the \textsc{TwoLevel} model}

To motivate our \textsc{TwoLevel} model, consider what the \textsc{Congress} 
data ``looks like'' if one ``squints'' at it, \emph{i.e.}, in 
``coarse-grained'' sense.
In this case, most edges are between different members of a single 
Congress, \emph{i.e.}, they are temporally-local at a single time-slice; and 
the remainder of the edges are between a single individual in two 
consecutive Congresses, \emph{i.e.}, they are still fairly temporally-local.
That is, there is some structured graph (structured depending on the details 
of the voting pattern in any particular Congress) for which the 
temporally-local connections are reasonably strong (assuming that the 
connection parameter between individuals in successive Congresses is not 
extremely small or extremely large) that is ``evolving'' along a 
one-dimensional temporal scaffolding.
Thus, if one ``zooms in'' and looks locally at a single Congress, then the 
properties of that Congress should be apparent.
For example, the best partition computed from a spectral clustering 
algorithm for any single Congress is typically strongly correlated with 
party affiliation~\cite{PR97,WPFMP09_TR,multiplex_Mucha}.
On the other hand, if one ``zooms out'' and looks at the entire graph, 
then the linear time series structure should be apparent and the properties 
of any single Congress should be less important.
For example, the best partition computed from a spectral clustering
algorithm for the entire data set split the data into the first temporal 
half and the second temporal half and thus fails to see party affiliations.

In cases such as this, where there are two different ``size scales'' to 
the interactions, a zero-th order model for the data may be given by the 
following tensor product structure.
Let $W$ be a ``base graph'' representing the structure of ``local'' 
interactions at local or small size scales.
For example, this could be a simple model for the voting patterns within a 
single Congress; or this could represent the inter-county migration patterns 
within a single state or geopolitical region.
In addition, let $N$ be an ``interaction model'' that governs the ``global''
interaction between different base graphs $W$.
For example, this could be a ``banded'' or ``tridiagonal'' matrix, in which 
the nonzero components above and below the diagonal represent the connection 
links between two Congresses at adjacent time steps; or this could be a 
discretization of a low-dimensional manifold representing the geographical 
connections in a nation, if spatially-local couplings are most important; 
or this could even be a more general noise model in which edges are added 
randomly between every pair of nodes (if, \emph{e.g.}, the connections 
between different base graphs are much less structured, as in social and 
information networks~\cite{LLDM09_communities_IM}).
Then, a simple a zero-th order model, which we will denote the 
\textsc{TwoLevel} model, is given by 
\begin{eqnarray*}
G &=& H + N  \mbox{ , where}  \\
H &=& I \otimes W ,
\end{eqnarray*}
where $I$ is the identity matrix and $H = I \otimes W$ denotes the tensor 
product between $I$ and~$W$. 

In what follows, we will illustrate the properties of the
\textsc{TwoLevel} model in several idealized settings.
To do so, we will consider the base graph $W$ to be either ``structured'' or 
``unstructured,'' and we will also consider the interaction model $N$ to be 
either ``structured'' or ``unstructured.''
\begin{itemize}
\item
For the base graph, $W$, we will model the unstructured case by a single
unstructured Erd\H{o}s-R\'{e}nyi random graph~\cite{Bollobas85}, $G_{np}$, 
on some number $n$ of nodes, where the connection probability between each 
pair of nodes is $p$;
and we will model the structured case by a so-called $2$-module.
By a ``$2$-module,'' we mean two Erd\H{o}s-R\'{e}nyi random graphs, where 
intra-module nodes are randomly connected with probability $p_1$ and 
inter-module nodes are connected with some much lower probability $p_2$.
(This $2$-module is structured in the sense that the top eigenvector of the 
$2$-module graph is the Fiedler vector that would clearly separate the two 
modules.) 
\item
For the interaction model, $N$, we will model structured noise as a ``path 
graph,'' \emph{i.e.}, a tree with two or more vertices that is not branched 
at all and which thus has a ``banded'' adjacency matrix; and we will model 
unstructured noise by randomly connecting any two nodes in different modules 
with some small probability, \emph{i.e.}, by an Erd\H{o}s-R\'{e}nyi random 
graph with some small connection probability $p$.
\end{itemize}
%%%
Clearly, for both the base graph and for the interaction model, these are 
limiting cases.  
For example, rather than consider a $2$-module as the base graph, one could 
consider a $3$-module to model the existence of a good tri-partition of the
base graph, a $4$-module, etc.
Similarly, rather than just considering interactions along a one-dimensional 
scaffolding, one could consider it along a two-dimensional scaffolding, etc.
Unpublished empirical results indicate that, for both the base graph and for 
the interaction model, by considering these weaker forms of structure
(in particular, $3$-modules rather than $2$-modules or a two-dimensional 
scaffolding rather than a one-dimensional scaffolding), we obtain results 
that are similar to but intermediate between the structured and unstructured 
results that we report below.
Formalizing this more generally and understanding the theoretical and 
empirical implications of perturbations of tensor product matrices is an 
open problem raised by our observations.

\subsection{Empirical properties of the \textsc{TwoLevel} model}

Here, we will examine the behavior of the \textsc{TwoLevel} model for 
various combinations of structured and unstructured graphs for the base 
graph and the the interaction model.
Our goal will be to reproduce qualitatively some of the properties we 
observed in Section~\ref{sxn:empirical} and to understand their behavior 
in terms of the parameters of the \textsc{TwoLevel} model.

\begin{figure}[t] 
\begin{center}
\includegraphics[width=0.32\columnwidth]{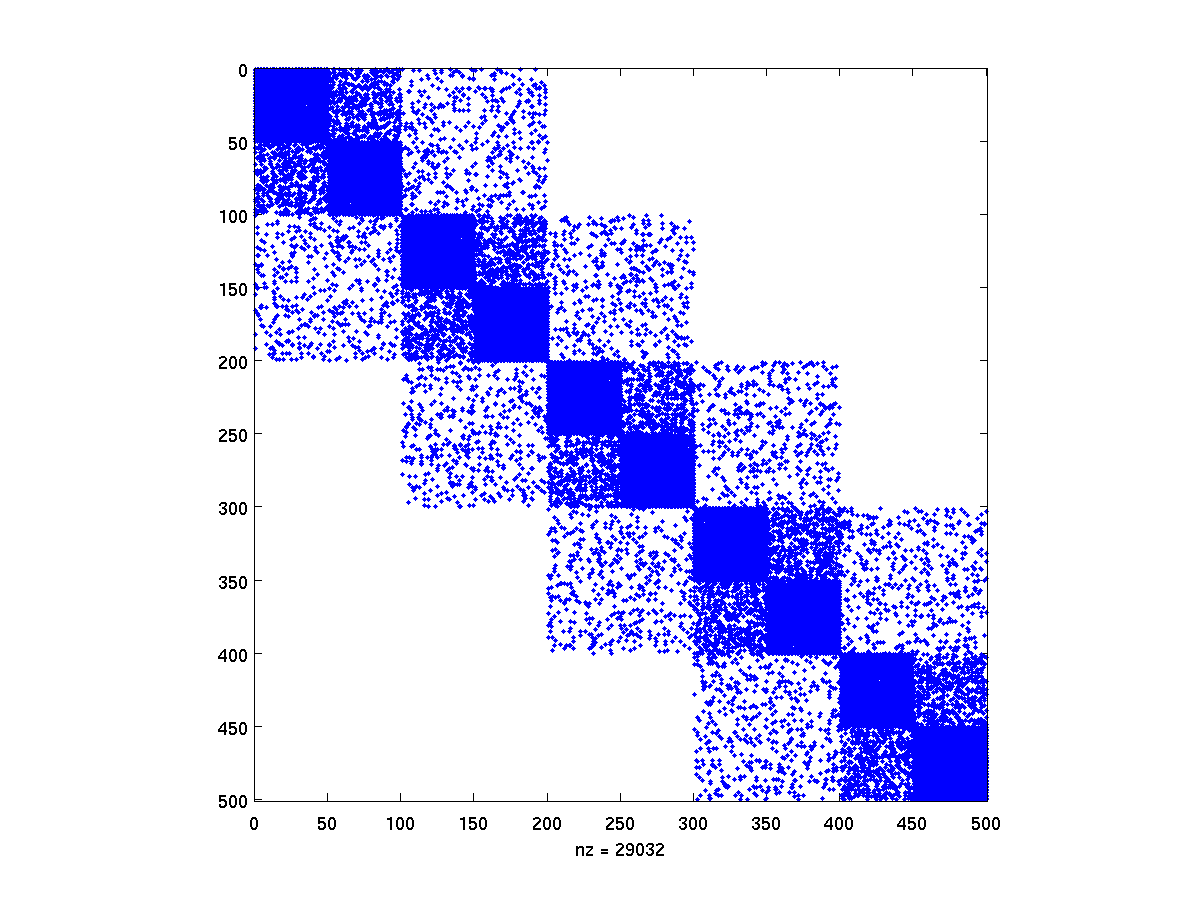}
\includegraphics[width=0.32\columnwidth]{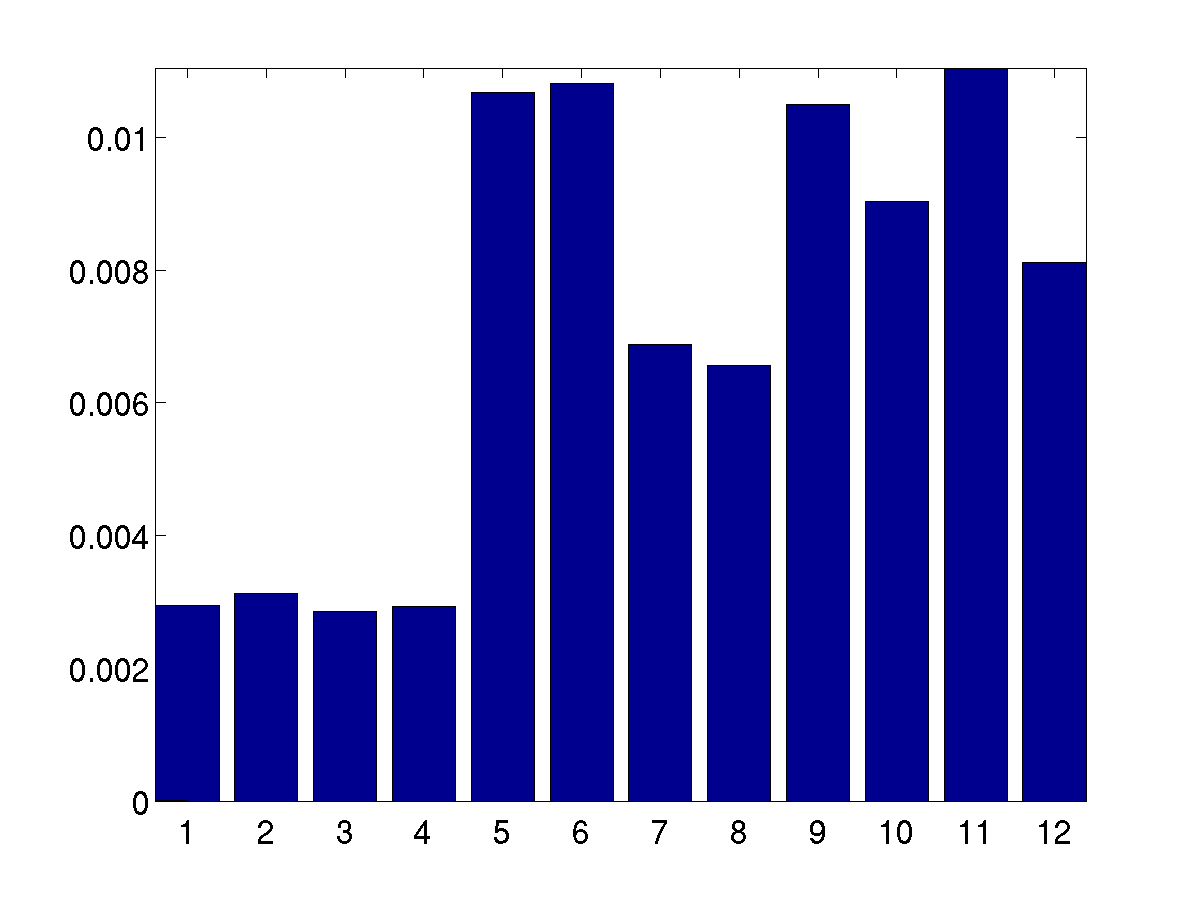}
\includegraphics[width=0.32\columnwidth]{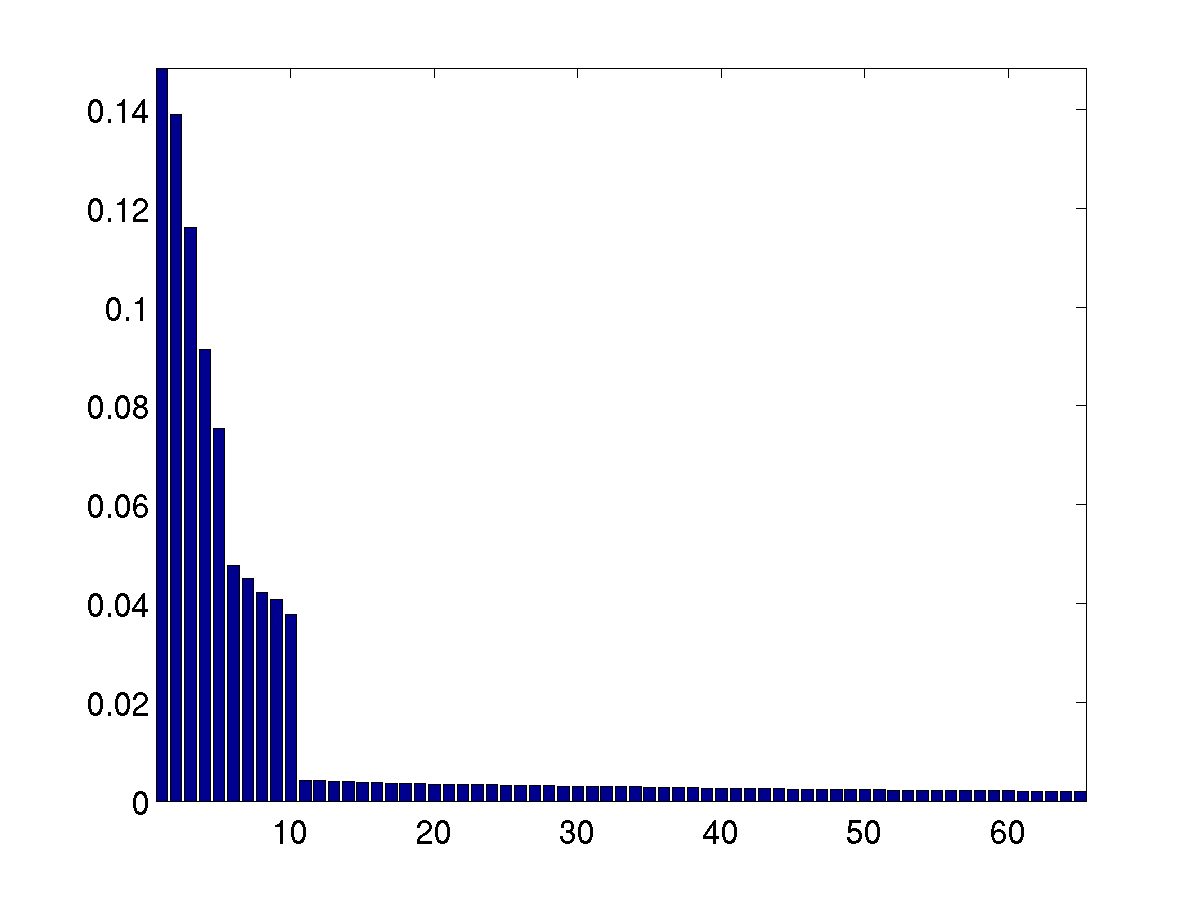}
\includegraphics[width=0.16 \columnwidth]{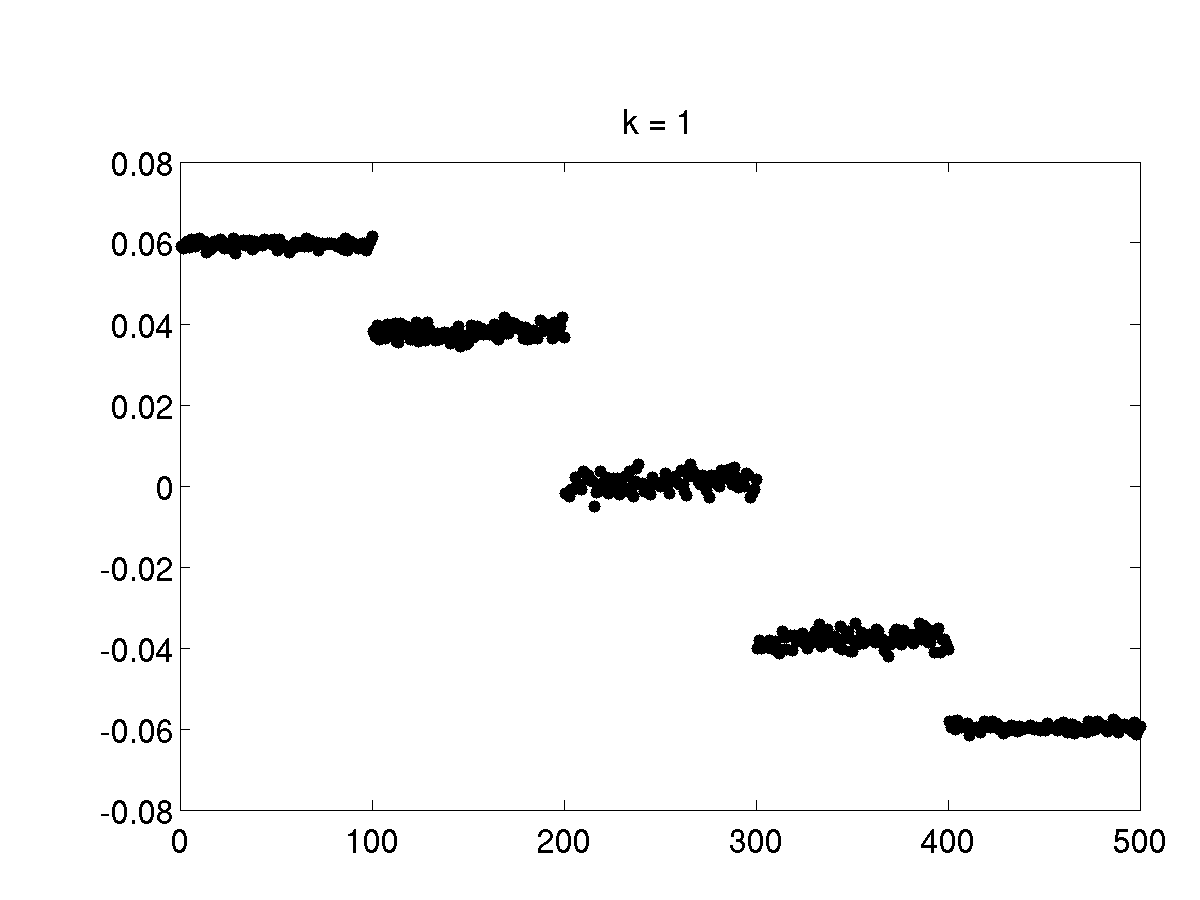}
\includegraphics[width=0.16 \columnwidth]{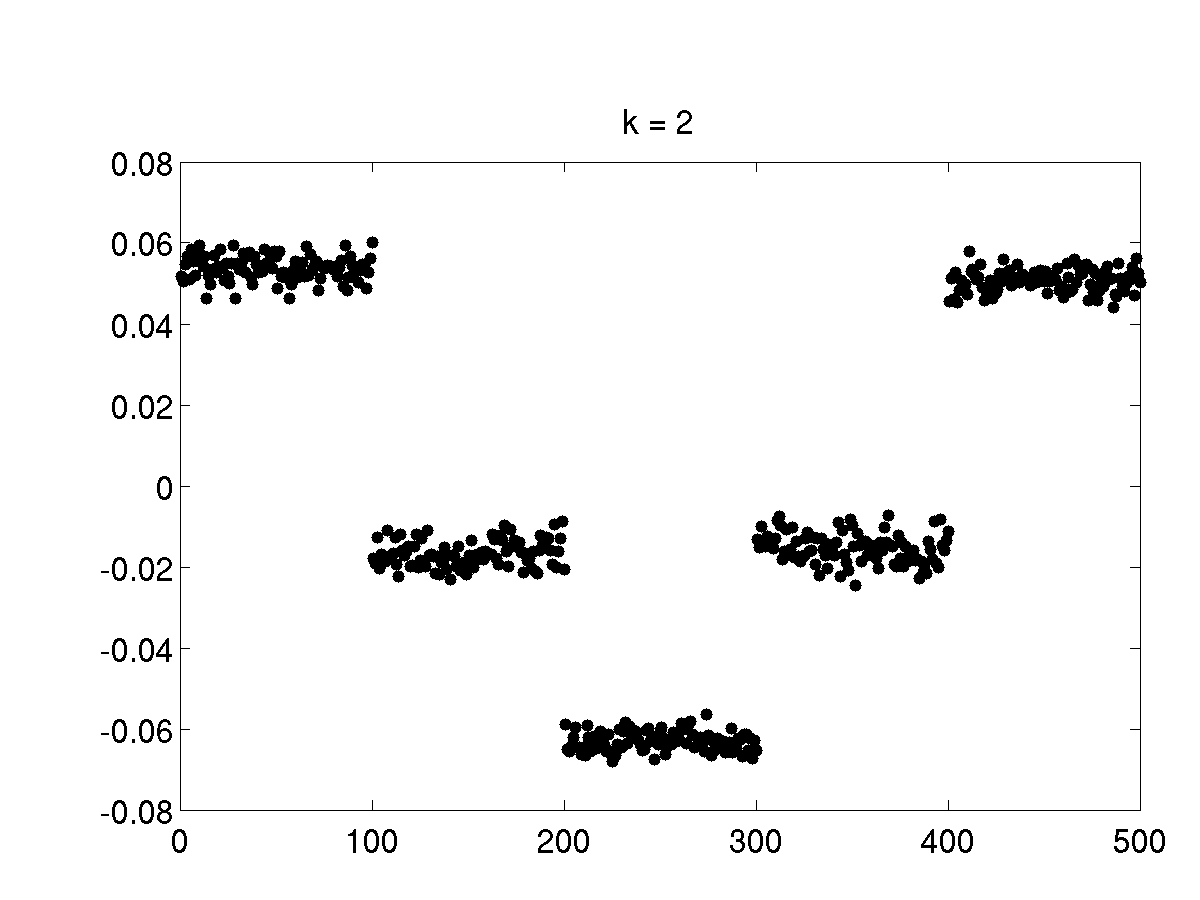}
\includegraphics[width=0.16 \columnwidth]{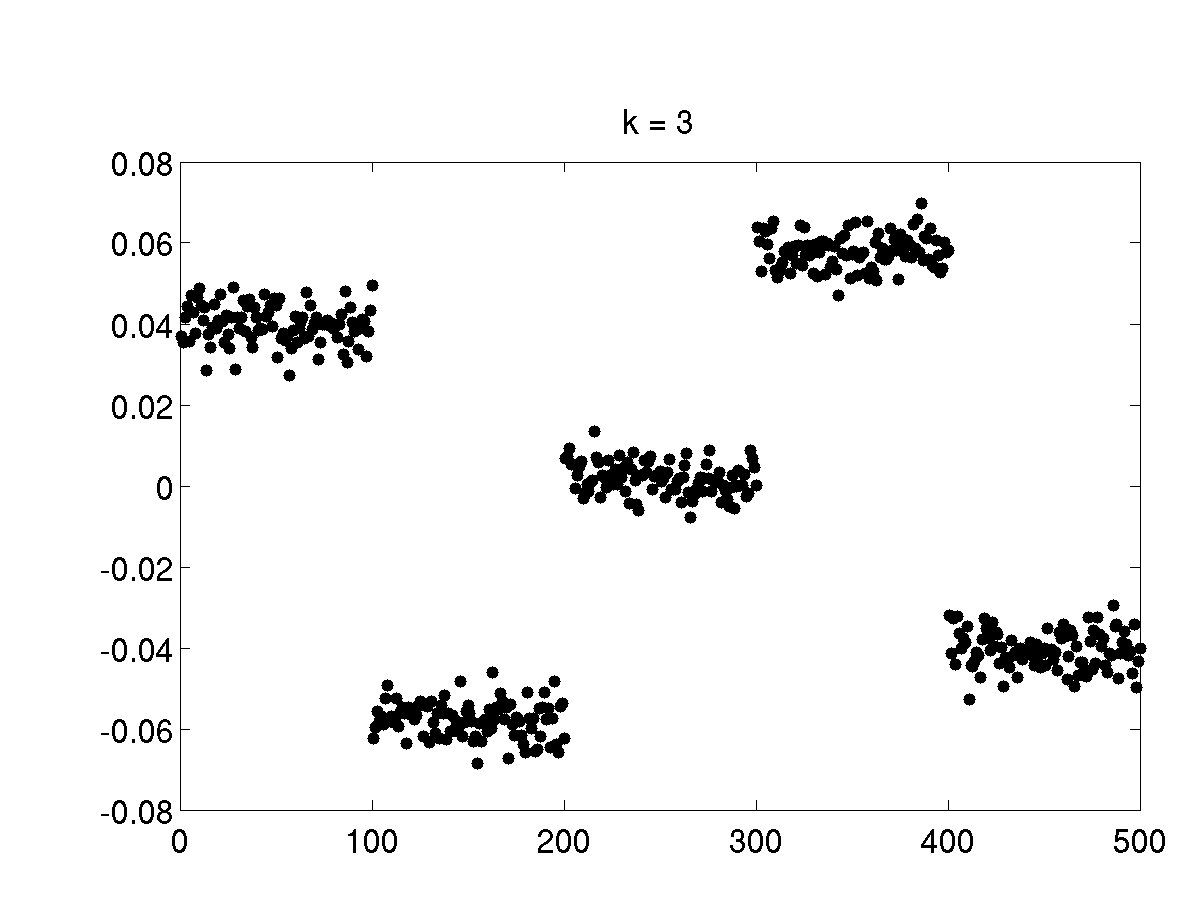}
\includegraphics[width=0.16 \columnwidth]{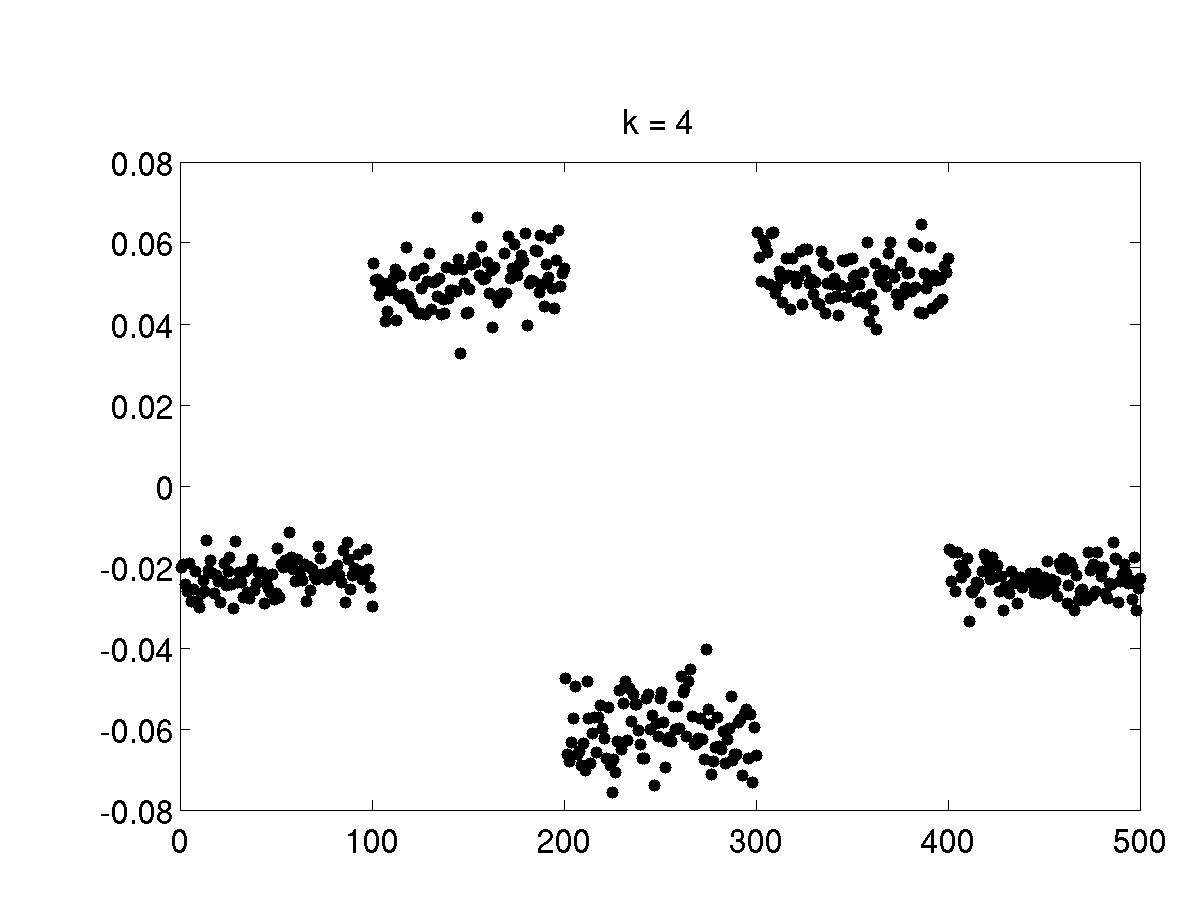}
\includegraphics[width=0.16 \columnwidth]{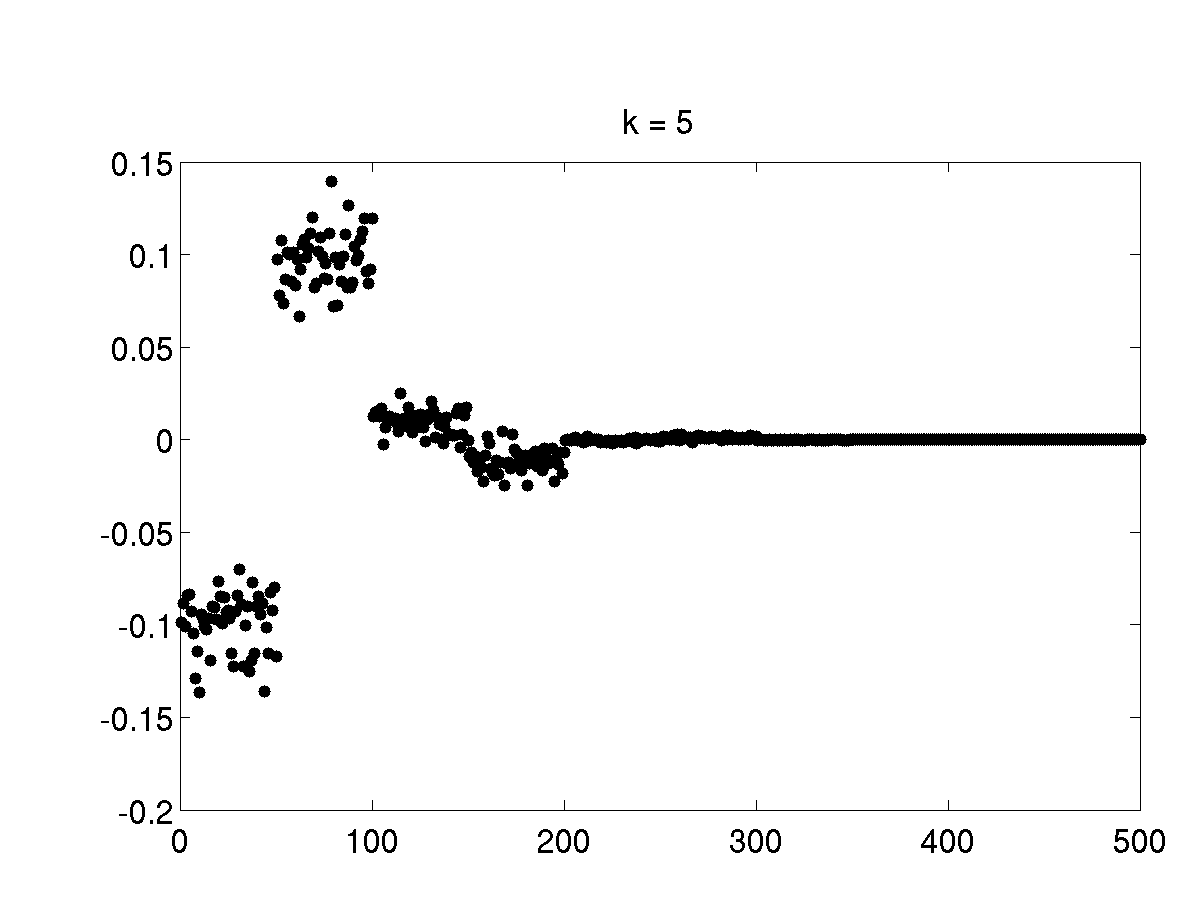}
\includegraphics[width=0.16 \columnwidth]{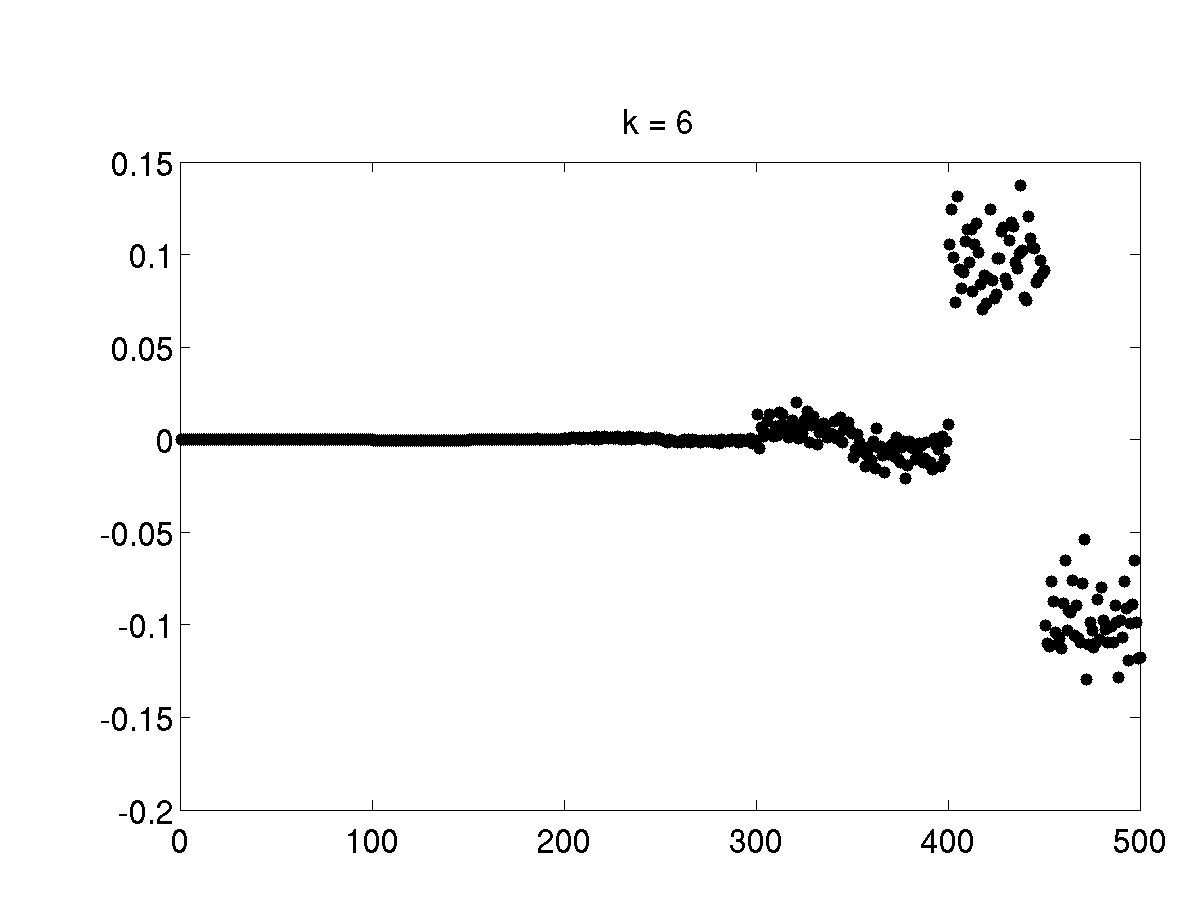} 
\includegraphics[width=0.16 \columnwidth]{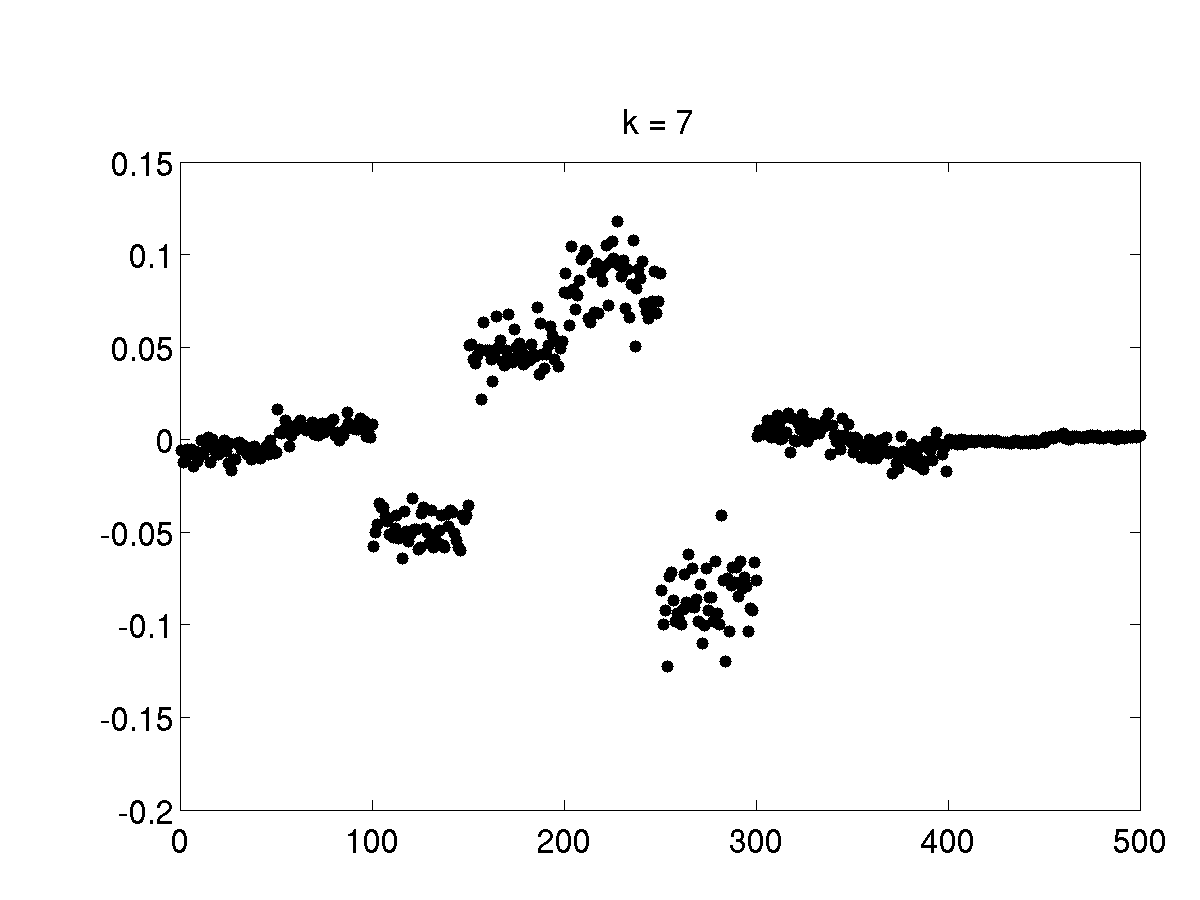}
\includegraphics[width=0.16 \columnwidth]{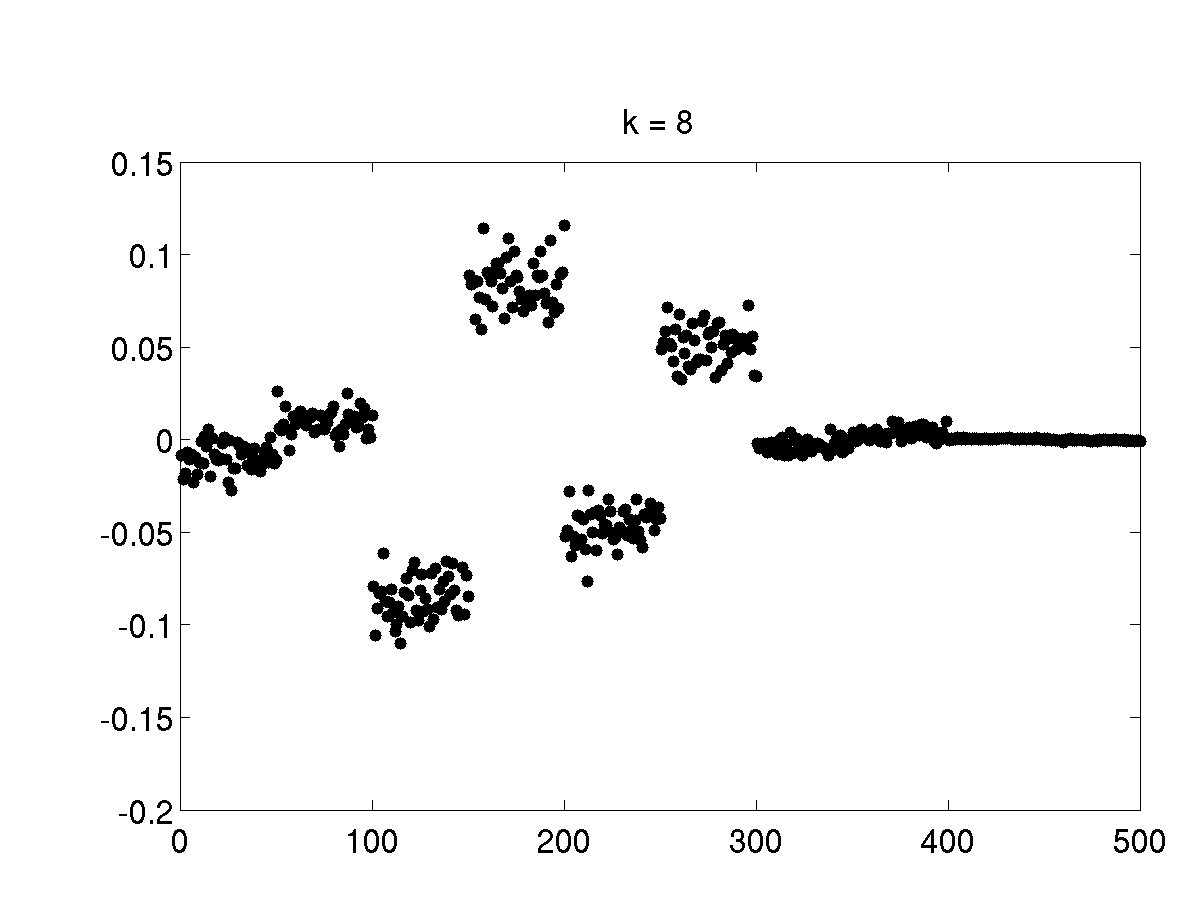}
\includegraphics[width=0.16 \columnwidth]{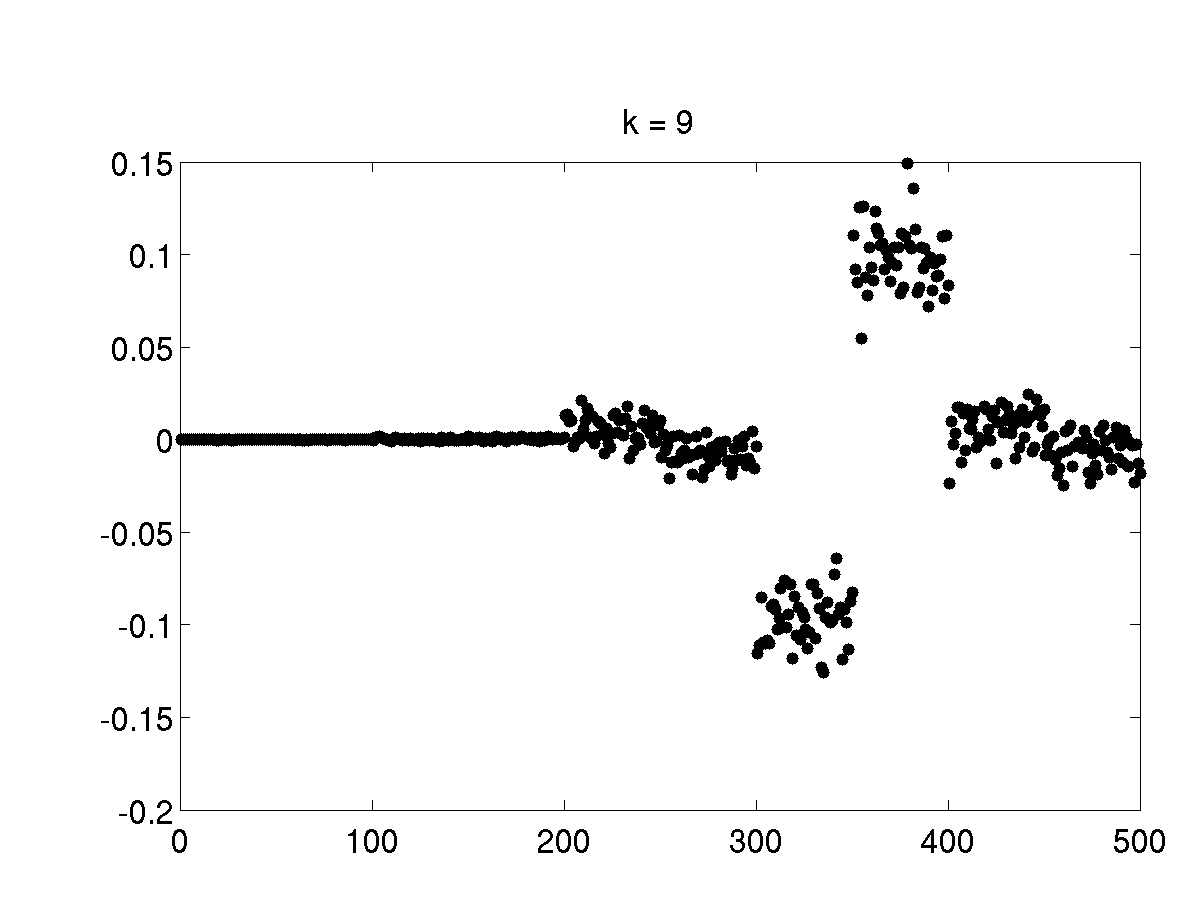}
\includegraphics[width=0.16 \columnwidth]{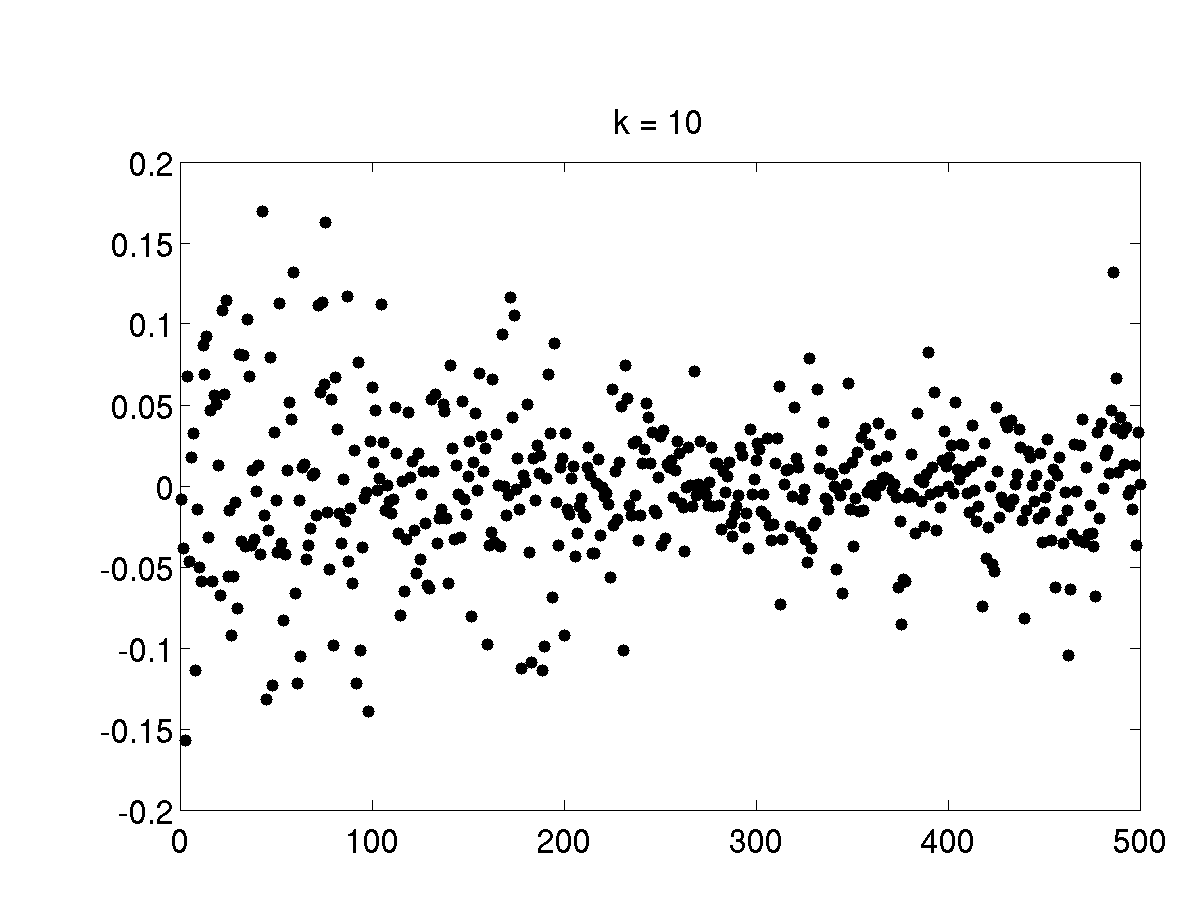}
\includegraphics[width=0.16 \columnwidth]{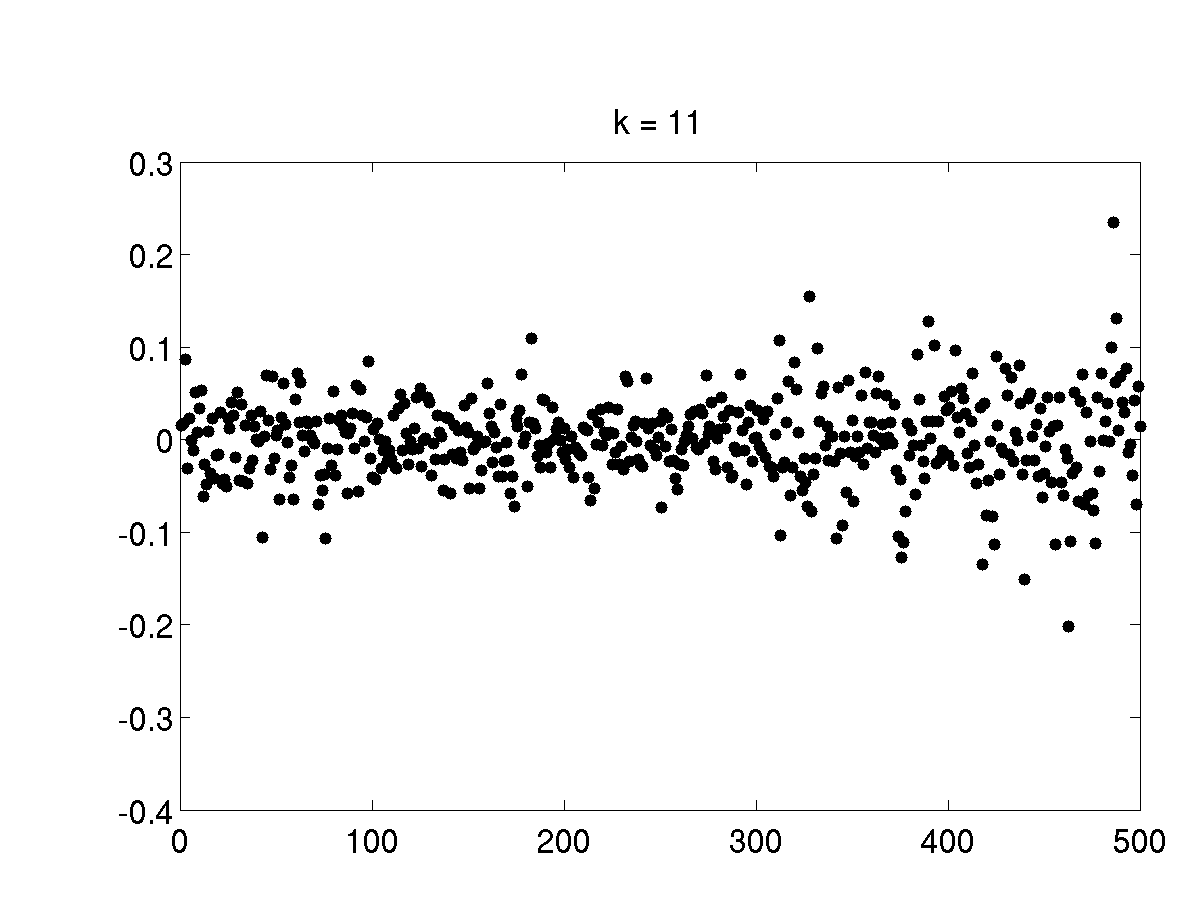} 
\includegraphics[width=0.16 \columnwidth]{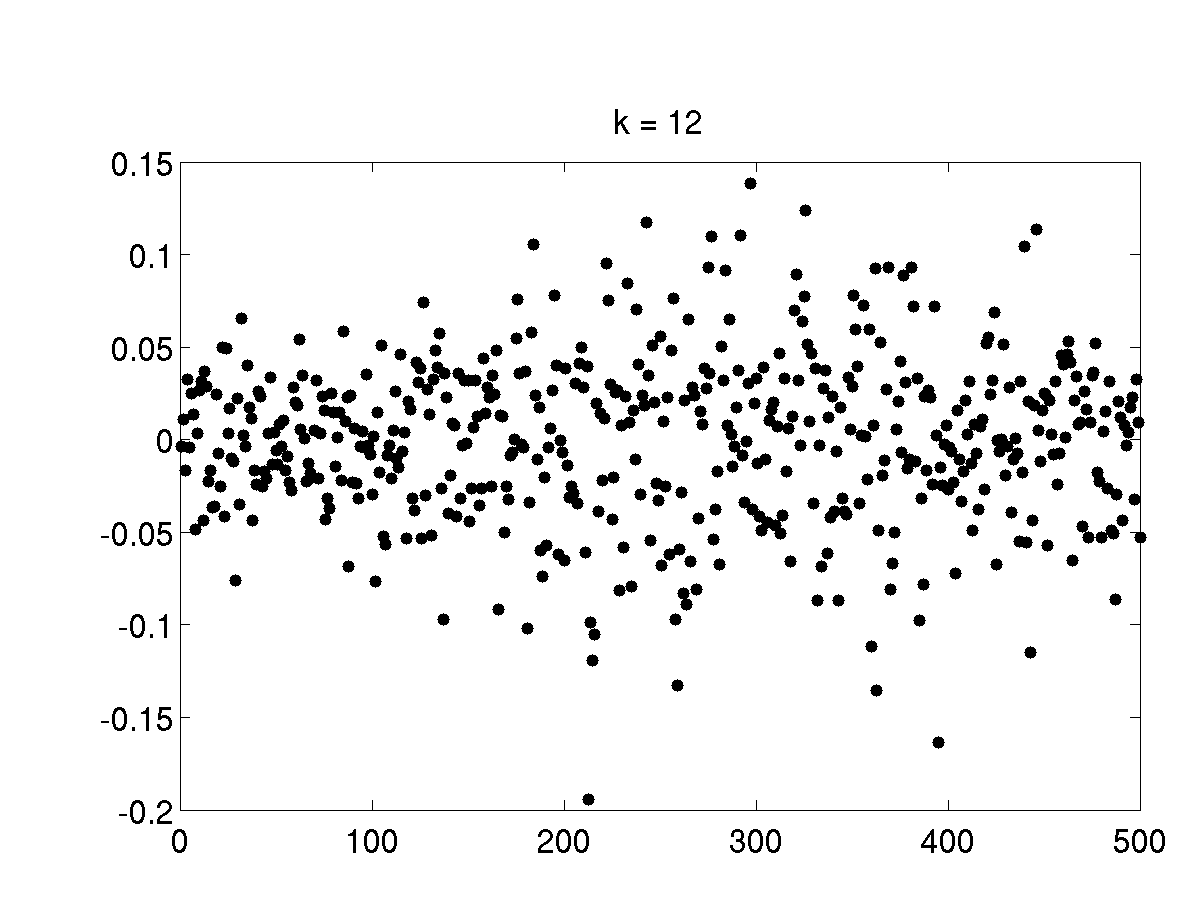} \\
\includegraphics[width=0.16 \columnwidth]{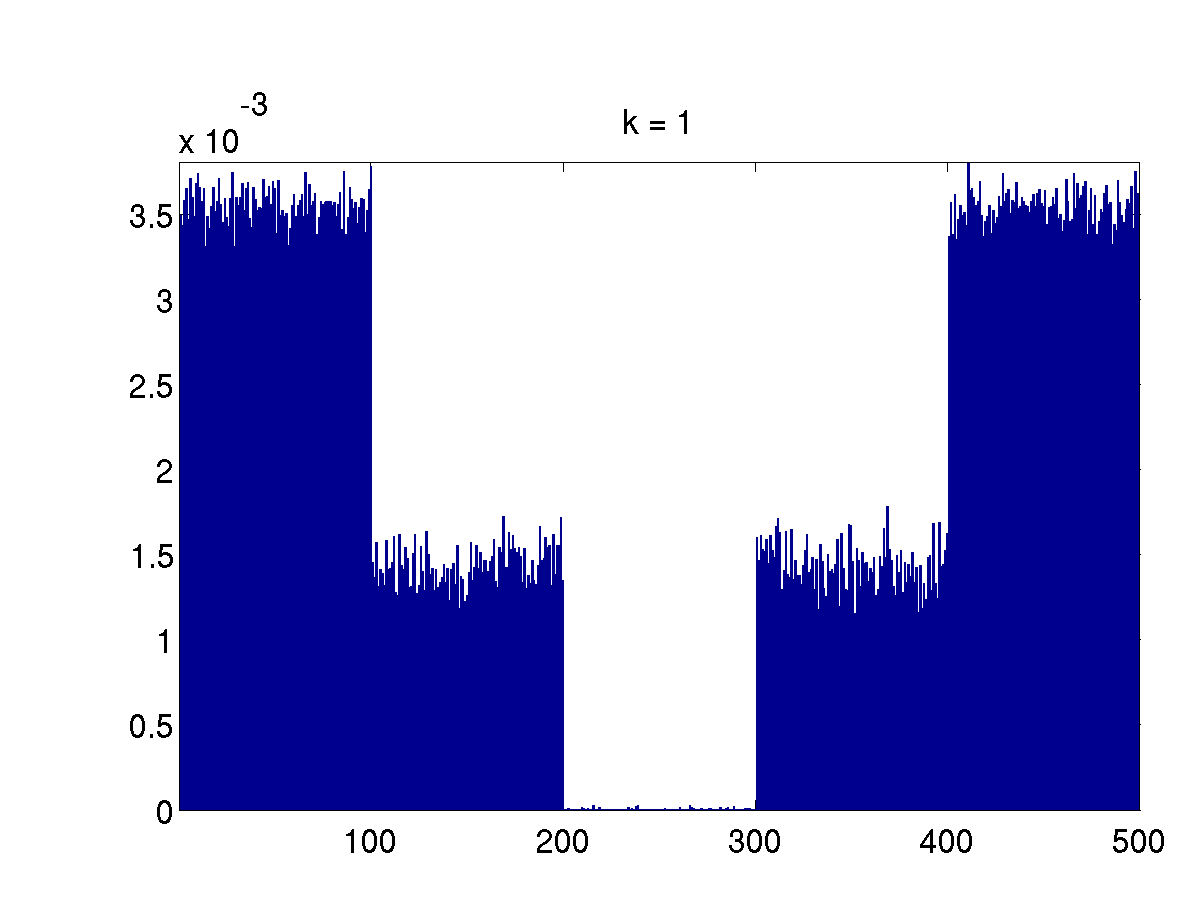}
\includegraphics[width=0.16 \columnwidth]{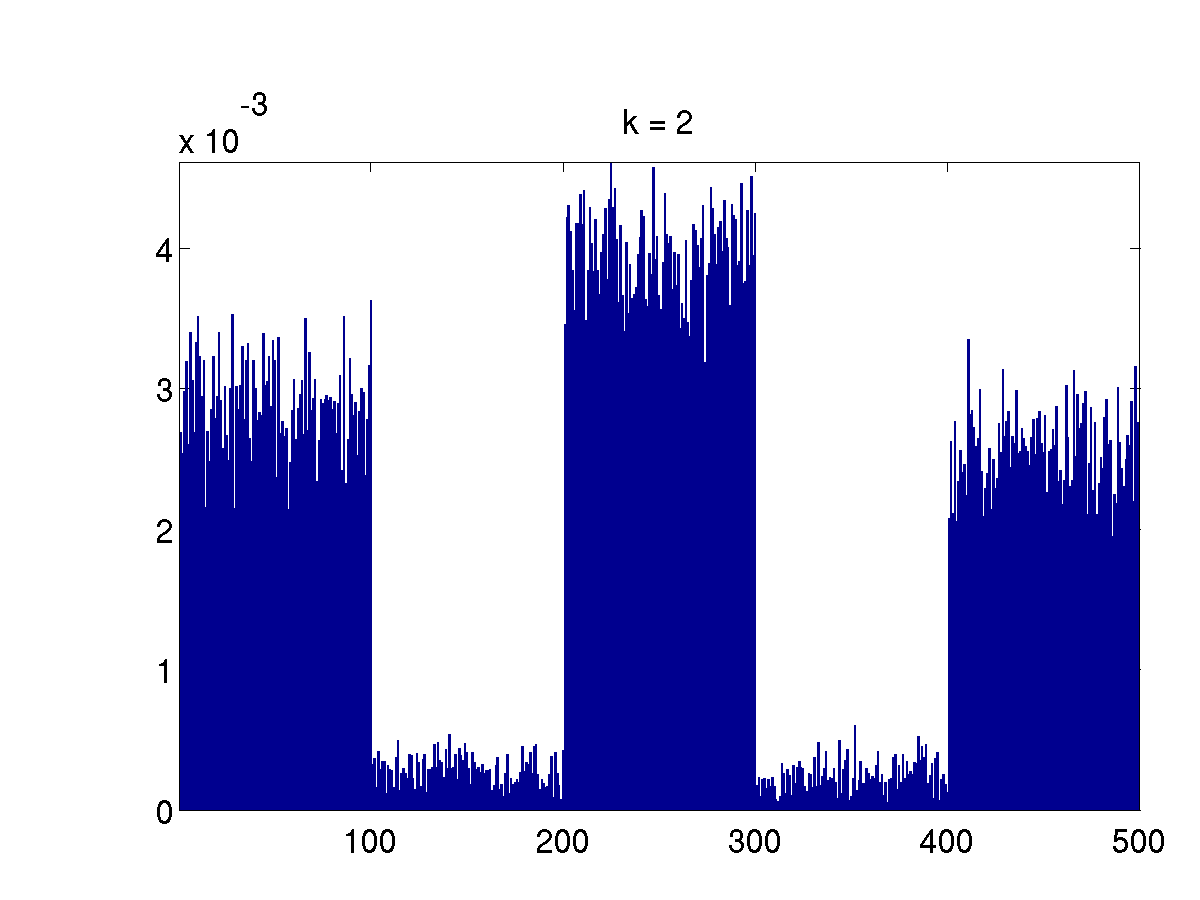}
\includegraphics[width=0.16 \columnwidth]{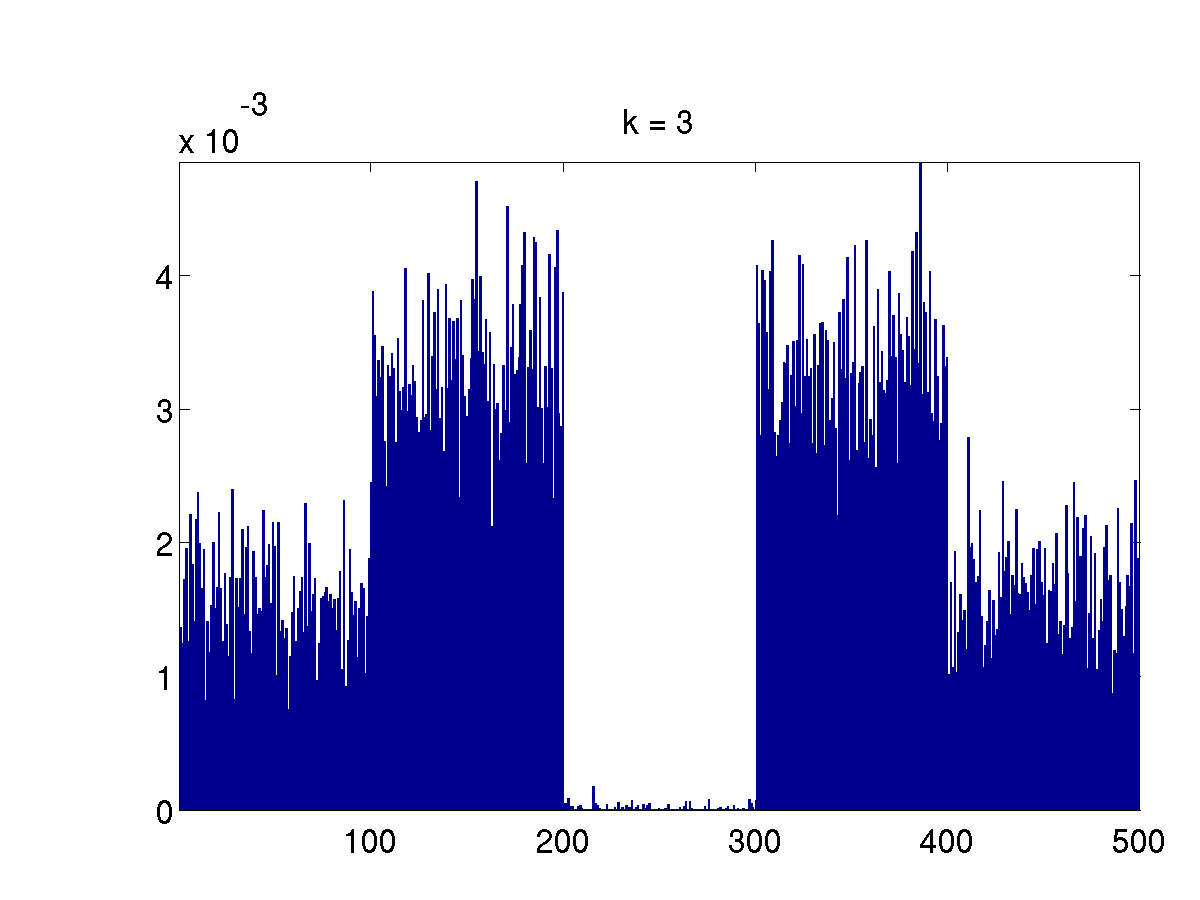}
\includegraphics[width=0.16 \columnwidth]{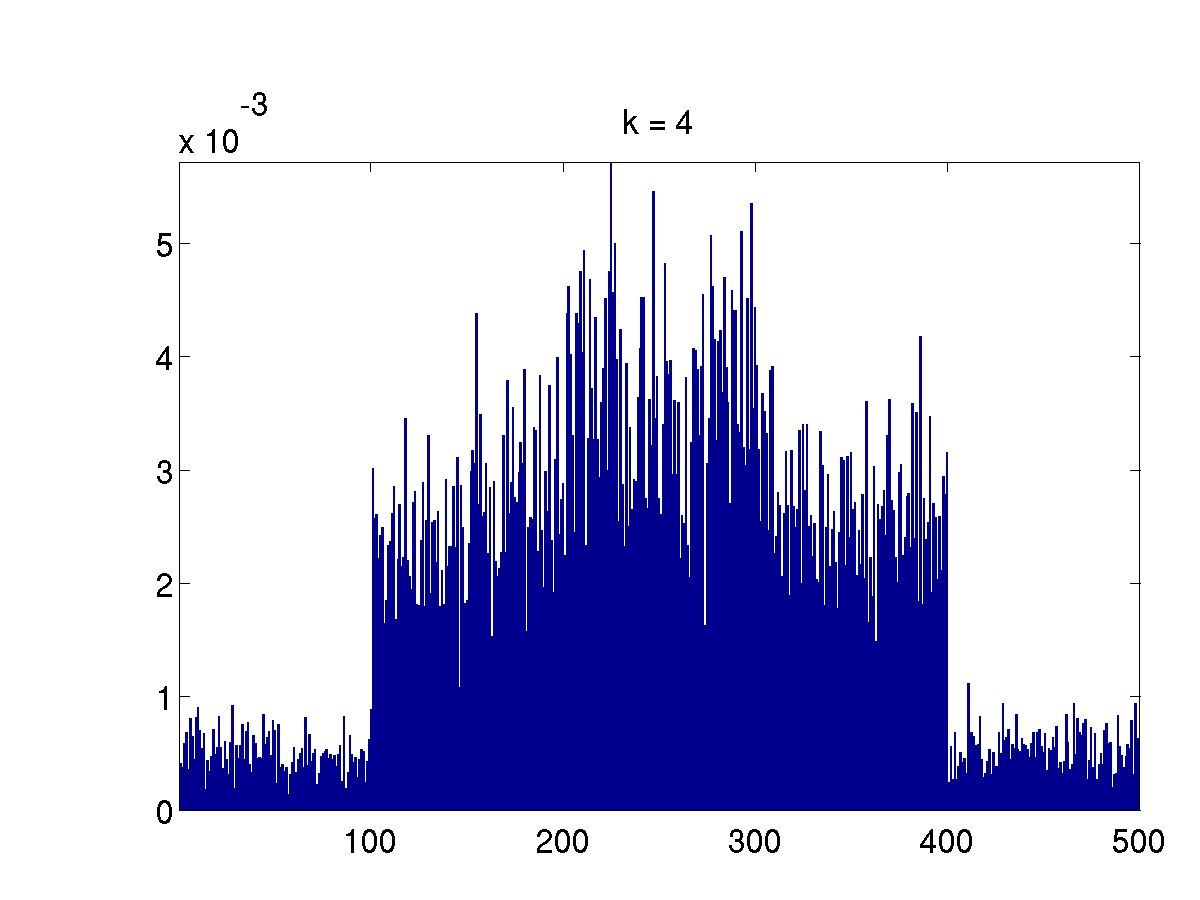}
\includegraphics[width=0.16 \columnwidth]{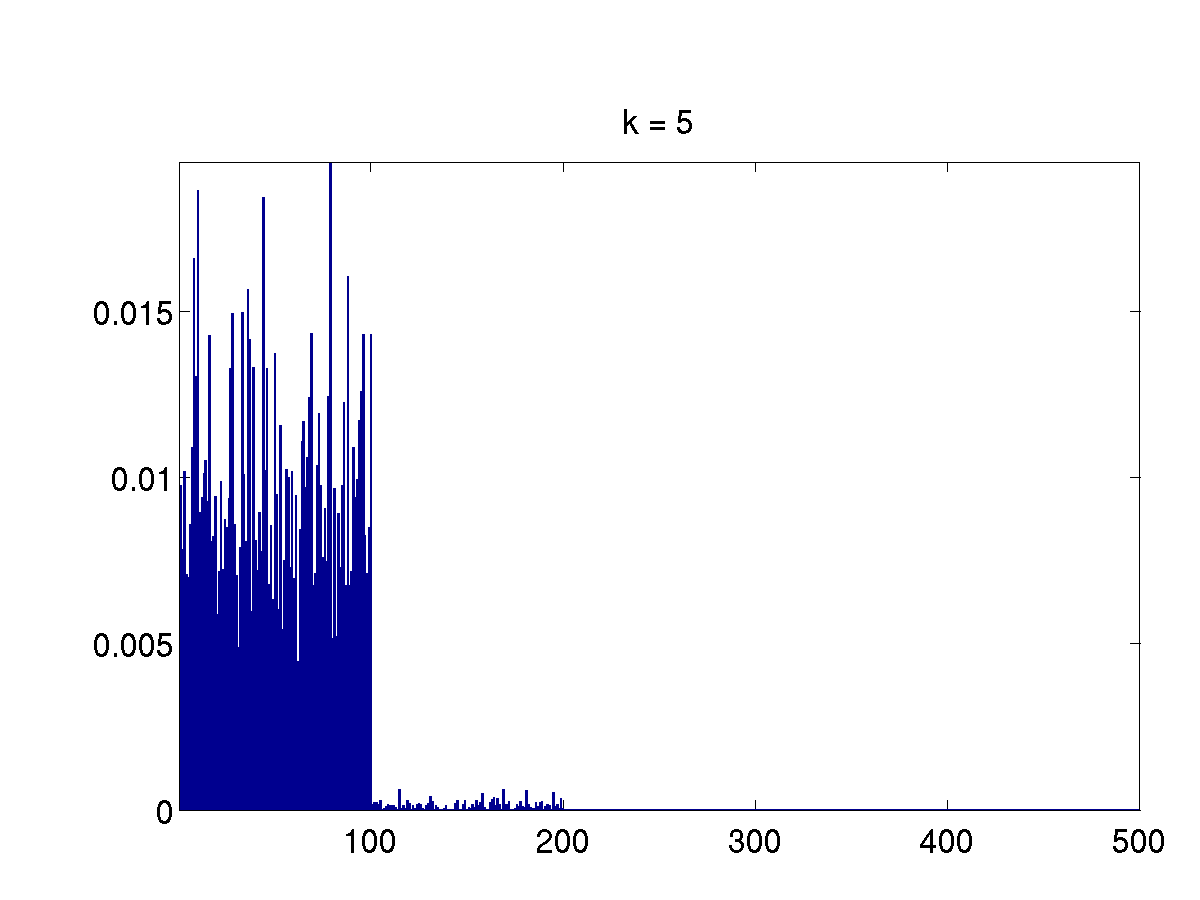}
\includegraphics[width=0.16 \columnwidth]{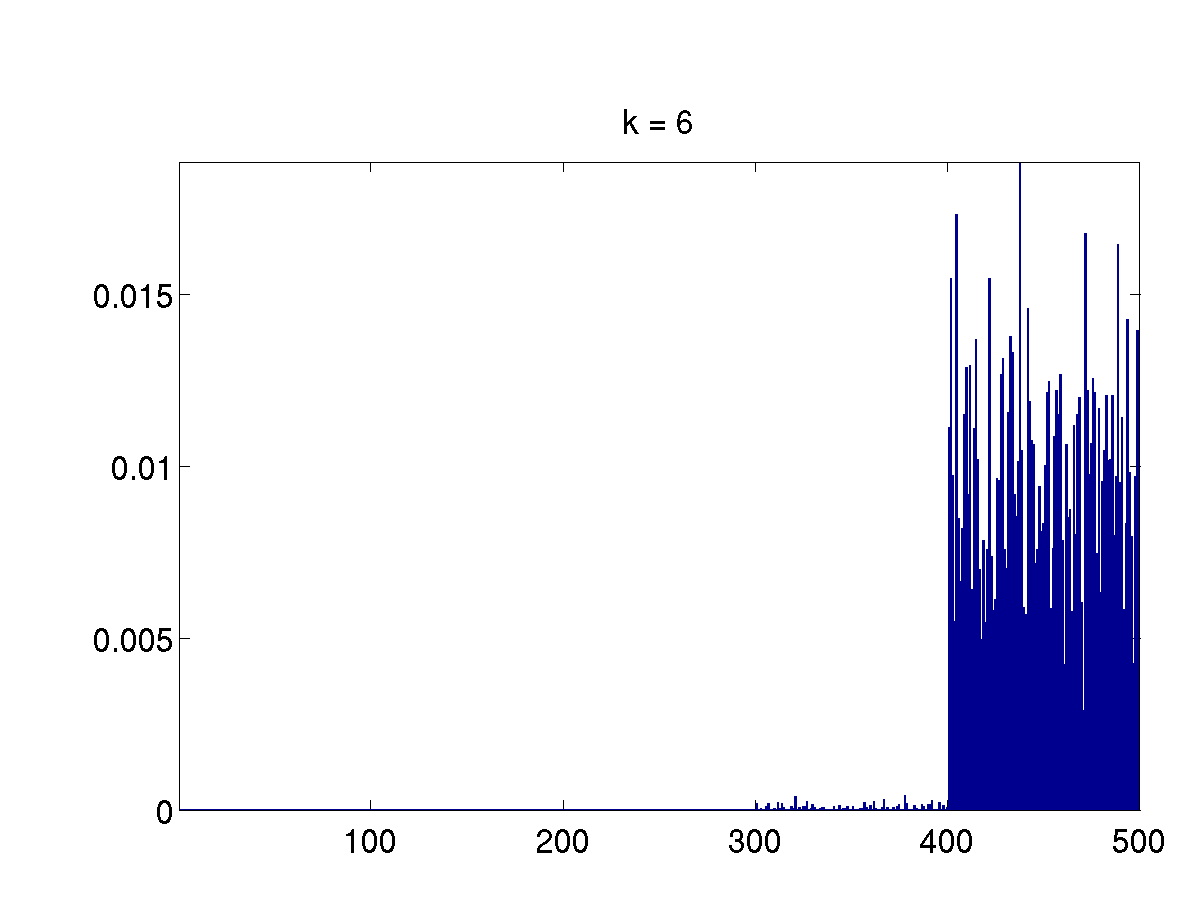} 
\includegraphics[width=0.16 \columnwidth]{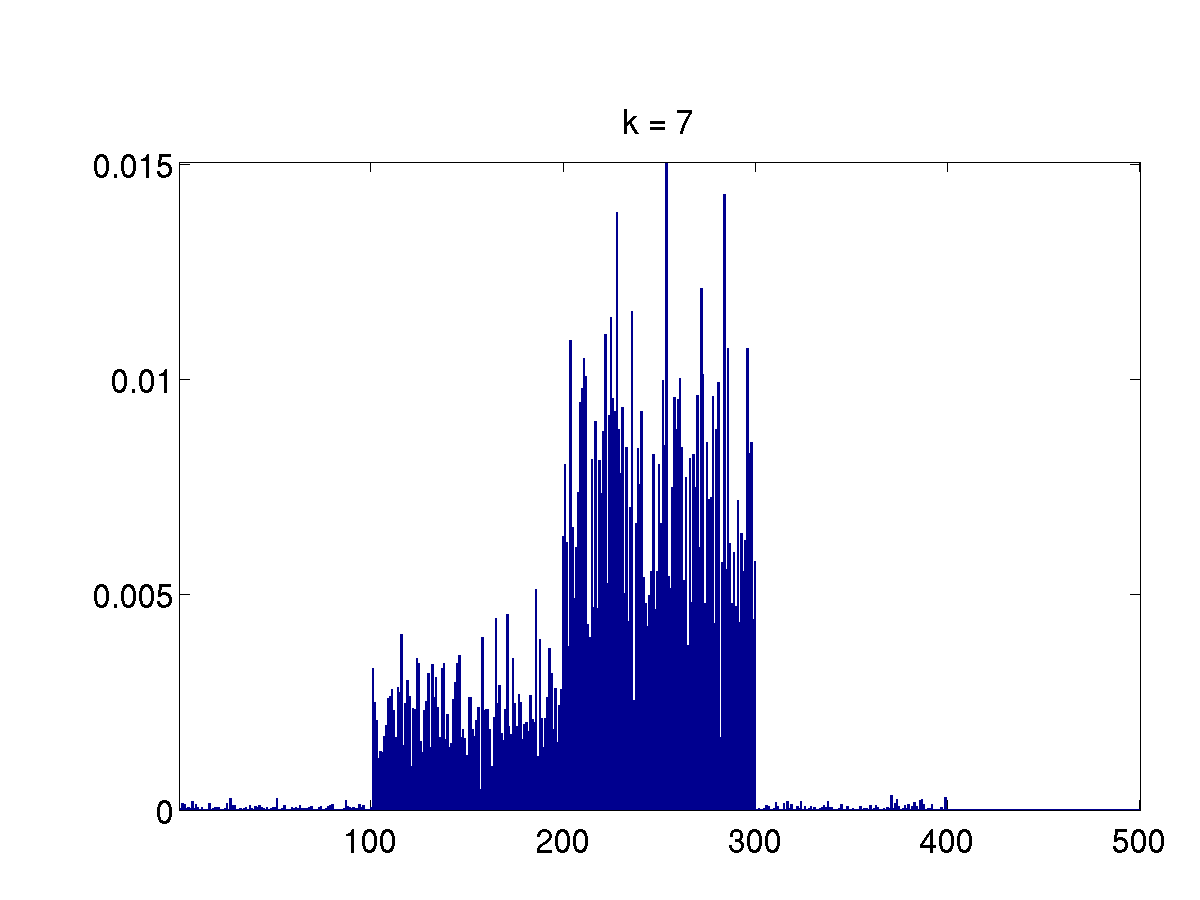}
\includegraphics[width=0.16 \columnwidth]{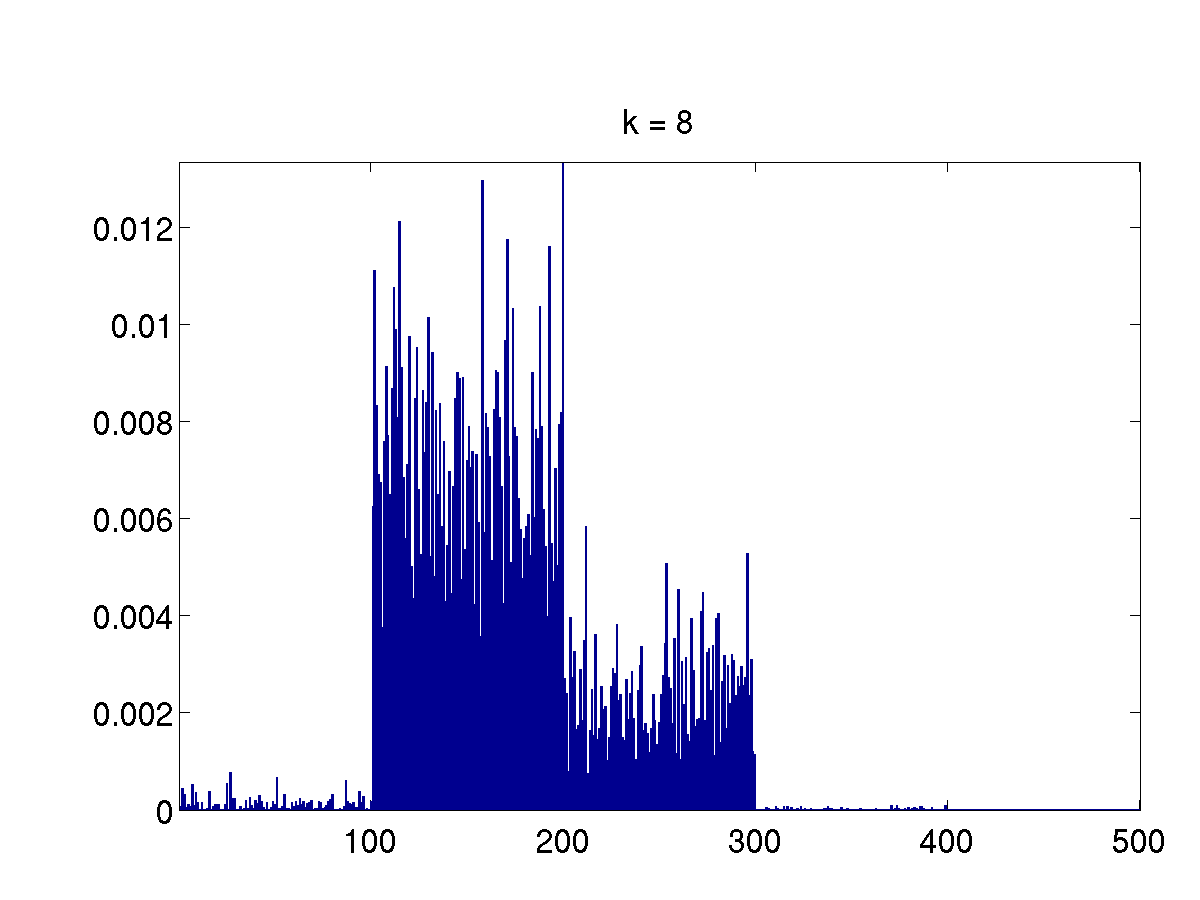}
\includegraphics[width=0.16 \columnwidth]{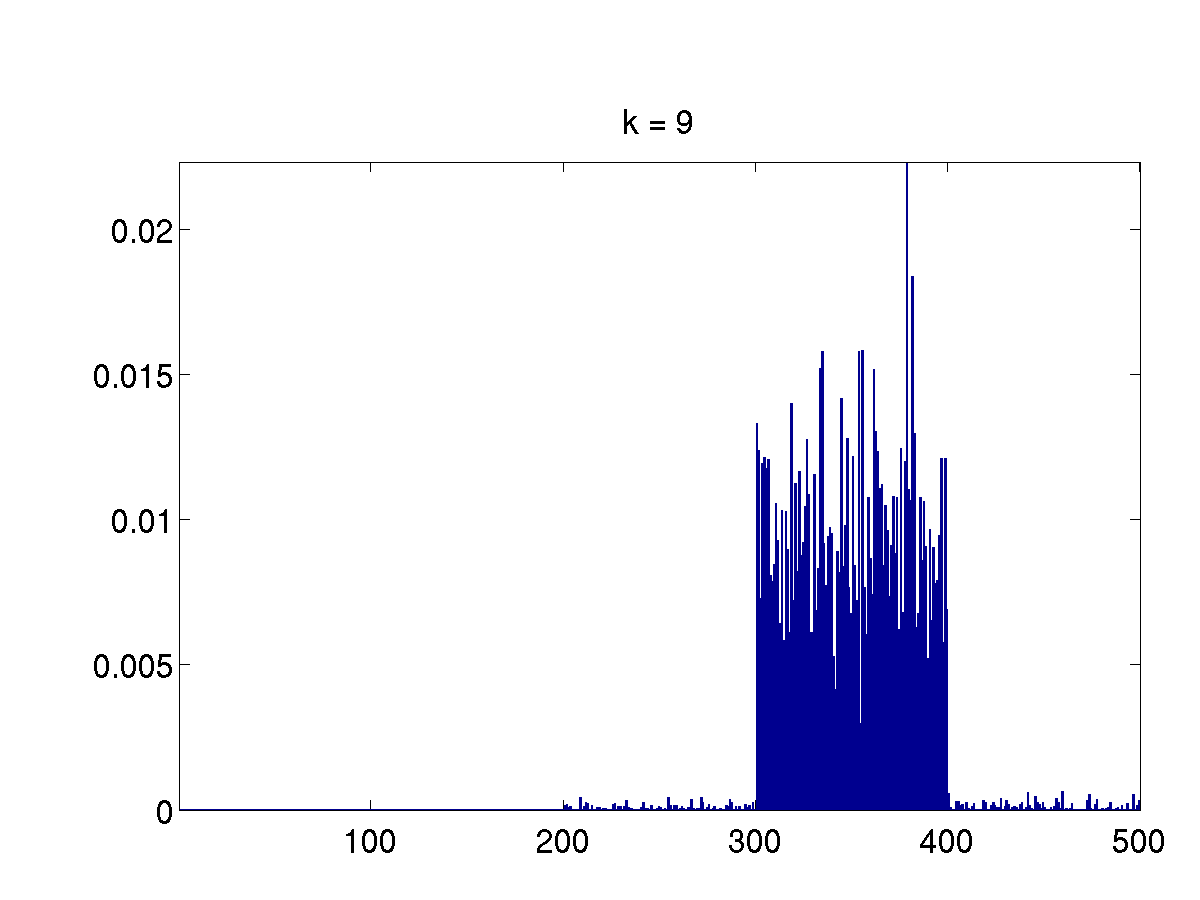}
\includegraphics[width=0.16 \columnwidth]{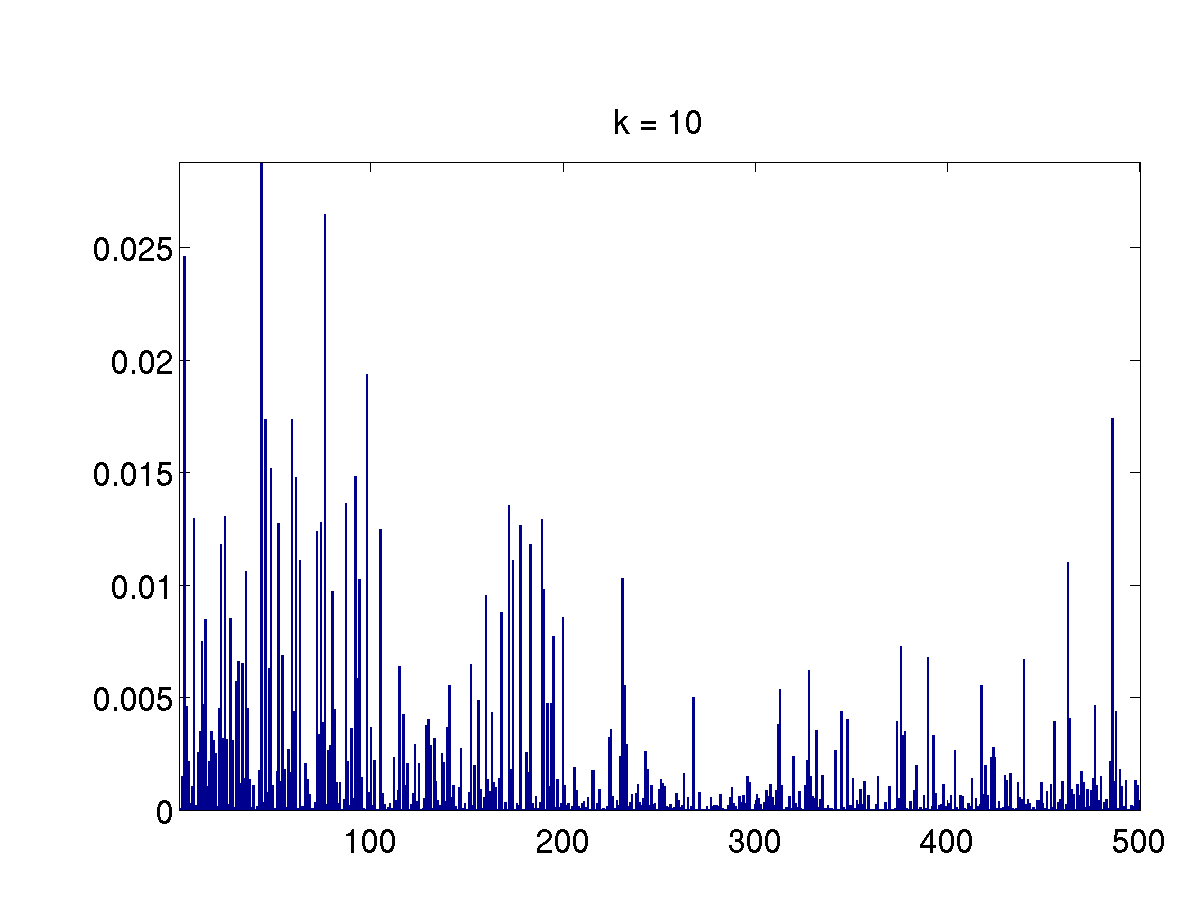}
\includegraphics[width=0.16 \columnwidth]{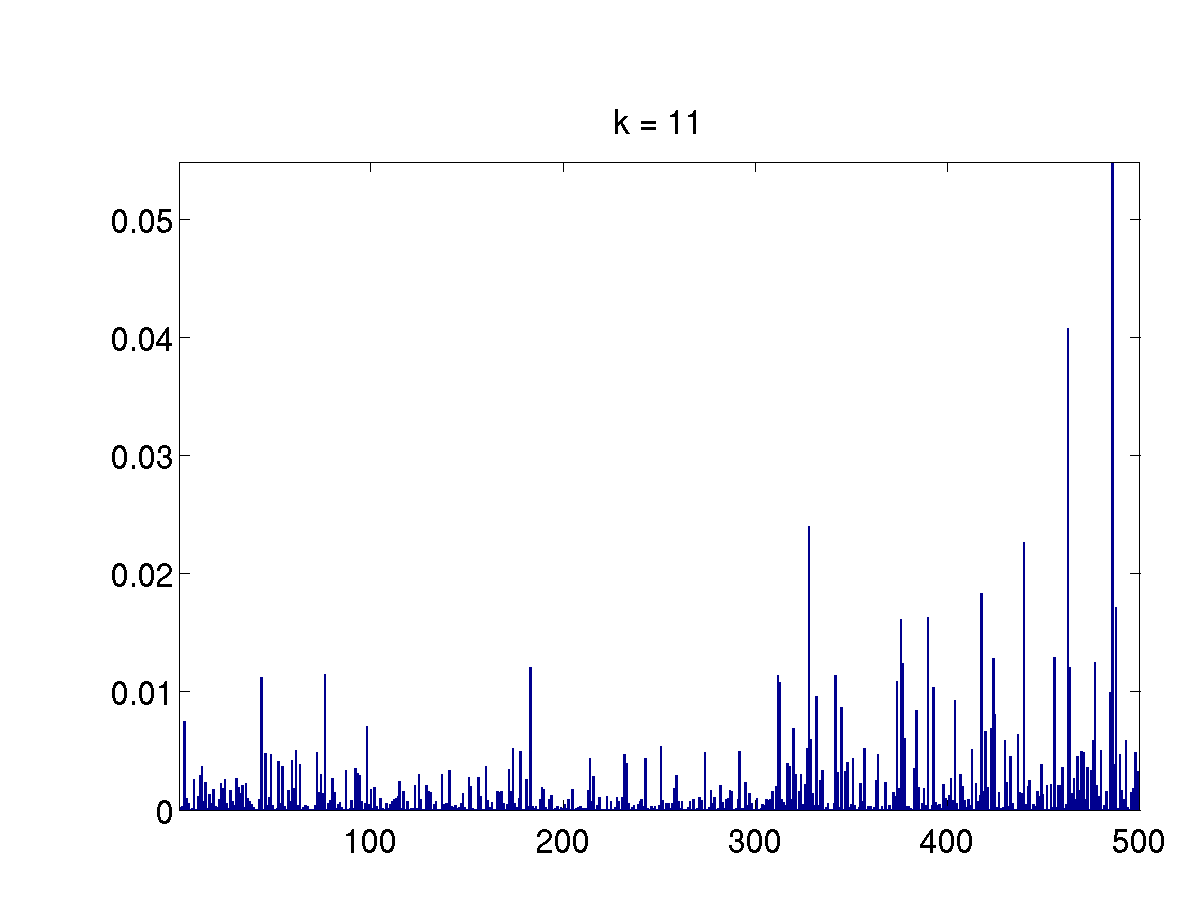} 
\includegraphics[width=0.16 \columnwidth]{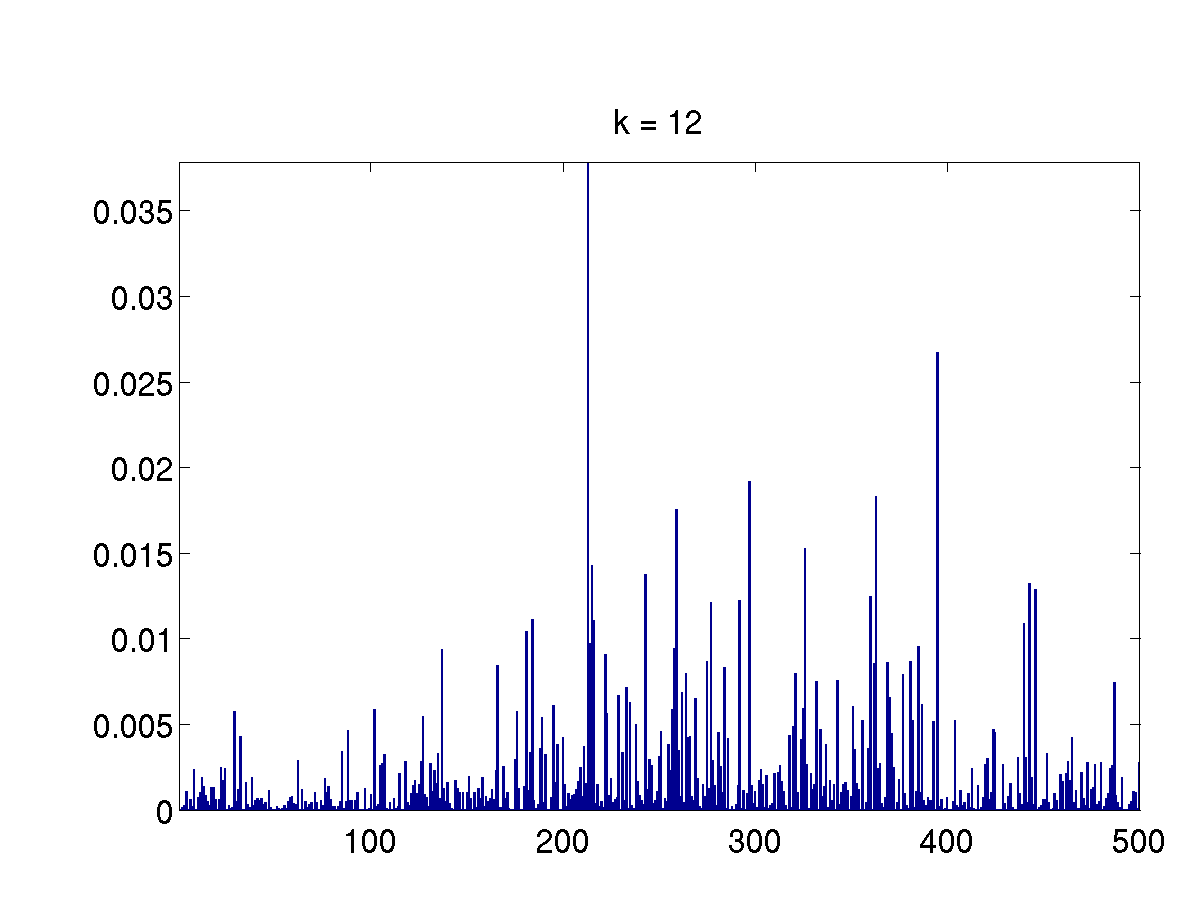} \\
\end{center}
\caption{Results from the \textsc{TwoLevel} model, where the parameters have 
been set as a ``path graph of $2$-modules,'' with edge densities $p_1=0.8$, 
$p_2=0.2$, and where a pair of nodes from consecutive $2$-modules are 
connected with probability $p=0.05$. 
Top left is a pictorial illustration of the graph in the form of a ``spy'' plot.
Top middle is the IPR scores, as a function of the rank of the eigenvector.
Top right is a barplot of the normalized square spectrum, \emph{i.e.}, 
$\frac{\lambda_i^2}{\sum_{j=1}^{n} \lambda_j^2}$ for $i=1,\ldots,65$. 
Next two rows are the top $12$ eigenvectors.
Last two rows are the corresponding statistical leverage scores.}
\label{fig:model1-2mod_on_line}
\end{figure}

To begin, Figure~\ref{fig:model1-2mod_on_line} illustrates a graph 
consisting of several hundred nodes organized as a ``path graph of 
$2$-modules''; that is, it consists of five $2$-modules connected together 
as beads along a one-dimensional scaffolding.  
(All of the figures for the behavior of the \textsc{TwoLevel} model contain 
a subset of: a pictorial illustration of the graph in the form of a ``spy'' 
plot; the IPR scores, as a function of the rank of the eigenvector; a 
barplot of the normalized square spectrum; plots of several of the 
eigenvectors; and the corresponding statistical leverage scores.
Figure~\ref{fig:model1-2mod_on_line} plots all of these quantities.)
The first four nontrivial eigenvectors in 
Figure~\ref{fig:model1-2mod_on_line} are fairly constant along each of the 
beads; and they exhibit the characteristic sinusoidal oscillations that one 
would expect from eigenfunctions of the Laplacian on the continuous line or 
a discrete path graph.
The next five eigenvectors are much more localized; and they tend to be 
localized either on a single bead at the endpoints of the path or on a small 
number of nearby beads in the middle of the path.
In addition, on the fifth and sixth eigenfunction, which are localized on a 
single $2$-module, there is a natural partition of that $2$-module based on 
the sign of that eigenvector, and that partition splits the $2$-module into 
the two separate~modules.
Later eigenvectors are still more localized than leading-order 
eigenvectors, at least by the IPR measure, but they do not seem to be 
localized in such a way as to yield insight into the data.
 
%TMP% 

\begin{figure}[t] 
\begin{center}
\includegraphics[width=0.32\columnwidth]{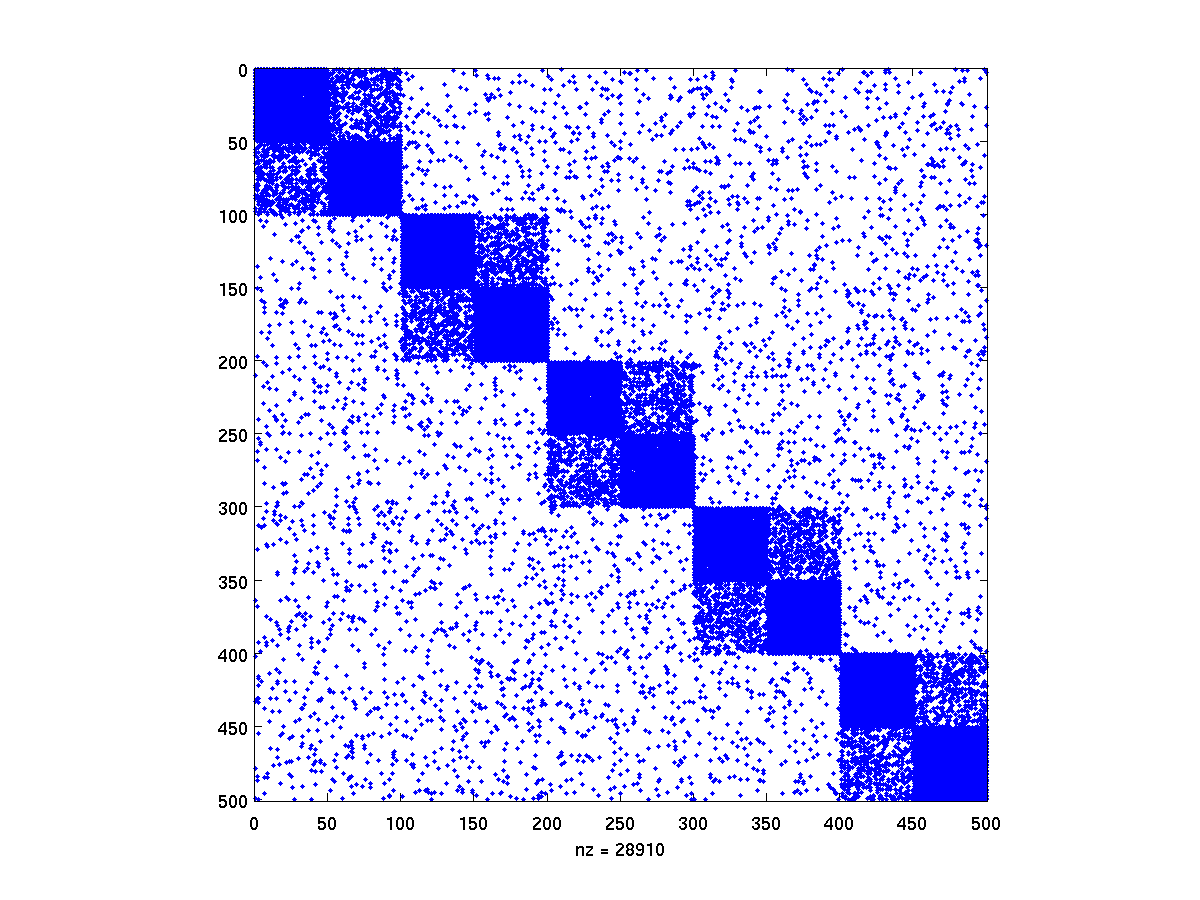}
\includegraphics[width=0.32\columnwidth]{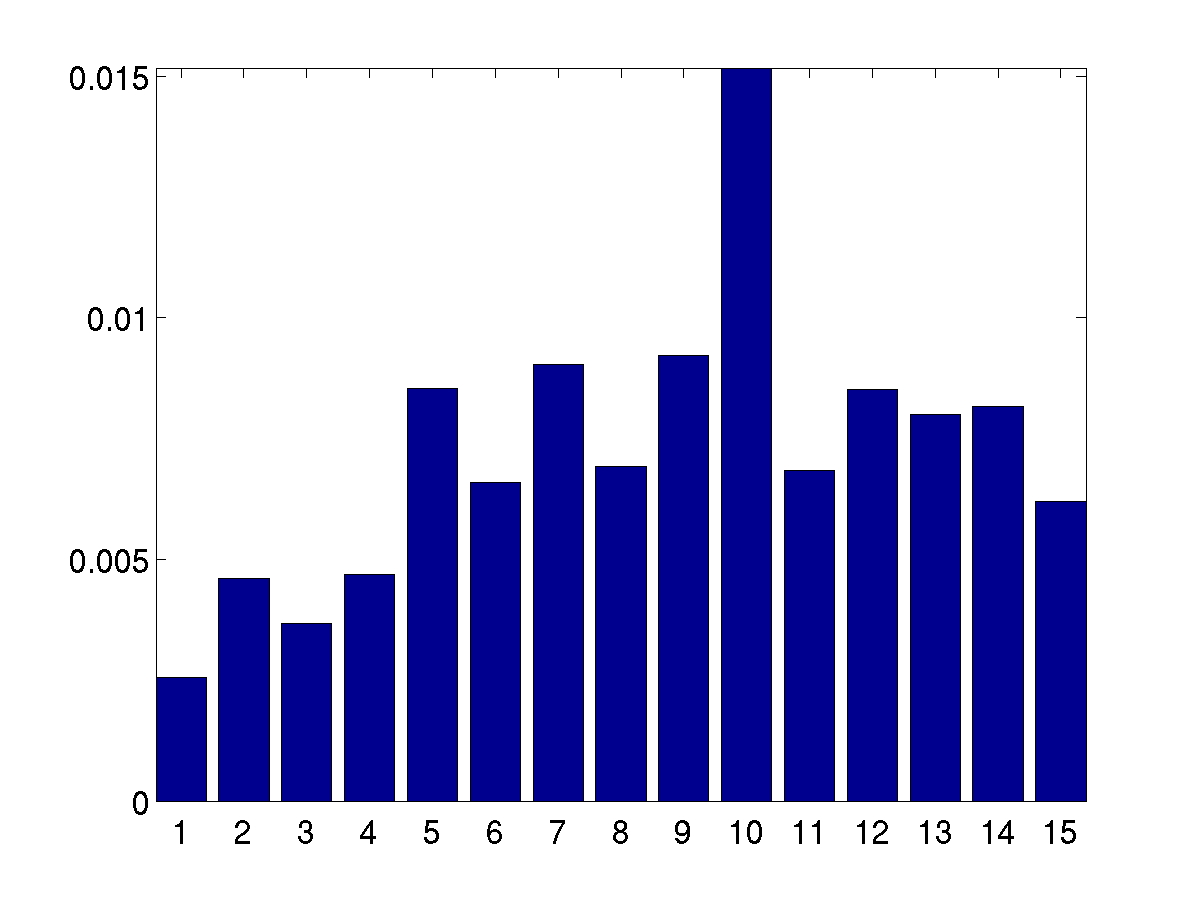}
\includegraphics[width=0.32\columnwidth]{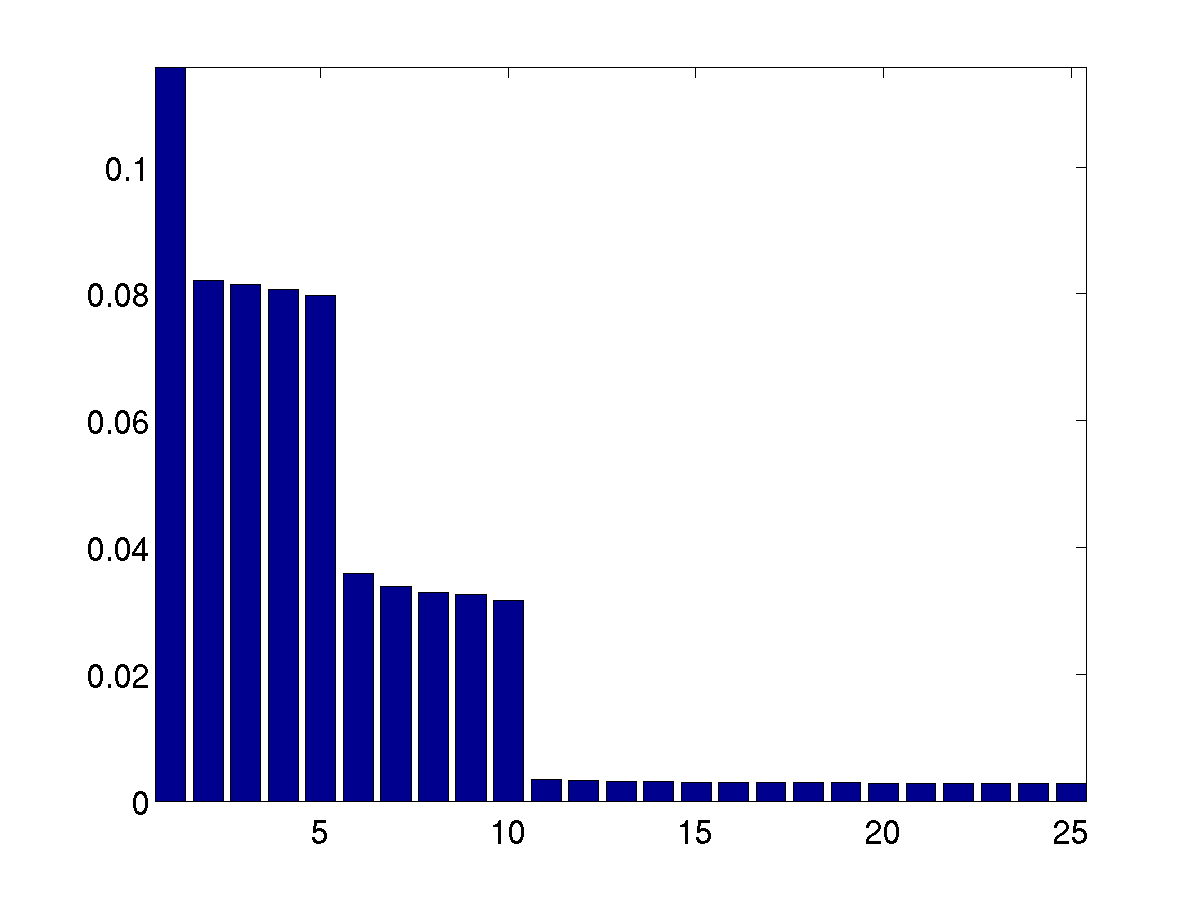}
\includegraphics[width=0.19 \columnwidth]{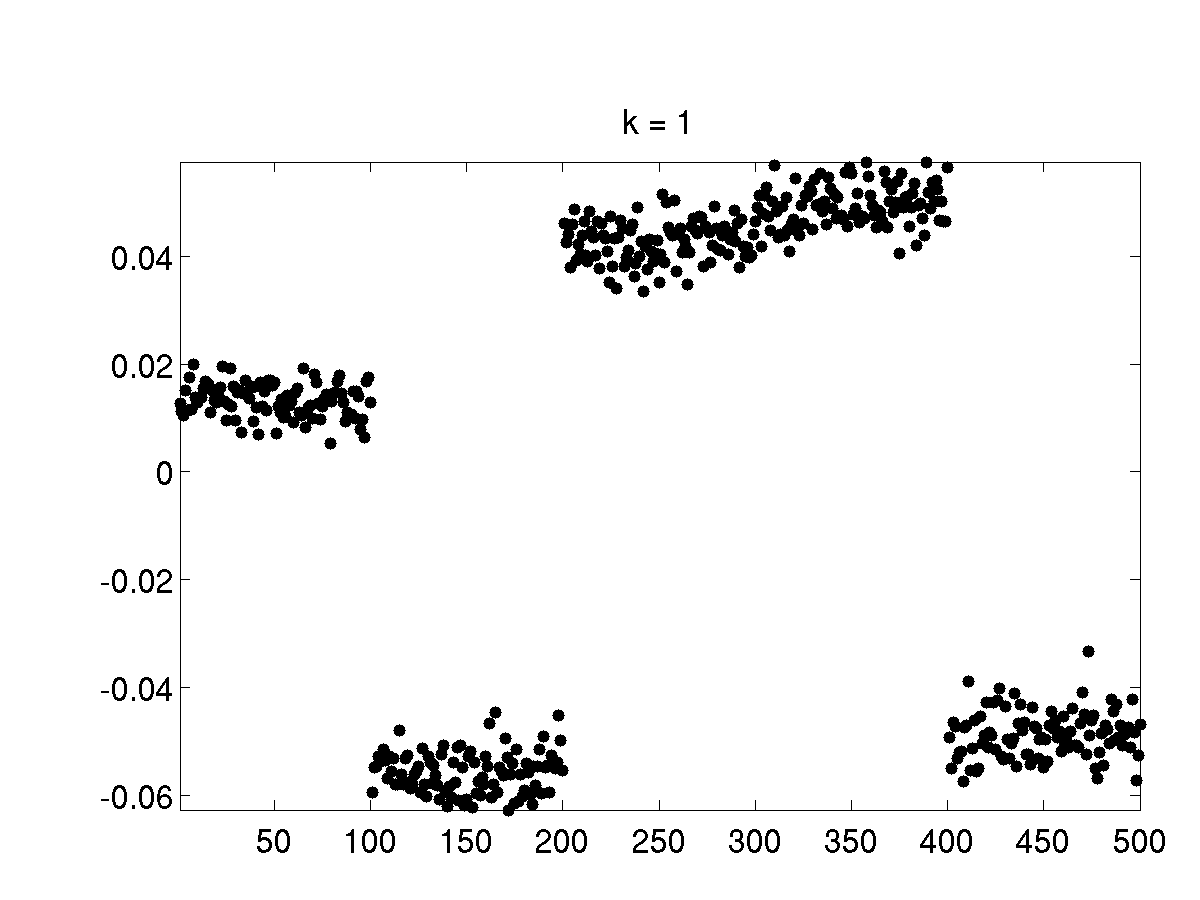}
\includegraphics[width=0.19 \columnwidth]{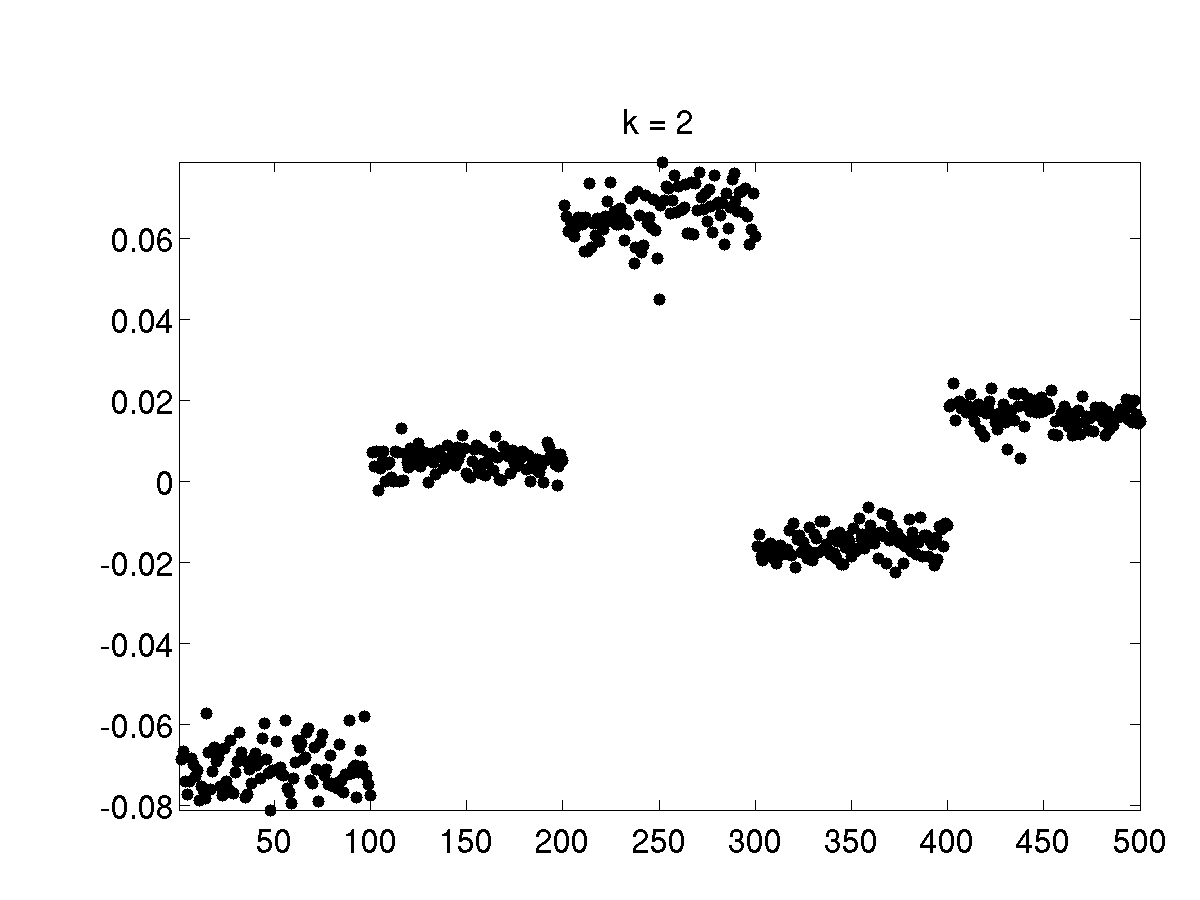}
\includegraphics[width=0.19 \columnwidth]{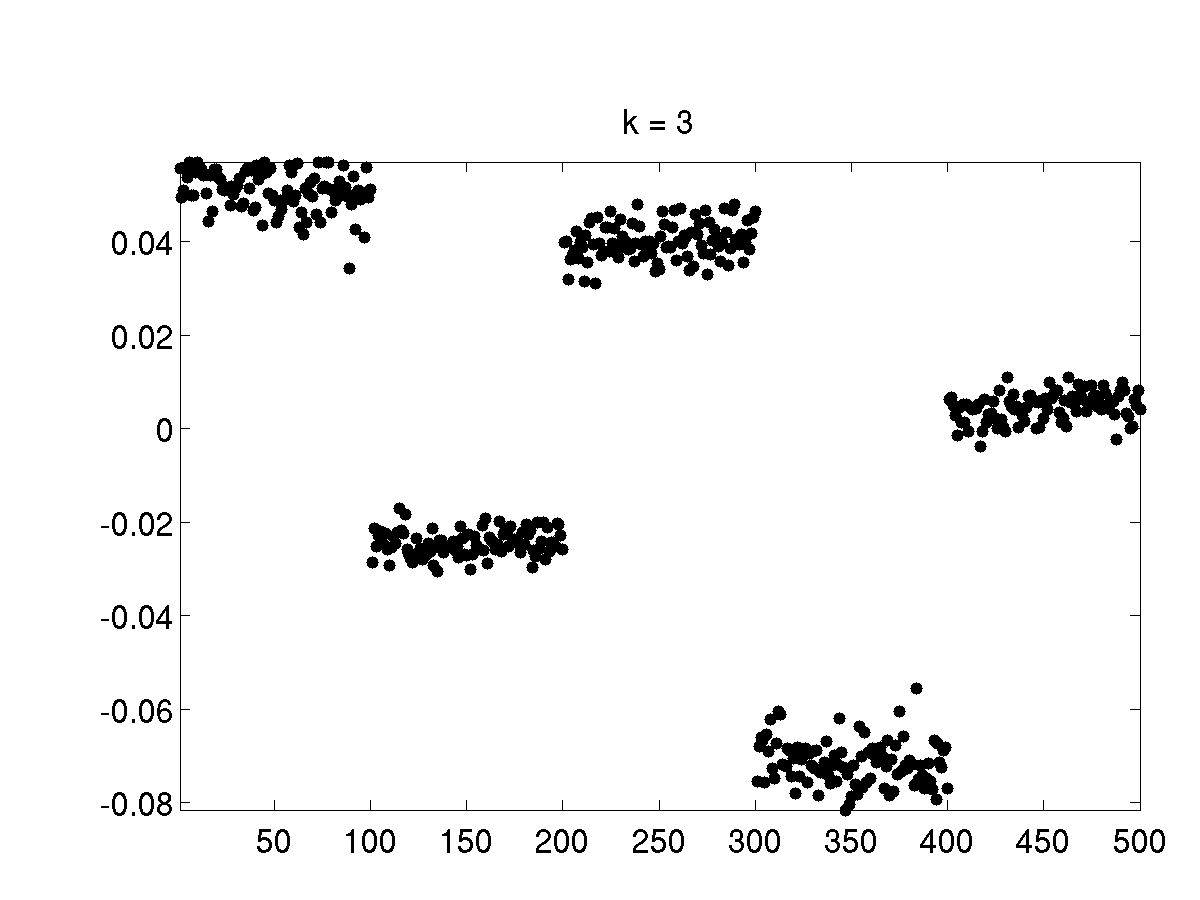}
\includegraphics[width=0.19 \columnwidth]{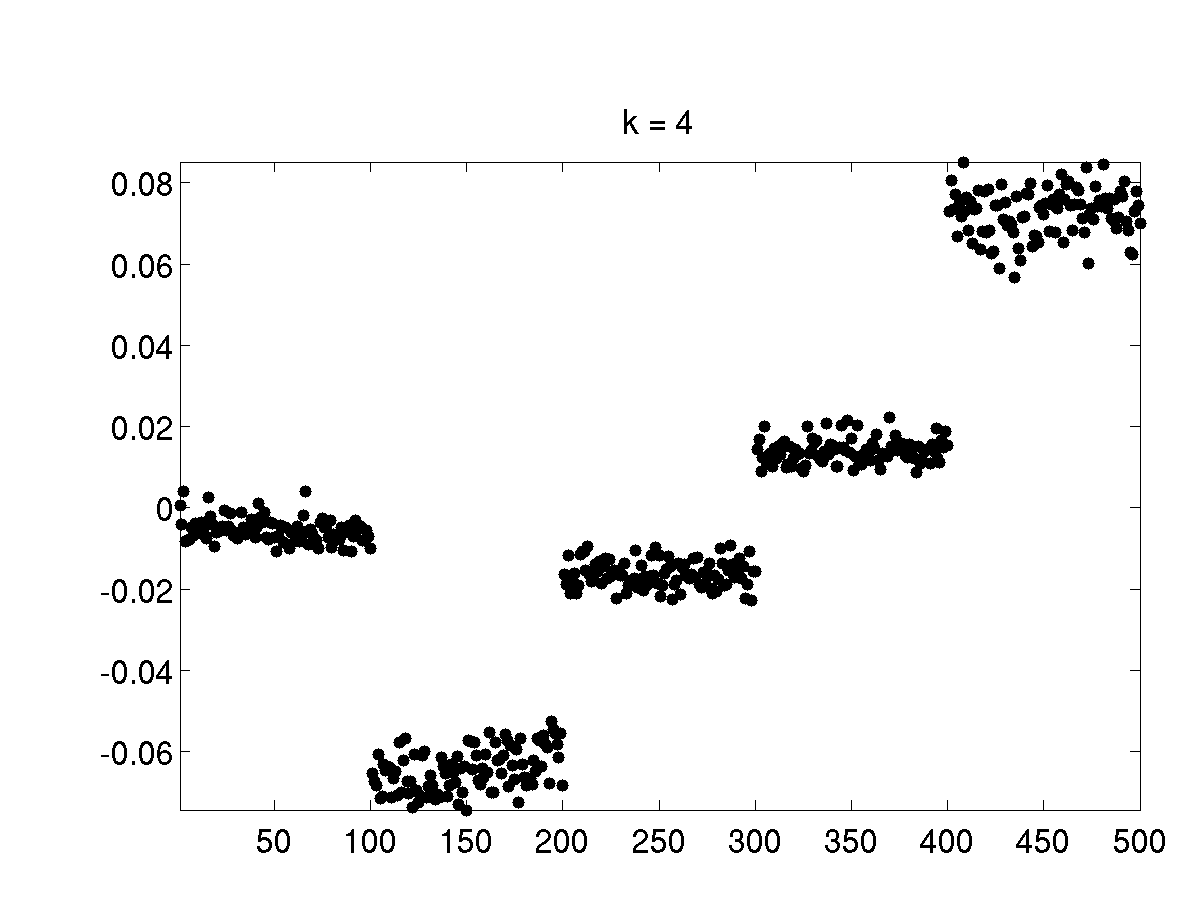}
\includegraphics[width=0.19 \columnwidth]{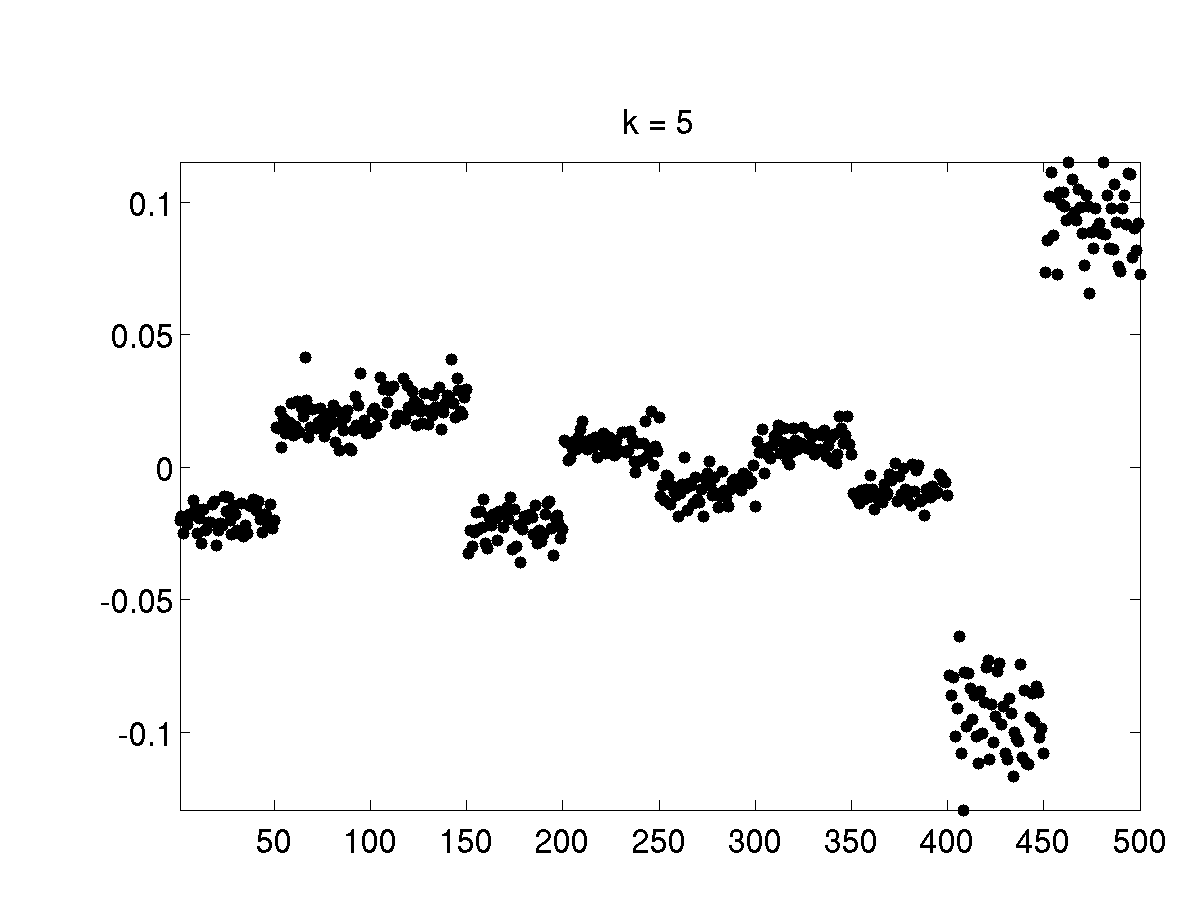}
\includegraphics[width=0.19 \columnwidth]{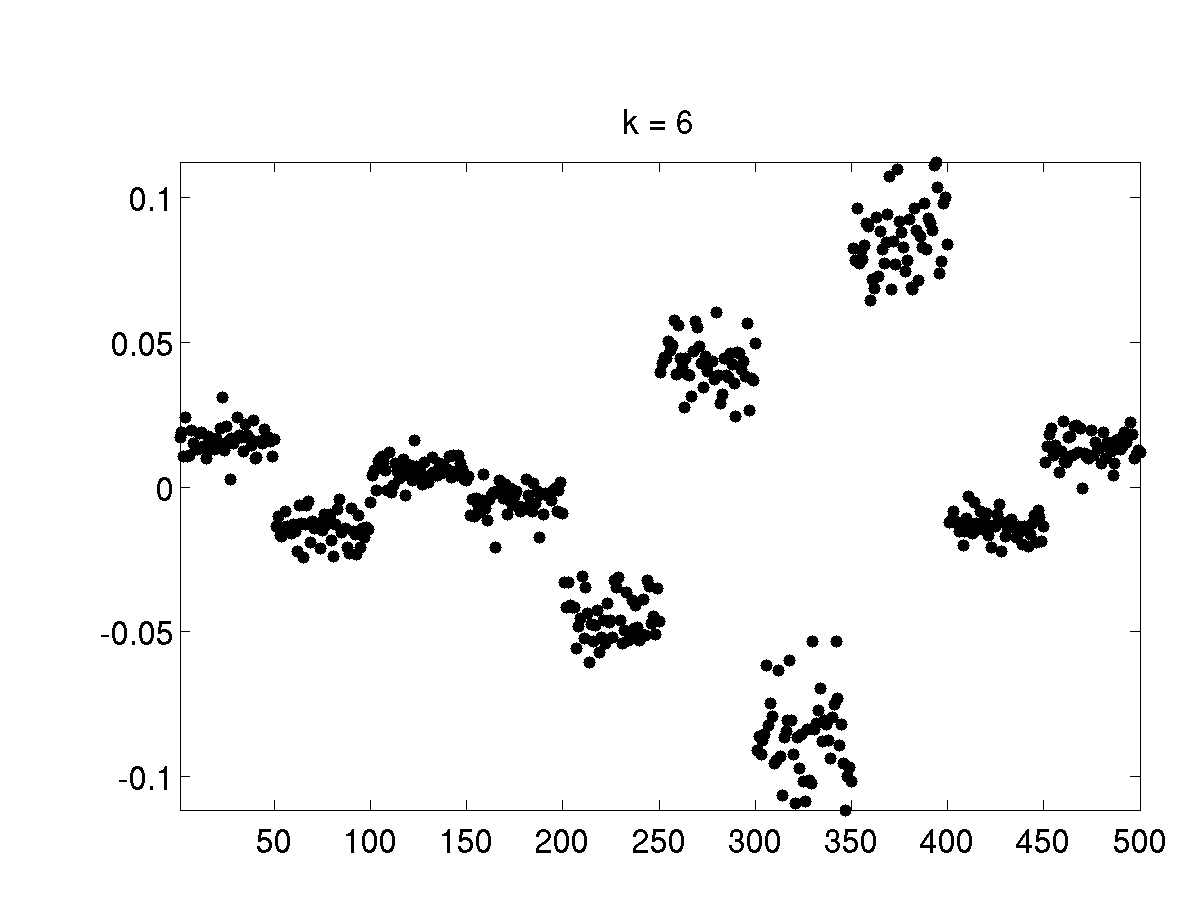}
\includegraphics[width=0.19 \columnwidth]{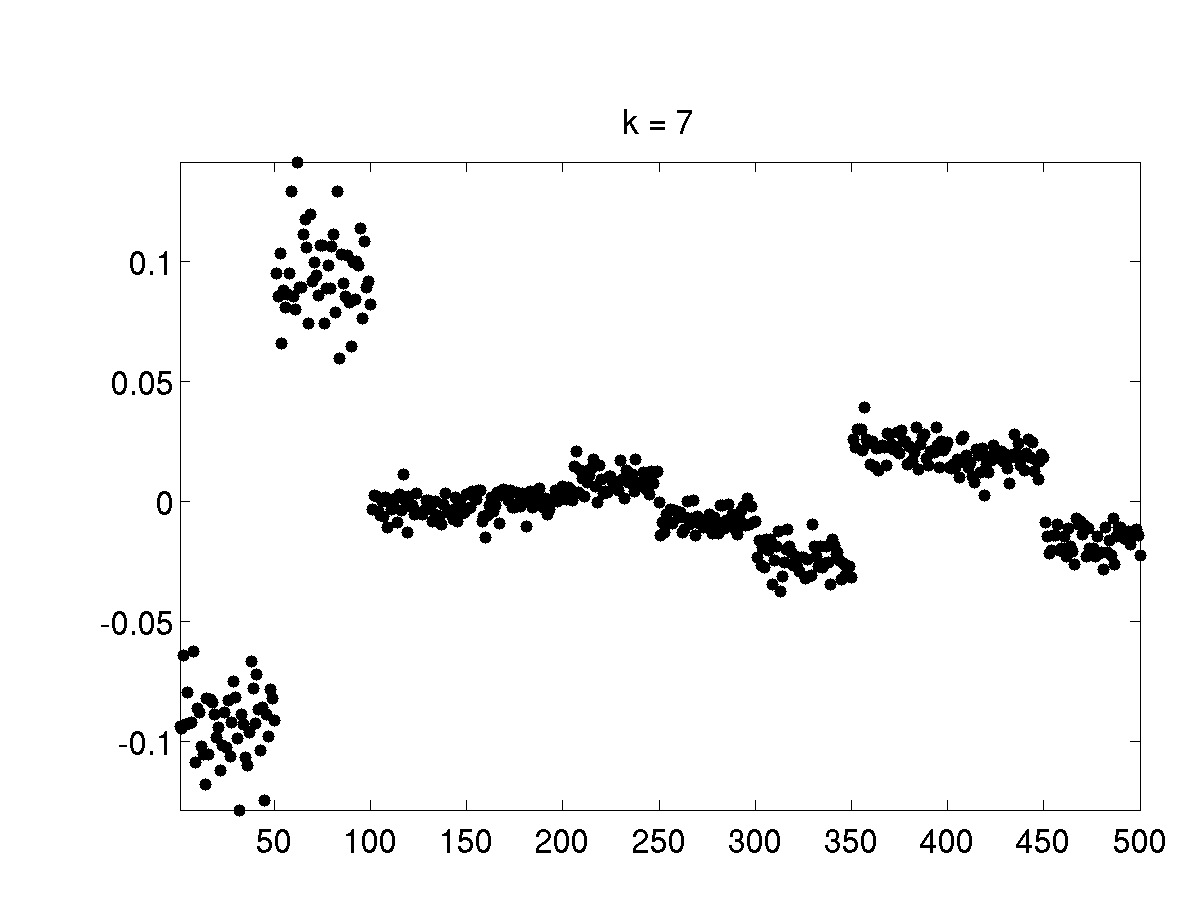}
\includegraphics[width=0.19 \columnwidth]{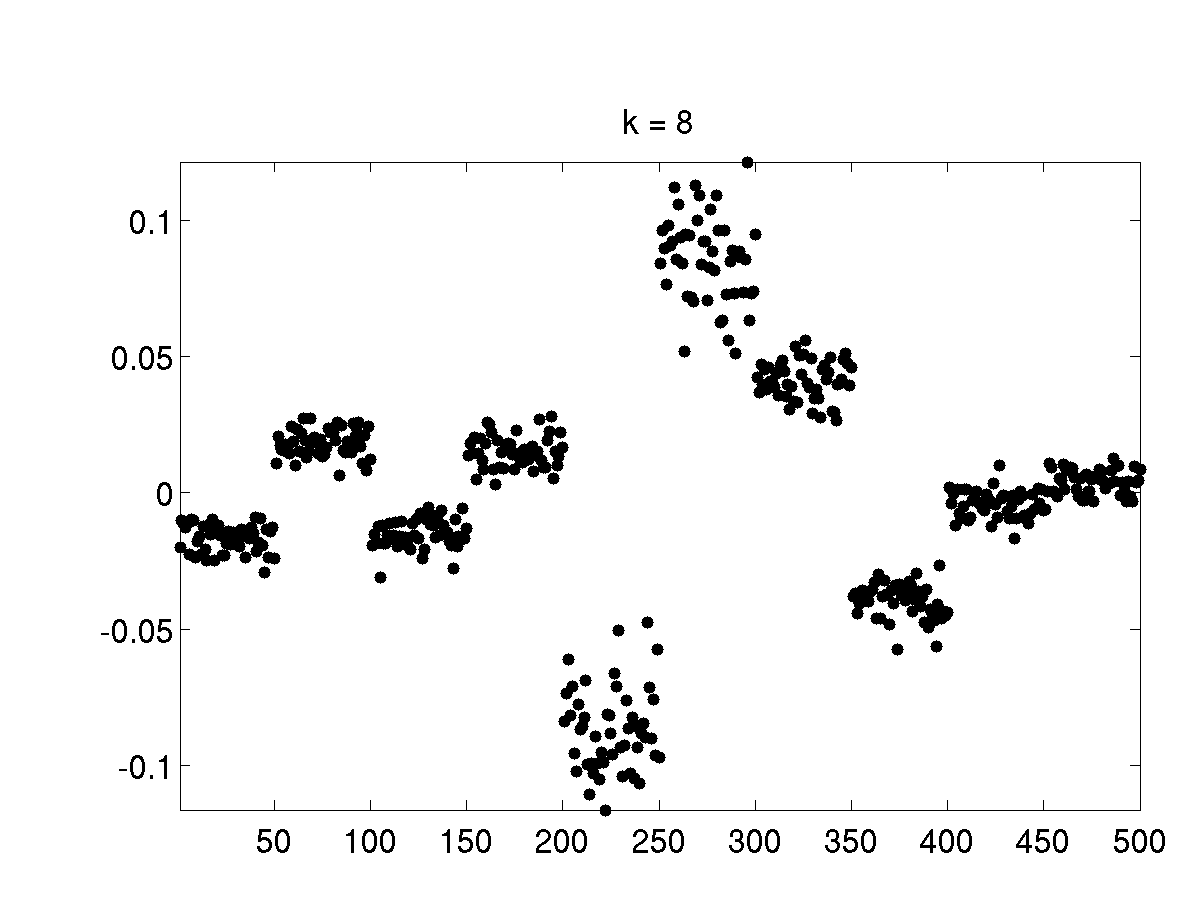}
\includegraphics[width=0.19 \columnwidth]{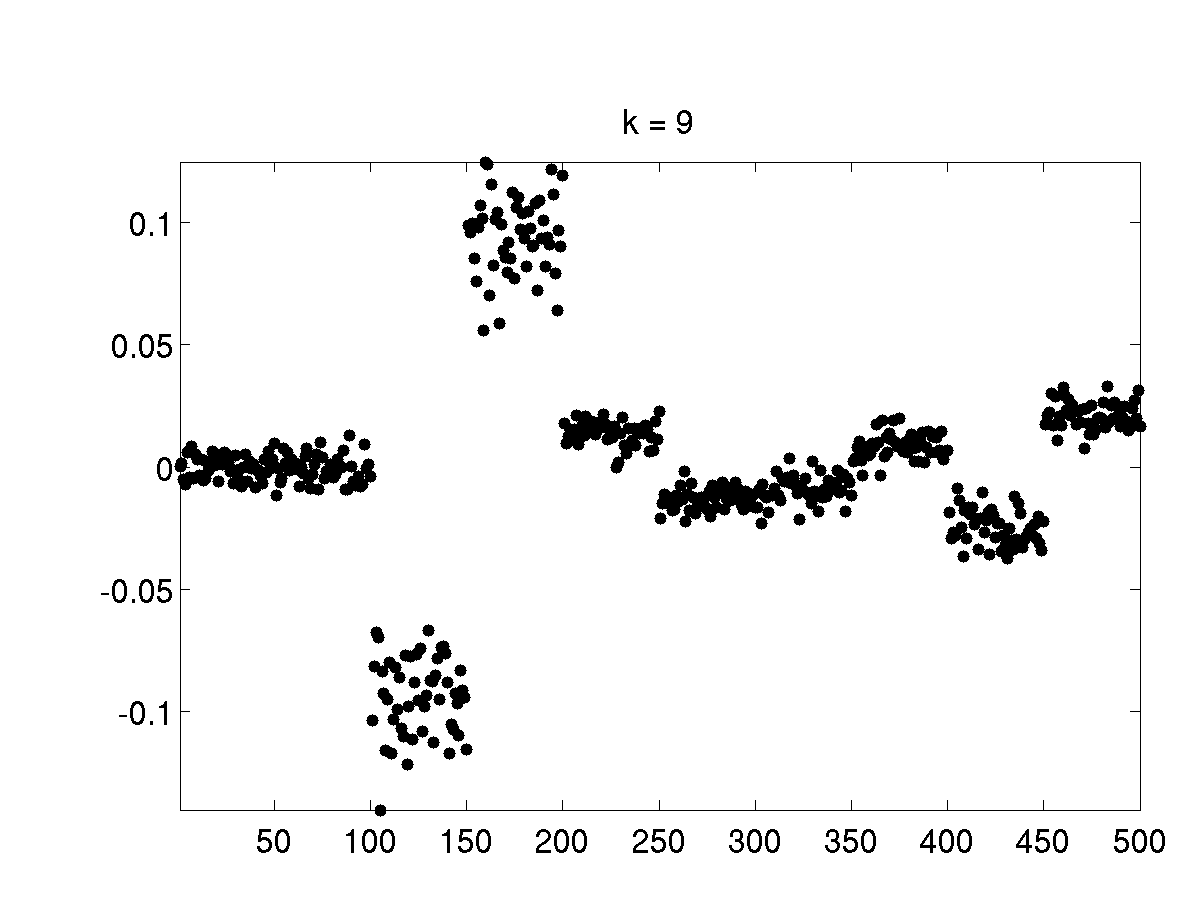}
\includegraphics[width=0.19 \columnwidth]{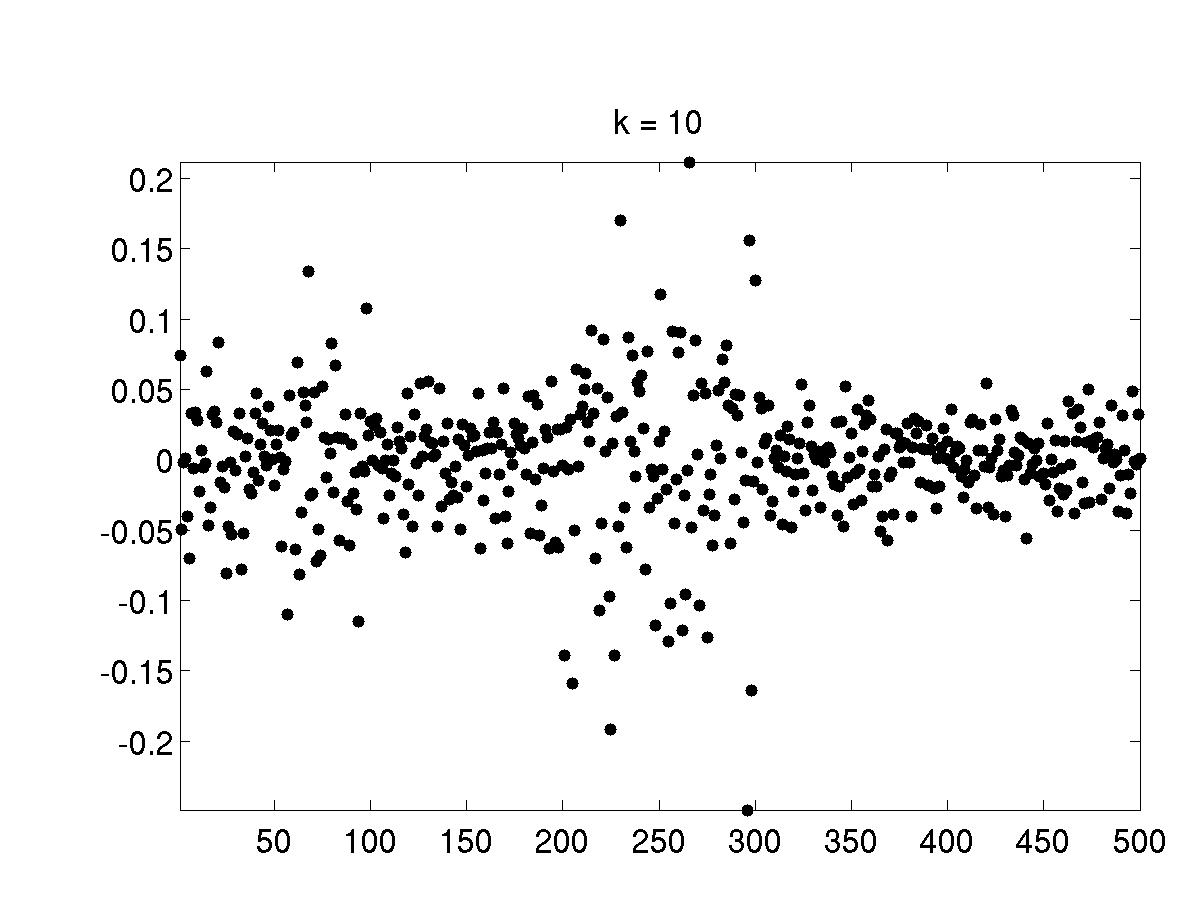}
\end{center}
\caption{Results from the \textsc{TwoLevel} model, where the parameters have 
been set as a ``unstructured graph of $2$-modules,'' where each $2$-module 
$W$ has edge densities $p_1=0.8$ and $p_2=0.2$, and where the unstructured 
$N$ is a random grpah with $p= 0.02$. 
%%% and number of noise edges $ 100 \times 100 * 0.02 *  {5\choose 2} = 2000 $.
Shown are:
a pictorial illustration of the graph;
the IPR scores;
the normalized square spectrum; and
the top $10$ eigenvectors.
  }
\label{fig:model2-2mod_on_noise}
\end{figure}

\begin{figure}[t] 
\begin{center}
\includegraphics[width=0.32\columnwidth]{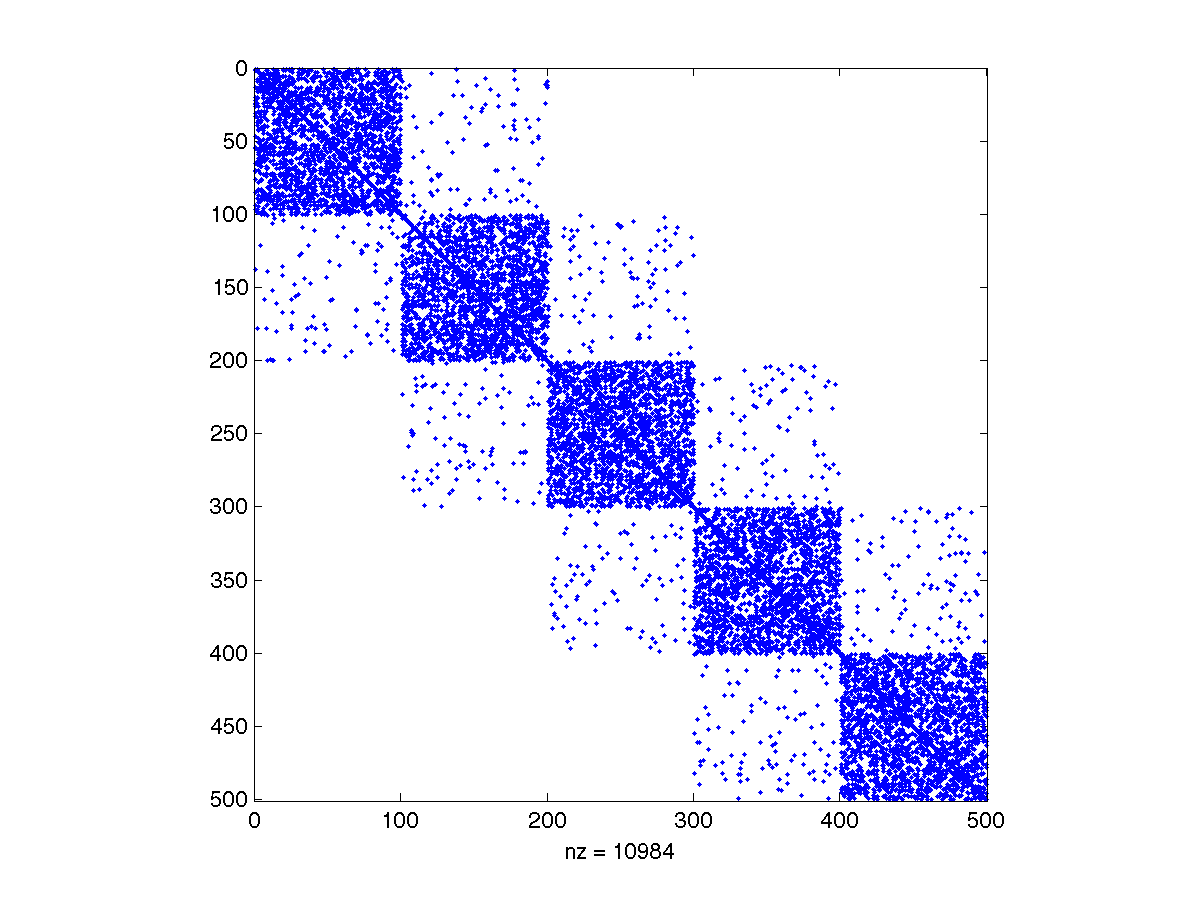}
\includegraphics[width=0.32\columnwidth]{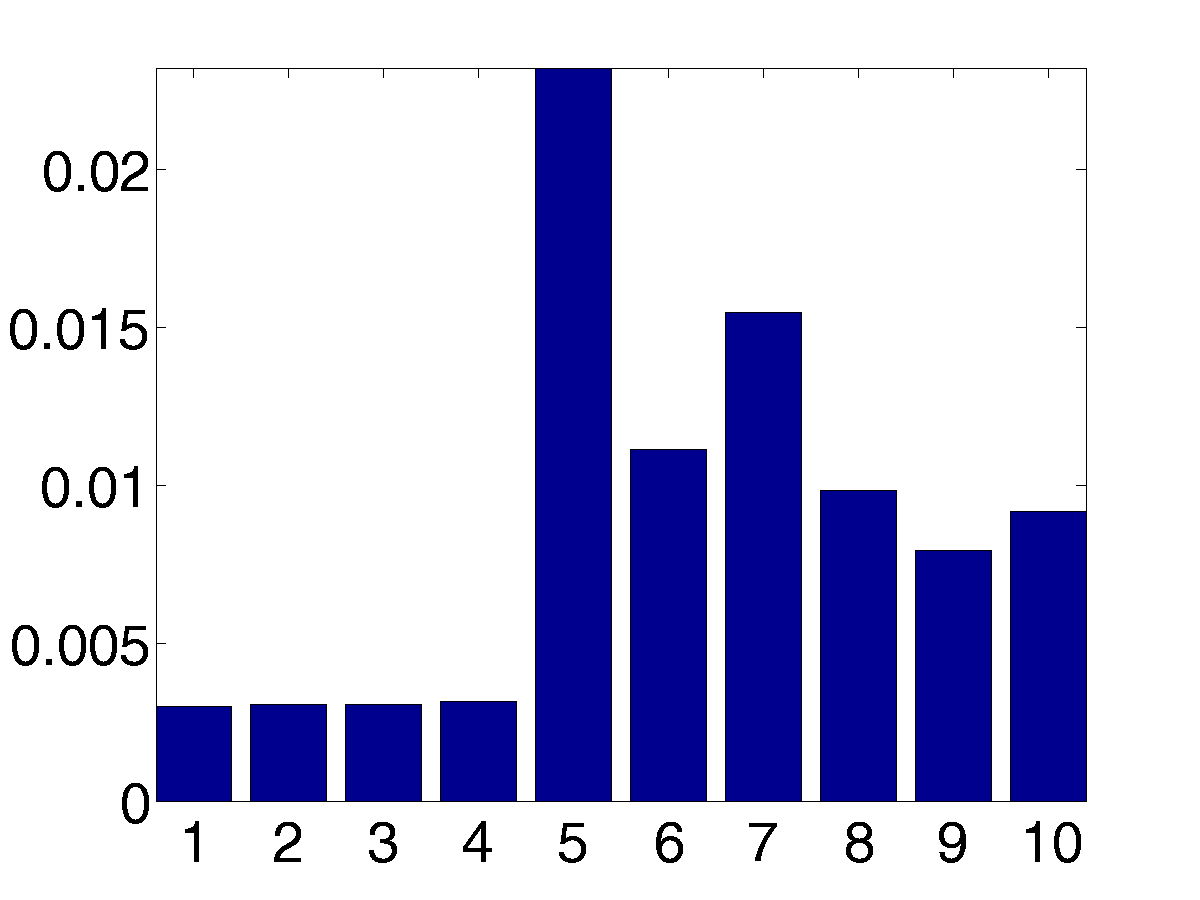}
\includegraphics[width=0.32\columnwidth]{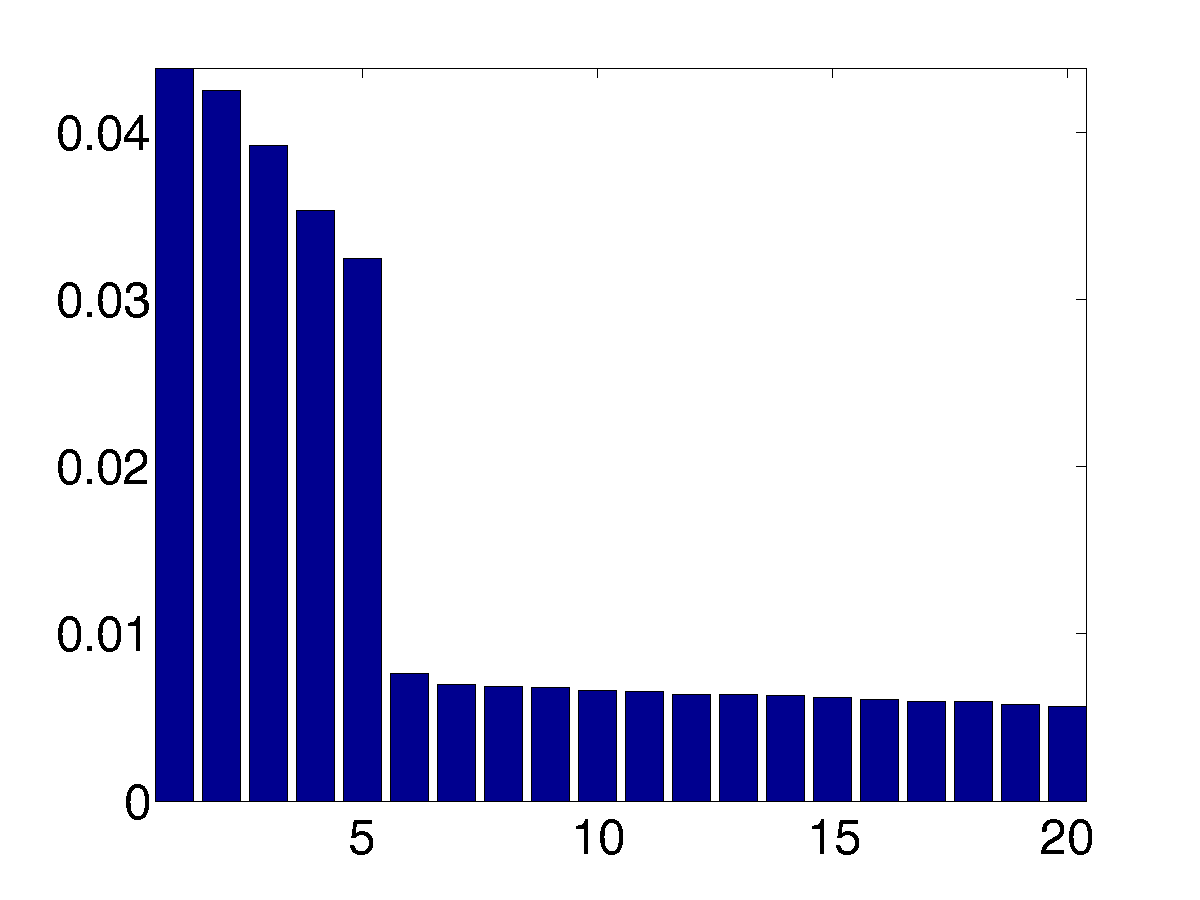}
\includegraphics[width=0.19 \columnwidth]{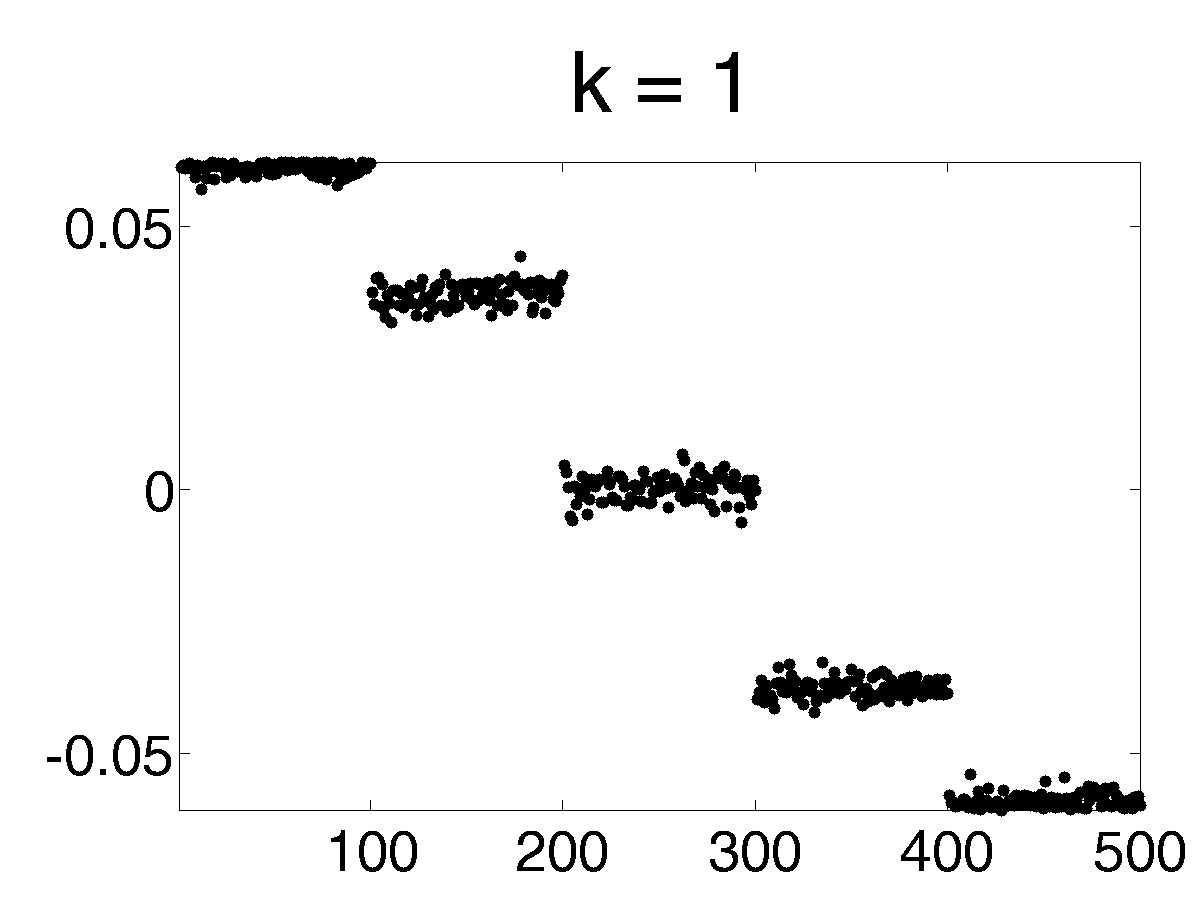}
\includegraphics[width=0.19 \columnwidth]{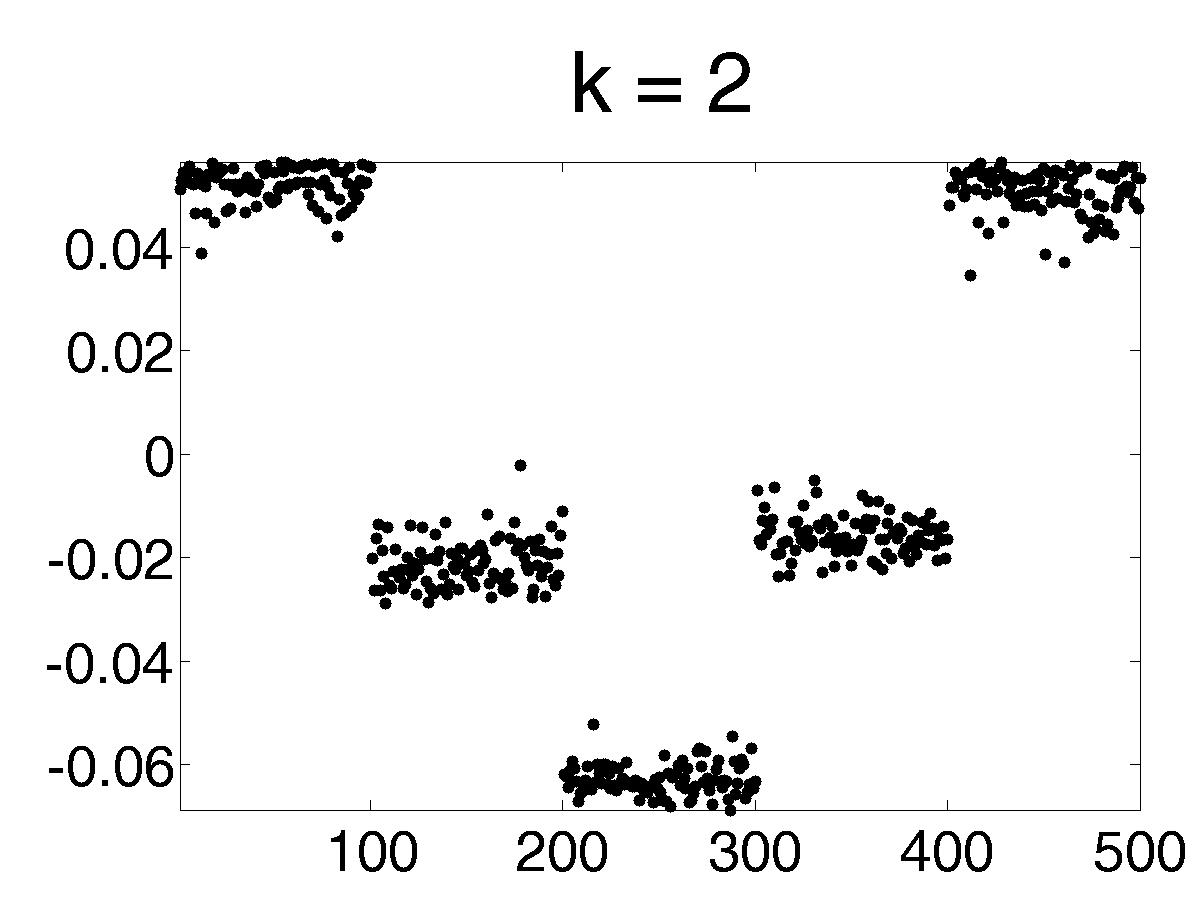}
\includegraphics[width=0.19 \columnwidth]{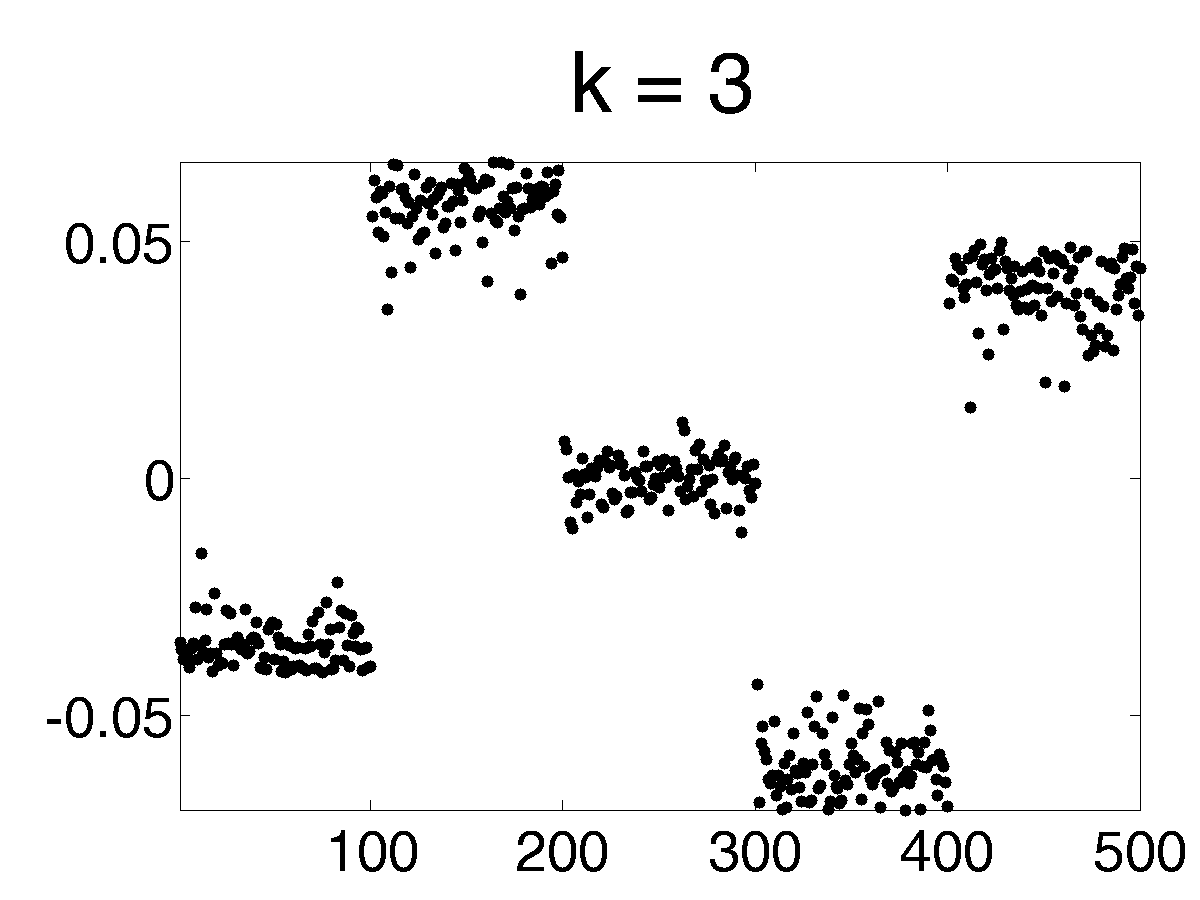}
\includegraphics[width=0.19 \columnwidth]{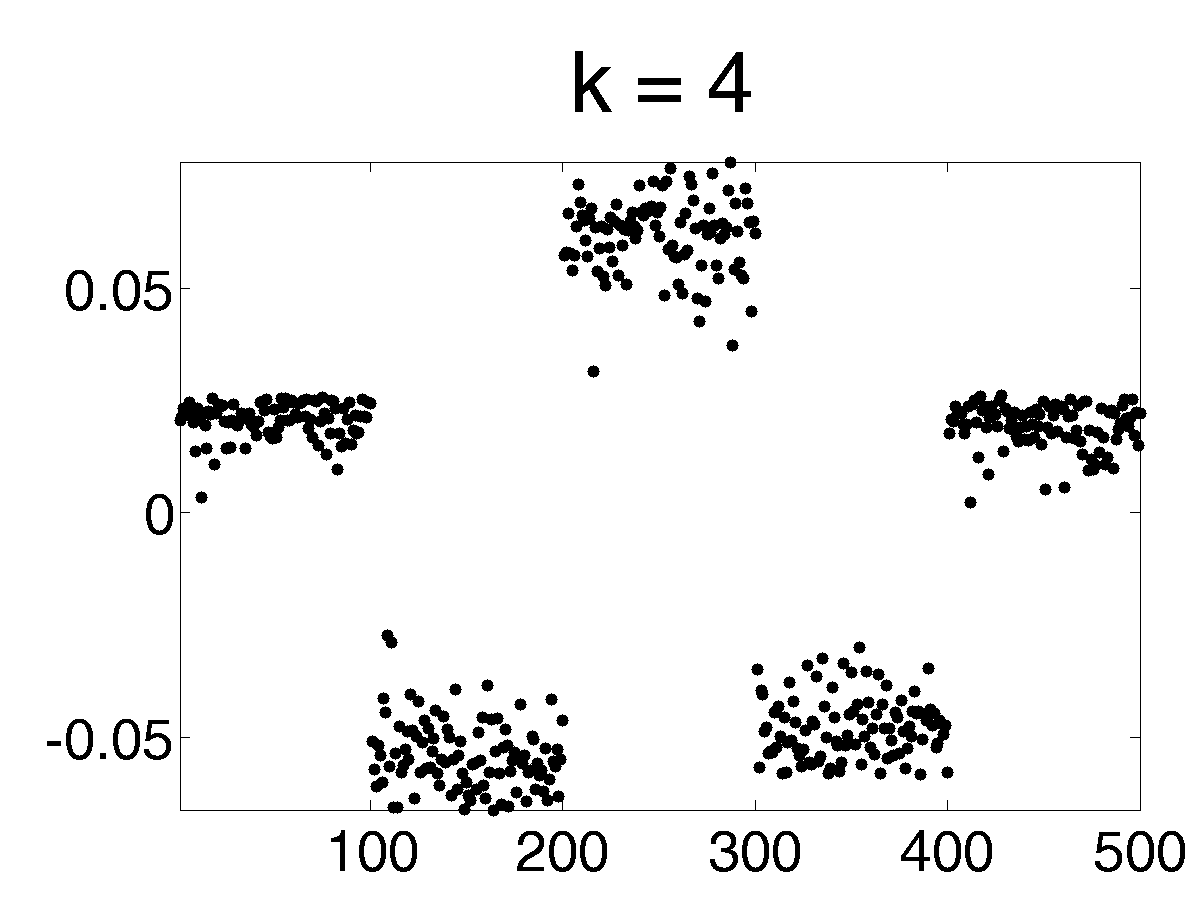}
\includegraphics[width=0.19 \columnwidth]{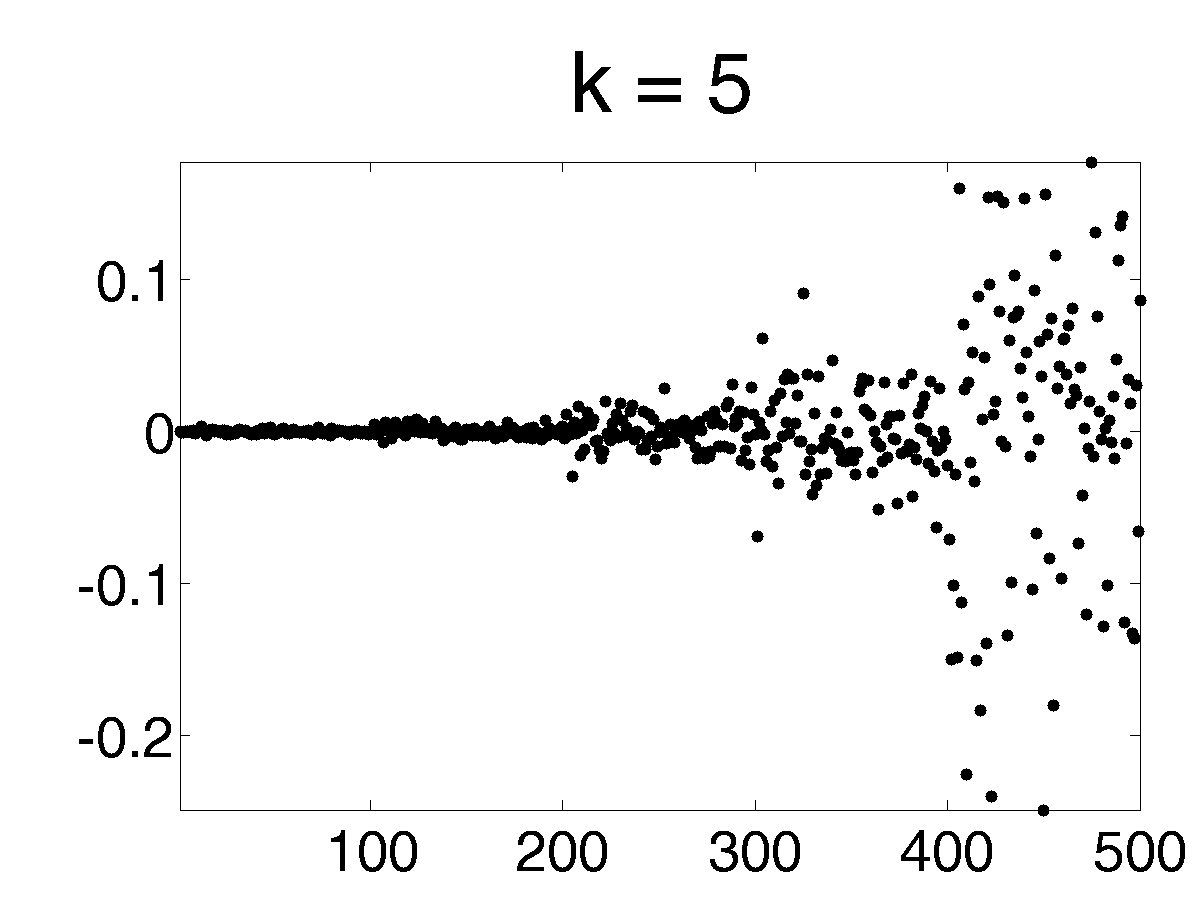}
\includegraphics[width=0.19 \columnwidth]{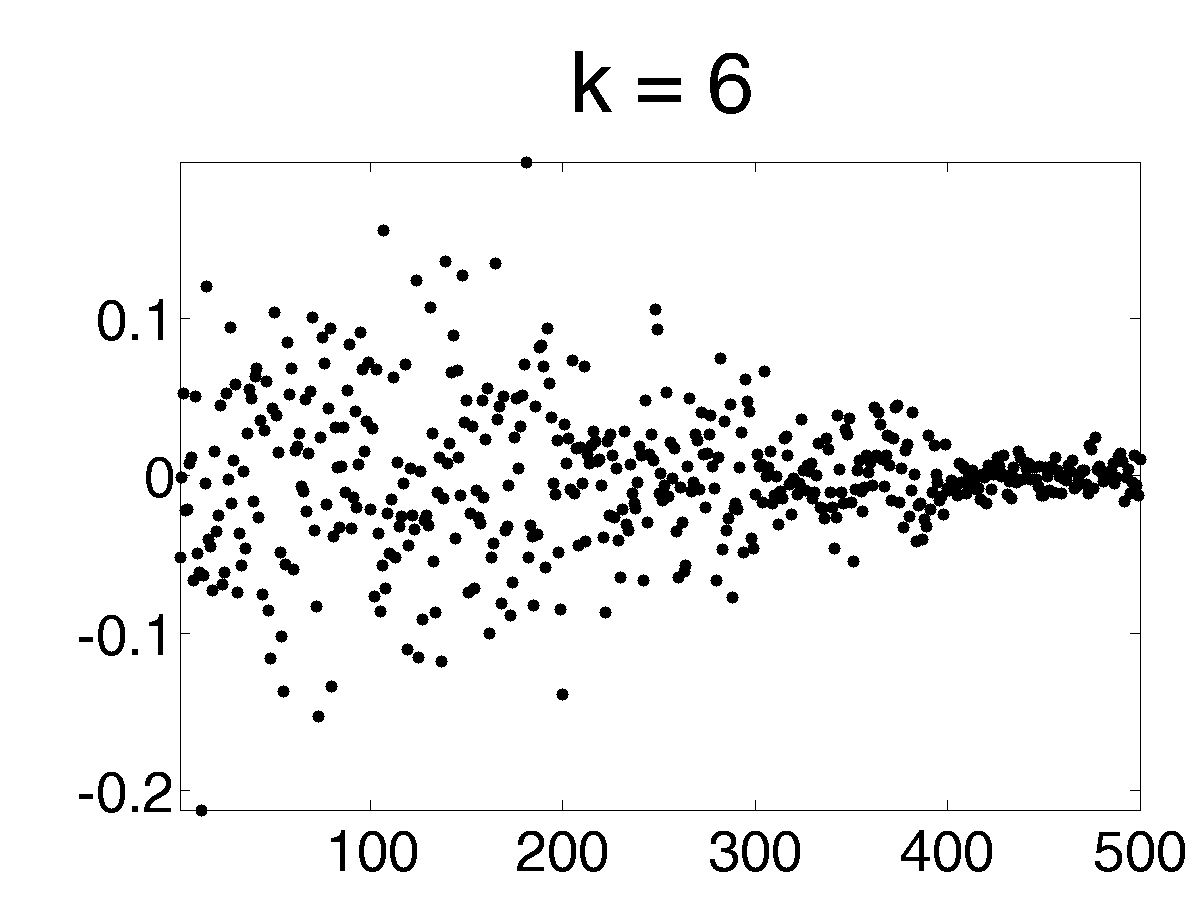}
\includegraphics[width=0.19 \columnwidth]{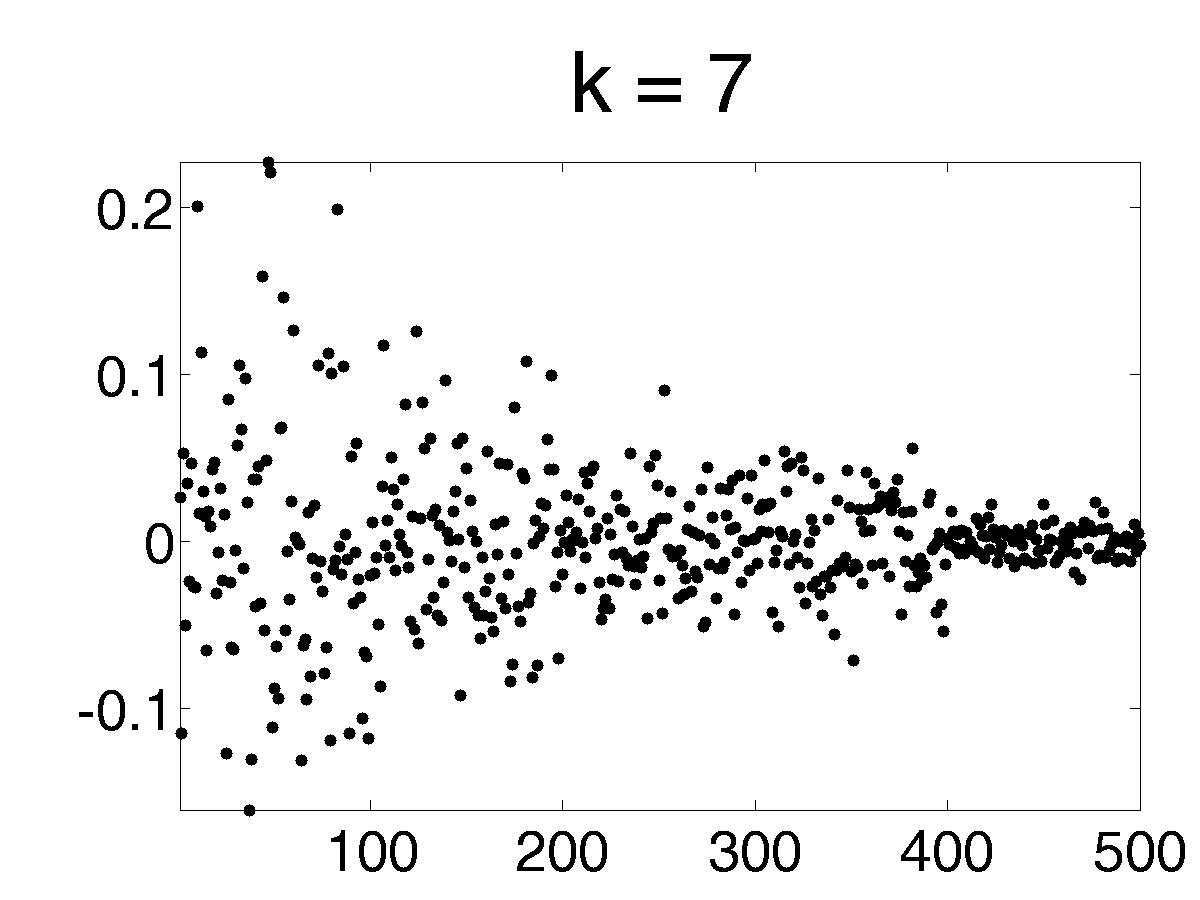}
\includegraphics[width=0.19 \columnwidth]{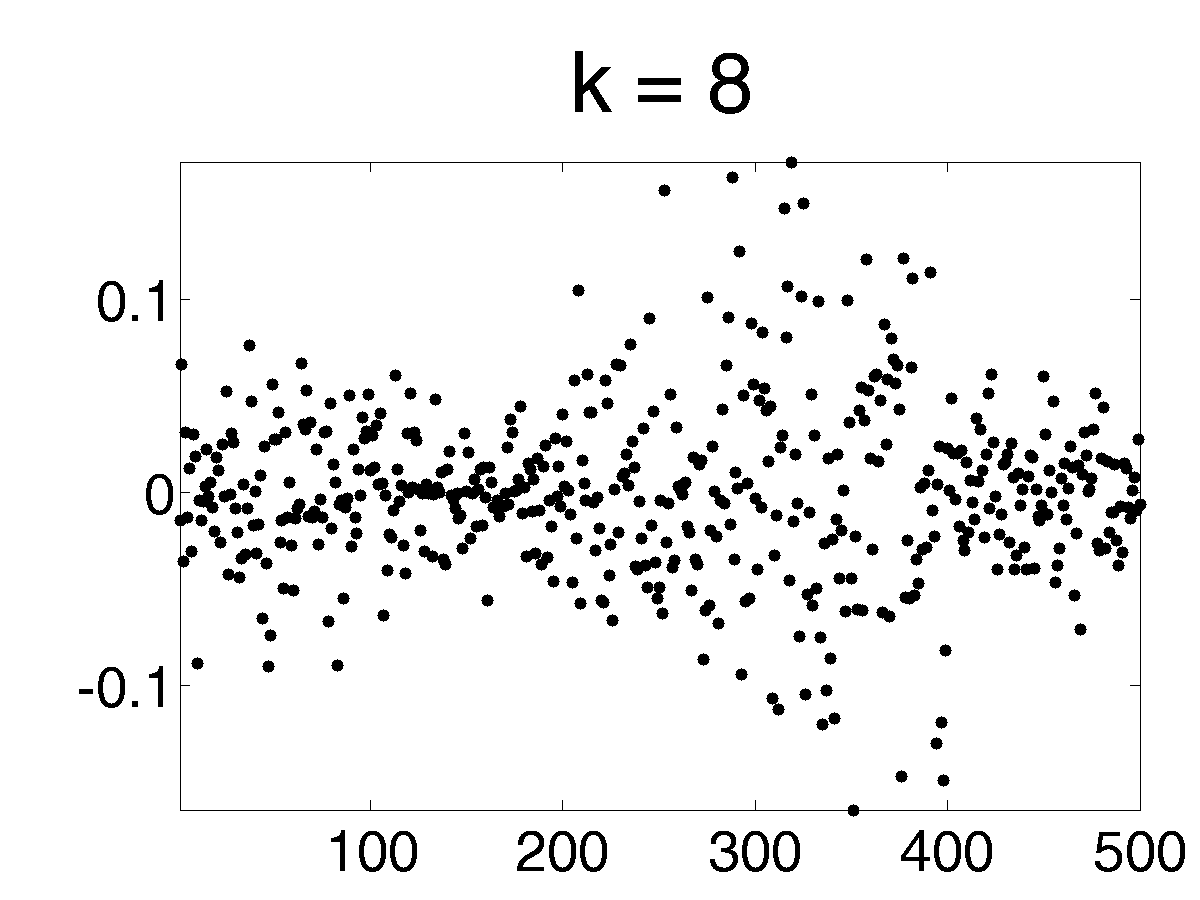}
\includegraphics[width=0.19 \columnwidth]{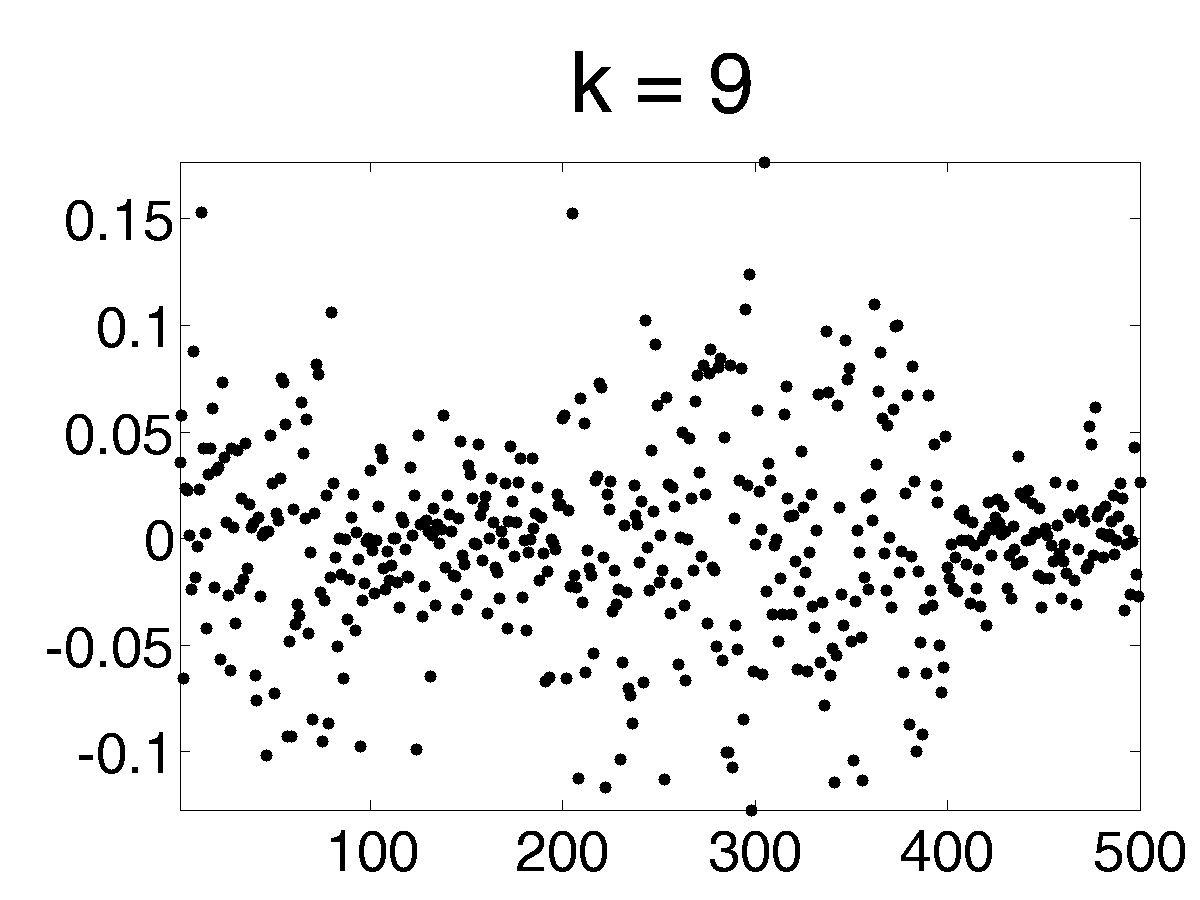}
\includegraphics[width=0.19 \columnwidth]{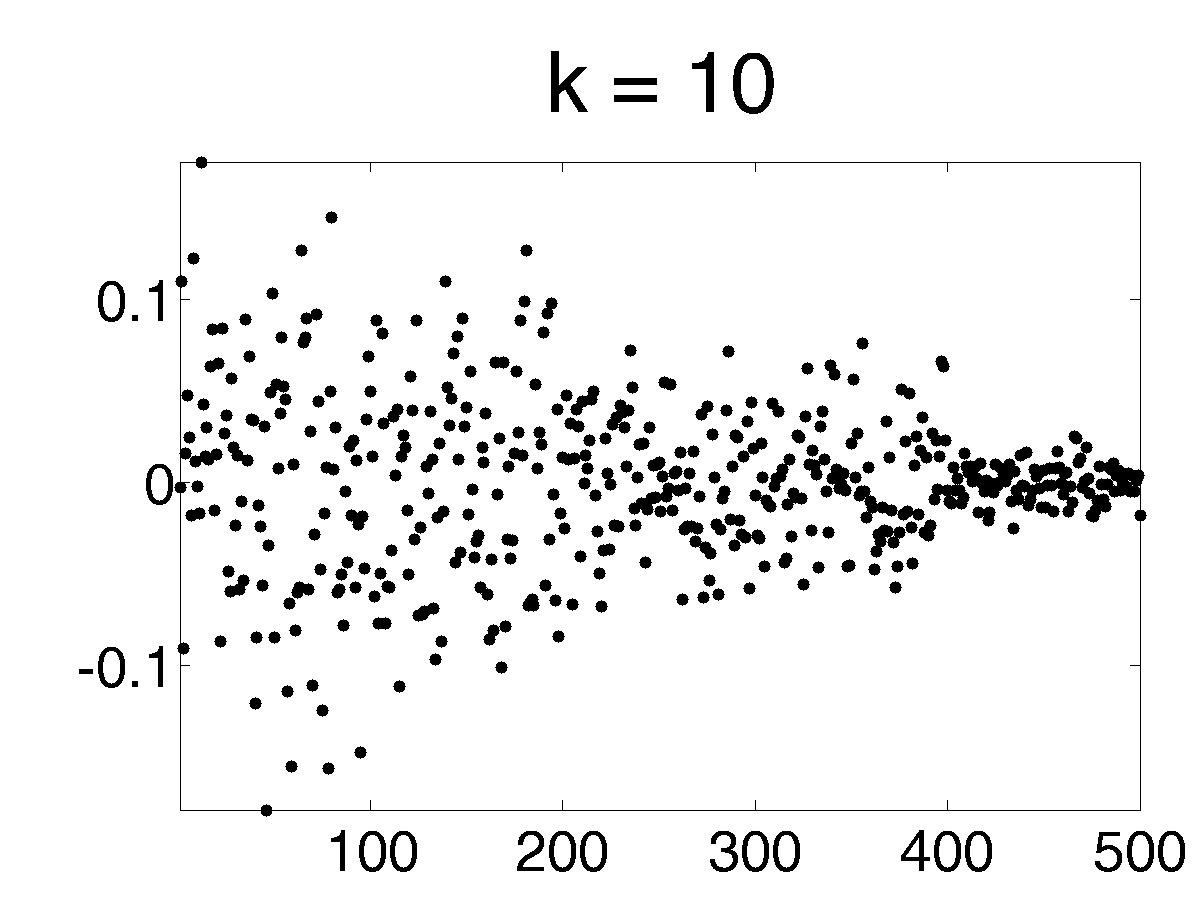} 
\end{center}
\caption{Results from the \textsc{TwoLevel} model, where the parameters have 
been set as an ``path graph of unstructured graphs,'' where each base graph 
$W$ is a random graph $G(n=100,p=0.2)$, and where a pair of nodes from 
consecutive base graphs are connected with  probability $0.01$. 
%%% (400 bad edges).
Shown are:
a pictorial illustration of the graph;
the IPR scores;
the normalized square spectrum; and
the top $10$ eigenvectors.
}
\label{fig:model3-exp_on_line}
\end{figure}

Next, Figures~\ref{fig:model2-2mod_on_noise} and~\ref{fig:model3-exp_on_line}
present the same results for two modifications of this basic setup.  
Figure~\ref{fig:model2-2mod_on_noise} does it for an ``unstructured graph 
of $2$-modules,'' \emph{i.e.}, for five $2$-modules connected with random 
interactions.
In this case, low-order eigenvector localization is still present, but it 
is much less prominent by the IPR measure, and it is significantly more 
noisy when the eigenvectors themselves are visualized.
Also, and not surprisingly, the situation becomes noisier still if the 
off-diagonal noise is increased.
Figure~\ref{fig:model3-exp_on_line} presents results for a ``path graph of 
unstructured graphs,'' \emph{i.e.}, several random unstructured graphs 
organized as beads along a one-dimensional scaffolding.
Again, low-order eigenvector localization is still present, but again the 
situation is significantly more noisy.
Note, though, that although the localization does not lead to most of the 
mass on low-order eigenvectors being localized on a single bead, there is 
still a tendency for localization to occur at the endpoints of the path.

%TMP% 

%TMP% 

\begin{figure}[t] 
\begin{center}
\includegraphics[width=0.32\columnwidth]{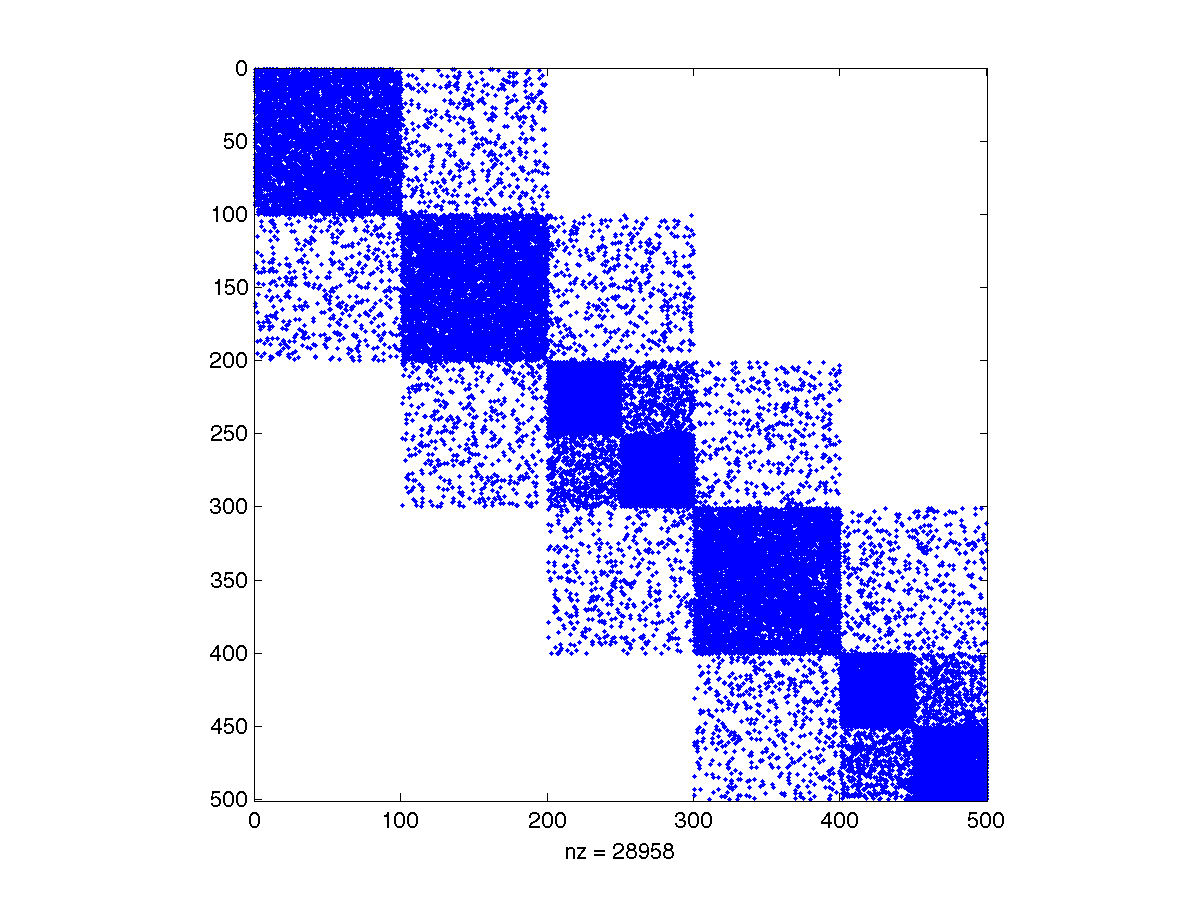}
\includegraphics[width=0.32\columnwidth]{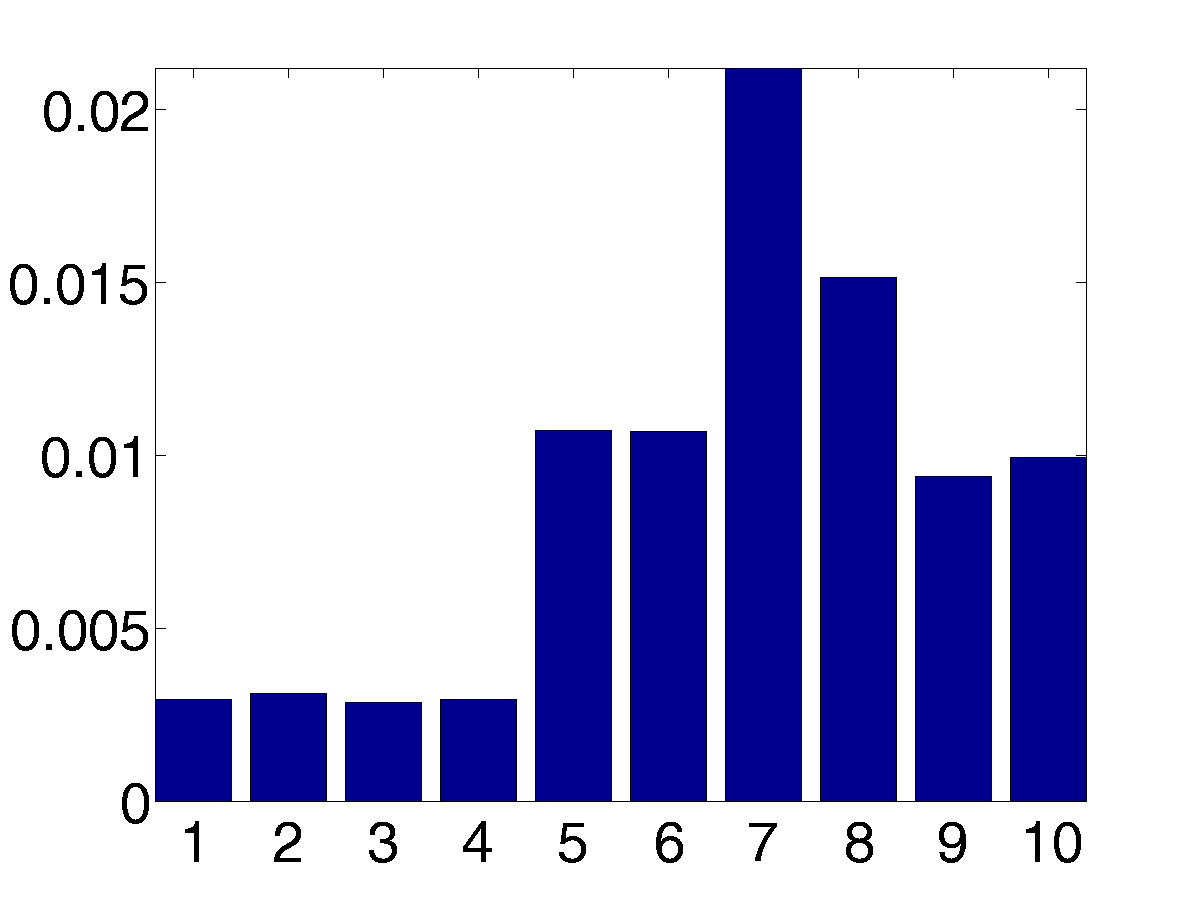}
\includegraphics[width=0.32\columnwidth]{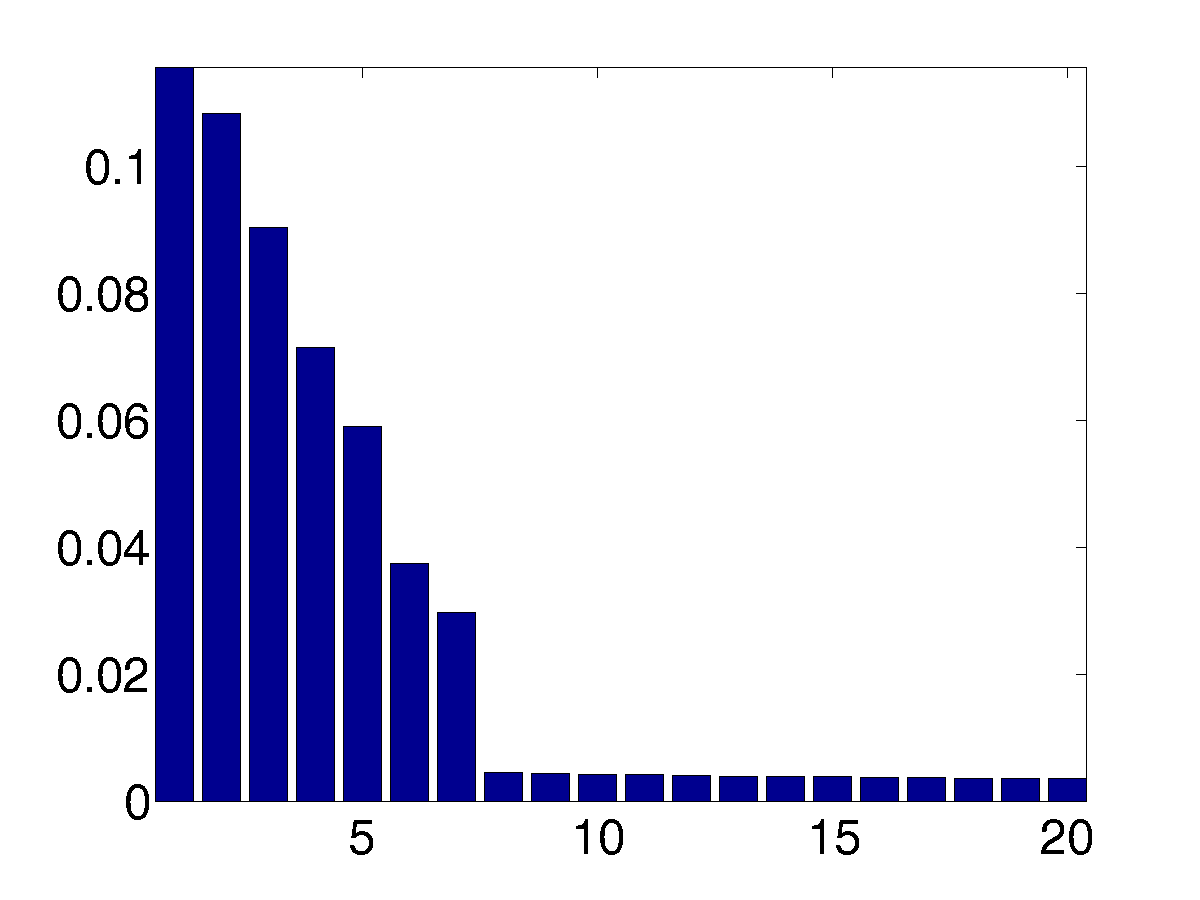}
\includegraphics[width=0.19 \columnwidth]{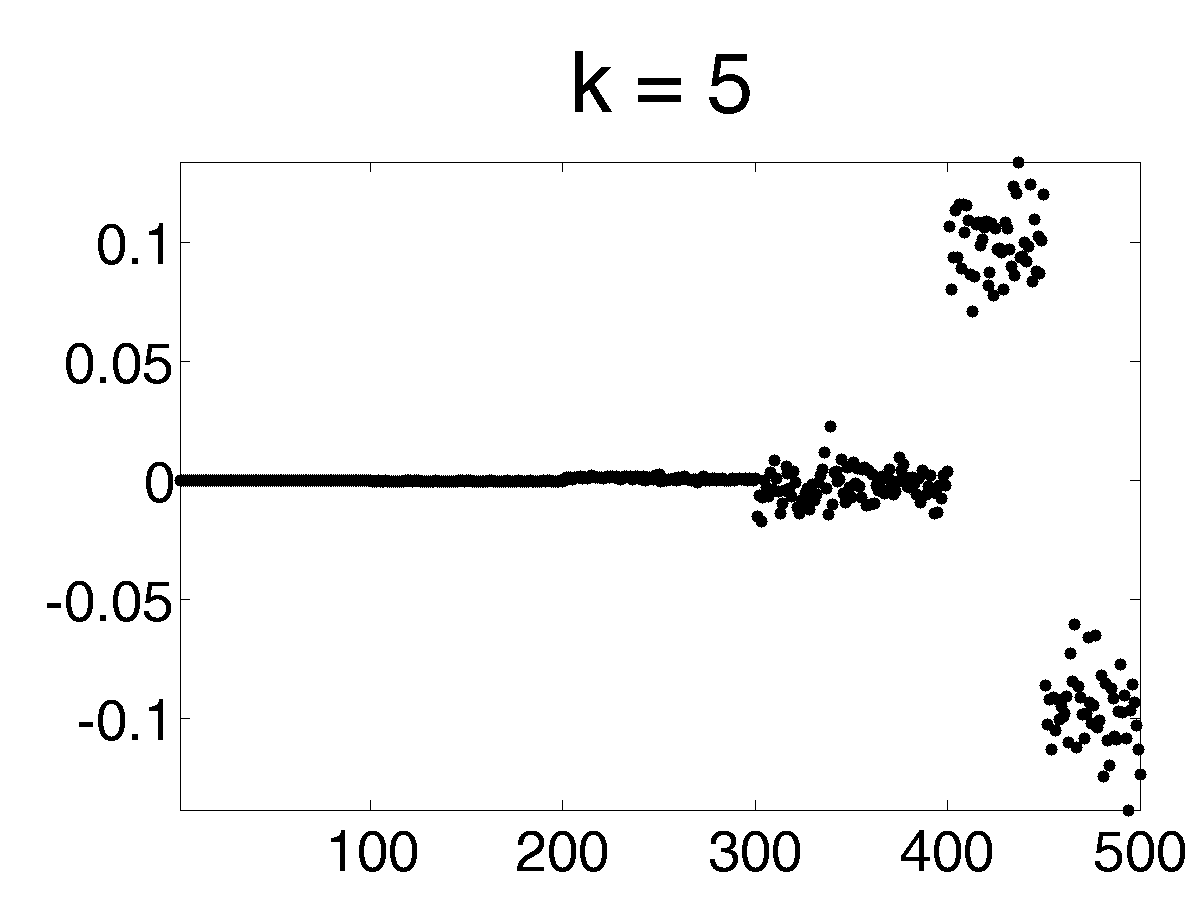}
\includegraphics[width=0.19 \columnwidth]{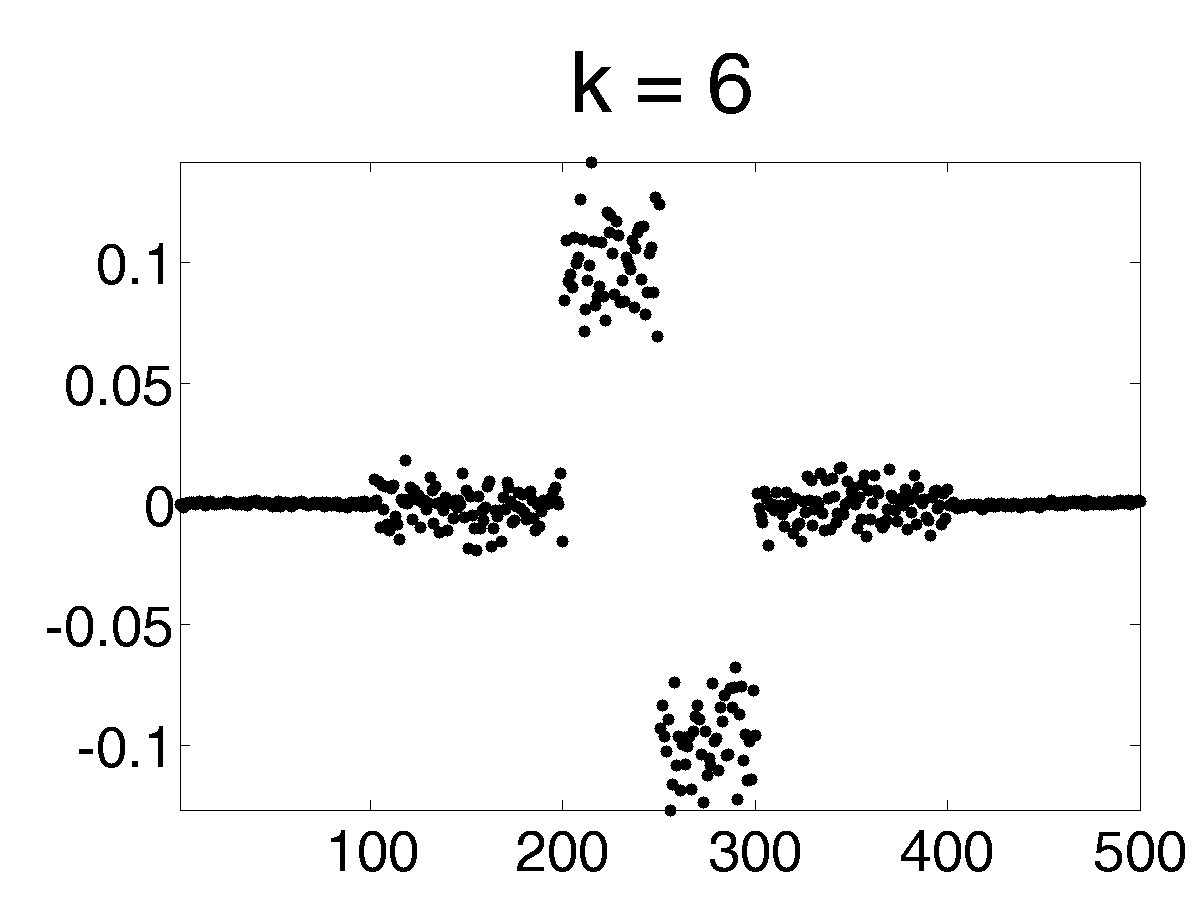}
\includegraphics[width=0.19 \columnwidth]{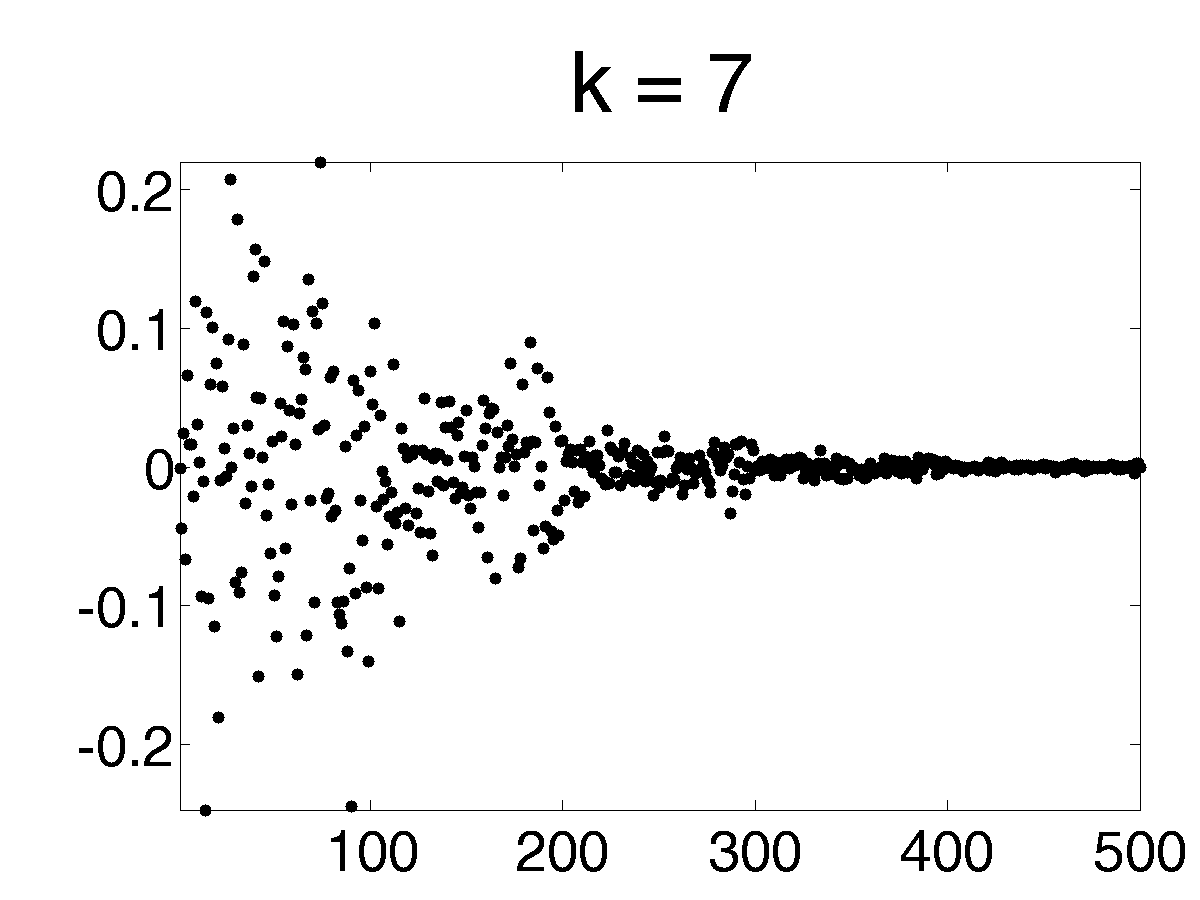}
\includegraphics[width=0.19 \columnwidth]{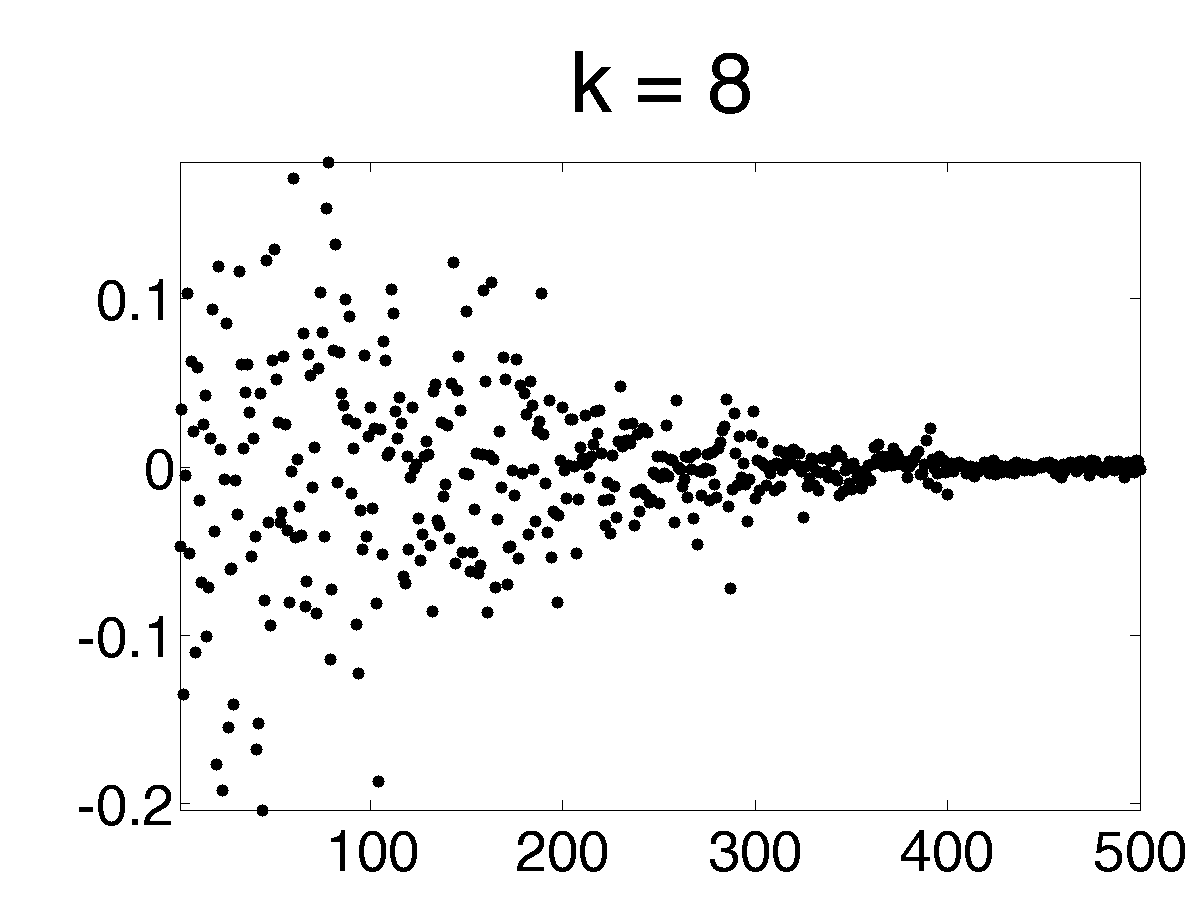}
\includegraphics[width=0.19 \columnwidth]{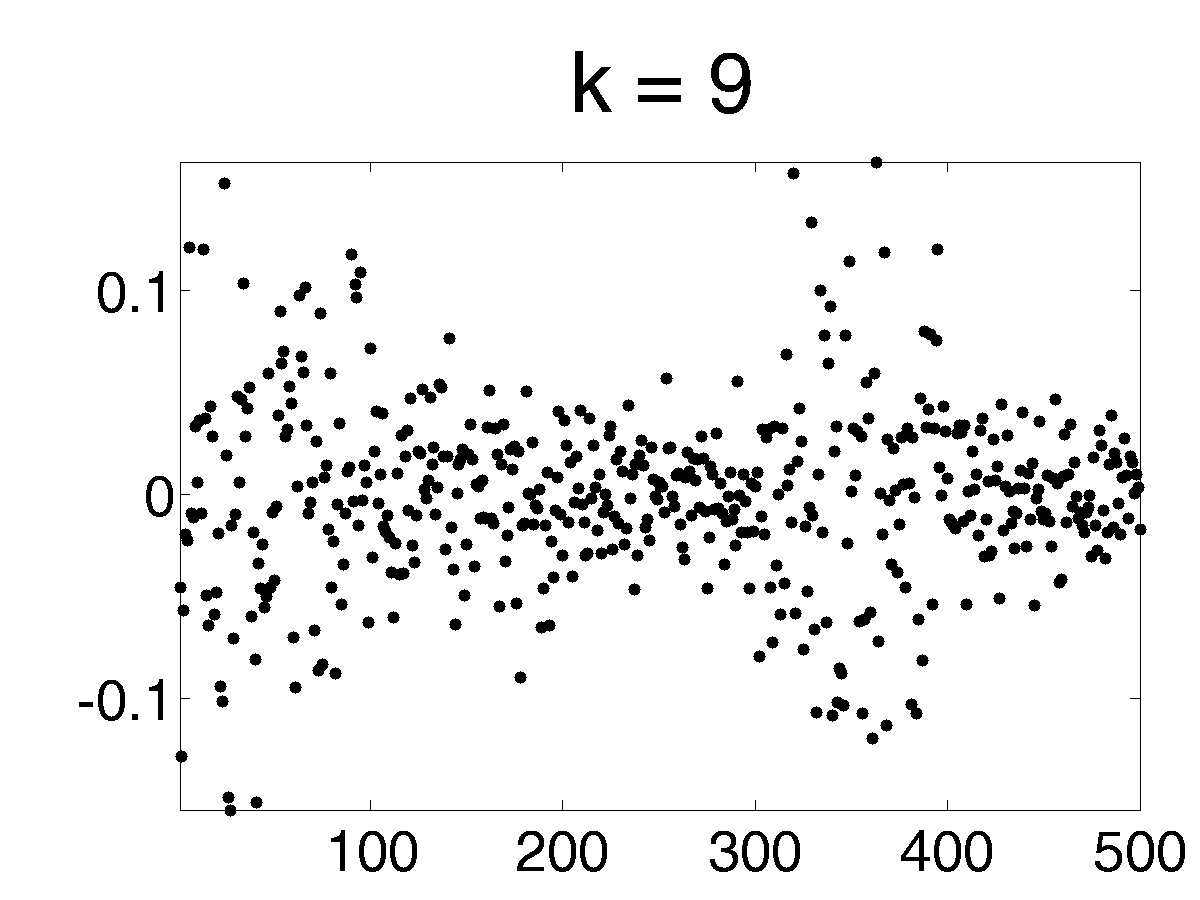}
\end{center}
\caption{Results from the \textsc{TwoLevel} model, with two different types 
base graphs organized as a path graph; the order of the base graphs is EE2E2.
Each ``2'' has edge densities $p_1=0.8$ and $p_2=0.2$; each ``E'' is a 
random graph with $p$ to match the edge densities inside the beads; 
and nodes between successive beads are connected with probability $0.05$.
Shown are:
a pictorial illustration of the graph;
the IPR scores;
the normalized square spectrum; and
the $5^{th}$ through $9^{th}$ eigenvectors.
}
\label{fig:model4-ee2e2}
\end{figure}

\begin{figure}[t] 
\begin{center}
\includegraphics[width=0.32\columnwidth]{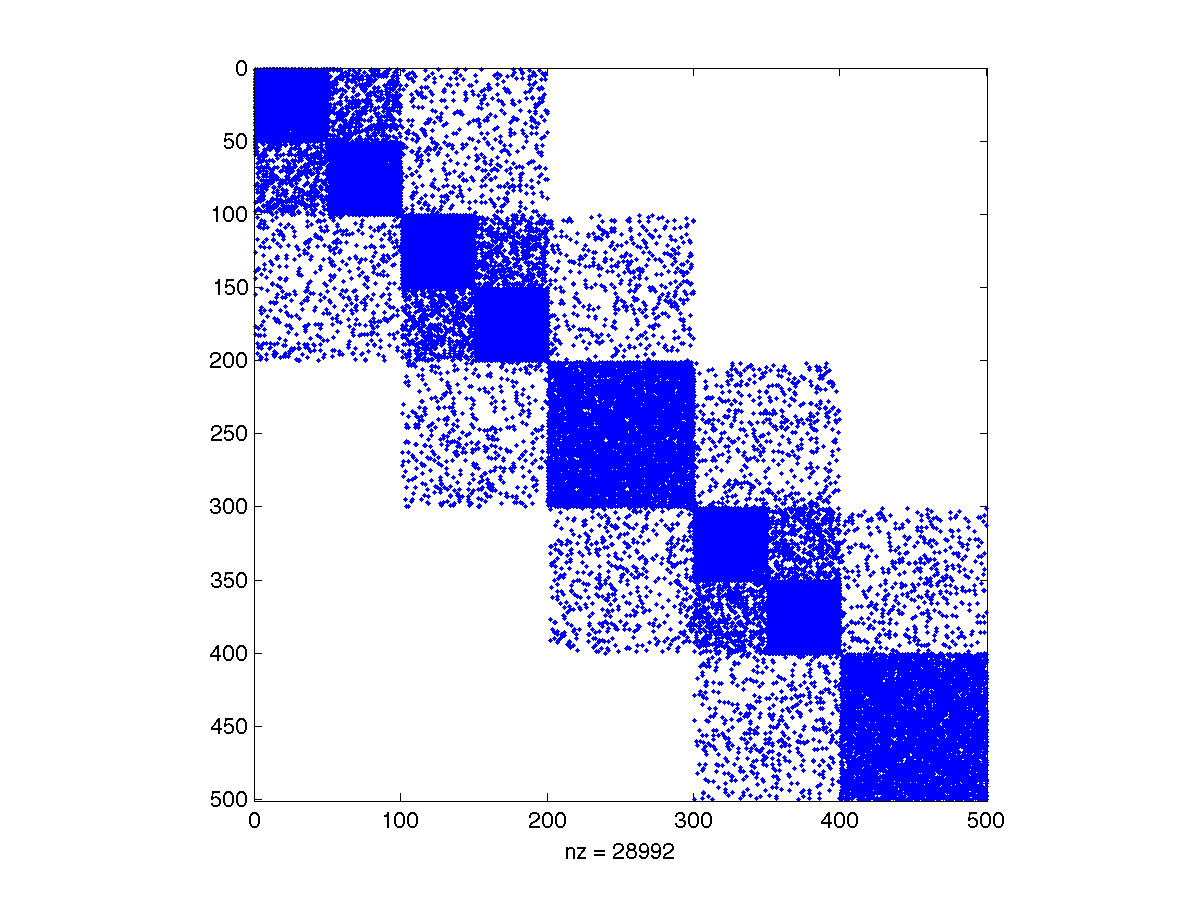}
\includegraphics[width=0.32\columnwidth]{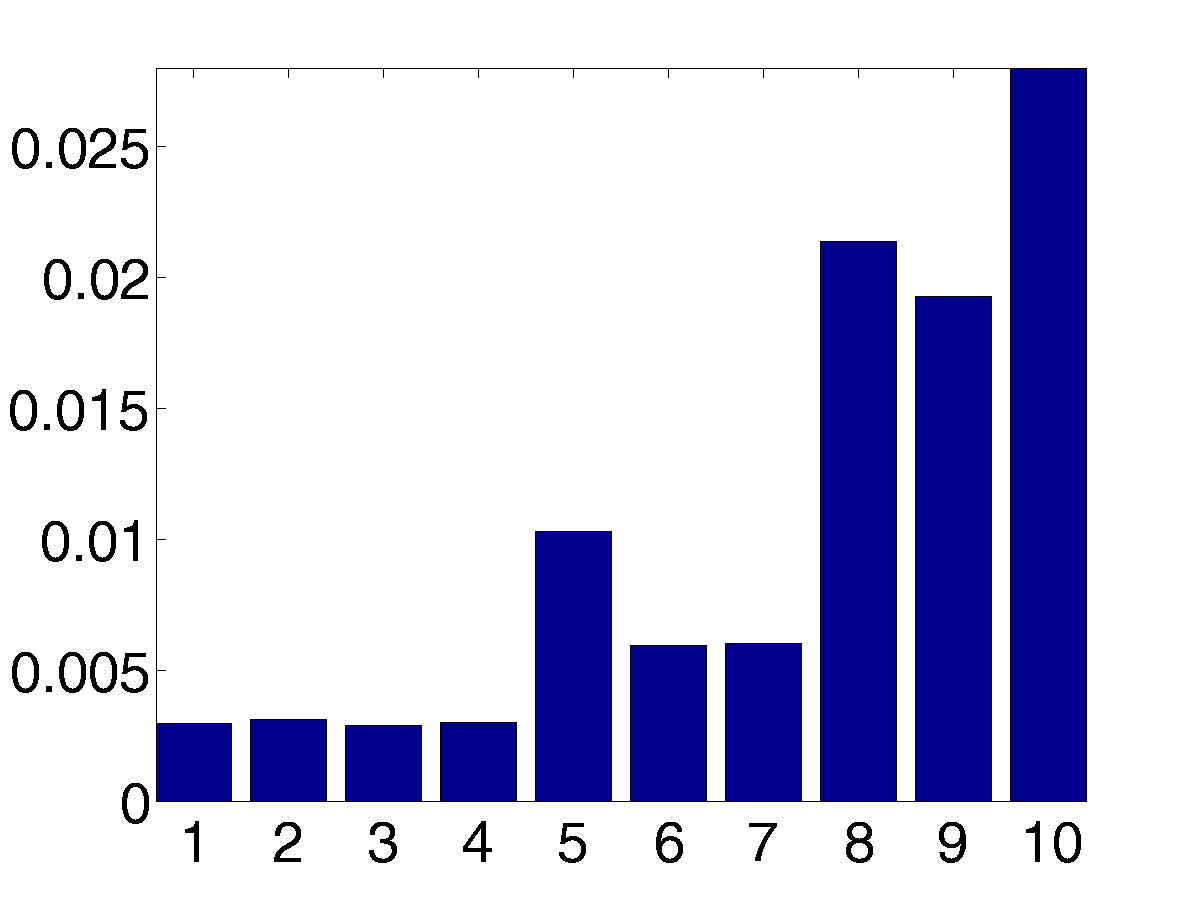}
\includegraphics[width=0.32\columnwidth]{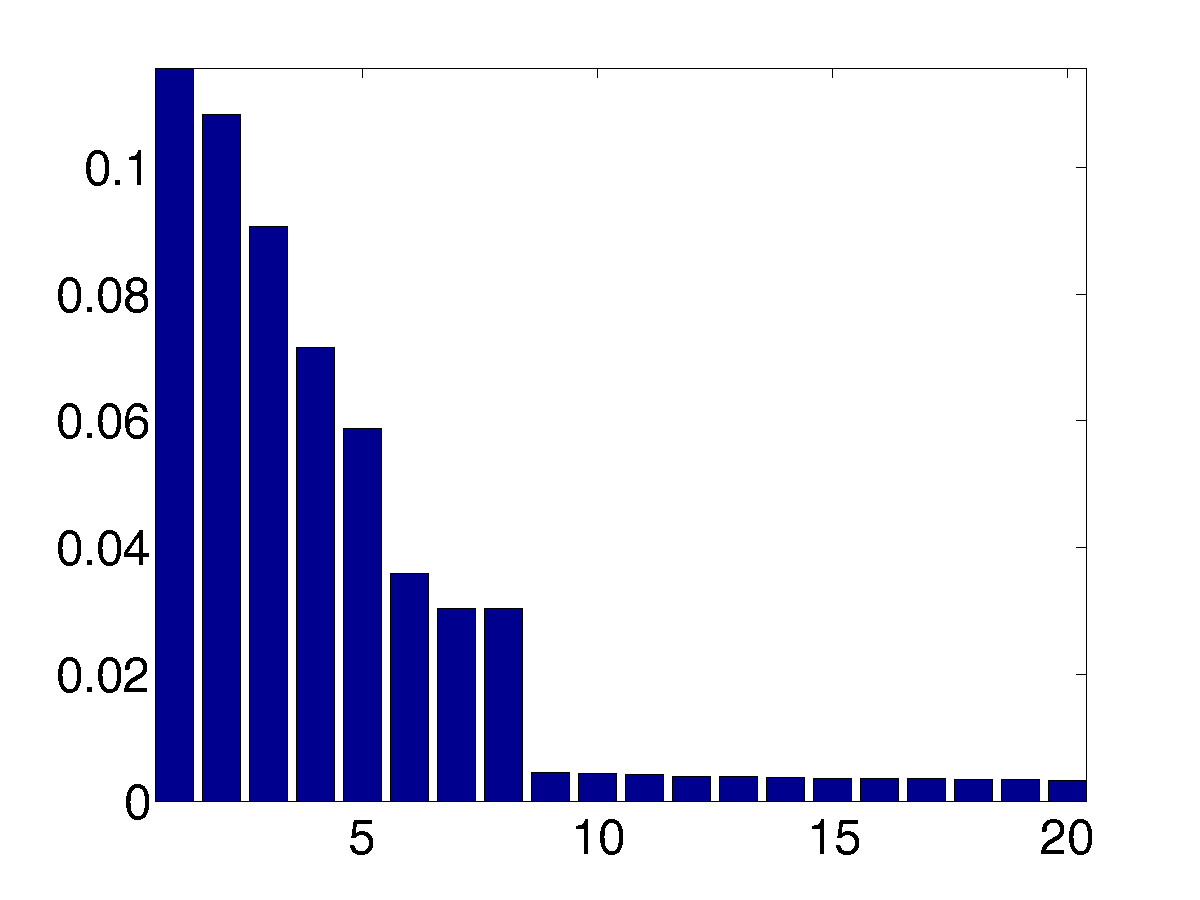}
\includegraphics[width=0.19 \columnwidth]{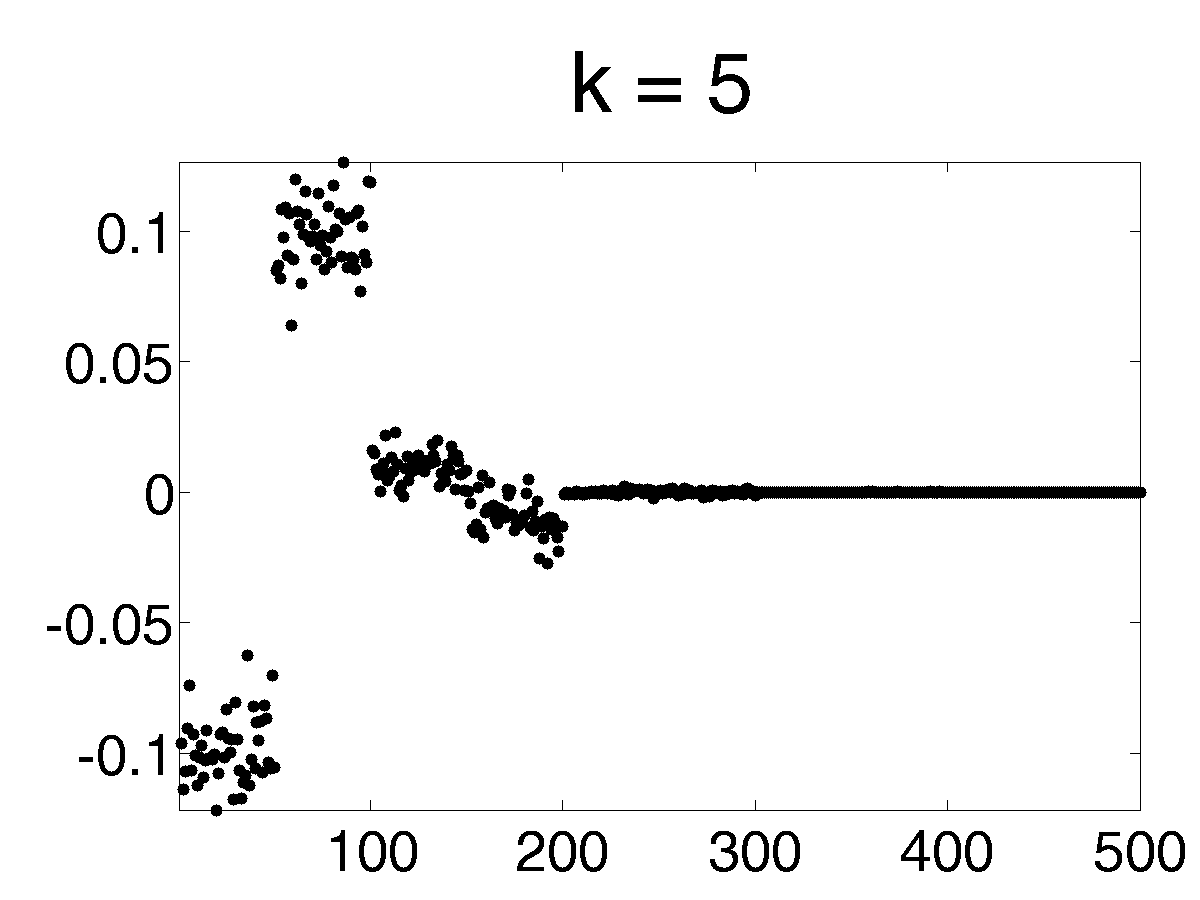}
\includegraphics[width=0.19 \columnwidth]{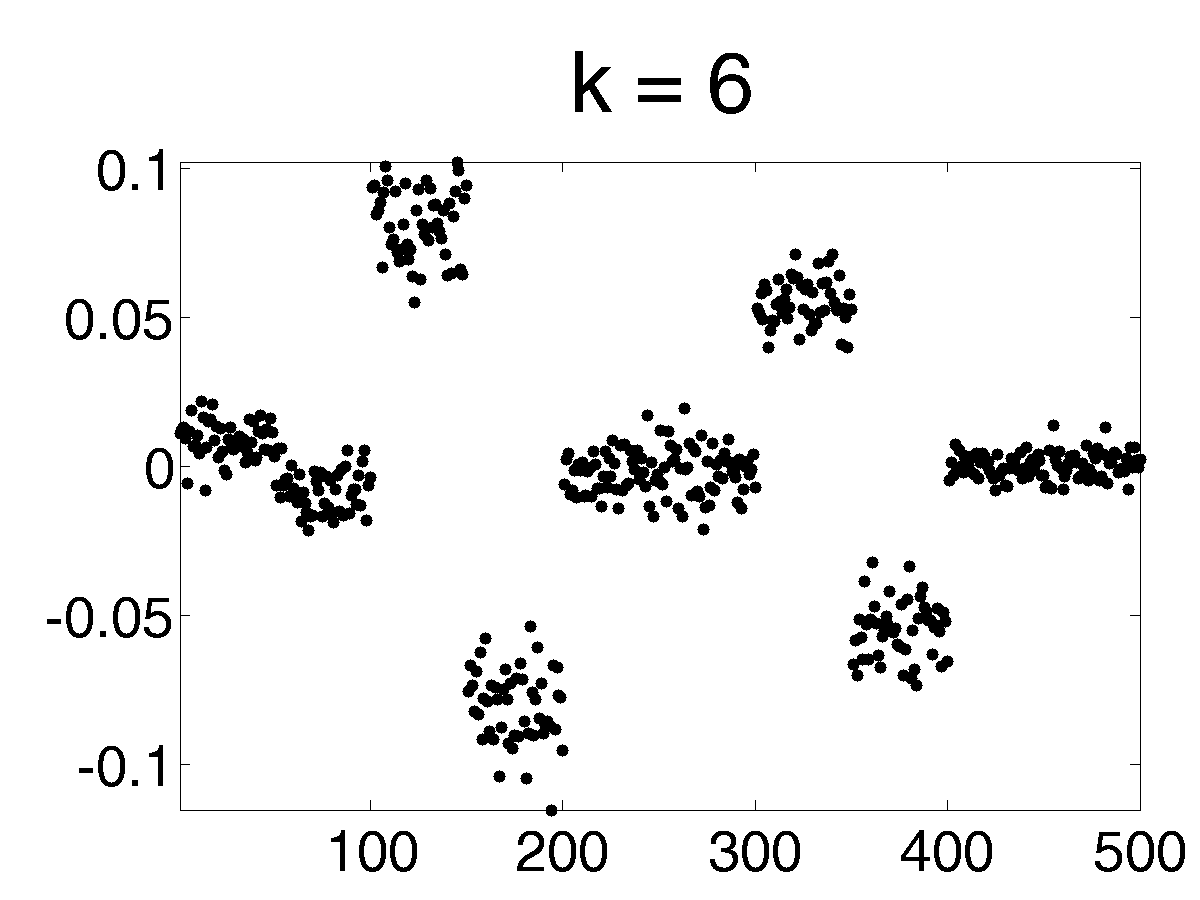}
\includegraphics[width=0.19 \columnwidth]{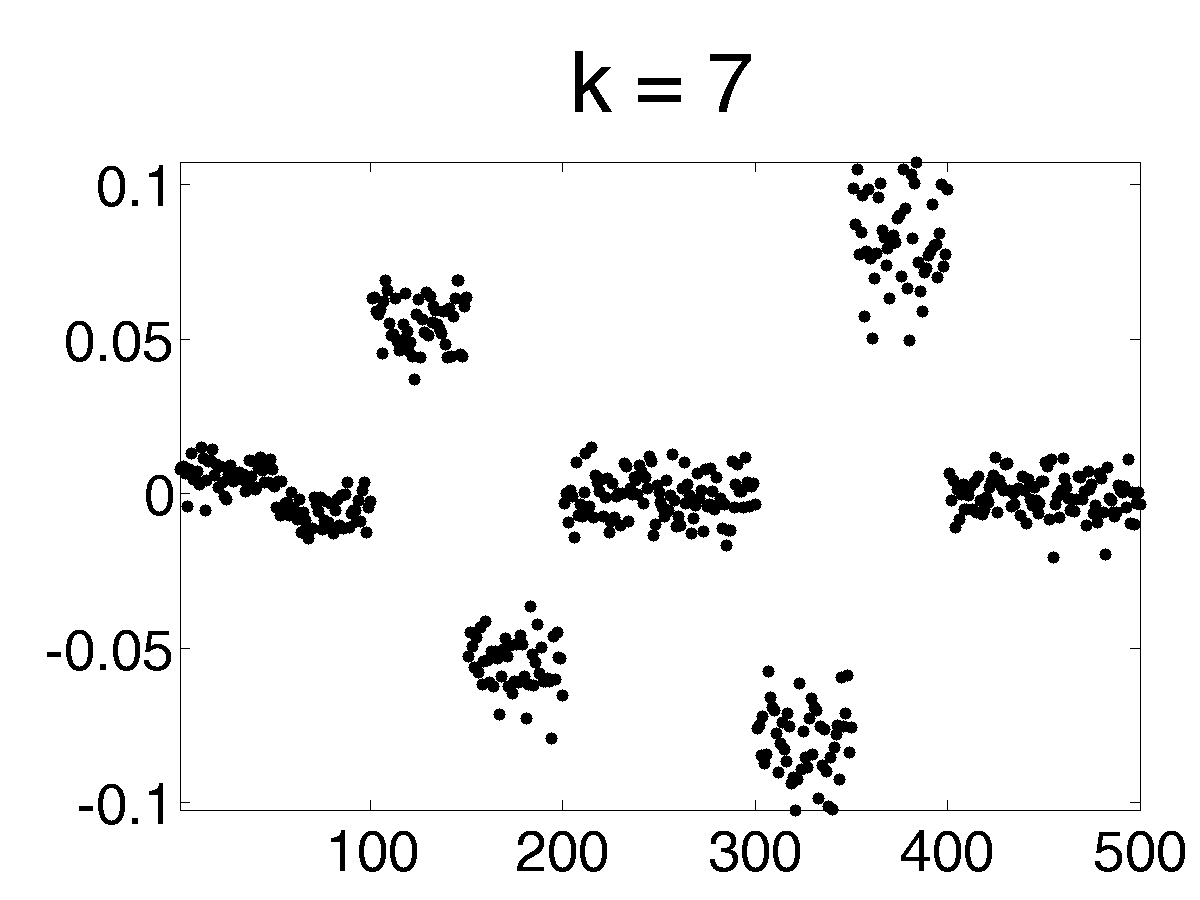}
\includegraphics[width=0.19 \columnwidth]{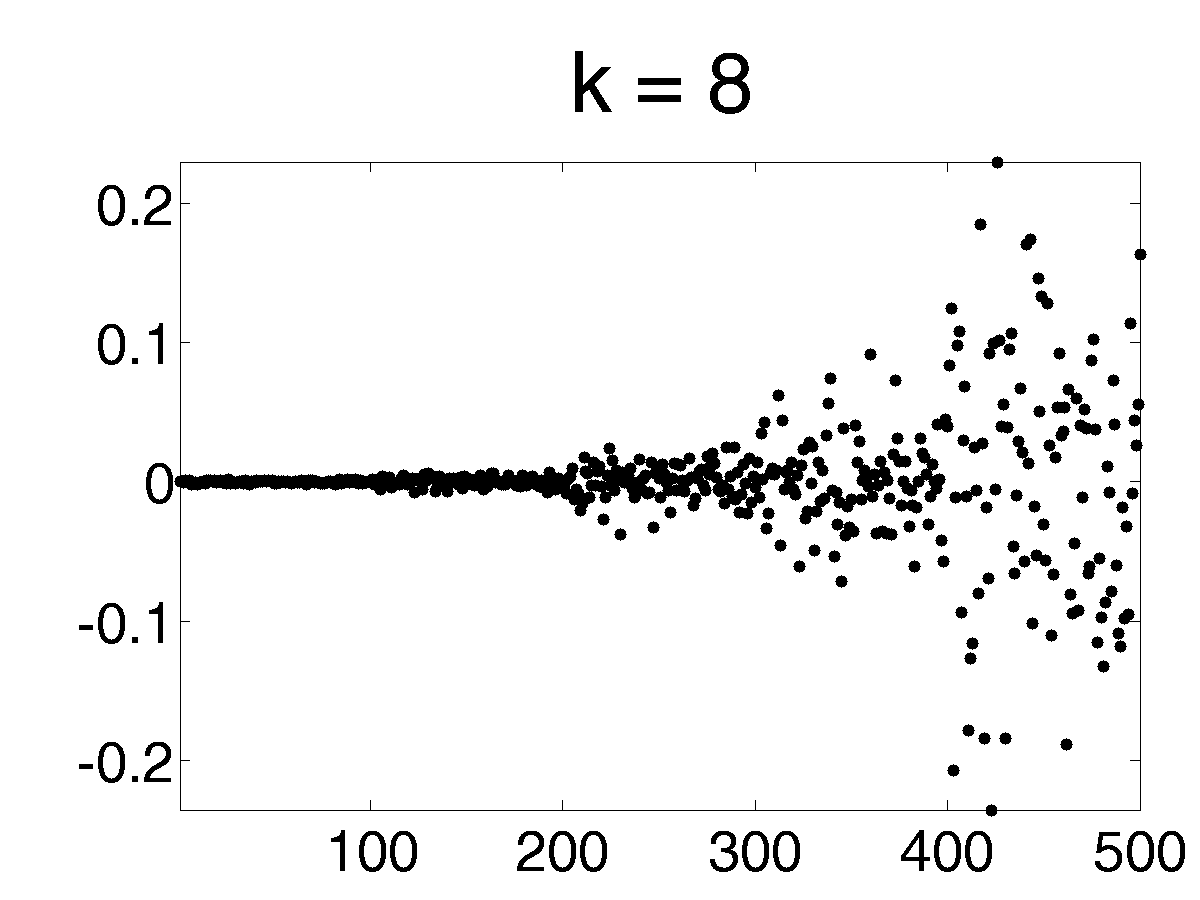}
\includegraphics[width=0.19 \columnwidth]{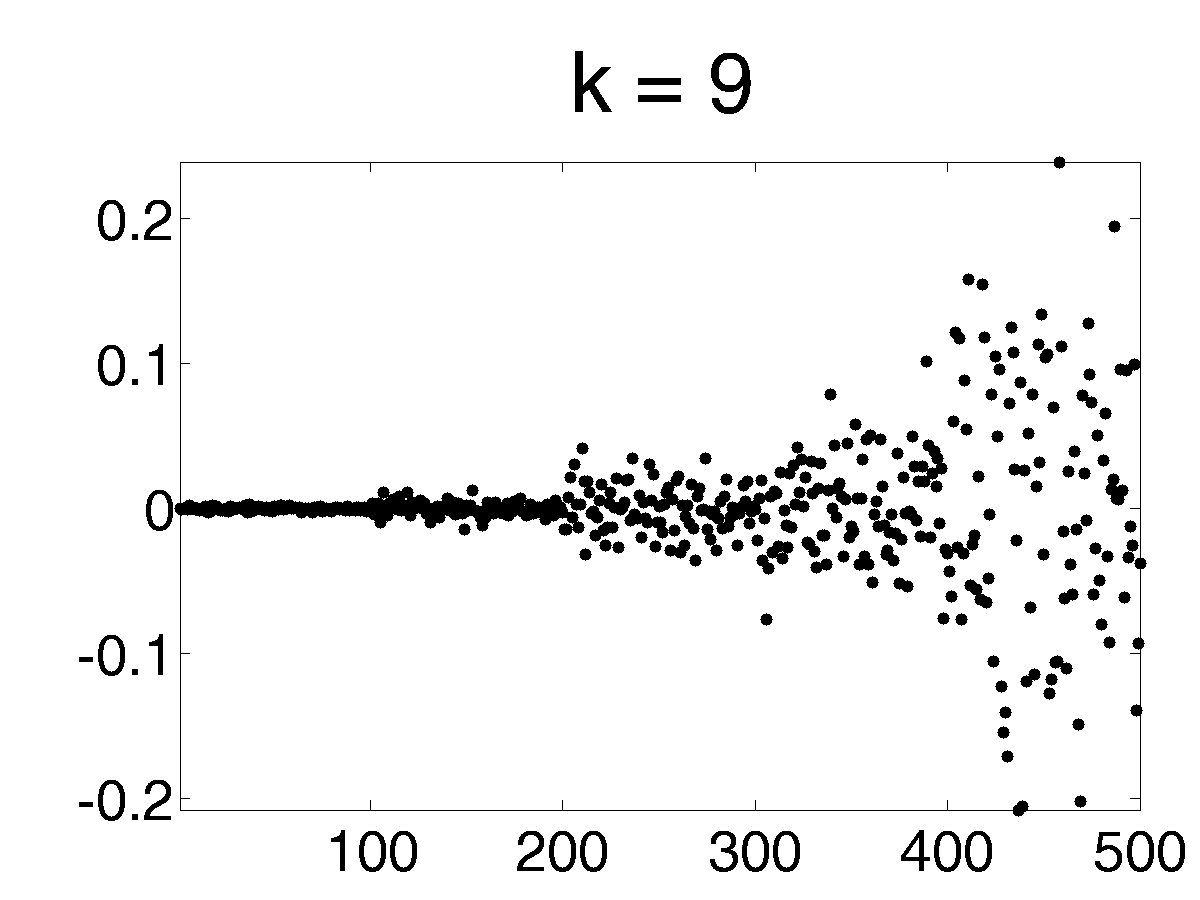}
\end{center}
\caption{Results from the \textsc{TwoLevel} model, with two different types 
base graphs organized as a path graph; the order of the base graphs is 22E2E.
Each ``2'' has edge densities $p_1=0.8$ and $p_2=0.2$; each ``E'' is a 
random graph with $p$ to match the edge densities inside the beads; 
and nodes between successive beads are connected with probability $0.05$.
Shown are:
a pictorial illustration of the graph;
the IPR scores;
the normalized square spectrum; and
the $5^{th}$ through $9^{th}$ eigenvectors.
}
\label{fig:model5-22e2e}
\end{figure}

Finally, in order to understand the effect of varying the structure of the 
base modules on the localization properties of the eigenvectors, 
Figures~\ref{fig:model4-ee2e2} and~\ref{fig:model5-22e2e} illustrate the 
situation when the beads of the path graph are of two different types: 
unstructured Erd\H{o}s-R\'{e}nyi random graph (to be denoted by ``E''); and
structured $2$-modules (to be denoted by ``2'').
Combining beads in this way is of interest since may be thought of as a 
zero-th order model of, \emph{e.g.},  a more-or-less polarized Congress.
The former figure illustrates the case when most of the beads are 
unstructured (in the order EE2E2), while the latter illustrates the case 
when most of the beads are structured and a few are less-structured (in 
the order 22E2E).
For the EE2E2 situation, the low-order eigenvectors highlight the two 
relatively more-structured $2$-modules, starting with the one at the 
endpoint, although there is some residual structure highlighted by low-order 
eigenvectors on the unstructured E beads.
Conversely, for the 22E2E case, the $2$-modules tend to be highlighted; the
E beads tend to be lost, but they do tend to make the localization on 
nearby $2$-modules less pronounced.

%TMP% 

%TMP% 

\subsection{Theoretical considerations}

The empirical results on the \textsc{TwoLevel} model demonstrate that a very 
simple tensor product construction can shed light on some of the empirical 
observations for the \textsc{Congress} data and the \textsc{Migration} data
that were made in Section~\ref{sxn:empirical}. 
More generally, the \textsc{TwoLevel} model may be used as a diagnostic 
tool to help extract insight that is useful for a downstream analyst from 
data graphs when such low-order eigenvector localization is present. 
To help gain insight into ``why'' our empirical observations hold, here we 
will provide some insight that is guided by theory.
A detailed theoretical understanding of the \textsc{TwoLevel} 
model is beyond the scope of this paper, as it would require a matrix 
perturbation analysis of the tensor product of structured 
matrices.
This is a technically-involved topic, in part since a straightforward 
application of matrix perturbation ideas tends to ``wash out'' the bottom 
part of the spectrum~\cite{Stewart90}.

Instead of attempting to provide this, we will illustrate how many of the 
empirical results can be ``understood'' as a consequence of several
rules-of-thumb that are well-known to practitioners of eigenvector-based 
machine learning and data analysis tools.

\begin{itemize}
\item
First, recall that tensor product constructions lead to separable eigenstates.
In particular, for the \textsc{TwoLevel} model, the spectrum of $H$ is related 
in a simple way to those of $I$ and $W$: its eigenvalues are just the direct 
products of the eigenvalues of $I$ and $W$, and the corresponding eigenvectors 
of $W$ are the tensor products of the eigenvectors of $I$ and $W$. 
For example, if $1$ and $v_1$ are the top eigenvalue/eigenvector of $I$, 
and $5$ and $u_1$ of $W$, then the eigenvector of $W$ corresponding to the 
top eigenvalue $1 \otimes 5=5$  is $(v_1)\otimes(u_1)$; and so on.
Assuming that the perturbation caused by the interaction model $N$ is 
``sufficiently weak'' relative to the base graph $W$, this suggests two things: 
first, that the top eigenvectors of the full graph will not ``see'' the 
internal structure of the base graph $W$; 
second, that the number of these top eigenvectors will equal the number of 
base graphs (minus one, if the trivial eigenvector is not counted); and 
third, that properties of the eigenvectors of the base graph $W$ may 
manifest themselves in subsequent low-order eigenvectors of the full graph.
All of these phenomena are clearly observed in the empirical results for 
the \textsc{TwoLevel}, as well as for the \textsc{Congress} data when the 
inter-Congress couplings are small to moderate. 
When the inter-Congress couplings become larger, the interaction model is 
less weak, in which case the situation is much noisier and more complex.
Similarly, for the \textsc{Migration} data, there is some geographically-local 
structure illustrated in the low-order eigenvectors, but the situation is 
much noisier, suggesting that the interaction model $N$ is more complex or
that a simple separation of scales in a tensor product construction is less 
appropriate for these data.
%%%
\item
Second, recall that eigenvectors have strong connections with diffusions.
For example, the power method can be used to compute the top eigenvector 
of certain matrices, and random walks can be used to compute vectors 
which find good partitions of the data.
Empirical results on the \textsc{TwoLevel} data illustrate that when the 
base graph $W$ is structured (a $2$-module with a good bipartition, as 
opposed to an unstructured random graph) and/or when the interaction model 
$N$ is structured (a path graph, as opposed to an unstructured random graph) 
then the low-order localization is most pronounced.
(This may be seen as a consequence of the implicit ``isoperimetric capacity 
control'' associated with diffusing in very low-dimension spaces or when 
there are very good bipartitions of the data.
Formalizing these trade-offs would provide a precise but nontrivial sense 
in which perturbation caused by the interaction model $N$ is ``sufficiently 
weak'' relative to the base graph $W$.)
Relatedly, below the localization-delocalization transition, there is a 
fairly strong tendency in the \textsc{Congress} data for localization to 
occur on very early Congresses or very late Congresses, \emph{i.e.}, at 
early or late but not at intermediate times.
A similar but somewhat weaker tendency is seen for localization to occur 
in the \textsc{Migration} data at the boundaries or geographic borders of 
the data, suggesting that an explanation for this has to do with random 
walks ``getting stuck'' at ``corners'' of the configuration space. 
Relatedly, in the \textsc{Congress} data, on low-order eigenvectors for 
which the localization is somewhat less pronounced, there is often but not 
always substantial mass on several temporally-adjacent Congresses.
%%%
\item
Third, recall that higher-variance eigenvectors occur earlier in the spectrum.
As a consequence of this, the conventional wisdom is that the top 
eigenvector is relatively smooth and that subsequent eigenvectors exhibit 
characteristic higher-frequency sinusoidal oscillations; and, indeed, this 
is observed in both the real and synthetic data.
More interestingly, one should observe that in the \textsc{Congress} data, 
eigenvectors with localization on recent, \emph{i.e.}, temporally-later, 
Congresses tend to occur earlier in the spectrum, \emph{i.e.}, account for a 
larger fraction of the variance, than eigenvectors with localization on much 
older Congresses. 
An explanation for this is given by the observation that more recent 
Congresses are substantially more ``polarized'' than earlier 
Congresses~\cite{PR97,WPFMP09_TR,multiplex_Mucha}.
Since the variance associated with a more polarized base graph should be 
larger than that associated with a less polarized base graph, one would 
expect that (assuming that eigenvectors with localization on both earlier 
and on later Congresses are observed in the data) eigenvectors with 
localization on recent (and thus more polarized) Congresses should be seen 
before eigenvectors with localization on older (and less polarized) 
Congresses.
This explanation is given clear support by considering the order in which 
localized low-order eigenvectors appear when more-structured and 
less-structured base graphs are combined; see Figures~\ref{fig:model4-ee2e2}
and~\ref{fig:model5-22e2e}.
%%%
\item
Fourth, recall that lower-order eigenvectors are exactly orthogonal to 
earlier eigenvectors.
Since the requirement of exact orthogonality is typically unrelated to the
processes generating the data, this often manifests itself in denser 
eigenvectors that often have weaker localization properties and that are 
largely uninterpretable in terms of the domain from which the data are 
drawn.
This is the conventional wisdom, and (although not presented pictorially) 
this is also seen in some of the lower-order eigenvectors in the data 
sets we have been discussing.
\end{itemize}
%%%
Although these rule-of-thumb principles do not explain everything that a 
rigorous perturbation analysis of the tensor product of structured matrices 
might hope to provide, they do help to understand many of the observed 
empirical results that are seemingly arbitrary or simply artifacts of noise 
in the data.
In addition, they can be used to understand the properties of 
eigenvector-based methods more generally.
As a trivial example, recall that the \textsc{Congress} data from 
Section~\ref{sxn:empirical} was for a time period when the number of U.S. 
states and thus U.S. senators did not change substantially and thus when 
the size of the Congress was roughly constant, suggesting that fixed-sized
beads evolving along a one-dimensional scaffolding might be appropriate. 
If, instead, one was interested in using eigenvector-based methods to 
examine Congressional voting data from $1789$ to the 
present~\cite{PR97,WPFMP09_TR,multiplex_Mucha}, then one must take into 
account that the number of senators changed substantially over time.
In this case, an ``ice cream cone'' model, where the beads along the 
one-dimensional scaffolding grow in size with time, would be more 
appropriate.

\section{Discussion and conclusion}
\label{sxn:conc}

We have investigated the phenomenon of low-order eigenvector localization 
in Laplacian matrices associated with data graphs.
Our contributions are threefold: 
first, we have introduced the notion of low-order eigenvector localization; 
second, we have described several examples of this phenomenon in two real 
data sets, illustrating that the localization can in some cases highlight 
meaningful structural heterogeneities in the data that are of potential 
interest to a downstream analyst; and 
third, we have presented a very simple model that qualitatively reproduces 
several of the empirical observations.
Our model is a very simple two-level tensor product construction, in which 
each level can be ``structured'' or ``unstructured.''
Although simple, this model suggests certain structural similarities among 
the seemingly-unrelated applications where we have observed low-order 
eigenvector localization, and it may be used as a diagnostic tool to help 
extract insight from data graphs when such low-order eigenvector 
localization is present.
At this point, our model is mostly ``descriptive,'' in that it can be used 
to describe or rationalize empirical observations.
We will conclude this paper with a discussion of our results in a more 
general context.

Recall that the idea behind nonlinear dimensionality reduction methods such 
as Laplacian eigenmaps~\cite{BN03} and the related diffusion 
maps~\cite{CLLMNWZ05a} is to use eigenvectors of a Laplacian matrix 
corresponding to the coarsest modes of variation in the data matrix to 
construct a low-dimensional representation of the data.
The embedding provided by these top eigenvectors is is often interpreted 
in terms of an underlying low-dimensional manifold that is ``nice,''
\emph{e.g.}, that does not have pathological curvature properties or other 
pathological distributional properties that would lead to structural 
heterogeneities that would lead to eigenvector localization; and 
this embedding is used to perform tasks such as classification, clustering, 
and regression.
Our results illustrate that meaningful low-variance information will often 
be lost with such an approach.
Of course, there is no reason that general data graphs should look like 
limiting discretizations of nice manifolds, but it has been our experience 
that the empirical results we have reported are very surprising to 
practitioners of eigenvector-based machine learning and data analysis 
methods.

Far from being exotic or rare, however, a two-level structure such as that
posited by our \textsc{TwoLevel} model is quite common---\emph{e.g.}, time 
series data have a natural one-dimensional temporal ordering, DNA 
single-nucleotide polymorphism data are ordered along a one-dimensional 
chromosome along which there is correlational or linkage disequilibrium 
structure, and hyperspectral data in the natural sciences have a natural 
ordering associated with the frequency.
Not surprisingly, then, we have observed similar qualitative properties to 
those we have reported here on several of these other types of data sets, 
and we expect observations similar to those we have made to be made in 
many other~applications.

In some cases, low-order eigenvector localization has similarities with 
localization on extremal eigenvectors.
In general, though, drawing this connection is rather tricky, especially if 
one is interested in extracting insight or performing machine learning when 
low-order eigenvector localization is present.
Thus, a number of rather pressing questions are raised by our observations.
An obvious direction has to do with characterizing more broadly the 
manner in which such localization occurs in practice.
It is of particular interest to understand how it is affected by 
smoothing and preprocessing decisions that are made early in the data 
analysis pipeline.
A second obvious direction has to do with providing a firmer theoretical 
understanding of low-order localization.
This will require a matrix perturbation analysis of the tensor product of 
structured matrices, which to the best of our knowledge has not been 
considered yet in the literature.
This is a technically-involved topic, in part since a straightforward 
application of matrix perturbation ideas tends to ``wash out'' the bottom 
part of the spectrum.
A third direction has to do with understanding the relationship between
the low-order localization phenomenon we have reported and 
recently-developed local spectral methods that implicitly construct 
local versions of eigenvectors~\cite{Spielman:2004,andersen06local,MOV09_TR}. 
A final direction that is clearly of interest has to do with understanding 
the implications of our empirical observations on the applicability of 
popular eigenvector-based machine learning and data analysis tools.

%-----------------------------------------------------------------------
\textbf{Acknowledgments:}
We would like to acknowledge SAMSI and thank the members of its
2010-2011 Geometrical Methods and Spectral Analysis Working Group 
for helpful discussions.

%-----------------------------------------------------------------------
%-----------------------------------------------------------------------
%%% \bibliographystyle{plain}
%%% %\bibliography{mwmbib_jrnl,mwmbib_proc,mwmbib_book,mwmbib_misc,mwmbib_drft,communities}
%%% \bibliography{communities,mwmbib_jrnl,mwmbib_proc,mwmbib_book,mwmbib_misc,mwmbib_drft,communities}
%-----------------------------------------------------------------------

\end{document}